\documentclass[review]{elsarticle}

\usepackage{amsmath}
\usepackage{amssymb}
\usepackage{subcaption}
\usepackage{float}
\usepackage{siunitx}
\usepackage{tikz}
\usepackage{array}
\usepackage{upgreek}
\usepackage{setspace}
\usepackage{upgreek}
\usepackage{relsize}
\usepackage[margin=2.5cm]{geometry}
\newcolumntype{M}[1]{>{\centering\arraybackslash}m{#1}}
\journal{Acta Materialia}

\DeclareSIUnit{\atom}{atom}

\biboptions{sort&compress}

\begin{document}
\doublespacing
\begin{frontmatter}

\title{Splitting Instability in Superalloys: A Phase-Field Study}



\author[mymainaddress,mysecondaryaddress3]{Tushar Jogi\corref{mycorrespondingauthor}}
\ead{tushar.jogi@rub.de}
\author[mymainaddress,mysecondaryaddress4]{Saurav Shenoy}
\author[mysecondaryaddress1]{R. Sankarasubramanian}
\author[mysecondaryaddress2]{Abhik Choudhury}

\author[mymainaddress]{Saswata Bhattacharyya\corref{mycorrespondingauthor}}
\cortext[mycorrespondingauthor]{Corresponding authors}
\ead{saswata@msme.iith.ac.in}

\address[mymainaddress]{Department of Materials Science and Metallurgical
    Engineering, Indian Institute of Technology Hyderabad, Kandi, Sangareddy,
    Telangana, India - 502285}
\address[mysecondaryaddress1]{Defence Metallurgical Research Laboratory, Kanchanbagh, Hyderabad, India - 500058}
\address[mysecondaryaddress2]{Department of Materials Engineering, Indian
    Institute of Science, Bangalore, Karnataka, India - 560012}
\address[mysecondaryaddress3]{Materials Informatics and Data Science, Interdisciplinary Centre for Advanced Materials Simulations, Ruhr-Universit\"{a}t Bochum, Universit\"{a}tstrasse 150, 44809 Bochum, Germany}
\address[mysecondaryaddress4]{Department of Materials Science and Engineering and the Materials Research Institute, The Pennsylvania State University, University Park, PA 16802, USA}

\begin{abstract}
Precipitation-strengthened alloys, such as Ni-base, Co-base and Fe-base superalloys, show the development of dendrite-like precipitates in the solid state during aging at near-$\gamma^{\prime}$ solvus temperatures. 
These features arise out of a diffusive instability wherein, due to the point effect of diffusion, morphological perturbations over a growing sphere/cylinder are unstable. 
These dendrite-like perturbations exhibit anisotropic growth resulting from anisotropy in interfacial/elastic energies. 
Further, microstructures in these alloys also exhibit ``\emph{split}'' morphologies wherein dendritic precipitates fragment beyond 
a critical size, giving rise to a regular octet or quartet pattern of near-equal-sized precipitates separated by thin matrix channels. 
The mechanism of formation of such morphologies has remained a subject of intense investigation, and multiple theories have been proposed to explain their occurrence. Here, we developed a phase-field model incorporating anisotropy in elastic and interfacial energies to investigate the evolution of these split microstructures during growth and coarsening of dendritic $\gamma^{\prime}$ precipitates. 
Our principal finding is that the reduction in elastic energy density drives the development of split morphology, albeit a concomitant increase in the surface energy density. We also find that factors such as supersaturation, elastic misfit, degree of elastic anisotropy and interfacial energy strongly 
modulate the formation of these microstructures. We analyze our simulation results in the light of classical theories of elastic stress effects on coarsening and prove that negative elastic interaction energy leads to the stability of split precipitates.

\end{abstract}


\end{frontmatter}
\section{Introduction}
\label{intro}
 Solid-state precipitation of elastically coherent phases gives rise to interesting dynamical patterns with an inherent self-organization of the precipitates during microstructure evolution. This phenomenon is due to the coupling between the diffusional and elastic interactions. In addition, the shape of individual precipitate is modulated by the elastic misfit, the anisotropy in intrinsic elastic properties as well as the inhomogeneity in elastic moduli. Since these parameters control the total elastic energy of the precipitate, elastic stress strongly affects the morphological evolution of precipitates. 
For example, elastic stresses can lead to microstructural characteristics such as shape transitions (sphere to cuboid and cuboid to rod or plate) with size, shape-instability such as solid-state dendrites and split morphologies~\cite{doi1992Coarsening,DOI199679,su1996dynamics,gururajan2016elastic,yeon2005phase,fratzl1999modeling,johnson1984elastically}. 
In particular, elastic stresses can give rise to split morphology in Ni-base and Co-base superalloys where a larger cuboidal precipitate appears to disintegrate into two, four or eight smaller cuboidal precipitates. 
Several groups have proposed the splitting process as an instance of inverse Ostwald ripening where the average particle size decreases with time~\cite{johnson1984elastic,johnson1990coarsening,VOORHEES20041,su1996dynamics}.
Hence, particle splitting in high-temperature materials can be beneficial to mechanical properties.

\begin{figure}[!htb]
    \centering
    \begin{subfigure}{0.3\textwidth}
    \centering
    \includegraphics[width=4.6cm]{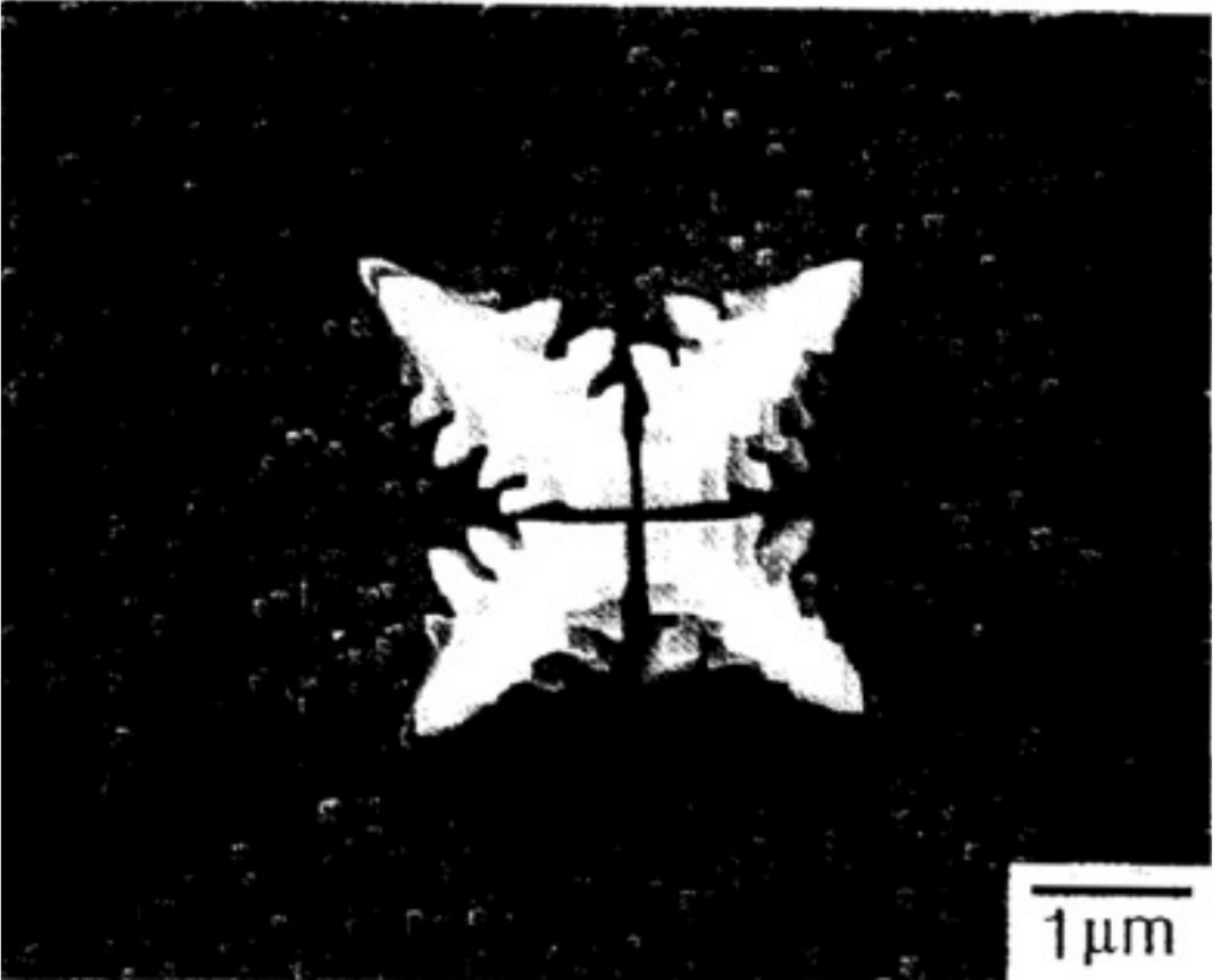}
    \caption{Split dendrite \cite{yoo1995effect}.}
    \label{subfig:split_dendrite}
    \end{subfigure}%
    \begin{subfigure}{0.3\textwidth}
    \centering
    \includegraphics[width=4.6cm]{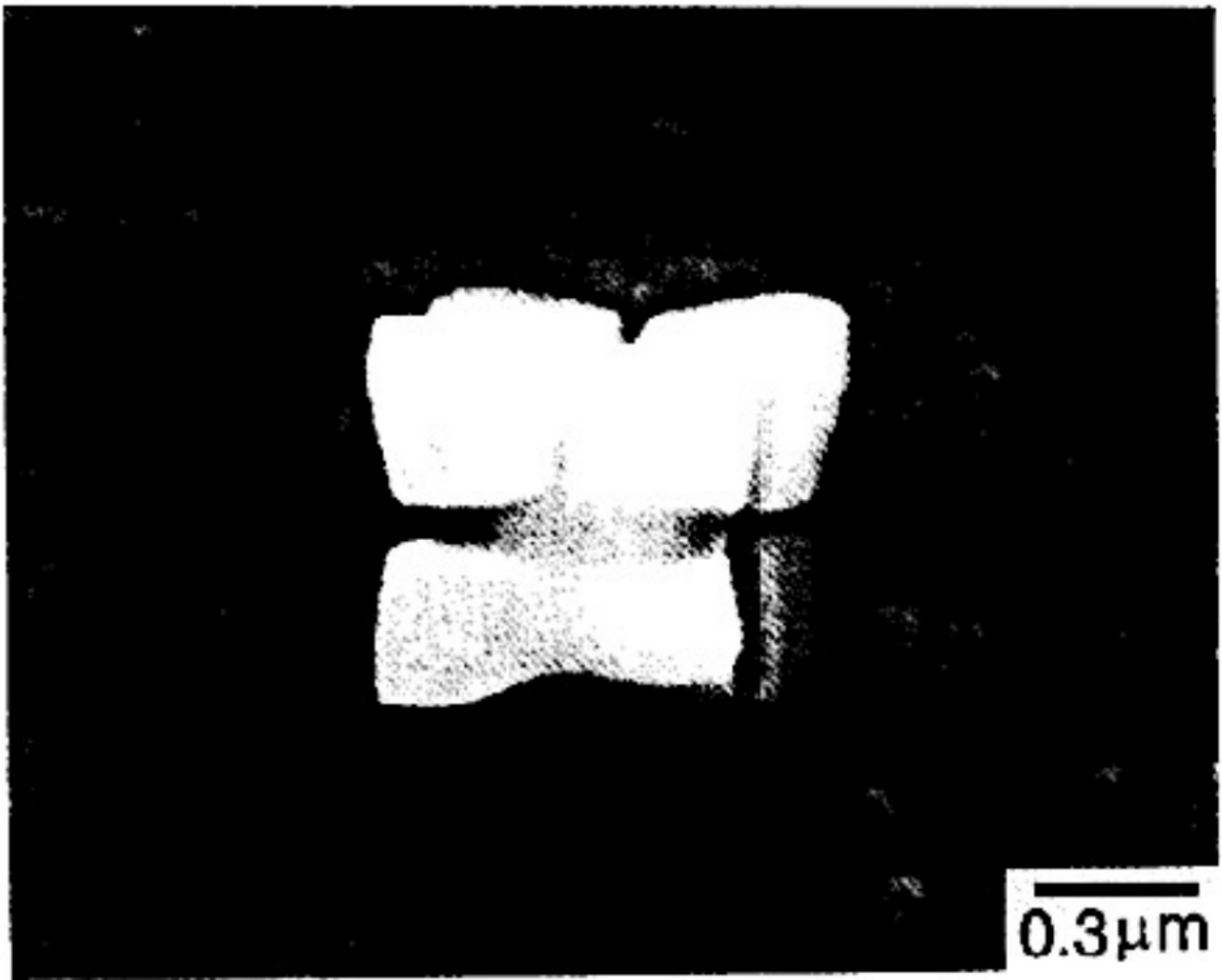}
    \caption{ Octodendritic structure \cite{yoo1995effect}.}
    \label{subfig:octodendrite}
    \end{subfigure}%
    \begin{subfigure}{0.3\textwidth}
    \centering
    \includegraphics[width=4.8cm]{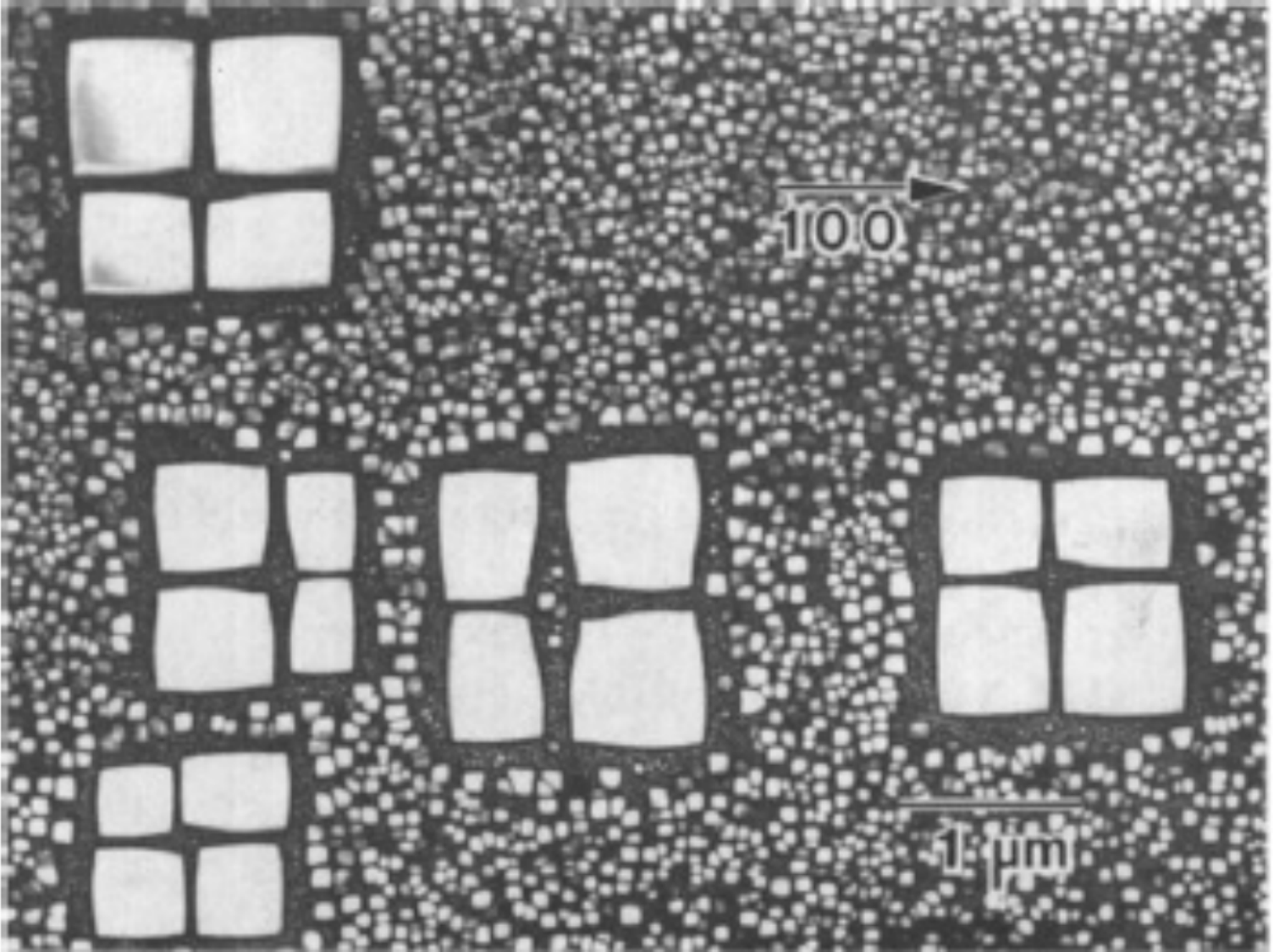}
    \caption{Split pattern~\cite{Kaufman1989}}
    \label{subfig:kaufman}
    \end{subfigure}
    \caption{Experimental observations of split patterns (a) Formation of 
    split dendritic structure when undercooled just below solvus temperature in model Ni-base 
    superalloy~\cite{yoo1995effect}. (b) High magnification micrograph showing fissure planes which 
    appeared to grow on the surface of $\gamma^{\prime}$ precipitate in 
    model Ni-base superalloy~\cite{yoo1995effect}. (c) Transmission electron micrograph of Ni-Al alloy aged 
    at $1100$\si{\celsius} for $1$h. Here, several larger cuboidal particles split into smaller octets of 
    cuboids~\cite{Kaufman1989}.}
    \label{fig:exp_pics}
\end{figure}

Apparently, for the first time, Westbrook observed the split patterns in Ni-base alloys, where he described them as ``ogdoadically diced cubes''~\cite{westbrook1958precipitation}. 
Subsequently, a large body of experimental observations reveals split morphologies in Ni-base and Co-base superalloys~\cite{miyazaki1982formation,doi1984effect,doi1985effects,Kaufman1989,yoo1995effect,qiu1996retarded,qiu1998splitting,grosdidier1998precipitation}, Fe-base alloys~\cite{DOI199679,doi1996transmission,calderon1997coarsening}, Cu-base alloys~\cite{castro1998isothermal}, Ir-Nb alloys~\cite{yamabe2002formation}, and Pt-base alloys~\cite{cornish2007overview}. 
Many electron microscopy studies observe these morphologies during isothermal heat treatment just below the solvus temperature~\cite{miyazaki1982formation,Kaufman1989,yoo1995effect} or during the slow continuous cooling from the single-phase region~\cite{grosdidier1998precipitation,doi1985effects}.
Evidently, in the same regime, the microstructural observations show dendrite-like morphologies during the solid-state phase transformation~\cite{yoo1995effect,ricks1983growth} (see Figs.~\ref{subfig:split_dendrite} and \ref{subfig:octodendrite}). 
The formation of dendrite-like morphology during solid-state phase transformation can be explained based on the theory of morphological instability~\cite{leo1989effect}, where the point effect of diffusion predominates over the restoring capillary forces~\cite{Bhadak2020formation,yeon2005phase}. 
On the other hand, based on the thermodynamic argument, the splitting of a particle is favourable at a critical size where elastic interaction energy between the split precipitates minimizes the energy of the system~\cite{DOI199679,wang1995shape}. 
However, debate still persists over explanatory mechanisms for the formation of split patterns. 
In the literature, three distinct mechanisms for the formation of split patterns exist: (i) Re-nucleation of matrix phase at the centre of precipitate (hollowing mechanism) (ii) Particle aggregation/coalescence where interactions among anti-phase domains can lead to organization of precipitates resembling split patterns (iii) Fissioning mechanism where concavities on surface of the precipitate develop into grooves which eventually deepens to the centre and splits the precipitate. 
In the following, we summarize the experimental reports, theoretical and simulation studies done till date that base their findings on one of the proposed mechanisms.

Doi et al.~explained the formation of doublet as well as octet morphologies in Ni-based alloys based on the hollowing mechanism~\cite{doi1984effect}. 
As per this mechanism, the matrix phase renucleates at the centre of the precipitate and grows along $\langle001\rangle$ directions leading to disintegration of the precipitate. 
Using two-dimensional phase-field simulations, Wang et al.~showed the formation of a split pattern via hollowing mechanism~\cite{wang1993kinetics}. 
They described the eigen-strain field in the system as a function of local composition field. 
During the precipitate growth, the composition field in the precipitate becomes inhomogeneous due to higher strain energy contribution, and a further doublet pattern forms by renucleating the matrix phase at the centre of the precipitate. 
Following a similar approach, Li et al.~showed the formation of a quartet pattern in two-dimensions~\cite{zhang_li_chen_1997}.
However, in the work of Wang et al.~and Li et al.~, the chosen values of elastic misfit for which splitting occurs are physically unrealistic. 
Moreover, with the help of phase-field simulations, Li and Chen also showed the formation of doublets under the influence of applied strain fields in an elastically inhomogeneous system~\cite{li1998shape}.
In another work, Liu et al.~reported the formation of block of split particles in a single as well as multiparticle systems using two-dimensional phase-fieldsimulations~\cite{liu2016elastic}. 
Further, they extended their work to show split dendrite using two-dimensional phase-field simulations under the influence of anisotropic interfacial energy and isotropic elastic energy~\cite{liu2017split}. 
Liu and co-workers obtained the split pattern from the initial configuration where the precipitate is highly under-saturated. 
However, according to the classical nucleation theory, the critical nucleus size of the matrix phase that can nucleate for such under-saturations can be comparable to that of the precipitate. 
Moreover, the physical basis for considering undersaturation of precipitates under isothermal conditions cannot be explained.

Banerjee et al.~suggested an alternative mechanism of formation of split patterns in Ni-base alloys via particle aggregation. They used a coherent phase-field model that incorporates the evolution of anti-phase domains of $\mathrm{L}1_{2}$ ordered $\upgamma^{\prime}$ particles in a disordered $\upgamma$ matrix phase~\cite{banerjee1999formation}. 
When the anti-phase boundary energy is higher than interfacial energy, the interactions between the four different ordered domains in a disordered matrix can result in a quartet and octet pattern in two dimensions and three dimensions, respectively. 
High-resolution electron microscopy (HREM) studies of Ni-base alloys corroborate the particle aggregation mechanism where different translational domains can migrate to form split patterns~\cite{calderon2000direct,Calderon2005high,kisielowski2007statistical}. 
Using HREM, the statistical analysis of the phase relationships between the neighboring pairs of $\upgamma^{\prime}$ particles show that 72\% of the particle pairs are out-of-phase relationship, i.e., one translational domain of $\upgamma^{\prime}$ particle surrounded by another translational domain of $\upgamma^{\prime}$ particle~\cite{Calderon2005high,kisielowski2007statistical}. 
Since all the four variants of $\gamma^{\prime}$ have equal thermodynamic possibility to randomly nucleate in the $\gamma$ matrix, nearly three quarters of particle pairs surrounding a given particle will statistically have out-of-phase relationship. 
However, this evidence does not suggest that the particle can aggregate to form a quartet or octet of the precipitate while isothermal aging. 
On the other hand, in a Cu-Zn-Al alloy, the $\upgamma$ brass phase has a complex $\mathrm{D}8_2$structure, which nucleates in the $\mathrm{B}2$ ordered $\upbeta$-brass matrix phase and shows a quartet pattern of $\upgamma$-brass precipitates~\cite{castro1998isothermal}. 
Unlike four variants describing the $\upgamma^{\prime}$ precipitates in Ni-base alloys, the $\upgamma$-brass precipitate phase is described by 36 translational variants~\cite{mehl2017aflow}; it is unclear to explain comprehensively how different translational variants lead to the formation of split patterns in Cu-Zn-Al alloy. Hence, particle aggregation mechanism may not explain the split patterns observed in Cu-Zn-Al alloy where an ordered precipitate with lower symmetry nucleates in an ordered matrix phase. 

Kaufman et al.~proposed that the formation of split 
patterns in Ni-Al alloy occurs via interfacial 
instability over the precipitate faces~\cite{Kaufman1989}.
When the interfacial instability initiates on one or six 
faces of the particle, instability spreads through the 
particle towards the centre of the particle and leads to 
the formation of a split pattern. 
Using two-dimensional phase-field simulations, 
Cha et al.~show the formation of the quartet pattern 
via morphological instability in an elastically 
inhomogeneous system~\cite{cha2005phase}.
In addition, few theoretical studies reported the effect of applied 
fields and elastic incommensurity on the particle splitting.
Leo et al.~show that the presence of the deviatoric stresses can lead to 
the particle splitting in an elastically inhomogeneous 
alloy system~\cite{leo2001elastically}. In a discrete atom 
method study, Lee shows that elastically incommensurate 
system (i.e., elastically soft direction of the matrix aligns 
along elastically hard directions of the precipitate) can form 
quartet patterns. However, reports of elastically incommensurate 
elastic constants in an alloy system such as Ni-base alloys are 
not present. 

Amongst the previously discussed mechanisms, we rule out 
matrix renucleation and particle aggregation as the mechanisms 
to explain the splitting behaviour.
It is possible that the splitting instability may initiate 
at the concavities present on the faces of the 
cuboidal precipitates, as many experimental 
observations show the formation of concavities 
on the faces of cuboidal particles during growth which forms 
owing to the point effect of 
diffusion~\cite{miyazaki1982formation,yoo1995effect,doi1984effect, 
maheshwari1992elastic}. A sharp interface based equilibrium shape study
conforms to this possibility where the presence of 
a notch on the face of precipitate oriented along $[10]$ 
direction is a precursor for the splitting of precipitate 
to occur in an elastically inhomogeneous system~\cite{zhao2013effects}.
We hypothesize that kinetically 
driven morphological instability, e.g., dendrite-like structure, 
in an elastically anisotropic cubic alloy system can 
probably lead to the development of concavities at the 
centre of precipitate faces which further can develop grooves  
along $\langle 1 0 0 \rangle$ and $\langle 1 1 0 \rangle$ 
crystallographic directions ($\langle 10 \rangle$ directions in 
two-dimensions). These grooves continue to run toward the centre of
of the precipitate resulting in primary dendrite fragmentation.
We are motivated to draw this hypothesis based on the 
capillary-driven fragmentation of dendrite side-arms 
observed during the solidification of 
alloys~\cite{cool2017359,herlach2001grain}.  

In this work, we present the phase-field simulations of particle 
splitting instability in an elastically homogeneous 
and anisotropic alloy system which mimics the superalloys. 
We identify factors influencing the initiation of 
splitting instability; in particular, the 
effects of lattice misfit, the strength of anisotropy in elastic 
energy, and interactions between particles on 
the initiation of splitting instability in a modelled alloy system. 
Section~\ref{sec:model} 
presents the phase-field model followed by two-dimensional as well 
as three-dimensional simulation results and their discussion in 
section~\ref{sec:results}. We derive the conclusions in 
section~\ref{sec: conclusion}.

\section{Model}
\label{sec:model}
A phase-field model based on free energy functional, 
e.g., Kim-Kim-Suzuki (KKS) model, needs to impose continuity 
of diffusion potential across the interface to exhibit a 
quantitative nature. It is unlike phase-field models based on 
grand-potential functional wherein the grand potential energy 
naturally possesses quantitative character. Using the Legendre 
transform, one can show that the grand-potential formalism is 
equivalent to the KKS 
formalism~\cite{choudhury2012grand,choudhury2013quantitative}. 
Here, we employ the KKS formalism~\cite{kim1999,kim2007phase} to
perform the simulations of coherently misfitting two-phase 
modelled alloy system. In addition, we follow the extended 
Cahn-Hilliard model proposed by Abinandanan and 
Haider~\cite{abinandanan2001extended} to incorporate the 
anisotropy in interfacial energy by taking 
into account fourth derivatives in the free energy functional.

Sections~\ref{subsec:energetics} and~\ref{subsec:evolve} discuss 
the energetics and governing equations describing
the evolution of field variables of the system, respectively. 
Further, we present the numerical implementation of the governing 
equations in section~\ref{numerical}. 
\subsection{Energetics}
\label{subsec:energetics}
Eqn.~\eqref{eqn:free_energy1} represents the total free energy of 
a system  described in terms of order parameter 
field $\phi(\mathbf{r})$ and composition fields 
$c_{\upalpha}(\mathbf{r})$ and $c_{\upbeta}(\mathbf{r})$ of phases 
$\upalpha$ (matrix) and $\upbeta$ (precipitate) respectively. 
 \begin{eqnarray}
     \mathcal{F}\left(\phi(\mathbf{r}), \nabla \phi(\mathbf{r}), 
     c_{\upalpha}(\mathbf{r}),c_{\upbeta}(\mathbf{r})\right) &=& N_{\textrm{v}}
   \int_V  \Big[
        f^{*}_{\textrm{dw}} + f_{\textrm{grad}}^* + f^{*}_{\textrm{ch}} + f_{\textrm{el}}^* 
          \Big] \mathrm{d}V,\nonumber  \\ 
          \label{eqn:free_energy1}    
  &=& \int_V  \Big[ f_{\textrm{dw}} +  f_{\textrm{grad}} + 
  f_{\textrm{ch}} + f_{\textrm{el}}  
          \Big] \mathrm{d}V,        
  \label{eqn:free_energy2}
 \end{eqnarray}
where $f_{\textrm{dw}}$ represents the double-well potential 
energy density of the system, $f_{\textrm{grad}}$ is the 
gradient-energy density, $f_{\textrm{ch}}$ is the chemical 
free-energy density of the system, $f_{\textrm{el}}$ is the 
elastic free-energy density of the system, $N_{\textrm{v}}$ 
denotes the number of atoms per unit volume in the system, 
and $V$ represents the volume of the system. 
The double-well potential energy density $f_{\textrm{dw}}$ reads 
as
\begin{equation}
f_{\textrm{dw}}(\phi) = \eta_w \phi^2(1- \phi)^2,
\label{eqn:dw_energy}
\end{equation} 
where $\eta_w$ is the height of potential barrier  of  the double well potential. Here, $\phi = 1$ denotes the precipitate 
phase, whereas $\phi = 0$ is the matrix phase.
The gradient-energy density is discussed in~\ref{app:grad_energy_density}. 
We represent the bulk chemical free-energy density 
$f_{\textrm{ch}}$ of the system as the interpolation between 
the chemical free-energy densities of two phases:
\begin{equation}
f_{\textrm{ch}}(c_{\upalpha}, c_{\upbeta}) =  
h(\phi) f^{\upbeta}(c_{\upbeta}) + \left(1 - h(\phi)\right) f^{\upalpha}(c_{\upalpha}),
\label{eqn:ch_energy}
\end{equation}
where $f^{\upalpha}(c_{\upalpha})$ and $f^{\upbeta}(c_{\upbeta})$ are chemical free-energy 
densities of the designated phases, and $h(\phi)$ is an 
interpolation function which should ensure that $h(0)= 0$, $h(1)=1$, 
and $0<h(\phi)<1$ when $0<\phi<1$. We choose 
$h(\phi) = \phi^3 (6\phi^2 - 15\phi +10)$ which guarantees the contribution of the interpolated 
quantity to the interface to be relatively less. The chemical free-energy 
densities of the two phases are approximated as parabolic 
functions:
\begin{eqnarray}
f^{\upalpha}(c_{\upalpha}) = f_{0}^{\upalpha} 
(c_{\upalpha} - c_{\upalpha}^{\textrm{e}})^2\\
f^{\upbeta}(c_{\upbeta}) = 
f_{0}^{\upbeta} (c_{\upbeta} - 
c_{\upbeta}^{\textrm{e}})^2,
\label{eqn:ph_free_energy}
\end{eqnarray} 
where $\displaystyle f_{0}^{\upalpha, \upbeta}$ correspond to curvatures of the chemical 
free-energy densities of the designated phases, and $c_{\upalpha,\upbeta}^{\textrm{e}}$ 
denote equilibrium compositions of the designated phases.
The bulk elastic free-energy density 
$f_{\textrm{el}}$ of the system is given as
\begin{eqnarray}
f_{\textrm{el}} (\phi) &=& N_{\textrm{v}} C^*_{ijkl}(\mathbf{r}) 
      \epsilon_{ij}^{\textrm{el}}(\mathbf{r})  \epsilon_{kl}^{\textrm{el}}(\mathbf{r}), \\
      &=& C_{ijkl}(\mathbf{r}) \epsilon_{ij}^{\textrm{el}}(\mathbf{r})  
      \epsilon_{kl}^{\textrm{el}}(\mathbf{r}),
\end{eqnarray}
where $C_{ijkl}(\mathbf{r})$ represents the position 
dependent component of the elastic stiffness tensor, 
$\epsilon_{ij}^{\textrm{el}}(\mathbf{r})$ is the elastic strain 
field which is represented as
$\displaystyle \epsilon_{ij}^{\textrm{el}}(\mathbf{r}) = 
\bar{\epsilon}_{ij} + \delta \epsilon_{ij}(\mathbf{r}) 
- \epsilon_{ij}^0(\mathbf{r})$. Here, $\bar{\epsilon}_{ij}$ 
is the homogeneous strain in the system, 
$\delta\epsilon_{ij}(\mathbf{r})$ is the heterogeneous strain 
field which has no macroscopic effects on the system, and 
$\epsilon_{ij}^0 (\mathbf{r})$ is 
the eigen-strain field. The heterogeneous strain is the 
symmetric part of the displacement gradient, i.e., 
$\displaystyle \delta \epsilon_{ij}(\mathbf{r}) = \frac{1}{2} 
\left[ \frac{\partial u_i(\mathbf{r})}{\partial r_j} + 
\frac{\partial u_j(\mathbf{r})}{\partial r_i}
\right]$, where $u_i(\mathbf{r})$ represents the 
displacement field. 
We assume a dilatational eigen-strain field in the 
system which writes as $\epsilon_{ij}^0(\mathbf{r}) = \epsilon^*
\delta_{ij} h(\phi)$, where $\epsilon^*$ is the 
magnitude of eigen-strain and $\delta_{ij}$ is the Kronecker delta function. 

\subsection{Evolution Equations}
\label{subsec:evolve}
The local composition field $c(\mathbf{r})$ is 
represented as the interpolation of the phase compositions 
$c_{\upalpha}$ and $c_{\upbeta}$: 
\begin{equation}
 c(\mathbf{r}) = h(\phi) c_{\upbeta} + \left( 1 - h(\phi) \right) 
 c_{\upalpha}.
 \label{eqn:local_comp} 
\end{equation}
The diffusion equation governs the evolution of local 
composition field $c(\mathbf{r})$~\cite{kim1999}: 
\begin{equation}
    \frac{\partial c(\mathbf{r}) }{\partial t} = 
    M  \nabla^2 \widetilde{\mu} = D \left[ \nabla^2 c(\mathbf{r}) +  
    \boldsymbol{\nabla} \cdot \left( 
       h^{\prime}\left(\phi(\mathbf{r})\right)
       \left(c_{\upalpha} - c_{\upbeta}\right)
       \boldsymbol{\nabla} \phi \left(\mathbf{r} 
       \right)
    \right) \right],
\label{eqn:compeqn}
\end{equation}
where $D$ is the inter-diffusion coefficient.
In order to decouple the gradient and bulk energy 
contributions, we impose a sharp interface boundary 
condition of the equality of diffusion potentials 
across the interface~\cite{kim1999} which is  
 \begin{equation}
     \frac{\partial f^{\upalpha}(c_{\upalpha})}{\partial c_{\upalpha}}
     = \frac{\partial f^{\upbeta}(c_{\upbeta})}{\partial c_{\upbeta}}.
 \label{eqn:diffusionpotential}
 \end{equation}
The Allen-Cahn equation governs the evolution of order 
parameter field $\phi(\mathbf{r})$~\cite{allen1979microscopic}: 
\begin{equation}
    \frac{\partial \phi(\mathbf{r})}{\partial t} = -L \frac{\delta
    \mathcal{F}}{\delta \phi(\mathbf{r})} = L\left[ 2 \varepsilon^2 \nabla^2\phi - f^{\prime}_{\textrm{dw}} - f^{\prime}_{\textrm{ch}} - f^{\prime}_{\textrm{el}}\right],
 \label{eqn:phieqn}  
\end{equation}
where $L$ denotes the relaxation coefficient, 
primed quantities represent the derivatives 
with respect to $\phi$. The derivative of 
elastic driving force 
$f_{\mathrm{el}}^{\prime}$ is presented in~\ref{app:elast_driv_force}.

Since the elastic field relaxes faster than
$\phi$ or $c$, it is reasonable to assume the 
stress field $\sigma(\mathbf{r})$ in the system 
to be governed by static
equilibrium, given as 
\begin{equation}
\nabla_j \sigma_{ij} (\mathbf{r}) = 0,
\label{eqn:mecheqlm1}
\end{equation}
where $\sigma_{ij}(\mathbf{r})$ is the local stress field 
present in the system.
We implement the Khachaturyan's interpolation 
scheme~\cite{khachaturyan2013theory} to describe the 
local stress field in the system:
\begin{equation}
    \sigma_{ij}(\mathbf{r}) = C_{ijkl}(\mathbf{r}) \left( \bar{\epsilon}_{kl} + \delta 
    \epsilon_{kl}(\mathbf{r}) - \epsilon^0_{kl}(\mathbf{r}) \right).
\label{eqn:mecheqlm2}
\end{equation} 
The position dependent stiffness 
tensor reads as 
\begin{equation}
    C_{ijkl}(\mathbf{r}) = C_{ijkl}^{0}
                         + C_{ijkl}^{\prime}(\mathbf{r}),
    \label{eqn:Cijkl}
\end{equation}
where $C_{ijkl}^{0} = \frac{1}{2}(C_{ijkl}^{\upbeta} + C_{ijkl}^{\upalpha})$ 
is the homogeneous elastic stiffness tensor and 
$C_{ijkl}^{\prime}(\mathbf{r}) = 
\frac{1}{2}(C_{ijkl}^{\upbeta} - C_{ijkl}^{\upalpha}) 
(2 h(\phi) - 1)$ is the heterogeneous elastic stiffness tensor. 
$C_{ijkl}^{\upbeta}$ and
$C_{ijkl}^{\upalpha}$ are elastic stiffness tensors 
of the designated phases. Combining Eqns.~\eqref{eqn:mecheqlm1},~\eqref{eqn:mecheqlm2} 
and~\eqref{eqn:Cijkl}, and further simplifying, we can write the 
mechanical equilibrium equation as 
\begin{equation}
C_{ijkl}^0 \frac{\partial^2 u_k }{\partial r_j \partial r_l} = \nabla_j 
\left[
       C_{ijkl}(\mathbf{r}) \left( \epsilon_{kl}^0(\mathbf{r}) -
       \bar{\epsilon}_{kl} \right) + C_{ijkl}^{\prime}(\mathbf{r})
       \frac{\partial u_k}{\partial r_l}. 
\right]
\label{eqn:mecheqlm3}
\end{equation}
Eqn.~\eqref{eqn:mecheqlm3} represents one of the conditions of 
mechanical equilibrium given by Khachaturyan's micro-elasticity 
theory. One needs to impose the another necessary condition of 
mechanical equilibrium, i.e.,~\cite{khachaturyan2013theory}
\begin{equation}
\frac{\partial f_{\textrm{el}}}{\partial \bar{\epsilon}_{ij}} = 0.
\label{eqn:mecheqlm4}
\end{equation}
Further simplifying Eqn.~\eqref{eqn:mecheqlm4}, we get
\begin{equation}
\bar{\epsilon}_{ij} = \langle S_{ijkl} \rangle 
\left( \langle \sigma^0_{kl} \rangle -
       \langle \delta \sigma_{kl} \rangle
\right),
\label{eqn:hom_strain}
\end{equation}
where $\langle S_{ijkl} \rangle = \langle C_{ijkl} \rangle^{-1}$,
$\langle C_{ijkl} \rangle = 1/V \int_V C_{ijkl}(\mathbf{r}) \mathrm{d}V$,\\
$\langle \sigma_{ij}^0 \rangle = 1/V \int_V C_{ijkl}(\mathbf{r}) 
\epsilon^0_{kl} (\mathbf{r}) \mathrm{d}V$,
$\langle \delta \sigma_{ij} \rangle = 1/V \int_V C_{ijkl}(\mathbf{r}) 
\delta \epsilon_{kl}(\mathbf{r}) \mathrm{d}V$. Eqns.~\eqref{eqn:mecheqlm3} 
and~\eqref{eqn:hom_strain} 
constitute the set of governing equations 
which ensure that the mechanical equilibrium 
is attained in the system. 

\subsection{Numerical Implementation}
\label{numerical}
We implement semi-implicit Fourier spectral 
method to obtain the 
solution to Eqns.~\eqref{eqn:compeqn} 
and~\eqref{eqn:phieqn} at 
every time step~\cite{chen1998applications}. 
We seek the numerical solution of the 
displacement fields by spectral 
iterative perturbation 
method~\cite{hu2001phase,bhattacharyya2012spectral}. 
Here, we show the discretized form of 
equations which are evolved 
at each time step to get solutions of 
$c(\mathbf{r},t)$ and 
$\phi(\mathbf{r},t)$ in Fourier space. 
Quantities with tilde represent the Fourier 
transform of corresponding quantities. The 
evolution of the local composition field in 
Fourier space reads as
\begin{equation}
\widetilde{c}(\boldsymbol{\xi},t +\Delta t) 
= \frac{\widetilde{c}(\boldsymbol \xi ,t) - 
I D \Delta t \; \boldsymbol \xi \cdot 
\widetilde{\mathbf{p}}(\boldsymbol{\xi},t)}
{1 + D\xi^2 \Delta t},
\label{eqn:c_dft}
\end{equation}  
where $I=\sqrt{-1}$, $\Delta t$ is the time 
increment, $\boldsymbol \xi$ denotes the 
inverse space vector,  $\xi$ is the magnitude 
of $\boldsymbol \xi$, 
$\widetilde{\mathbf{p}}(\boldsymbol{\xi},t)$ 
represents the Fourier transform of 
$\mathbf{p}(\mathbf{r},t) = h'(\phi) 
(c_{\upalpha} - c_{\upbeta}) \boldsymbol{\nabla} \phi$.
The evolution of the order parameter $\phi$ in
Fourier space reads as
\begin{equation}
\widetilde{\phi}(\boldsymbol{\xi},t+\Delta t) 
= \frac{\widetilde{\phi}(\boldsymbol{\xi},t) -
L\widetilde{m}(\boldsymbol{\xi},t) \Delta t}
{1 + 2\varepsilon^2 \xi^2 \Delta t},
\label{eqn:phi_dft}
\end{equation}
where $\widetilde{m}(\boldsymbol{\xi},t)$ 
represents the Fourier 
transform of $\displaystyle m(\mathbf{r},t) = 
f^{\prime}_{\textrm{dw}} + 
f^{\prime}_{\textrm{ch}} + 
f^{\prime}_{\textrm{el}}$. 

The spectral iterative perturbation method 
involves the evaluation of displacement 
fields for a homogeneous system, i.e., 
zeroth-order approximation, and further 
higher-order approximations are obtained 
up to the prescribed tolerance level by 
substituting the previous order 
approximations. Eqns.~\eqref{eqn:zeroth_dft} 
and~\eqref{eqn:higher_dft} 
represent respectively the zeroth and higher-order 
approximations of the displacement fields in 
Fourier space.
\begin{equation}
\widetilde{u}^{(0)}_{k}(\boldsymbol{\xi}) = -I \xi_j G_{ik}(\boldsymbol{\hat{\xi}}) 
\widetilde{\sigma}^{0}_{ij}(\boldsymbol{\xi}),
\label{eqn:zeroth_dft}
\end{equation}
\begin{equation}
\widetilde{u}_{k}^{(\textrm{n})}(\boldsymbol{\xi}) = -I \xi_j G_{ik}(\boldsymbol{\hat{\xi}}) 
\left[ C_{ijpq}(\mathbf{r}) 
\left( \epsilon_{pq}^0(\mathbf{r}) -\bar{\epsilon}_{pq} \right) + 
C_{ijpq}^{\prime}(\mathbf{r}) \frac{\partial u_p^{(\mathrm{n}-1)}(\mathbf{r})}{\partial r_q} 
\right]_{\boldsymbol{\xi}},
\label{eqn:higher_dft} 
\end{equation}
where $G_{ik}$ represents the Green tensor 
whose inverse is given as 
$(G_{ik})^{-1} = C_{ijkl}^0 \hat{\xi}_j \hat{\xi}_l$, 
$\displaystyle \boldsymbol{\hat{\xi}} = \frac{\boldsymbol{\xi}}
{|\boldsymbol{\xi}|}$, $\xi_j$ denotes the 
components of inverse space vector,
$\widetilde{\sigma}_{ij}^0 (\boldsymbol{\xi})$
is the Fourier transform of 
$\sigma_{ij}^0 (\mathbf{r})$, 
$[\cdot]_{\boldsymbol \xi}$ represents the 
quantity in Fourier space. Previous 
phase-field studies~\cite{cha2005phase} 
evaluate displacement field only up to the third 
order that can result in appreciable 
error in the stress field of the system. 
Since we impose homogeneous moduli 
approximation in the system, 
Eqn.~\eqref{eqn:zeroth_dft} is sufficient to 
seek solution for the displacement field.

\subsection{Sharp-interface limit}
In this section, we show that in the sharp-interface limit, we 
recover the Gibbs-Thomson relation from our diffuse-interface 
model when the interface is curved and coherent. 
Here, we assume a coordinate system where normal ($\mathbf{n}$) and 
tangent vector ($\mathbf{t}$) to the interface constitute the basis 
vectors. The normal to the interface is given as
\begin{equation}
    \boldsymbol{\nabla} \phi = \mathbf{n} \frac{\partial \phi}{\partial n}.
    \label{eqn:normal}
\end{equation}
Upon taking divergence of~\eqref{eqn:normal},
\begin{equation}
    \nabla^2 \phi = \frac{\partial^2 \phi}{\partial n^2} + \frac{\partial \phi}{\partial n} \left( \boldsymbol{\nabla} \cdot \mathbf{n}\right).
    \label{eqn:div}
\end{equation}
The phase-field equation in the transformed coordinate is
\begin{align}
    \frac{1}{L}\frac{\partial \phi(\mathbf{r})}{\partial t} =&
     2 \varepsilon^2\left( \frac{\partial^2 \phi}{\partial n^2} + 
     \frac{\partial \phi}{\partial n} \left( \nabla \cdot 
     \mathbf{n}\right) \right) - 2 \eta_{\mathrm{w}} \phi(1-2\phi)(1-\phi) \nonumber \\ 
    &-h^{\prime}(\phi)\left[ f^{\upbeta}(c_{\upbeta})-f^{\upalpha}(c_{\upalpha})
    -(c_{\upbeta} - c_{\upalpha})\frac{\partial 
    f^{\upalpha}(c_{\upalpha})}{\partial c_{\upalpha}}\right]  
    -f_{\textrm{el}}^{\prime}.
    \label{eqn:trans_phi_eqn}
\end{align}
The leading order solution for the phase-field profile 
$\phi_0$ is:
\begin{equation}
2\varepsilon^2 \frac{\partial^2 \phi_0}{\partial n^2} - 
2\eta_{\mathrm{w}} \phi_0
(1 - 2\phi_0) (1 - \phi_0) = 0.
\label{eqn:lead_order_sol}
\end{equation}
After multiplying $\displaystyle \frac{\partial \phi_0}{\partial n}$ on 
both sides of Eqn.~\eqref{eqn:lead_order_sol} and integrating it along
the interface normal at the boundary such that in the bulk of matrix and precipitate $\displaystyle \frac{\partial \phi_0}{\partial n} = 0$,
\begin{equation}
    \varepsilon^2\left( \frac{\partial \phi_0}{\partial n} \right)^2 =
     \eta_{\mathrm{w}} \phi_0^2 (1-\phi_0)^2
     \label{eqn:lead_order_sol2}
\end{equation}
Solving Eqn.~\eqref{eqn:lead_order_sol2} gives a 
equilibrium phase-field profile without any driving force. After 
combining Eqns.~\eqref{eqn:trans_phi_eqn} and~\eqref{eqn:lead_order_sol}, 
the updated phase-field equation is
\begin{align}
    \frac{1}{L}\frac{\partial \phi}{\partial t} = &
    2 \varepsilon^2 \left( \frac{\partial \phi_0}{\partial n}(\nabla \cdot \mathbf{n}) \right) -h^{\prime}(\phi)\left[ f^{\upbeta}(c_{\upbeta})-f^{\upalpha}(c_{\upalpha})
    -(c_{\upbeta} - c_{\upalpha})\frac{\partial 
    f^{\upalpha}(c_{\upalpha})}{\partial c_{\upalpha}}\right]\nonumber \\  
    & -f_{\textrm{el}}^{\prime}
    \label{eqn:phi_eqn_update2}
\end{align}
In a sharp-interface limit, we can obtain the elastic driving
force $f_{\textrm{el}}^{\prime} = \frac{\partial f_{\textrm{el}}}{\partial \phi}$:

\begin{equation}
    \frac{\partial f_{\textrm{el}}}{\partial \phi} = 
    \frac{\mathrm{d}f_{\textrm{el}}}{\mathrm{d}\phi} - \frac{\partial 
    f_{\textrm{el}}}{\partial \epsilon_{nn}} \frac{\partial \epsilon_{nn}}{\partial \phi}
    -2\frac{\partial f_{\textrm{el}}}{\partial \epsilon_{nt}} \frac{\partial 
    \epsilon_{nt}}{\partial \phi} - 
    \frac{\partial f_{\textrm{el}}}{\partial \epsilon_{tt}} 
    \frac{\partial \epsilon_{tt}}{\partial \phi}
\end{equation}
At the sharp-interface limit, $\sigma_{tt}$ exhibits a jump and
$\sigma_{nn}$ and $\sigma_{nt}$ will be continuous across the 
interface, i.e., $\sigma_{nn}^{\upalpha} = \sigma_{nn}^{\upbeta} = \sigma_{nn}^0$ and $\sigma_{nt}^{\upalpha} = \sigma_{nt}^{\upbeta} = \sigma_{nt}^0$. Hence, $\sigma_{tt}$ vanishes in the sharp-interface 
limit. As a result, the leading order elastic driving force in 
the sharp-interface limit is expressed as
\begin{equation}
    \frac{\partial f_{\textrm{el}}}{\partial \phi} = 
    \frac{\mathrm{d}f_{\textrm{el}}}{\mathrm{d}\phi} - \sigma_{nn}^0 
    \frac{\partial \epsilon_{nn}}{\partial \phi}
    -2\sigma_{nt}^0 \frac{\partial 
    \epsilon_{nt}}{\partial \phi} = 
    \frac{\mathrm{d}}{\mathrm{d}\phi} 
    \left[ f_{\textrm{el}} - \sigma_{nn}^0 \epsilon_{nn} - 2 \sigma_{nt}^0 
    \epsilon_{nt}\right]
    \label{eqn:elast_driv_force}
\end{equation}
Inserting Eqn.~\eqref{eqn:elast_driv_force} in Eqn.~\eqref{eqn:phi_eqn_update2},
\begin{align}
    \frac{1}{L}\frac{\partial \phi}{\partial t} = &
    2 \varepsilon^2 \left( \frac{\partial \phi_0}{\partial n}(\nabla \cdot \mathbf{n}) \right) -h^{\prime}(\phi)\left[ f^{\upbeta}(c_{\upbeta})-f^{\upalpha}(c_{\upalpha})
    -(c_{\upbeta} - c_{\upalpha})\frac{\partial 
    f^{\upalpha}(c_{\upalpha})}{\partial c_{\upalpha}}\right]\nonumber \\  
    & - \frac{\mathrm{d}}{\mathrm{d}\phi} 
    \left[ f_{\textrm{el}} - \sigma_{nn}^0 \epsilon_{nn} - 2 \sigma_{nt}^0 \epsilon_{nt}\right].
    \label{eqn:phi_eqn_update3}
\end{align}
Multiplying both sides of Eqn.~\eqref{eqn:phi_eqn_update3} by 
$\displaystyle \frac{\partial \phi_0}{\partial n}$ and integrating from the bulk of the
precipitate to that of the matrix along normal to the interface, we derive

\begin{align}
    \frac{1}{L} \int_{\textrm{in}}^{\textrm{out}} \frac{\partial \phi}{\partial t} \frac{\partial \phi_0}{\partial n} \mathrm{d}n = 
    & 2 \varepsilon^2 \int_{\textrm{in}}^{\textrm{out}} \left( \frac{\partial \phi_0}{\partial n}\right)^2 (\nabla \cdot \mathbf{n}) \mathrm{d}n \nonumber \\
    & -\int_{1}^{0} h^{\prime}(\phi)\left[ f^{\upbeta}(c_{\upbeta})-f^{\upalpha}(c_{\upalpha})
    -(c_{\upbeta} - c_{\upalpha})\frac{\partial 
    f^{\upalpha}(c_{\upalpha})}{\partial c_{\upalpha}}\right] \mathrm{d}\phi \nonumber \\
    & - \int_{1}^{0} \frac{\mathrm{d}}{\mathrm{d}\phi} 
    \left[ f_{\textrm{el}} - \sigma_{nn}^0 \epsilon_{nn} - 2 \sigma_{nt}^0 \epsilon_{nt}\right] \mathrm{d}\phi
    \label{eqn:phi_eqn_update4}
\end{align}
The left-hand term in Eqn.~\eqref{eqn:phi_eqn_update4} can be simplified as
\begin{equation}
    \frac{\partial \phi}{\partial t} \frac{\partial \phi_0}{\partial n} = - \left( \frac{\partial n}{\partial t}\right)_{\phi} 
    \left(\frac{\partial \phi_0}{\partial n}\right)^2 = - 
    V_{\textrm{n}} \left(\frac{\partial \phi_0}{\partial n}\right)^2,
\end{equation}
where $V_n$ denotes the velocity of the interface. $\nabla \cdot 
\mathbf{n} = - \kappa$ denotes the mean curvature of the interface.
Hence, Eqn.~\eqref{eqn:phi_eqn_update4} is modified as
\begin{align}
    -\frac{V_{\textrm{n}}}{L} \int_{\textrm{in}}^{\textrm{out}} \left(\frac{\partial \phi_0}{\partial n}\right)^2 \mathrm{d}n = &-2 \varepsilon^2 \kappa \int_{\textrm{in}}^{\textrm{out}} \left(\frac{\partial \phi_0}{\partial n}\right)^2 \mathrm{d}n \nonumber \\
    &-\left[ f^{\upbeta}(c_{\upbeta})-f^{\upalpha}(c_{\upalpha})
    -(c_{\upbeta} - c_{\upalpha})\frac{\partial 
    f^{\upalpha}(c_{\upalpha})}{\partial c_{\upalpha}}\right] \Bigg. h(\phi)\Bigg|_1^0 
    \nonumber \\
    &- \left[ f_{\textrm{el}} - \sigma_{nn}^0 \epsilon_{nn} - 2 \sigma_{nt}^0 \epsilon_{nt}\right]_1^0
    \label{eqn:phi_eqn_update5}
\end{align}
The term $\displaystyle \int_{\textrm{in}}^{\textrm{out}} \left( \frac{\partial \phi_0}{\partial n}\right)^2 \mathrm{d}n$ corresponds to $\displaystyle \frac{\gamma}{2 \varepsilon^2}$. After 
invoking the equality of diffusion
potentials, i.e., $\displaystyle \frac{\partial f^{\upalpha}(c_{\upalpha})}{\partial c_{\upalpha}} = \frac{\partial f^{\upbeta}(c_{\upbeta})}{\partial c_{\upbeta}}$, we rewrite the Eqn.~\eqref{eqn:phi_eqn_update5} as
\begin{equation}
    V_{\textrm{n}} = M^{\textrm{eff}} \left[ \gamma \kappa + (\Psi^{\upalpha} - \Psi^{\upbeta}) + (\omega^{\upalpha} - \omega^{\upbeta})\right],
\end{equation}
where
\begin{eqnarray}
 M^{\textrm{eff}} &=& \frac{2\varepsilon^2 L}{\gamma},\\
\Psi^{\upalpha} &=& f^{\upalpha}(c_{\upalpha}) - c_{\upalpha} \frac{\partial f^{\upalpha}(c_{\upalpha})}{\partial c_{\upalpha}},\\
\Psi^{\upbeta} &=& f^{\upbeta}(c_{\upbeta}) - c_{\upbeta} \frac{\partial f^{\upbeta}(c_{\upbeta})}{\partial c_{\upbeta}},\\
\omega^{\upalpha} &=& f_{\textrm{el}}^{\upalpha} - \sigma_{nn}^0 \epsilon_{nn}^{\upalpha} - 2 \sigma_{nt}^0 \epsilon_{nt}^{\upalpha},\\
\omega^{\upbeta} &=& f_{\textrm{el}}^{\upbeta} - \sigma_{nn}^0 \epsilon_{nn}^{\upbeta} - 2 \sigma_{nt}^0 \epsilon_{nt}^{\upbeta},
\end{eqnarray}
and $\Psi^{\upalpha, \upbeta}$ represents the grand potential densities 
of the designated phases. 
We are interested in diffusion-controlled growth regime, therefore phase-field $\phi$ relaxes infinitely fast, i.e., 
$M^{\textrm{eff}} = \infty$, and as a result 
\begin{equation}
    \gamma \kappa + (\Psi^{\upalpha} - \Psi^{\upbeta}) + (\omega^{\upalpha} - \omega^{\upbeta}) = 0
\end{equation}
The driving force for 
phase transformation $\Delta \Psi = \Psi^{\upbeta} - \Psi^{\upalpha}$ 
balances the curvature and the elastic driving forces
\begin{equation}
    \Delta \Psi = \Psi^{\upbeta} - \Psi^{\upalpha} = \gamma \kappa + (\omega^{\upalpha} - \omega^{\upbeta}).
\end{equation}
Using the linear expansion of grand potential about the chemical 
potential~\cite{choudhury2012grand,choudhury2013quantitative}
\begin{equation}
    \Delta \Psi = \left( \frac{\partial \Psi^{\upbeta} }{\partial \mu} - 
    \frac{\partial \Psi^{\upalpha} }{\partial \mu} \right) \Delta \mu,
    \label{eqn:thermo_rel}
\end{equation}
where $\Delta \mu = \mu - \mu_{\textrm{eq}}$, and $\mu_{\textrm{eq}}$ represents the equilibrium 
chemical potential. After invoking thermodynamic relation 
$\displaystyle \frac{\partial \Psi}{\partial \mu} = - c$,
\begin{equation}
    (c_{\upalpha} - c_{\upbeta}) \Delta \mu = \gamma \kappa + (\omega^{\upalpha} - \omega^{\upbeta}).
\end{equation}
Performing linear expansion of chemical potential about the composition of $\upbeta$ phase, 
we derive
\begin{equation}
    (c_{\upalpha} - c_{\upbeta}) \frac{\partial \mu}{\partial c} (c_{\upbeta} - c_{\upbeta}^{\textrm{e}}) = (c_{\upalpha} - c_{\upbeta}) \frac{\partial^2 f^{\upbeta}}{\partial c^2_{\upbeta}} (c_{\upbeta} - c_{\upbeta}^e) = \gamma \kappa + (\omega^{\upalpha} - \omega^{\upbeta}).
\end{equation}
Thus, we obtain a Gibbs-Thomson relation for the
shift in the equilibrium composition of $\upbeta$ phase
\begin{equation}
    \Delta c_{\mathrm{\upbeta}} = \frac{\gamma \kappa + (\omega^{\upalpha} -
\omega^{\upbeta})}{\frac{\partial^2 f^{\upbeta}}{\partial c^2_{\upbeta}} (c_{\upalpha} - c_{\upbeta})}.
\label{eqn:gibbs_thomson}
\end{equation}
Eqn.~\ref{eqn:gibbs_thomson} provides the interfacial 
compositions for $\upalpha$ and $\upbeta$ phases. Further, the 
mass balance equation extends the normal velocity of the 
interface under the diffusion controlled regime, i.e.,

\begin{equation}
    V_{\mathrm{n}} (c_{\upbeta}^{\prime} - 
    c_{\upalpha}^{\prime}) = D (\boldsymbol{\nabla} c_{\upbeta}^{\prime} - 
    \boldsymbol{\nabla} c_{\upalpha}^{\prime}) \cdot \mathbf{n},
\end{equation}
Here $\displaystyle c_{\upbeta,\upalpha}^{\prime} =
c_{\upbeta,\upalpha}^e + \Delta c_{\upbeta,\upalpha}$ represent the
bulk compositions of $\upbeta$ and $\upalpha$ phases where
the curvature and elastic contributions are accounted.

\begin{figure}[!htb]
    \centering
    \centering
    \begin{subfigure}{0.49\textwidth}
    \centering
    \includegraphics[width=7cm]{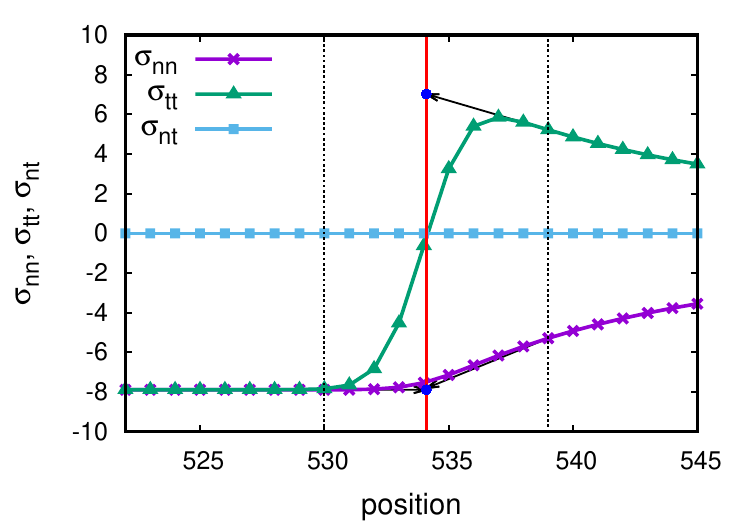}
    \caption{}
    \label{subfig:sts_prof}
    \end{subfigure}%
    \begin{subfigure}{0.49\textwidth}
    \centering
    \includegraphics[width=7cm]{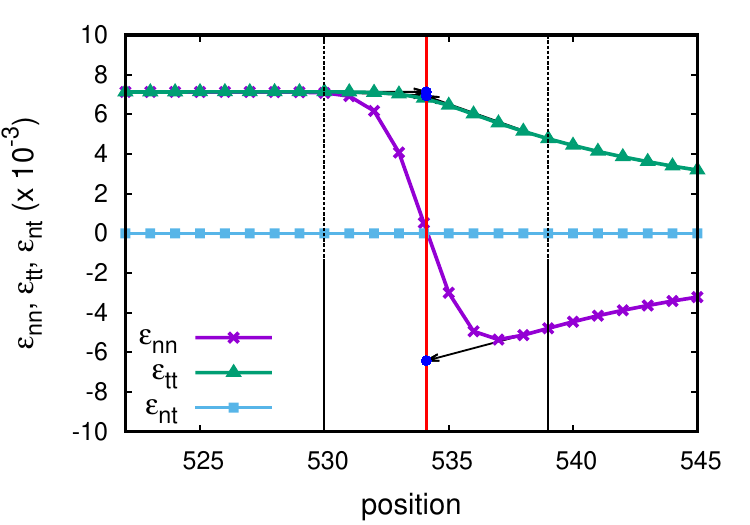}
    \caption{}
    \label{subfig:str_prof}
    \end{subfigure}%
    \caption{(a) Stress and (b) strain profiles across the 
    interface of the cylindrical precipitate growing in a 
    supersaturated matrix. The solid red line indicates the
    position of interface at $\phi = 0.5$. The asymptotic 
    extensions of the normal stress and transverse strain 
    are continuous at the interface. On the other hand, 
    transverse stress and normal strain fields exhibit jump across the interface.}
    \label{fig:sts_str_prof}
\end{figure}

\begin{figure}[!htb]
    \centering
    \includegraphics[width=0.75\textwidth]{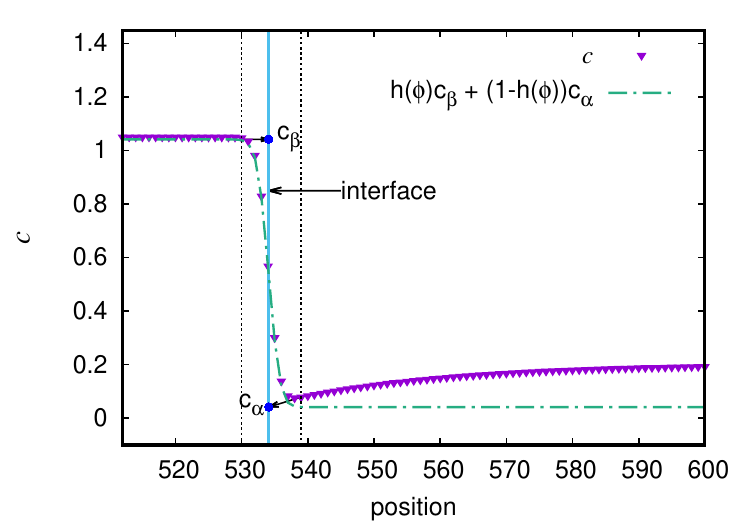}
    \caption{Composition profile of the circular precipitate
    in a supersaturated matrix ($c_{\infty} = 20\%$). Here, we assume isotropic 
    homogeneous moduli approximation with lattice misfit of $1\%$ and shear 
    modulus of $550$.}
    \label{fig:sharp_inter_comp}
\end{figure}

We compare the interfacial compositions from the 
simulations with that of the sharp-interface model
in the asymptotic limits.
We assume a circular precipitate ($\upbeta$) growing in a 
supersaturated matrix phase
($\upalpha$). 
Fig.~\ref{fig:sts_str_prof} shows that the 
asymptotic extensions of the bulk values of 
$\sigma_{nn}^{\upalpha},\sigma_{nn}^{\upbeta}$ and 
$\epsilon_{tt}^{\upalpha}, \epsilon_{tt}^{\upbeta}, 
\epsilon_{nt}^{\upalpha}, \epsilon_{nt}^{\upbeta}$ 
to the interface from diffuse interface simulation clearly 
captures the continuity of normal stresses and transverse 
strain fields. Other stress and strain fields exhibit jump across the interface. 
Fig.\ref{fig:sharp_inter_comp} shows the comparison of the interfacial 
compositions from sharp-interface model and diffuse interface. The shift in composition 
from the sharp-interface model comes out to be $0.0417$ and the corresponding shift in 
composition from the diffused interface simulations is $0.0506$. The interfacial conditions 
in the simulations compare well with sharp-interface model in the asymptotic limits.
\section{Results}
\label{sec:results}
We have performed two-dimensional as well as 
three-dimensional phase-field simulations of 
single and two precipitates in a 
supersaturated matrix. The morphological evolution is 
characterized by the quantification of Gibbs-Thomson 
effect. We investigate 
the effect of lattice misfit, anisotropy in elastic 
energy, and particle interactions on the morphological 
evolution.

\subsection{Generalized Gibbs-Thomson effect}
During the growth of coherently misfitting precipitate, 
the curvature and elastic effects shift the equilibrium
composition of the matrix which in turn shifts 
the chemical potential present in the matrix, i.e., 
Gibbs-Thomson 
effect~\cite{johnson1987precipitate,laraia1988growth}. 
Depending on the strengths of lattice misfit and 
anisotropy in elastic energy, the contribution of 
elastic effects varies in different directions. We 
evaluate the shift in chemical potential 
($\Delta \mu^*$) in the matrix phase ahead of interface 
as a function of an arc-length ($l_{\mathrm{arc}}$). The 
shift in chemical potential have two contributions, 
viz., curvature ($\Delta \mu^{\textrm{K}*}$) and elastic effects 
($\Delta \mu^{\mathrm{el}*}$): 
\begin{equation}
    \Delta \mu^* = \Delta \mu^{\textrm{K}*} + \Delta \mu^{\textrm{el}*},
    \label{eqn:Delta_mu}
\end{equation}
where $\displaystyle \Delta \mu^{\textrm{K}*} = \frac{\gamma \kappa}
{(c_{\upbeta}^{\textrm{e}} - c_{\upalpha}^{\textrm{e}})}$, 
$\kappa$ is the mean curvature and $\gamma$ is the 
interfacial energy. In a non-dimensional form, 
Eqn.~\eqref{eqn:Delta_mu} is rewritten as
\begin{equation}
    \Delta \mu = l_0 \kappa + \Delta \mu^{\textrm{el}},
\end{equation}
where $\displaystyle \Delta \mu = \frac{l_0 \Delta \mu^*}{\gamma}$, 
$\displaystyle \Delta \mu^{\textrm{el}} = \frac{l_0 \Delta \mu^{\textrm{el}*}}{\gamma}$, 
$l_0$ is the characteristic length which we choose as \SI{1}{\nano \meter}.

\%subsection{Topological characterization}
\subsection{Simulations Parameters}
\label{subsec:3D}

The thermodynamic and mobility parameters chosen in our study correspond to those of Ni-$17$ 
at.\% Al alloy aged at \SI{1000}{\celsius} where Kaufman et al.~\cite{Kaufman1989} reported the 
formation of split patterns of $\gamma^{\prime}$.
Table~\ref{table:3DsimParams} lists these parameters in 
dimensional as well as non-dimensional 
values. 
We use characteristic energy 
$\mathcal{E} = 10^{-20}$ \si{\joule}, 
characteristic length 
$\mathcal{L} = 0.25$ \si{\nano\meter}, 
and characteristic time 
$\mathcal{T} = 3 \times 10^{-3}$ \si{\second}.
\begin{table}[!htb]
	\caption{Simulation parameters used in
	simulations. Both dimensional and non-dimensional values of parameters are given.}
	\label{table:3DsimParams}
	\begin{center}
		\begin{tabular}{| c | c | c | c |}
			\hline
			\hline
			\textbf{Parameter} & \textbf{Non-dimensional value} & \textbf{Dimensional value} &
			\textbf{Formula}\\
			\hline
			\hline
			Interfacial energy ($\gamma$) & $0.12$ & $20$  \si{\milli \joule \per \square \meter} & $\gamma = \gamma^* (\mathcal{E}\mathcal{L}^{-2})$\\
			Interface width ($2\lambda$) & $1.6$ & $0.4$ \si{\nano \meter} &
			$2\lambda = (2\lambda)^* \mathcal{L}$\\
			Inter-diffusion coefficient ($D$) & $1.0$ & $10^{-15}$ \si{\meter \squared \per \second} & $D = D^* \mathcal{L}^2 \mathcal{T}^{-1}$\\
			Relaxation coefficient ($L$) & $1.615$ & $3.5 \times 10^{-5}$ \si{\meter \squared \per \newton \per \second} & 
			$L = L^* \mathcal{L}^3 \mathcal{E}^{-1} \mathcal{T}^{-1}$\\
			Time step ($dt$) & $0.1$ & $0.3$ \si{\milli \second} & $dt = dt^* \mathcal{T}$\\
			Shear modulus ($G$) & $550$ & $88$ \si{\giga \pascal} & $G = G^* \mathcal{E} \mathcal{L}^{-3}$\\
			Poisson's ratio ($\nu$) & -- & $0.3$ & --\\
			Zener anisotropy parameter ($A_{\textrm{z}}$) & -- & $4-2$ & -- \\
			Elastic misfit ($\epsilon^*$) & -- & $0.5-1\%$ & --\\
			supersaturation ($c_{\infty}$) & -- & $25-45\%$ & -- \\
			\hline
		\end{tabular}
	\end{center}
\end{table}
By choosing $2 \lambda =$ \SI{0.4}{\nano \meter} and 
$dx = dy = dz =$ \SI{0.05}{\nano \meter}, 
we have ensured a minimum of eight grid points at the 
interface of $\phi$. The inter-diffusion coefficient 
$D = 10^{-15}$ \si{\meter \squared \per \second}~\cite{LandoltBornstein1990} 
and relaxation coefficient $L = 3.5 \times 10^{-5}$ 
\si{\meter \squared \per \newton \per \second} 
assure the diffusion-controlled growth of precipitate in a 
supersaturated matrix. We employ relations given by Schmidt and 
Gross~\cite{schmidt1997equilibrium} to represent the components 
of the stiffness tensor in Voigt's notation for a system with 
cubic anisotropy in terms of average shear modulus ($G$), 
Poisson's ratio ($\nu$), and Zener anisotropy parameter ($A_{\textrm{z}}$).
\begin{align*}
    C_{11} &= G \left[ \frac{2(2+A_{\textrm{z}})}{1+A_{\textrm{z}}} - \frac{1-4\nu}{1-2\nu}\right];\\
    C_{12} &= G \left[ \frac{2 A_{\textrm{z}}}{1+A_{\textrm{z}}} - \frac{1-4\nu}{1+A_{\textrm{z}}}\right];\\
    C_{44} &= G \left[ \frac{2 A_{\textrm{z}}}{1+A_{\textrm{z}}} \right]
\end{align*}

Following sections discuss the effects of elastic misfit, anisotropy in elastic energy,
and degree of supersaturation on the particle splitting instability in three-dimensions. 
Since the three-dimensional simulations are memory intensive, we discuss the 
effect of particle interactions in two-particle 
settings using two-dimensional simulations. 
We perform both two-dimensional 
as well as three-dimensional simulations 
using NVIDIA Tesla V100 GPUs and utilize 
CUDA fast Fourier transform (cuFFT) 
library~\cite{cufft} 
to evaluate discrete 
Fourier transforms required for evolving 
Eqns.~\eqref{eqn:c_dft},~\eqref{eqn:phi_dft},~\eqref{eqn:zeroth_dft}, 
and~\eqref{eqn:higher_dft}. The simulation 
parameters, hereafter, render 
non-dimensional values. 

\subsection{Evolution of precipitate morphology to split patterns}
\label{sec:prec_evolve}
We begin with an isolated precipitate of size 
$3$ in a supersaturated matrix 
at the centre of the simulation box. 
Here, we choose supersaturation of $45\%$, lattice
misfit of $0.85\%$, and Zener anisotropy 
parameter of $4$. Table~\ref{tab:PrecipitateEvolve} 
shows the evolution of precipitate morphology in 
isosurface view, and composition and order parameter 
map in $(110)$ plane passing through the centre of 
the simulation box at different simulation times. 
Initially, the spherical precipitate transforms to a 
cuboidal shape and ears start developing along 
$\langle 111 \rangle$ directions. 
Later, the dendrite-like instability develops
which exhibits  
predominant primary arms along $\langle 111 \rangle$
directions (see precipitate morphology at $t=100$ in 
table~\ref{tab:PrecipitateEvolve}). The formation of
dendrite-like morphologies can be explained based on 
theory of morphological instability in coherent 
system given by Leo-Sekerka~\cite{leo1989effect}. 
The morphological instability occurs when the point 
effect of diffusion 
dominates over the capillary forces. 
Further, during the growth of dendrite-like 
instability, the concavities are developed along 
faces directed towards 
$\langle 100 \rangle$ and $\langle 110 \rangle$ 
directions. These concavities develop grooves that 
advance towards the centre of precipitate from the 
faces and lead to the split morphology consisting of 
eight smaller precipitates (ogdoad) (see precipitate 
morphologies at $t=1000$, $5000$, and $10000$ in 
table~\ref{tab:PrecipitateEvolve}).

\begin{table}[!htb]
    \centering
    \begin{tabular}{|M{1.8cm} | M{3cm} M{3cm} M{3cm} M{3cm} | }
    \hline
         &  $t=100$ & $t=1000$ & $t=5000$ & $t=10000$\\
    \hline         
    & & & &\\
    Isosurface view     & 
    \includegraphics[width=3cm]{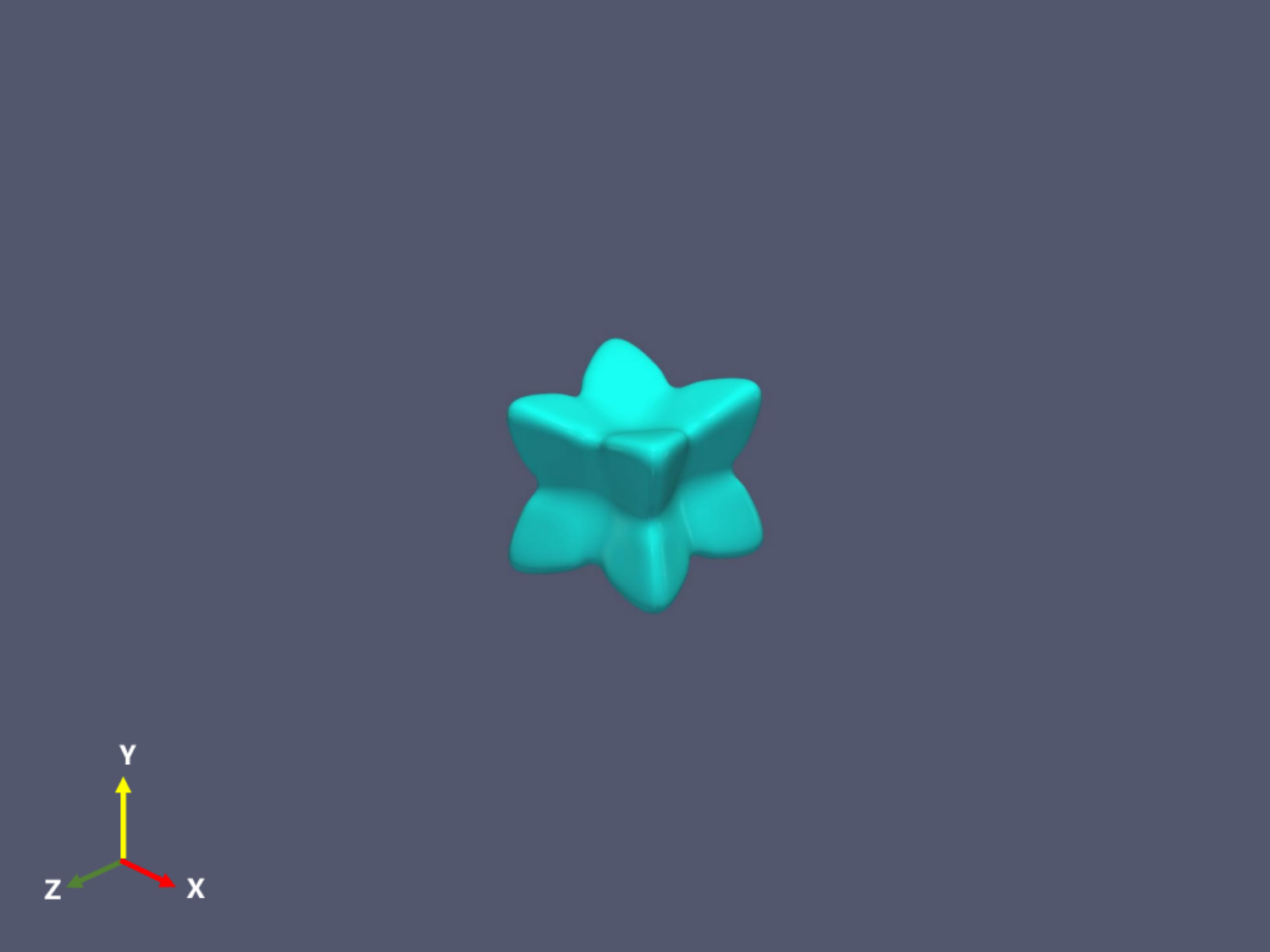} &
    \includegraphics[width=3cm]{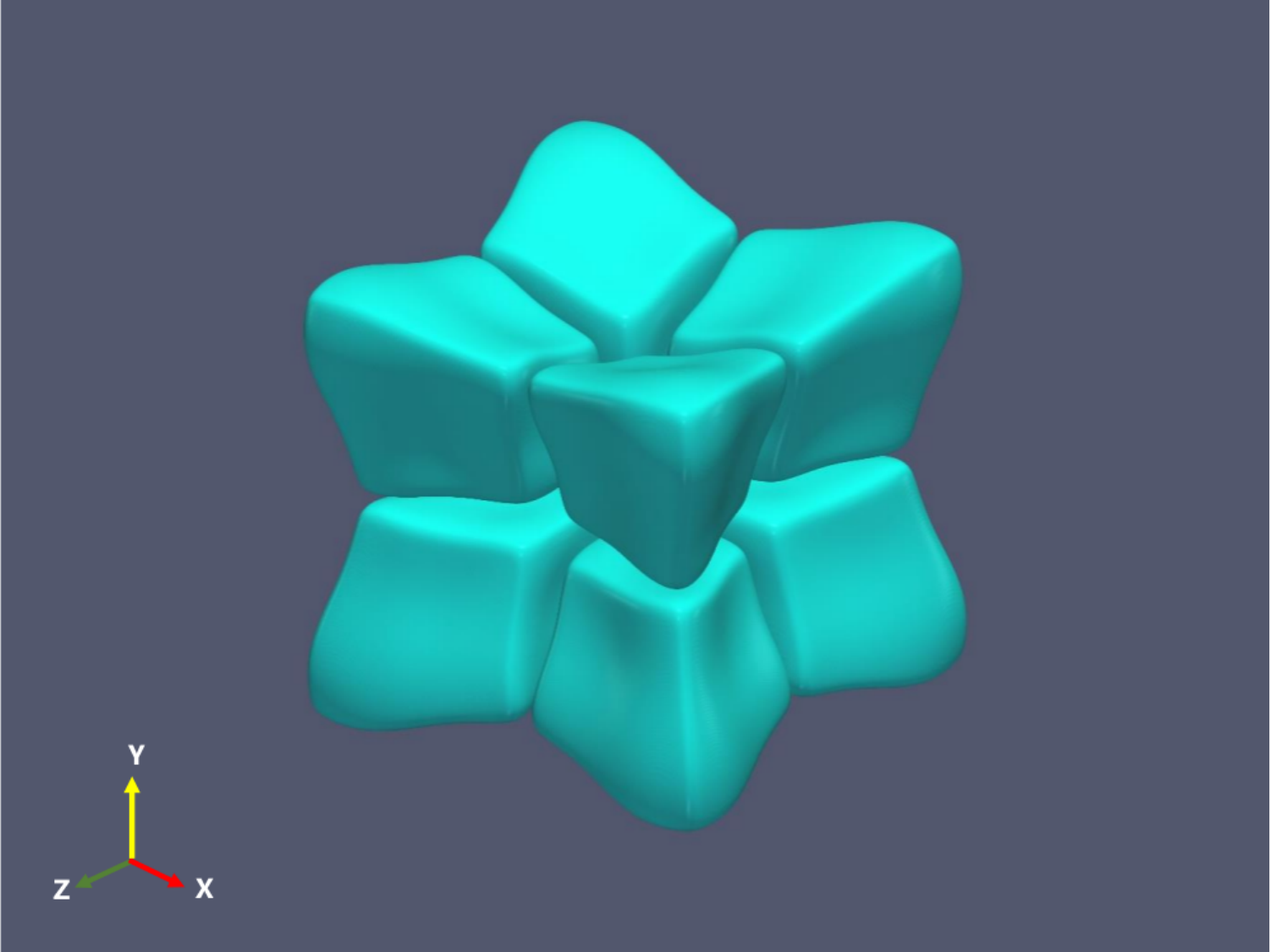} &
    \includegraphics[width=3cm]{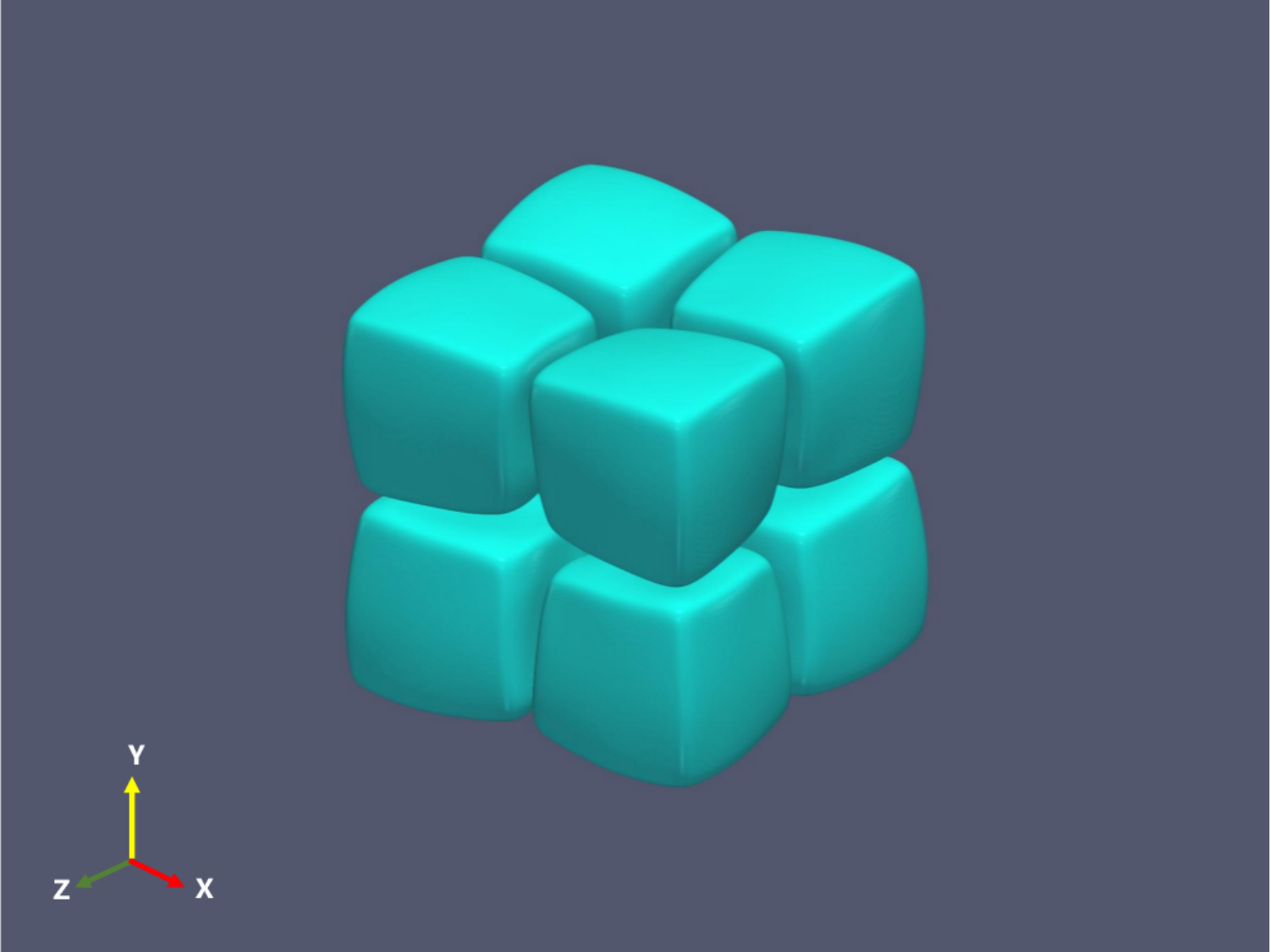} &
    \includegraphics[width=3cm]{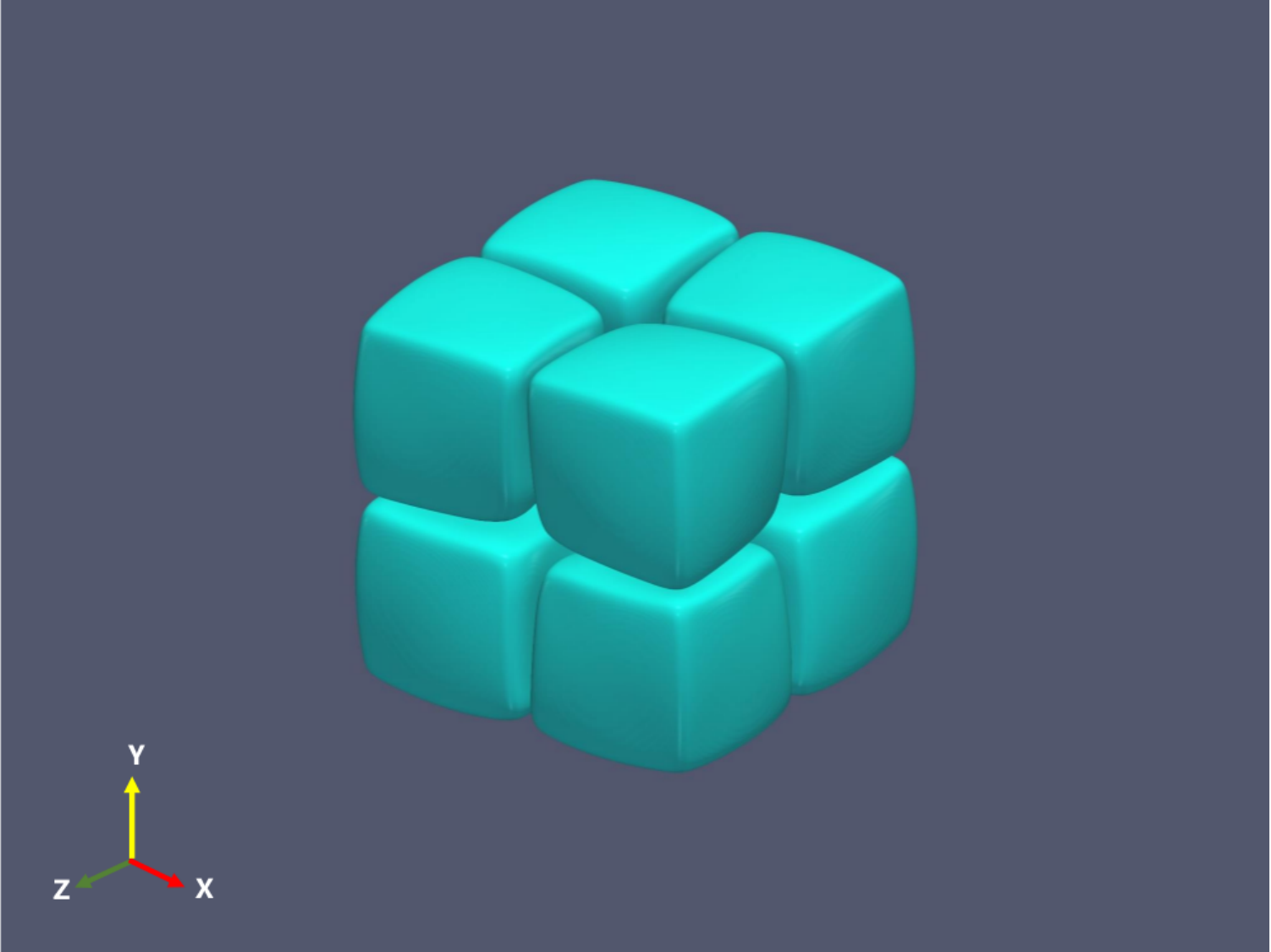}\\
    & & & &\\
    \hline
    & & & &\\
    Composition map &
    \includegraphics[width=3cm]{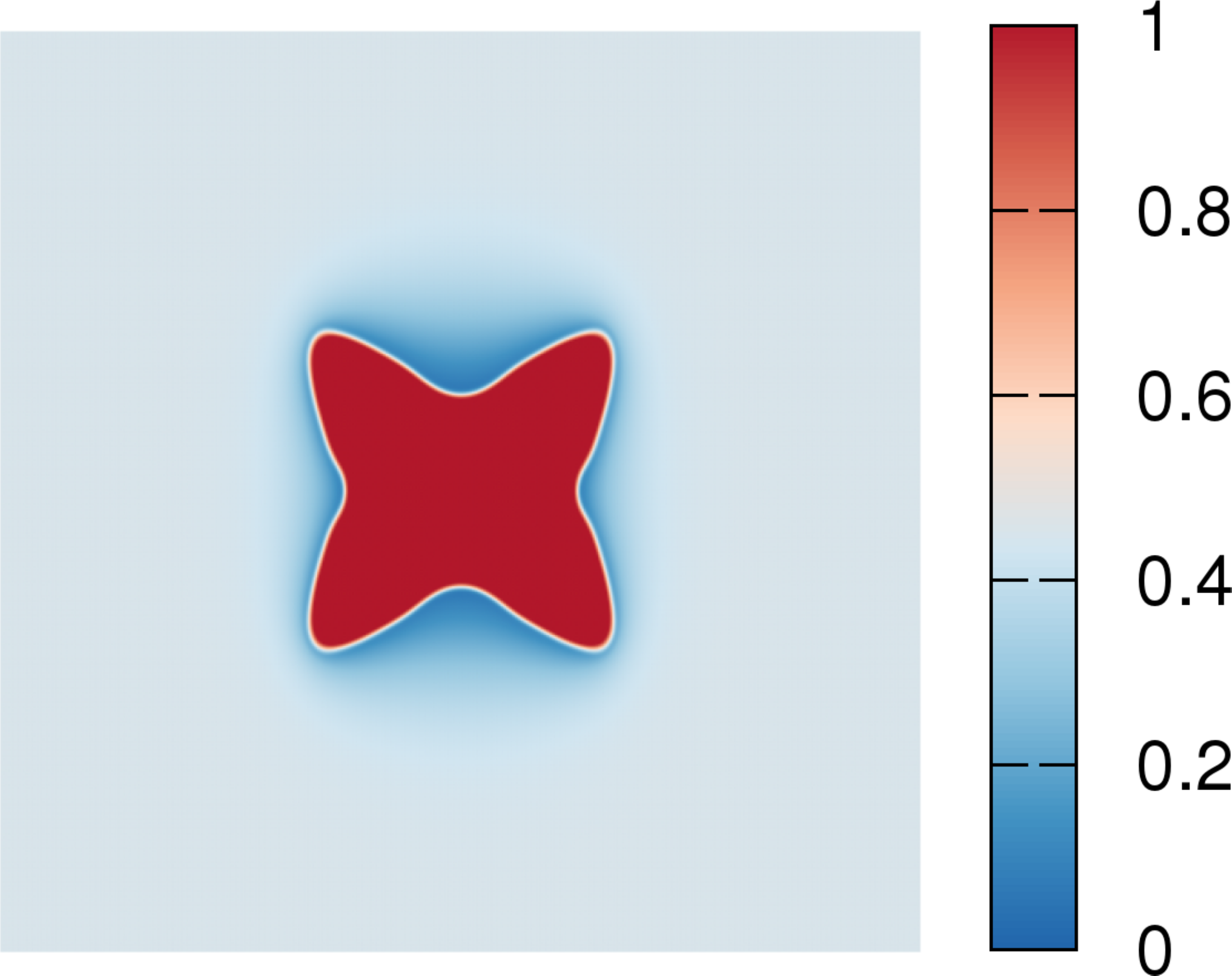} & 
    \includegraphics[width=3cm]{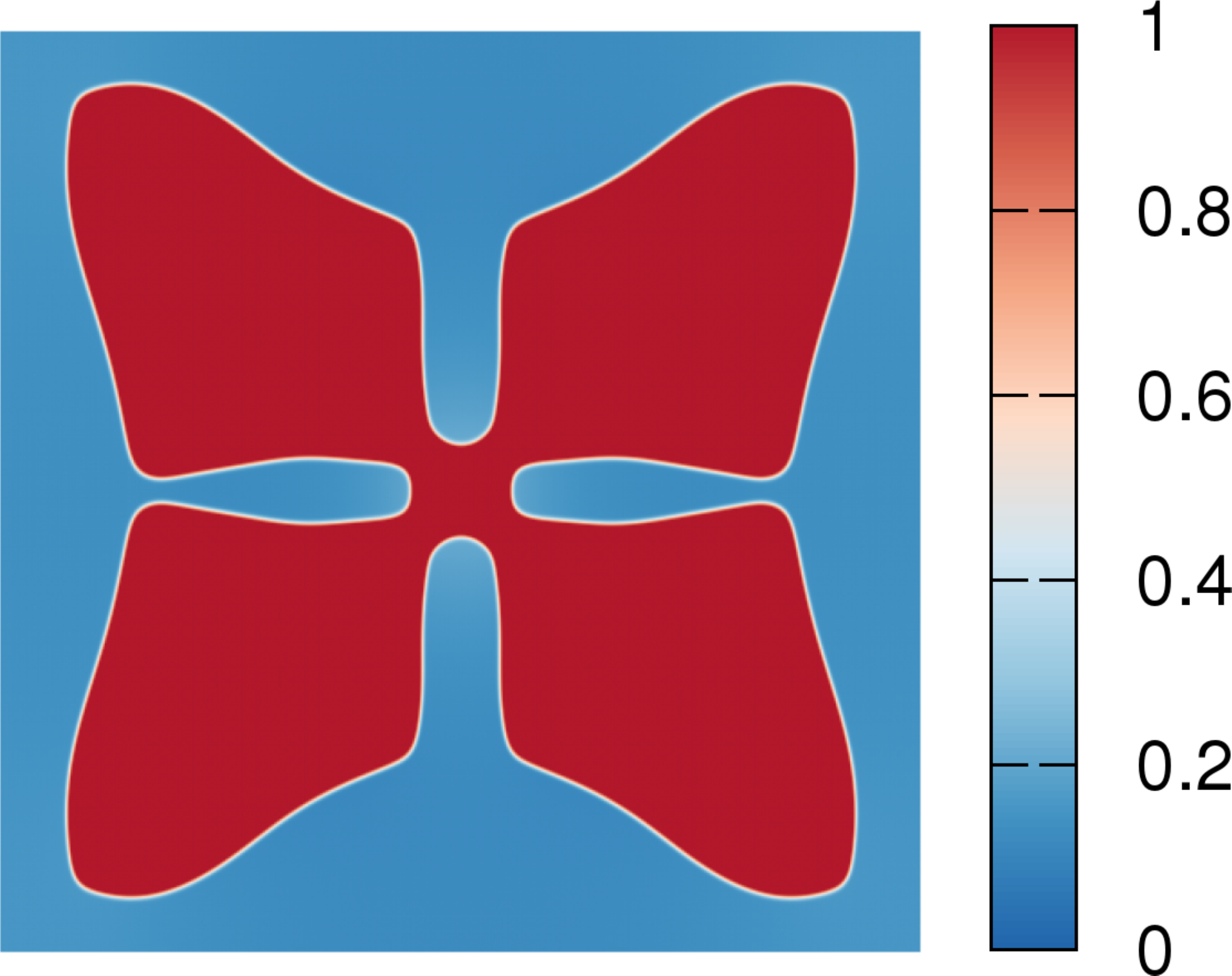} &
    \includegraphics[width=3cm]{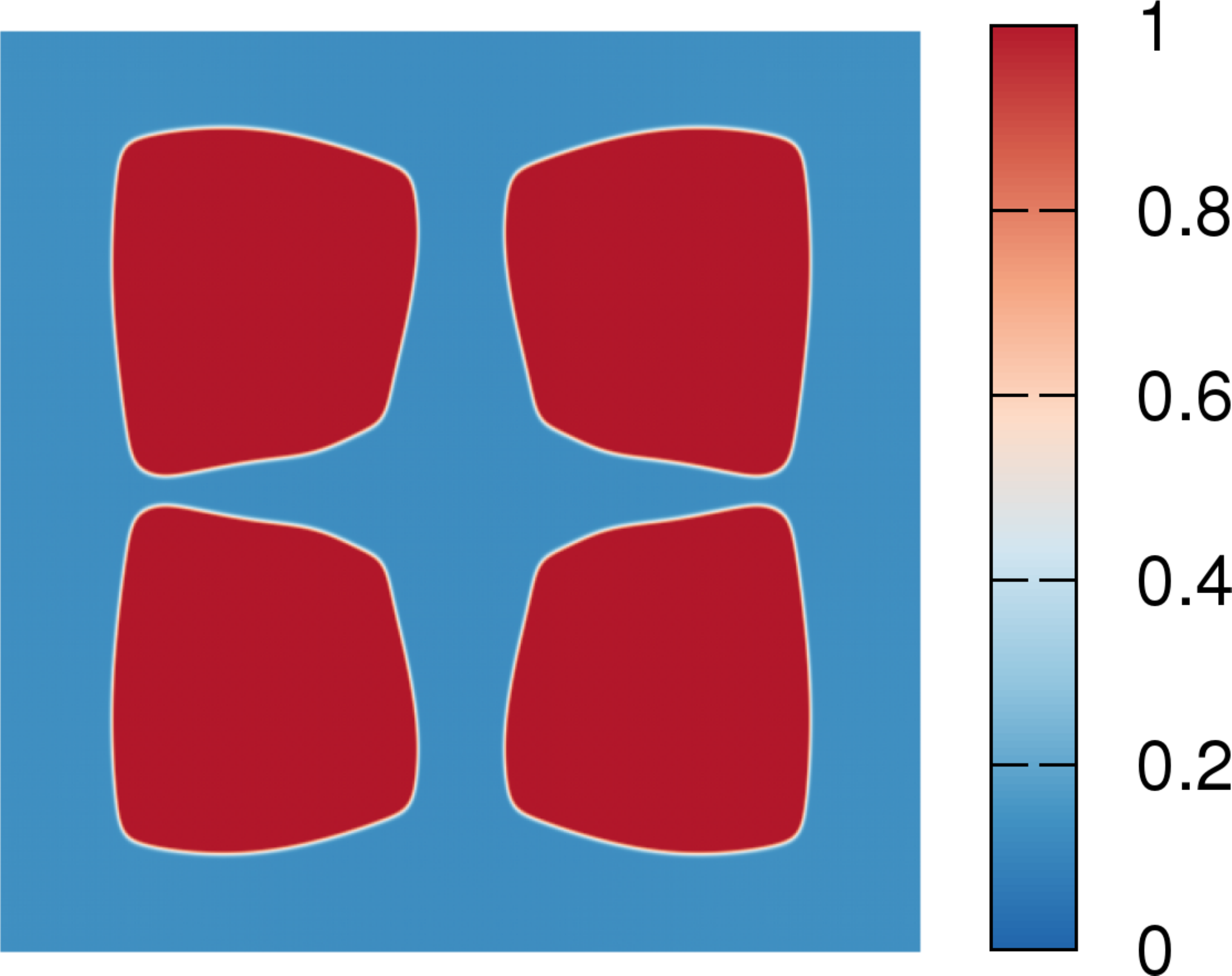} & 
    \includegraphics[width=3cm]{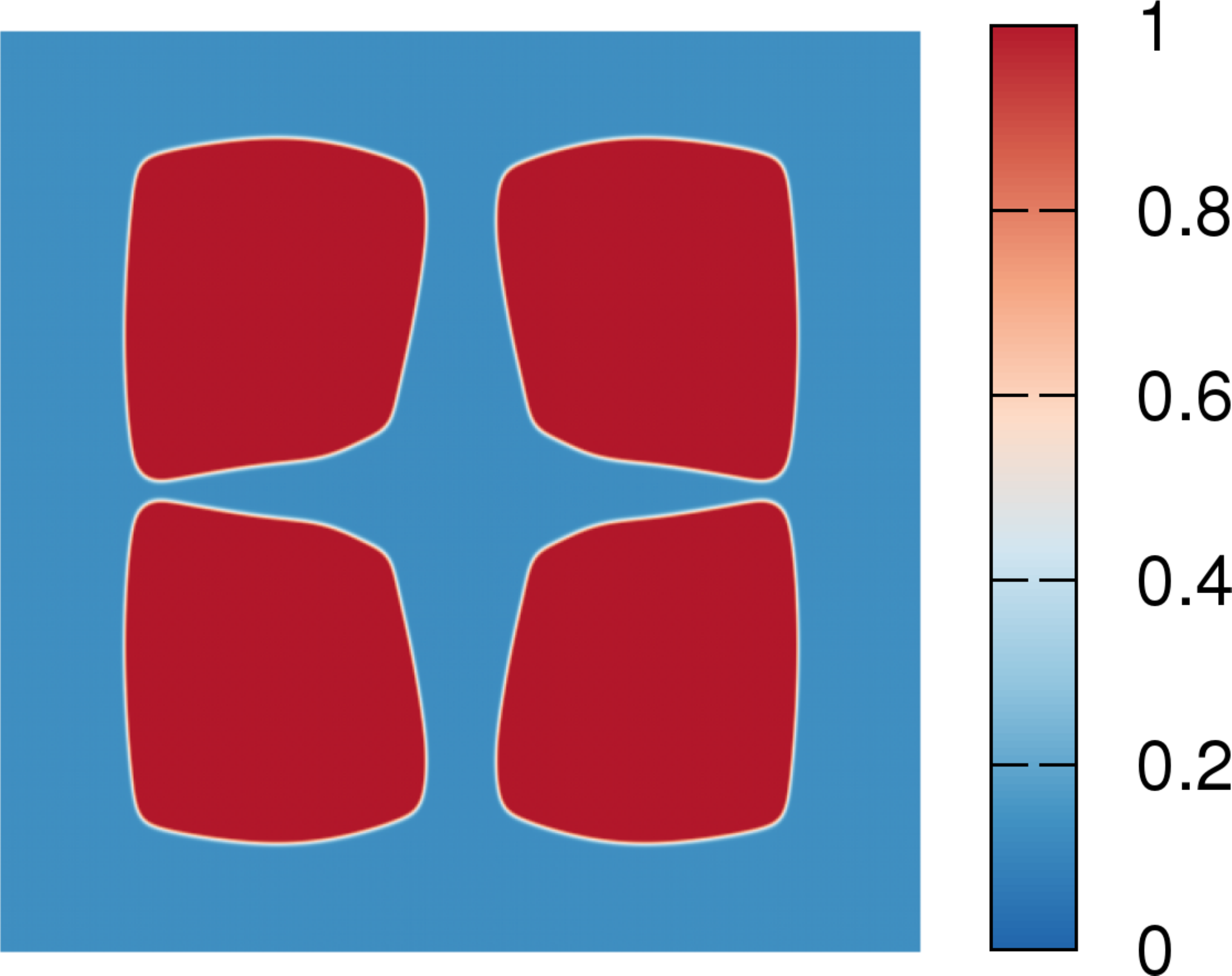} \\
    & & & &\\
    \hline
    & & & &\\
    Order Parameter map &
    \includegraphics[width=3cm]{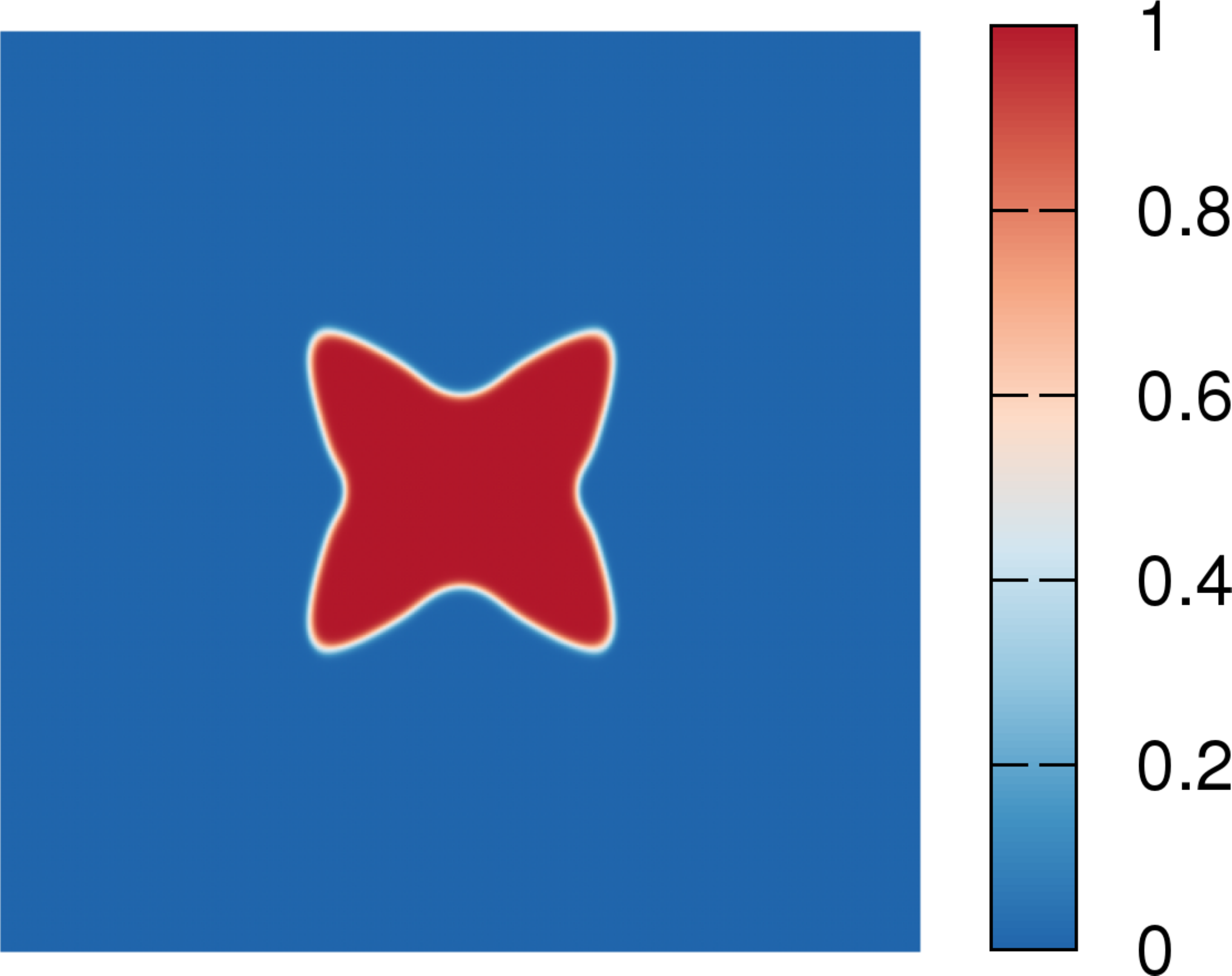} & 
    \includegraphics[width=3cm]{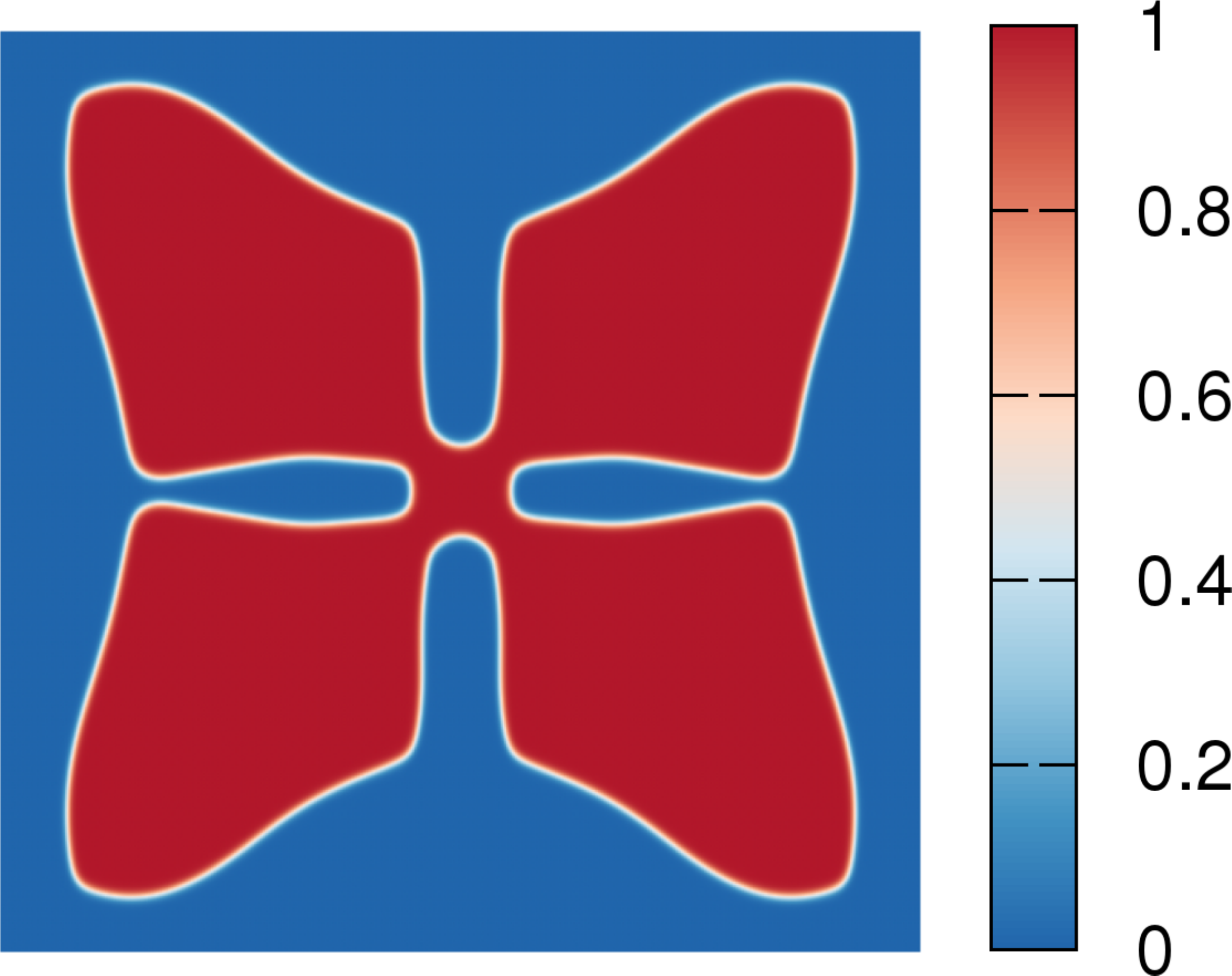} &
    \includegraphics[width=3cm]{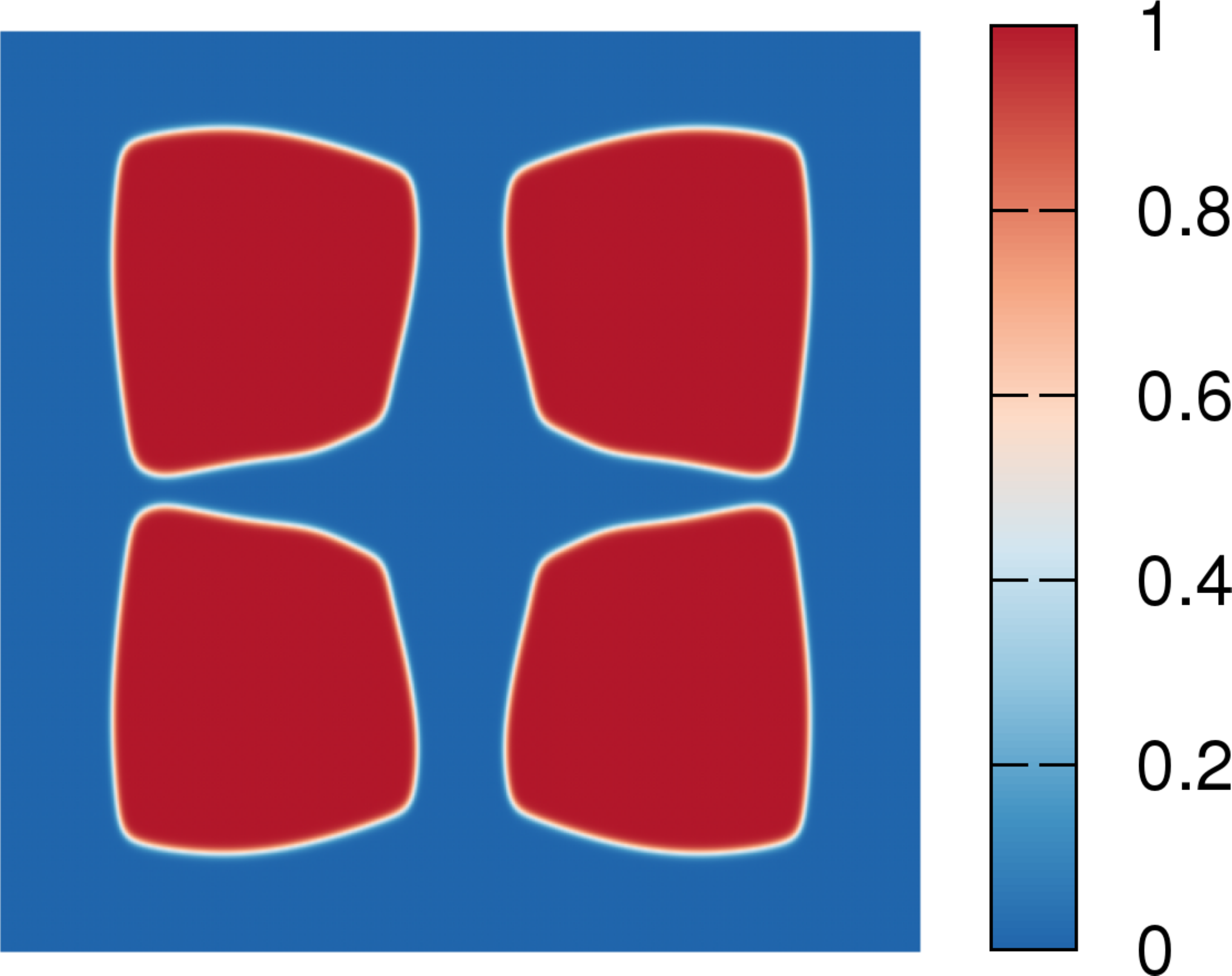} & 
    \includegraphics[width=3cm]{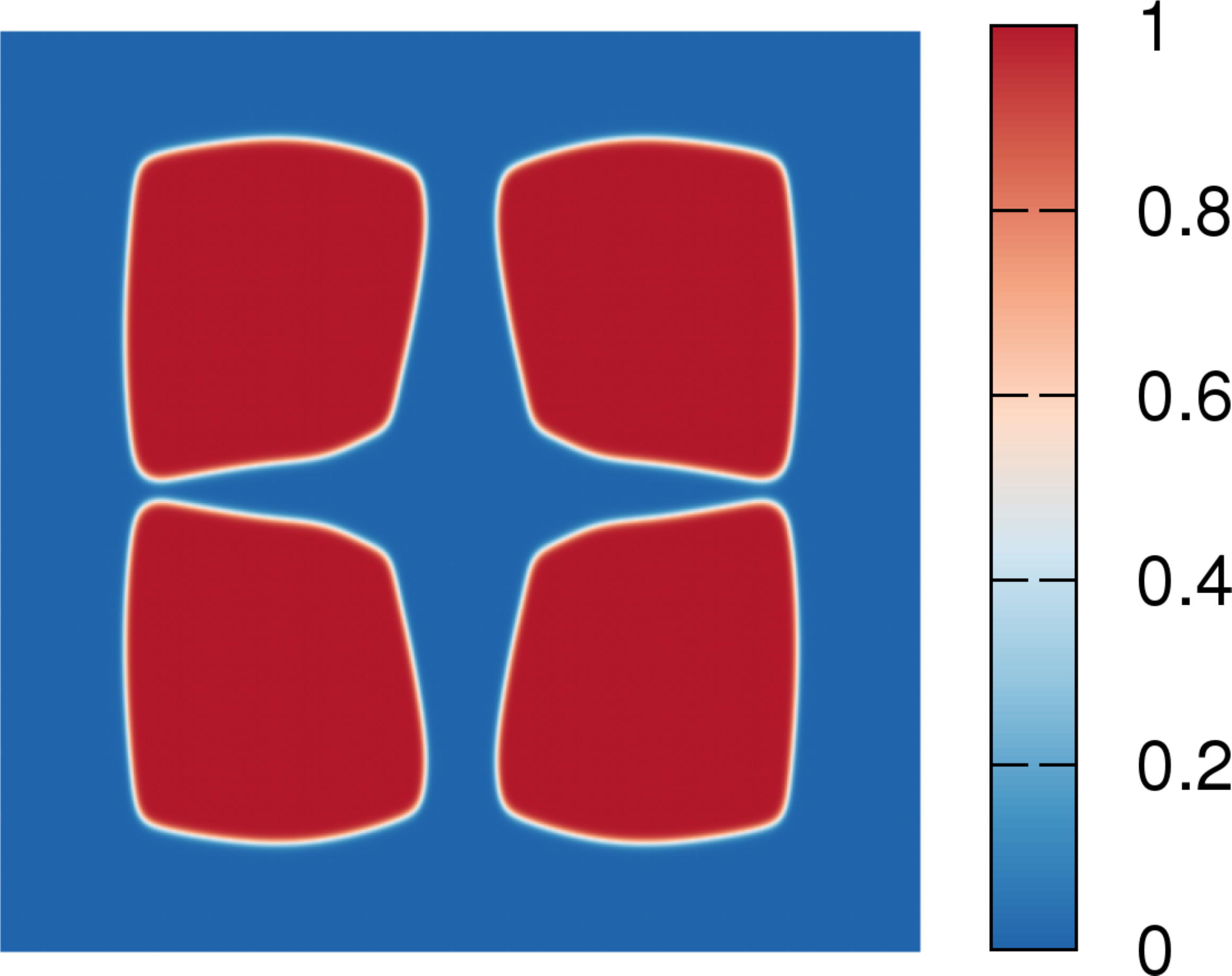} \\
    & & & &\\
    \hline
    \end{tabular}
    \caption{Temporal evolution of the precipitate morphology 
    at time $t = 100,\, 1000,\, 5000,\, 10000$ \si{\second}. First row shows the
    isosurface representation of precipitate morphology.
    Second row shows the composition map in $(110)$ plane 
    passing through the centre of the simulation box, and 
    third row shows the corresponding order parameter 
    map.}
    \label{tab:PrecipitateEvolve}
\end{table}

\begin{figure}[!htb]
    \centering
    \begin{subfigure}{0.33\textwidth}
    \centering
    \includegraphics[width=5cm]{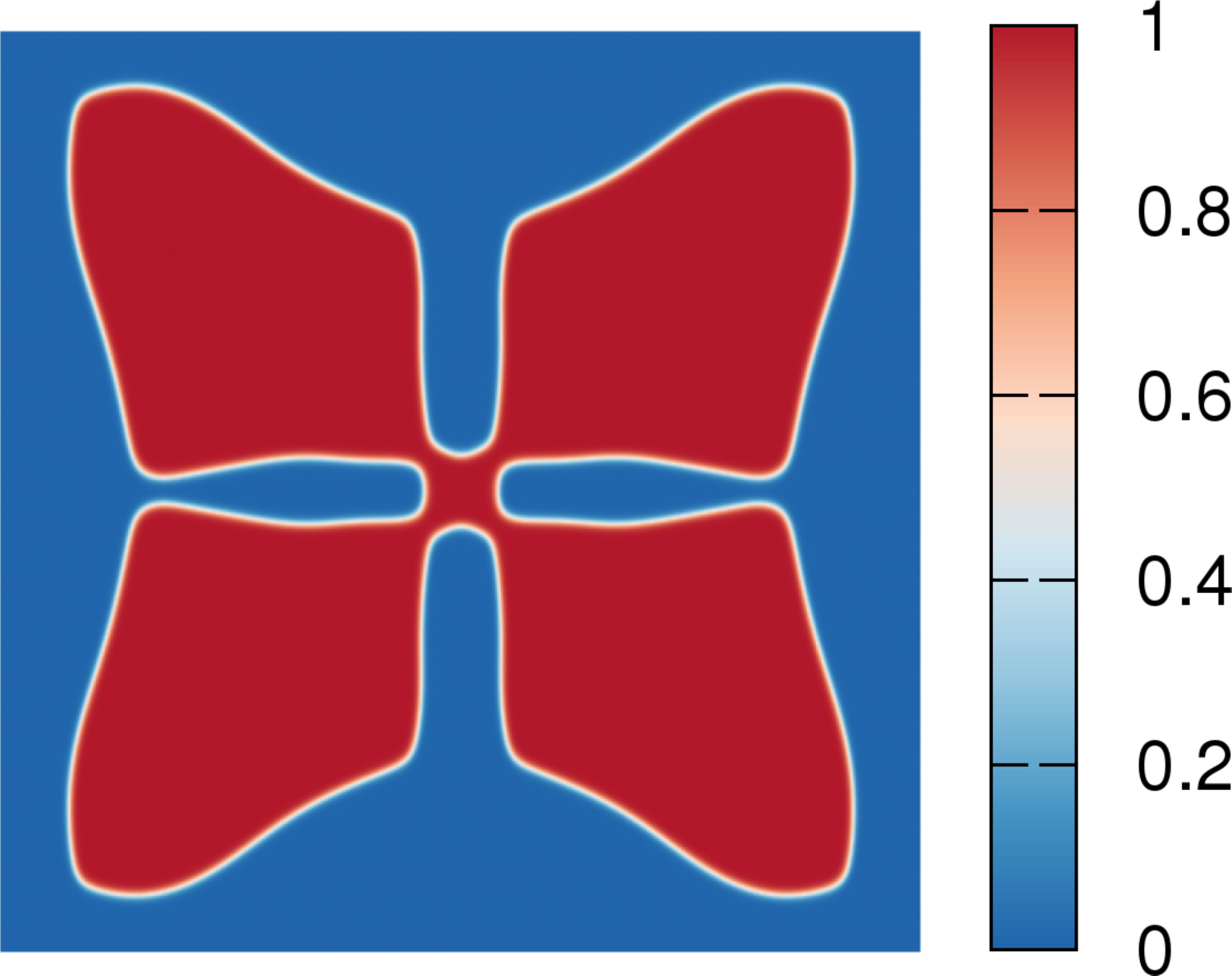}
    \caption{$t=1100$}
    \label{subfig:t11}
    \end{subfigure}%
    \begin{subfigure}{0.33\textwidth}
    \centering
    \includegraphics[width=5cm]{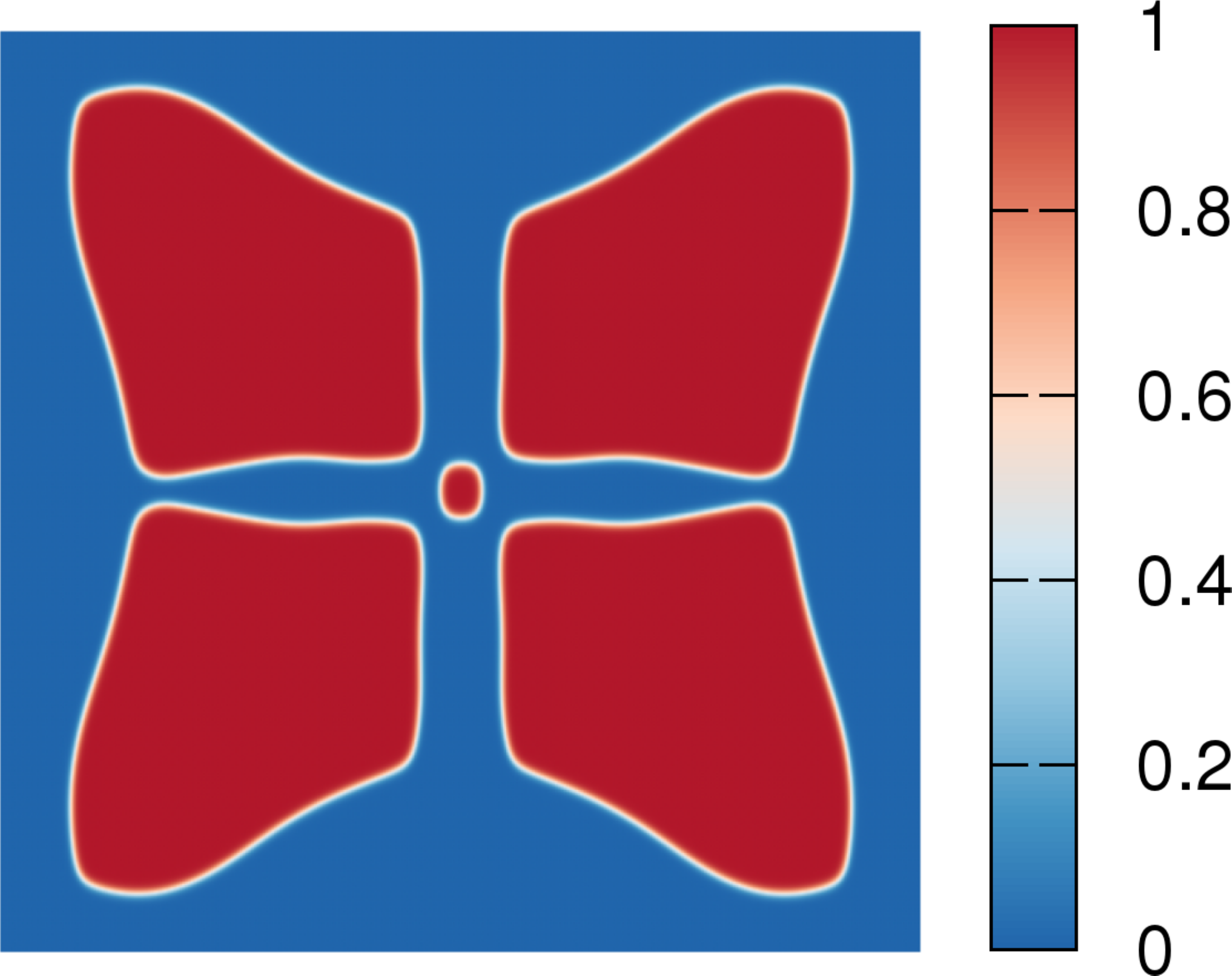}
    \caption{$t=1200$}
    \label{subfig:t12}
    \end{subfigure}%
    \begin{subfigure}{0.33\textwidth}
    \centering
    \includegraphics[width=5cm]{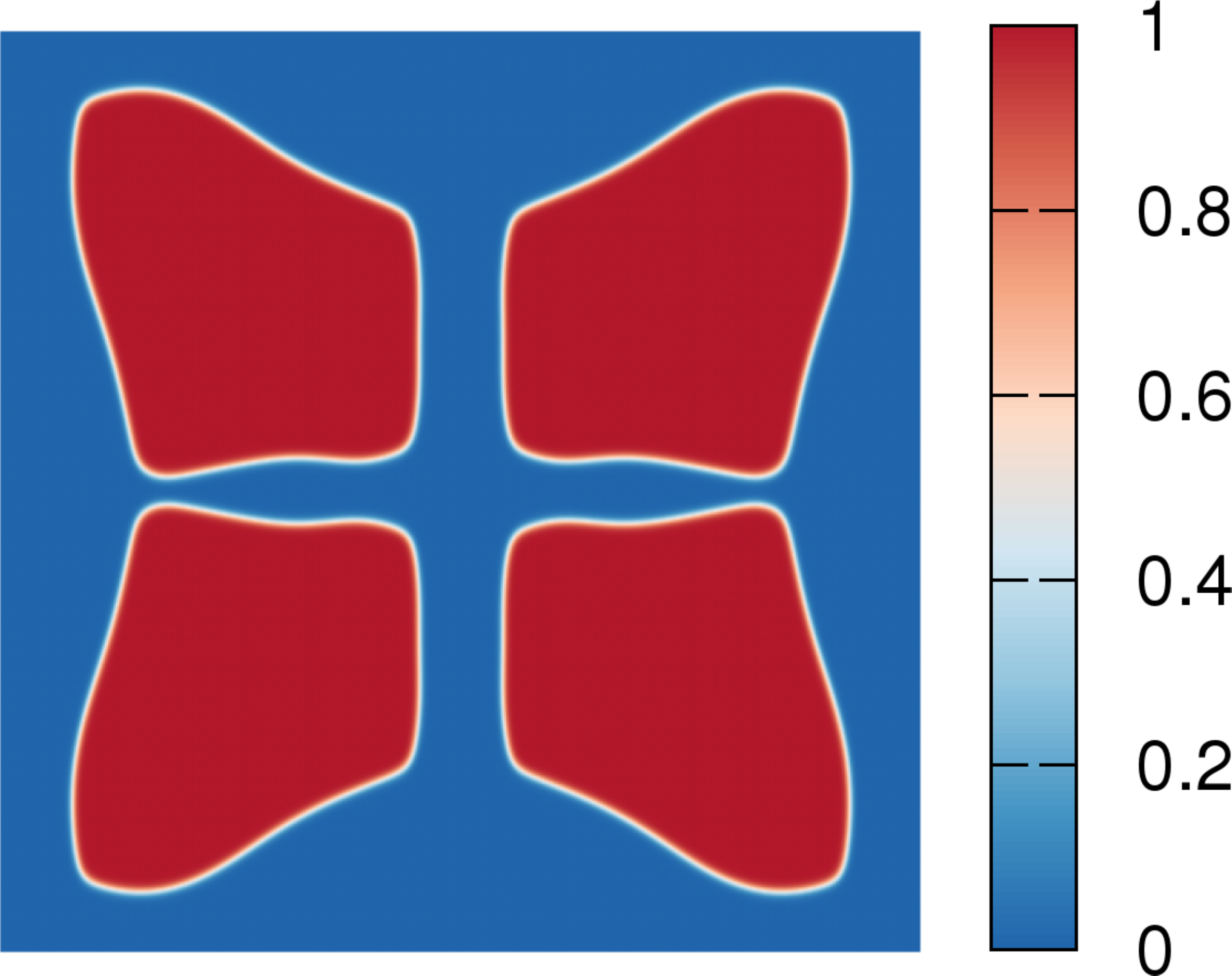}
    \caption{$t=1300$}
    \label{subfig:t13}
    \end{subfigure}
    \caption{Evolution of precipitate morphology 
    to split pattern in $(110)$ plane.
    (a) The precipitate morphology just before 
    pinch off at $t=1100$. (b) The primary arms pinch off from 
    the central precipitate core at time $t=1200$. 
    (c) The remaining central precipitate 
    core dissolves and a split pattern is formed. Here, $t=1300$.}
    \label{fig:pinch_off}
\end{figure}

By following the evolution of precipitate morphology 
with a finer time-steps, we observe that the primary 
arms of dendrite-like structure pinch off from 
the precipitate core. Fig.~\ref{fig:pinch_off} 
shows the evolution of precipitate morphology in 
the $(110)$ plane where primary arms pinch off from 
the precipitate core (see Fig.~\ref{subfig:t12}).
Further, the remaining precipitate core at the 
centre of the box dissolves completely.
The separation of primary arms from the precipitate
core is similar to secondary arms separating from the
primary arms during dendritic growth in 
solidification. The pinch-off of secondary arms 
from the primary arms during solidification is a 
curvature driven process, whereas the pinch-off of 
primary arms during solid-state transformation is an 
elastically induced instability. 

\begin{figure}[!htb]
    \centering
    \begin{subfigure}{0.49\textwidth} 
    \centering
    \fbox{ \includegraphics[width=7.5cm]{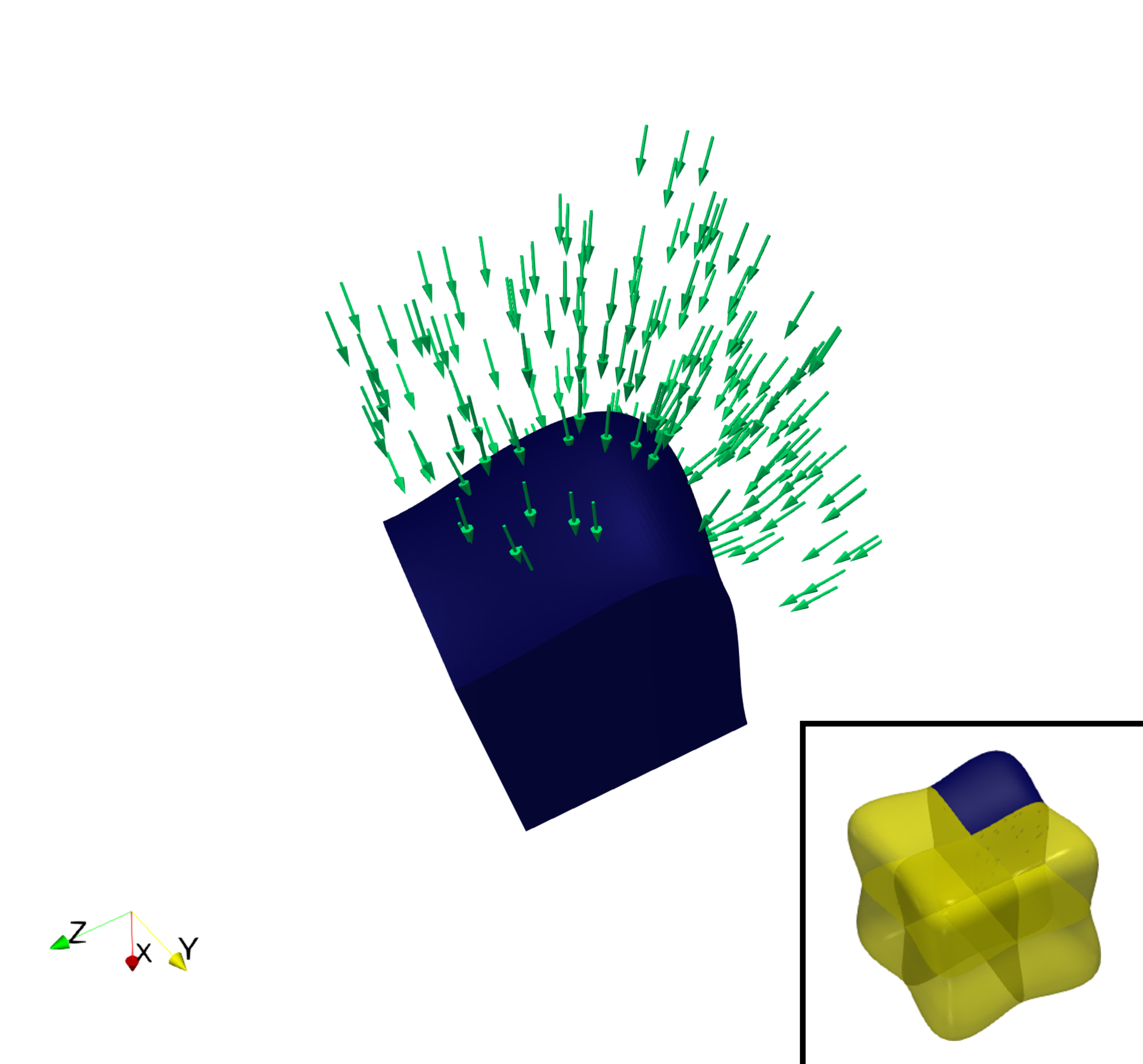}}
    \caption{$t=50$}
    \label{subfig:flux_map_t50}
    \end{subfigure}%
    \begin{subfigure}{0.49\textwidth}
    \centering
    \fbox{ \includegraphics[width=7.5cm]{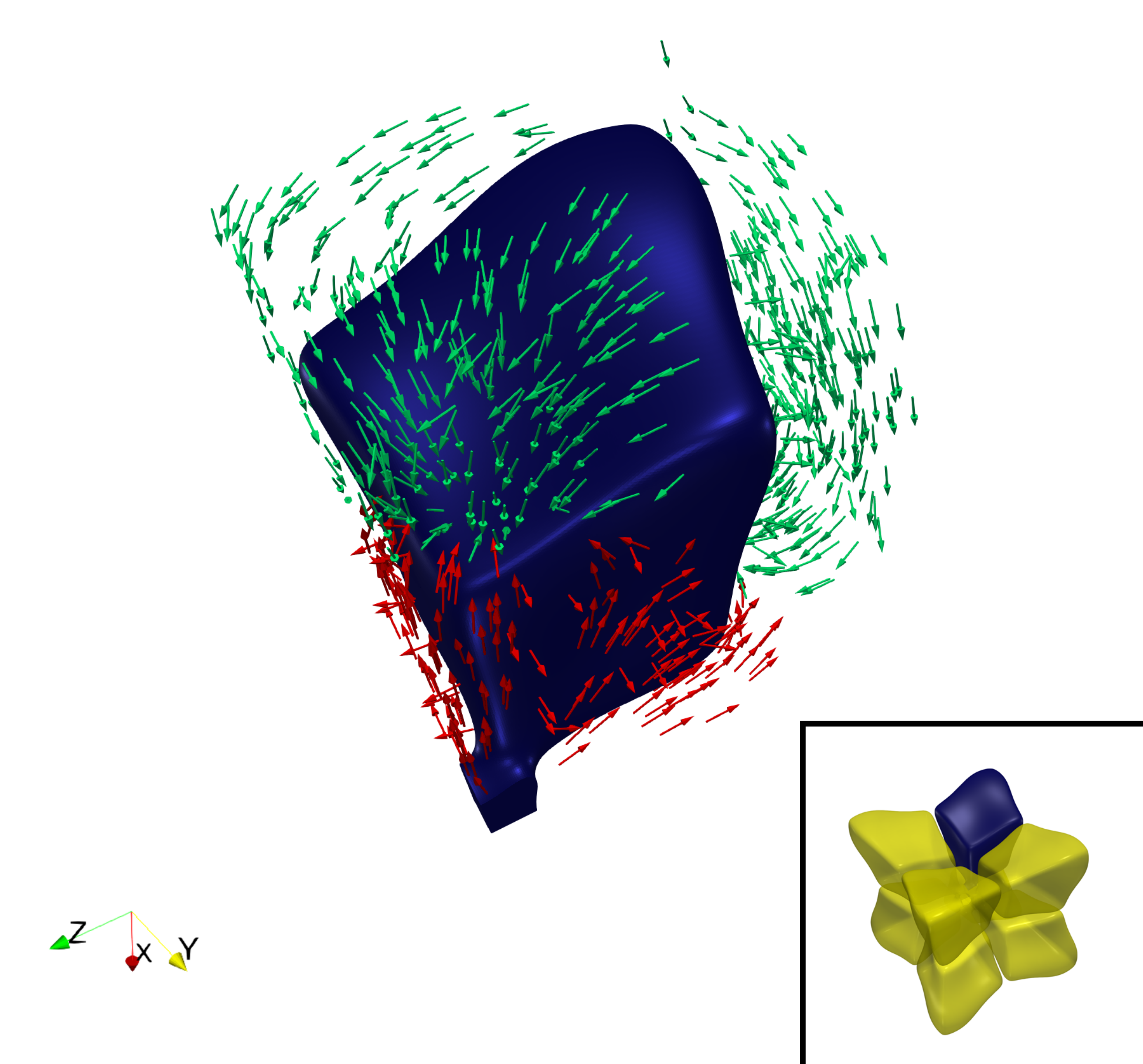}}
    \caption{$t=1000$}
    \label{subfig:flux_map_t1000}
    \end{subfigure}
    \caption{Compositional flux lines for time (a) $t = 50$
    (b) $t = 1000$. At $t = 50$, compositional flux lines are directed 
    towards the particle which suggests growth of precipitate.
    At $t = 1000$, some of the flux lines (red colored) are directed 
    away from the precipitate surface which leads to dissolution, whereas
    green colored flux lines exhibit composition flow towards the precipitate. }
    \label{fig:flux_map}
\end{figure}

Moreover, we plot composition fluxes to understand the flow of 
composition before the pinch-off of primary arms. Fig.~\ref{subfig:flux_map_t50}
depicts the compositional fluxes at time $t = 500$. The composition flow 
is predominantly towards the precipitate suggesting growth of precipitate.
On the other hand, Fig.~\ref{subfig:flux_map_t1000} shows the 
compositional fluxes at time $t = 1000$ where composition flows outward
from the precipitate in the groove regions (red colored flux lines) and 
solute deposits at the faces and tip of primary arms of the precipitate 
(green colored flux lines). The preferential dissolution in the grooves
will be explained later in this section.

\begin{figure}[!htb]
    \centering
    \begin{subfigure}{0.48\textwidth}
    \centering
    \includegraphics[width=6cm]{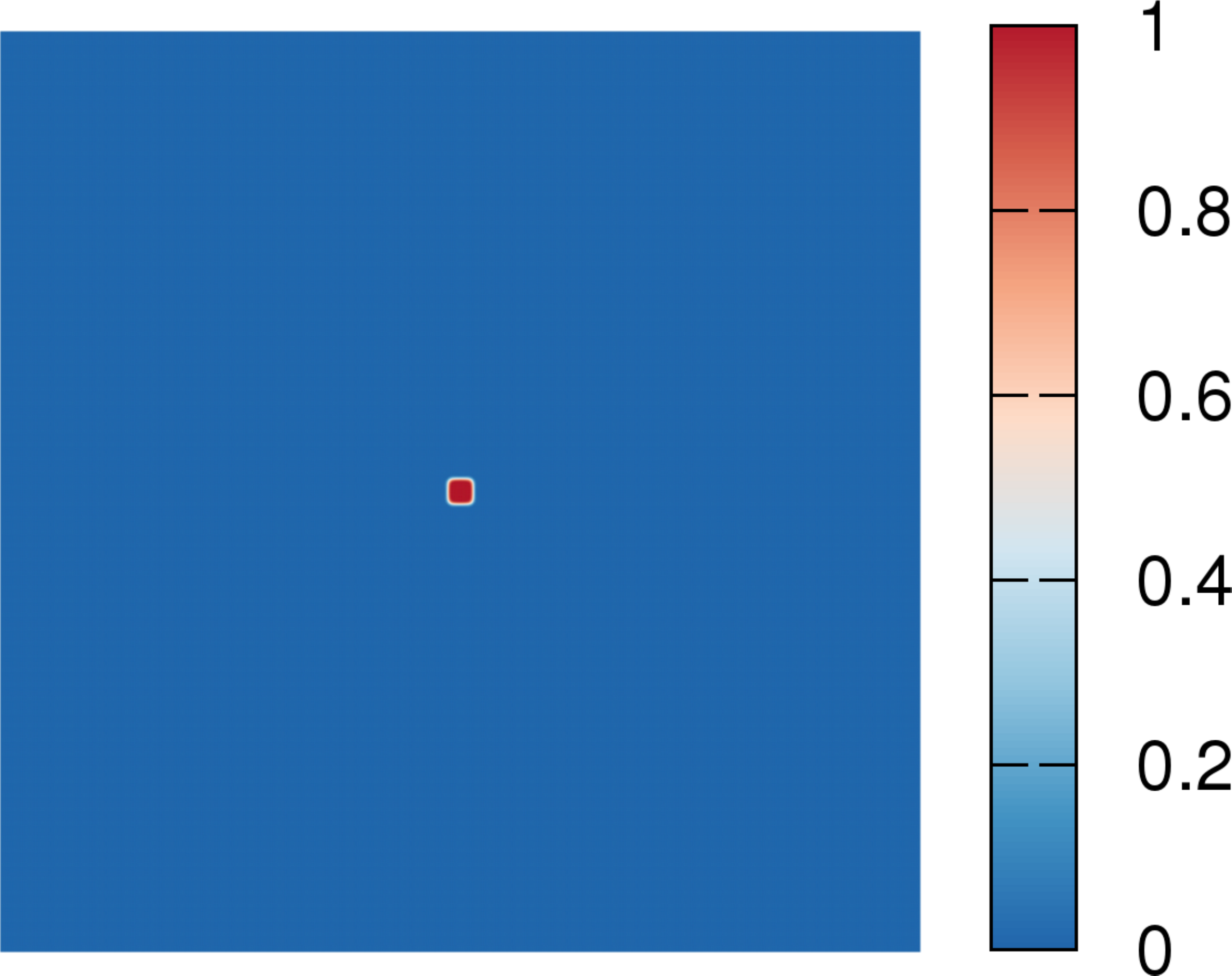}
    \caption{$t = 100$}
    \label{fig:int_aniso_t_.1s}
    \end{subfigure}%
    \begin{subfigure}{0.48\textwidth}
    \centering
    \includegraphics[width=6cm]{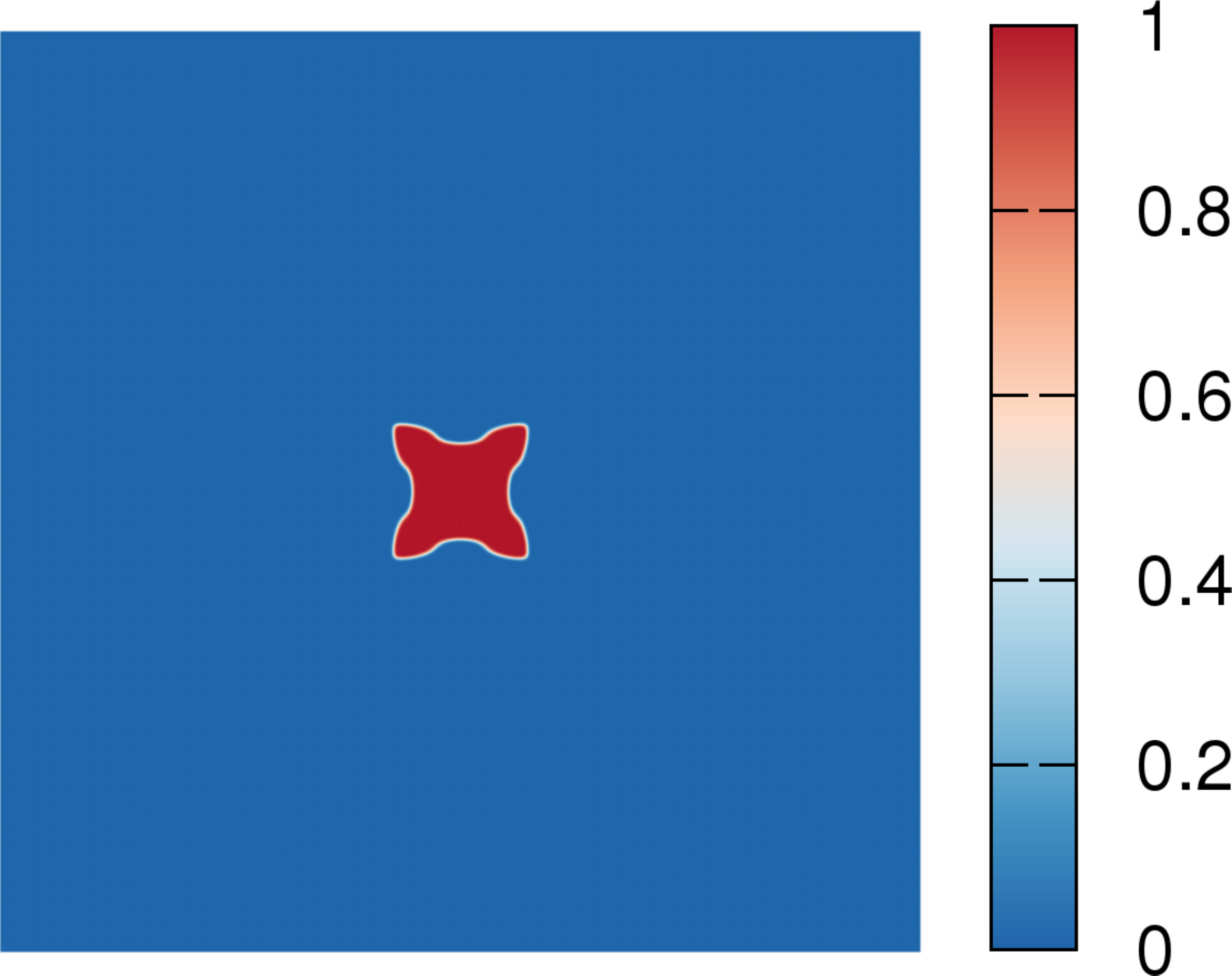}
    \caption{$t = 2000$}
    \label{fig:int_aniso_t_1s}
    \end{subfigure}
    \begin{subfigure}{0.48\textwidth}
    \centering
    \includegraphics[width=6cm]{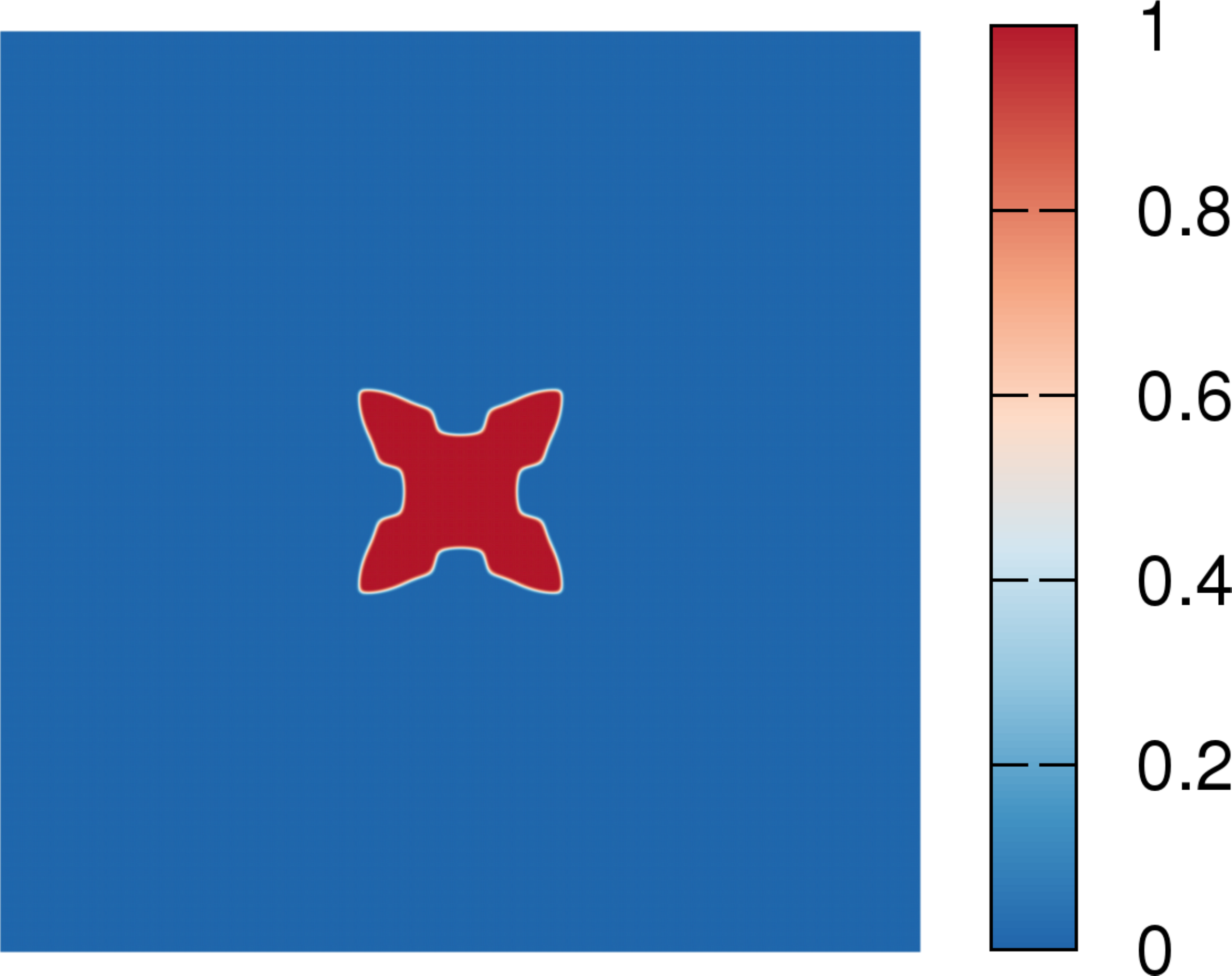}
    \caption{$t = 3000$}
    \label{fig:int_aniso_t_3s}
    \end{subfigure}%
    \begin{subfigure}{0.48\textwidth}
    \centering
    \includegraphics[width=6cm]{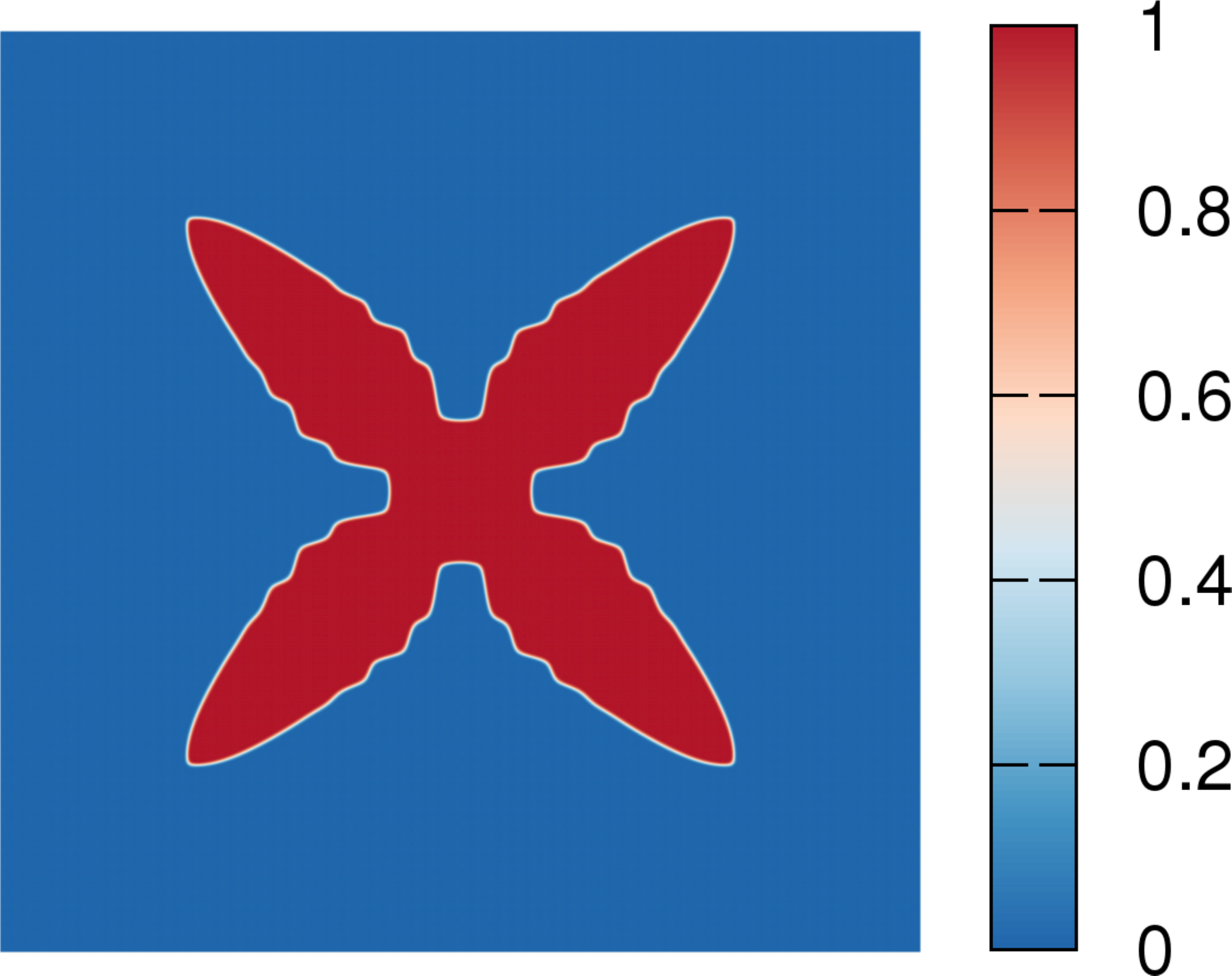}
    \caption{$t = 10000$}
    \label{fig:int_aniso_t_10s}
    \end{subfigure}
    \begin{subfigure}{0.48\textwidth}
    \centering
    \includegraphics[width=6cm]{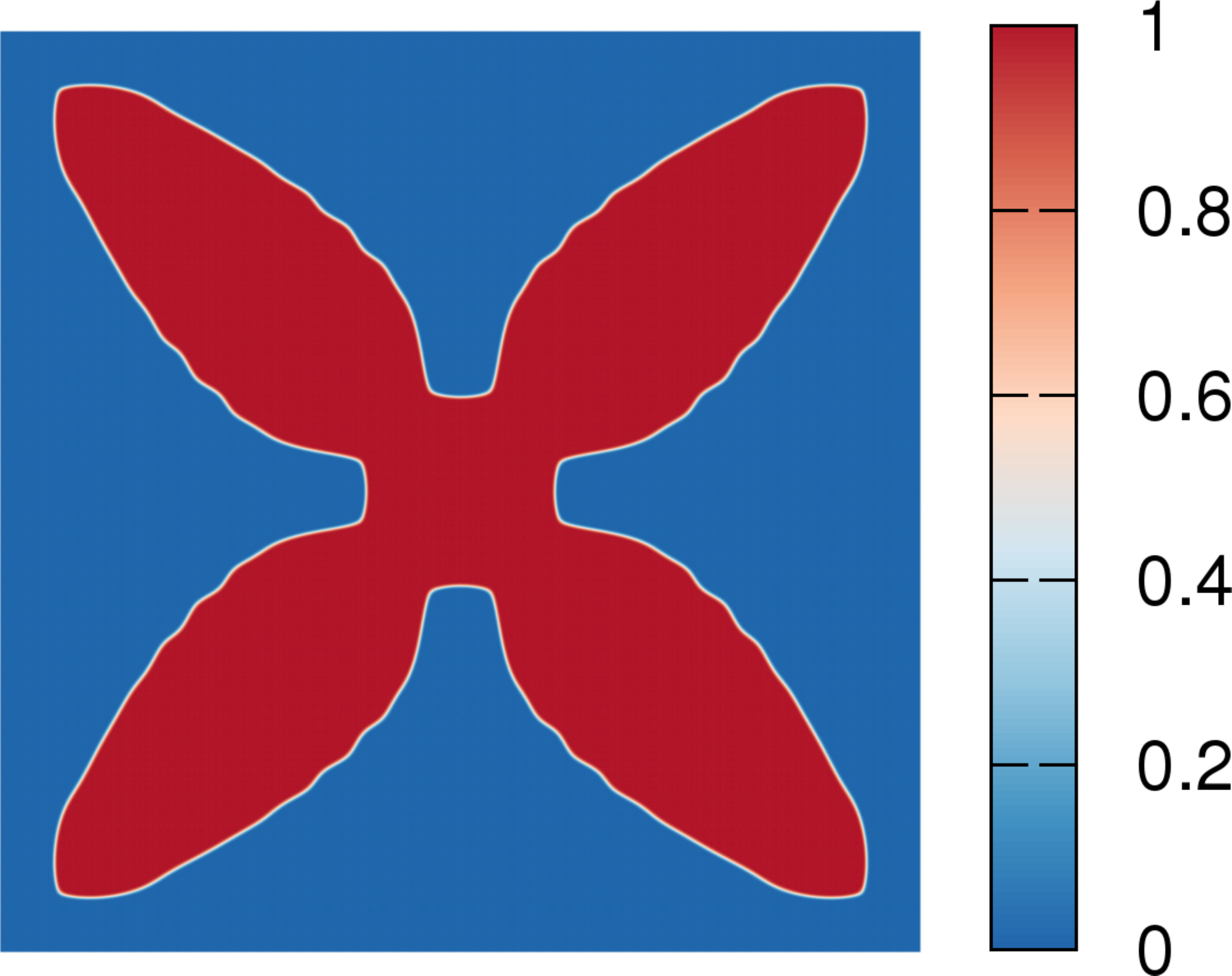}
    \caption{$t = 30000$ }
    \label{fig:int_aniso_t_30s}
    \end{subfigure}
    \caption{Temporal snapshots of precipitate morphology at time $t = 100$,
    $2000$, $3000$, $10000$, and $30000$. 
    Here, the elastic stresses are neglected and 
    interfacial energy along $\langle 11 \rangle$ 
    direction is $1.25$ times of that along $\langle 10 \rangle$ direction.}
    \label{fig:int_aniso_evolve}
\end{figure}

Previous reports show that particle splitting occurs
under the influence of interfacial energy anisotropy 
and isotropic elastic energy~\cite{liu2017split}. 
We investigate the effect of anisotropy in interfacial 
energy on the initiation of 
splitting of particle where the lattice 
misfit is ignored. To understand whether the 
anisotropy in interfacial energy can promote the 
splitting instability, we have performed 
two-dimensional simulations under the 
effect of anisotropy in interfacial energy. 
Fig.~\ref{fig:int_aniso_evolve} represents the 
temporal evolution of the precipitate morphology in 
two-dimensions where we choose interfacial energy 
along $\langle 11 \rangle$ directions 
($\gamma_{11}$) one and twenty-five hundredth of 
that along $\langle 10 \rangle$ directions 
($\gamma_{10}$), where $\gamma_{10}$ is $0.122223$. 
The simulation box size is 
$1024 \times 1024$ with the grid size of $0.5$. 
Here, we neglect the effect of elastic stresses on 
the growth of precipitate. 
In the presence of interfacial energy anisotropy, 
the initial circular precipitate develop ears, and 
subsequently prominent primary arms develop along 
$\langle 11 \rangle$ directions. 
Although the precipitate 
faces develop concavities, the grooves never advance 
towards the centre of the precipitate. Thus, this 
result implies that the presence of elastic stresses 
and the anisotropy in elastic energy is necessary 
for splitting to occur. 
However, different combinations of interfacial 
energy anisotropy and elastic energy anisotropy can 
either promote or suppress the splitting of 
precipitate. Here after, we only 
present the simulations where the interfacial energy 
is isotropic. 

Similar to three-dimensional simulations, we have 
performed two-dimensional simulations of the 
precipitate growth in the simulation box size of 
$809.6 \times 809.6$. We choose supersaturation of 
$25\%$, elastic misfit of $0.85\%$, and 
Zener anisotropy parameter of $4$. 
Table~\ref{tab:2d_evolve} depicts the temporal 
evolution of precipitate morphology in 
two-dimensions at different time steps.The initial 
circular precipitate transforms to square shape and 
gradually ears start developing along $[11]$ 
directions (see Fig.~\ref{tab:2d_evolve}). At 
subsequent stages, predominant primary arms develop 
along $[11]$ directions, and simultaneously 
concavities start building up along $[10]$ 
directions. Unlike three-dimensional simulation 
morphology, here the prominent secondary arms are 
observed. Further, the grooves developed along 
$[10]$ directions advance towards the centre of the 
precipitate and subsequent pinch-off takes place. 
Later, the secondary arms start disappearing and 
facets of the split particles emerge to gradually 
orient towards $[10]$ directions.

\begin{table}[!htb]
    \centering
    \begin{tabular}{| M{3.6cm} M{3.6cm} M{3.6cm} M{3.6cm} |}
    \hline 
     $t=20000$ & $t=40000$ & $t=70000$ & $t=300000$ \\
    \hline
     & & & \\
    \fbox{\includegraphics[width=3.45cm]{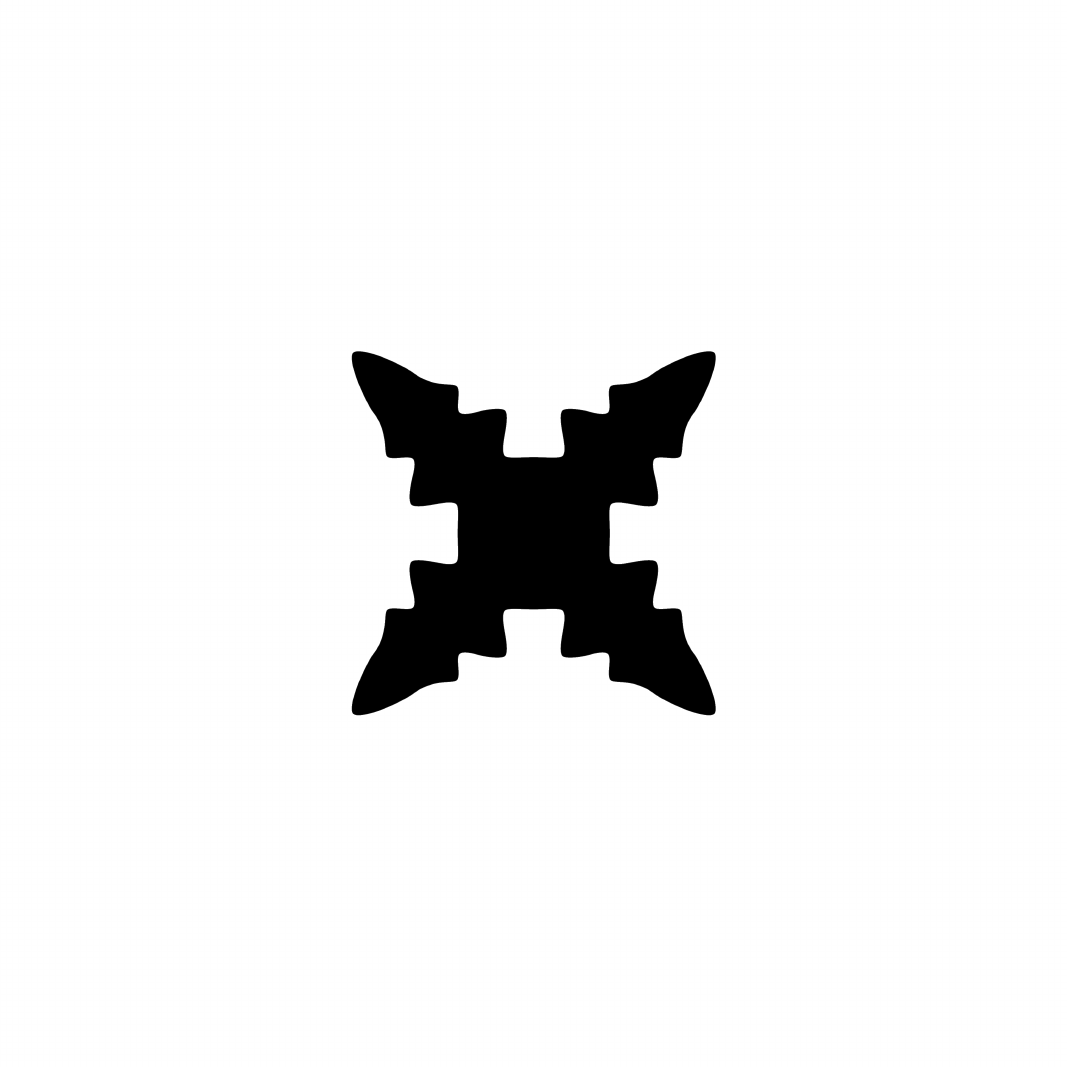}}
     &
    \fbox{\includegraphics[width=3.45cm]{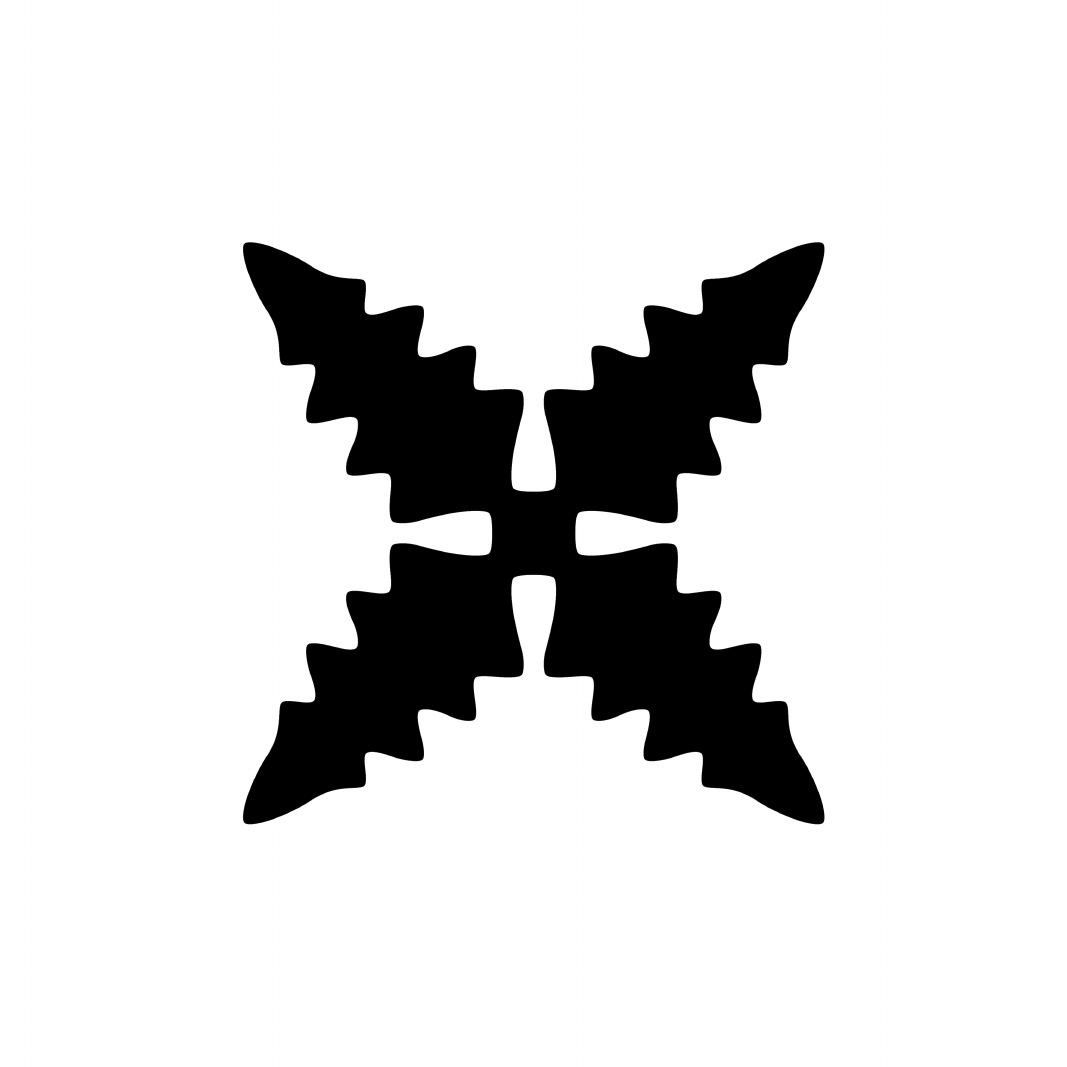}} 
     &
    \fbox{\includegraphics[width=3.45cm]{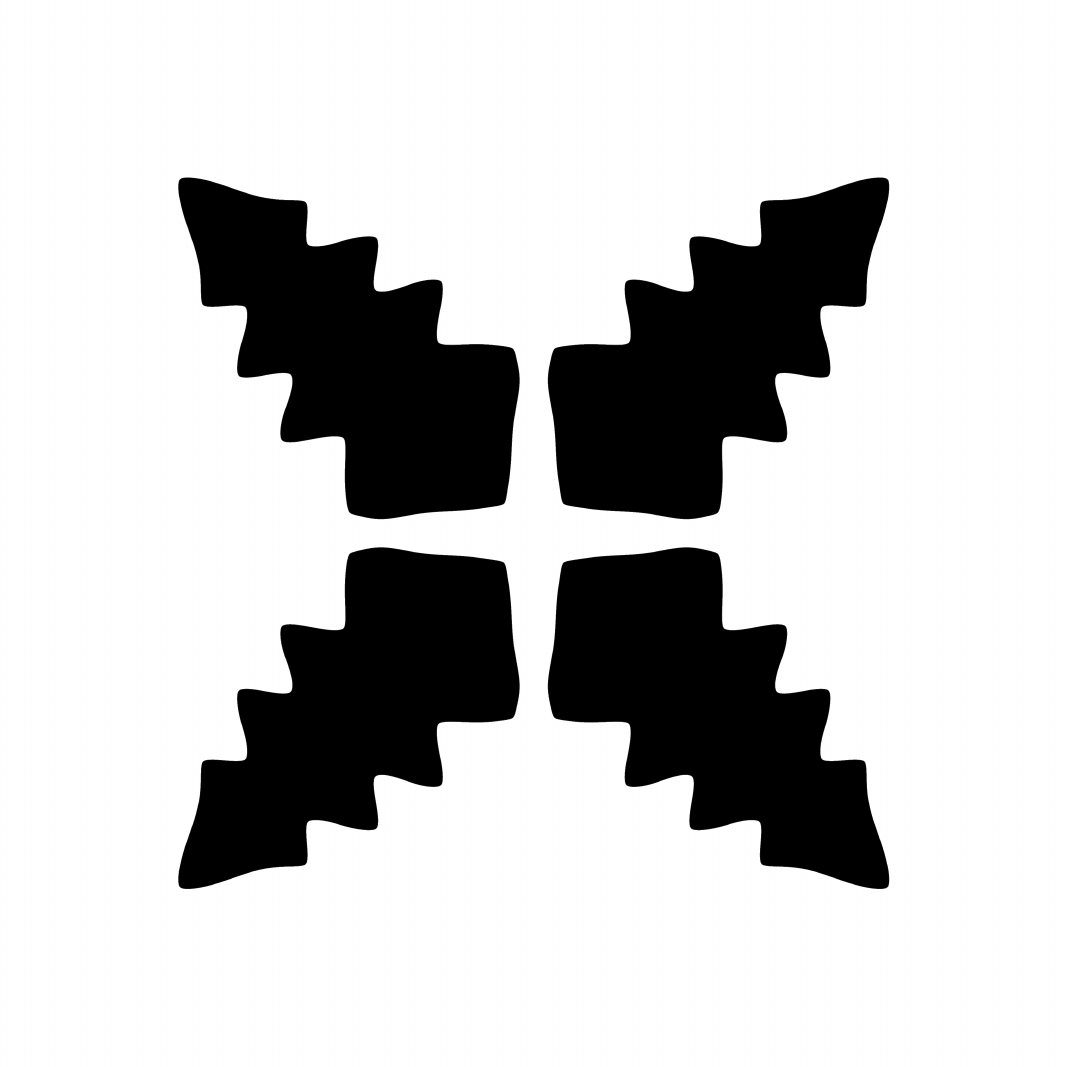}} 
     &
    \fbox{\includegraphics[width=3.45cm]{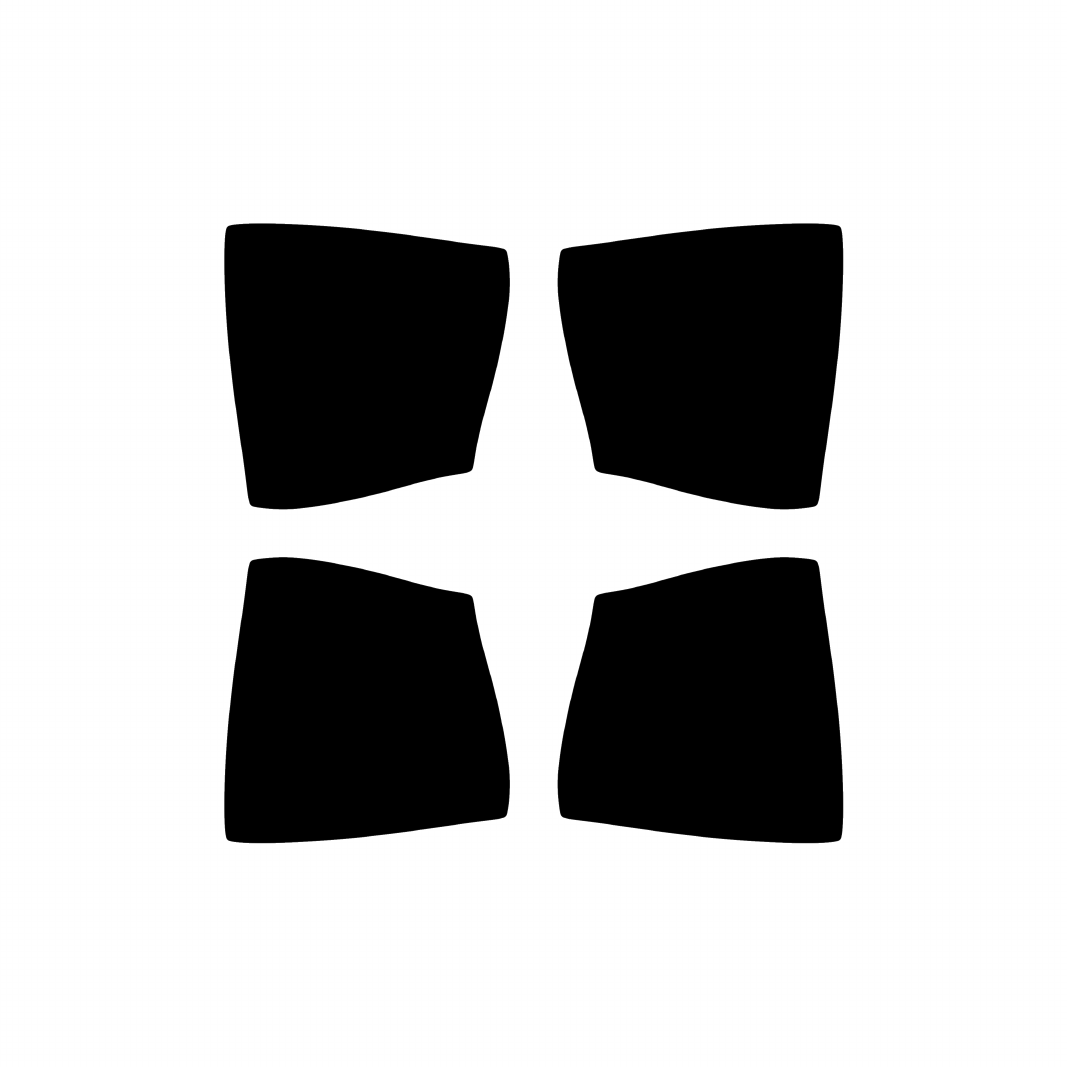}} \\
    & & & \\
    \hline
    \end{tabular}
    \caption{Temporal 
    snapshots of precipitate morphology in 
    two dimensions at $t = 2000$, $20000$, 
    $40000$, $70000$, $300000$. Here, 
    supersaturation is $25\%$, misfit is $1\%$, 
    and Zener anisotropy parameter is $4$.}
    \label{tab:2d_evolve}
\end{table}

\begin{figure}[!htb]
    \centering
    \begin{subfigure}{0.5\textwidth} 
    \centering
    \includegraphics[width=6.5cm]{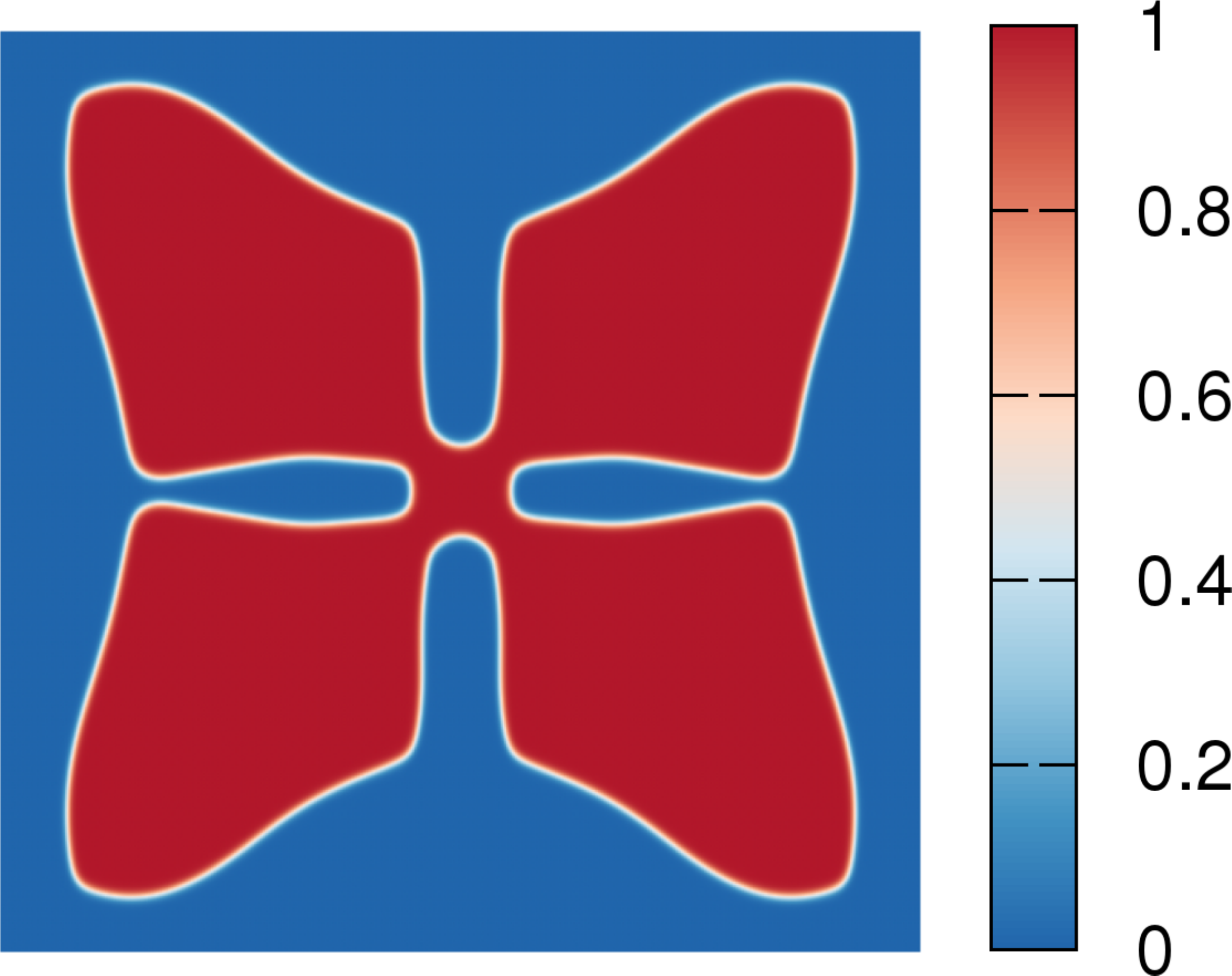}
    \caption{}
    \label{subfig:phi_map}
    \end{subfigure}%
    \begin{subfigure}{0.5\textwidth}
    \centering
    \includegraphics[width=6.5cm]{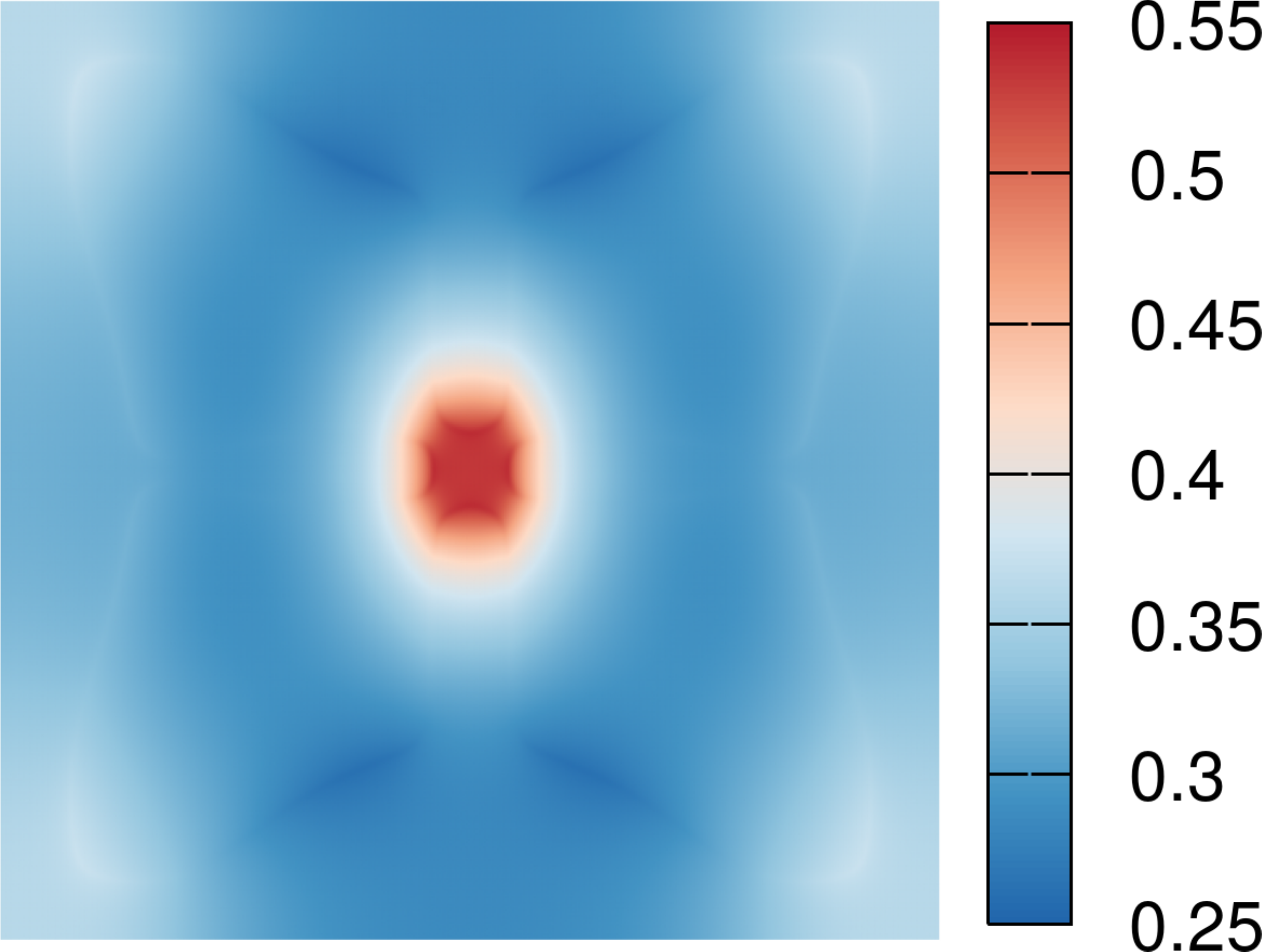}
    \caption{}
    \label{subfig:deltamu_map}
    \end{subfigure}
    
    \begin{subfigure}{0.5\textwidth} 
    \centering
    \includegraphics[width=6.5cm]{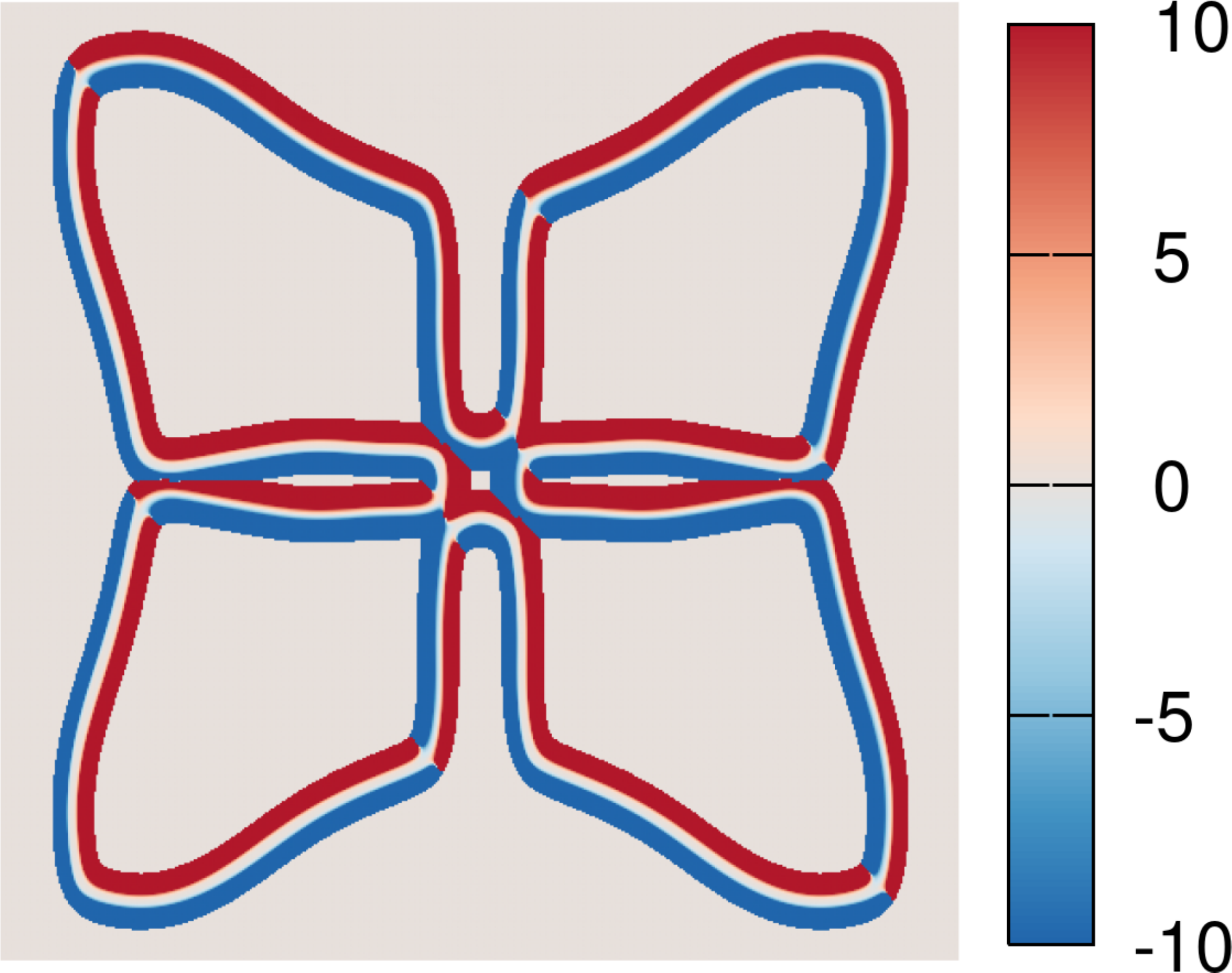}
    \caption{}
    \label{subfig:contour}
    \end{subfigure}%
    \begin{subfigure}{0.5\textwidth} 
    \centering
    \includegraphics[width=6.5cm]{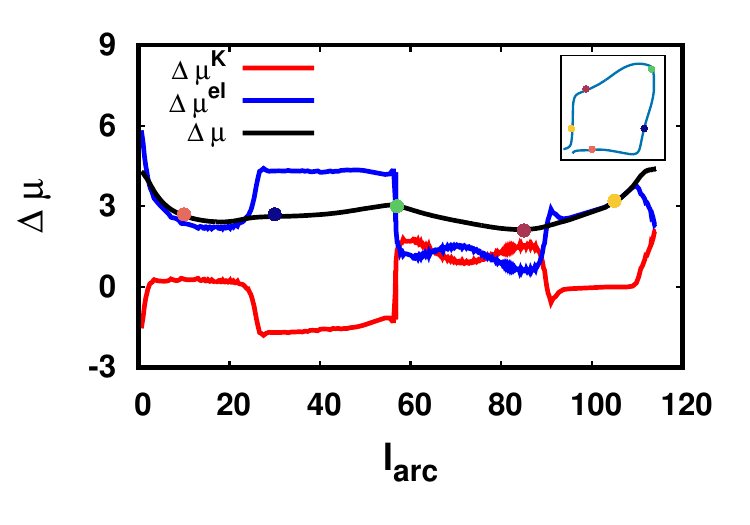}
    \caption{}
    \label{subfig:larc_vs_deltamu}
    \end{subfigure}
    \caption{(a) Order parameter map in $(110)$ plane section 
    at $t=1100$ (b) Shift in chemical potential 
    ($\Delta \mu^*$) map corresponding to order 
    parameter map (c) curvature map corresponding 
    to the order parameter map (d) The variation 
    of shift in chemical potential ($\Delta \mu$) 
    as a function of arc-length. The inset shows 
    the one-fourth section of the contour profile 
    in $(110)$ plane. The coloured dots 
    on the interface contour are shown in the 
    inset, and the corresponding values of 
    $\Delta \mu$ are shown by the same coloured 
    dots.}
\end{figure}

\begin{figure}[!htb]
    \centering
    \begin{subfigure}{0.5\textwidth}
    \centering
    \includegraphics[width=7cm]{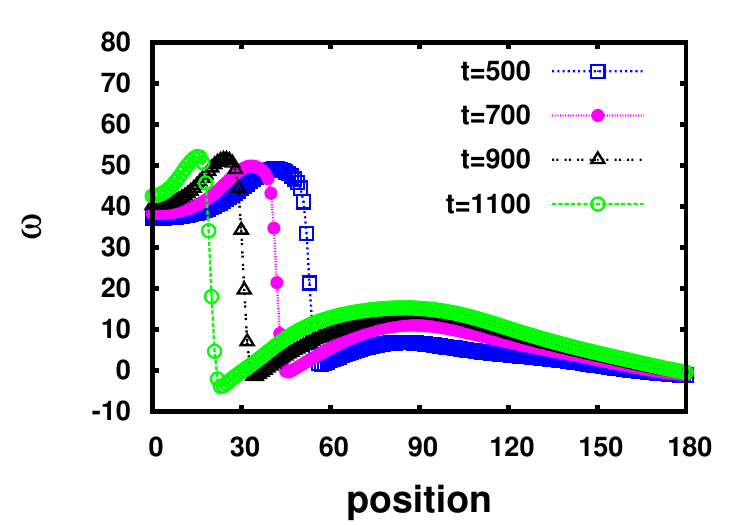}
    \caption{}
    \label{subfig:fel_evolve}
    \end{subfigure}%
    \begin{subfigure}{0.5\textwidth}
    \centering
    \includegraphics[width=7cm]{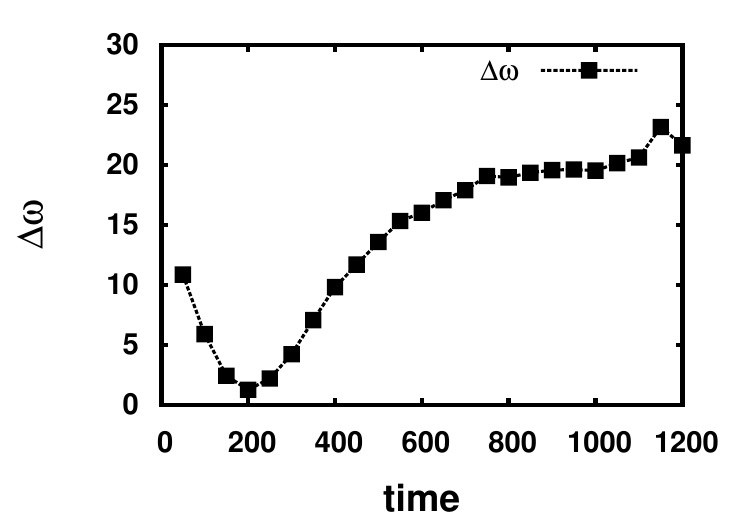}
    \caption{}
    \label{subfig:del_fel_evolve}
    \end{subfigure}
    \caption{(a) $\omega$ profiles
    along $[001]$ direction at simulation 
    times $t=500$, $700$, $900$, $1100$ 
    \si{\second}. The value of $\omega$ in
    the precipitate increases with time. 
    (b) The temporal evolution of  
    $\Delta \omega = \omega^{\upalpha} - 
    \omega^{\upbeta}$ across the interface 
    along $[001]$ direction. 
    The $\Delta \omega$ increases after 
    initial decrease which suggests 
    temporal increment in contribution to 
    the chemical potential. }
\end{figure}

The formation of a split pattern can be attributed 
to the higher contribution of elastic stresses 
to the shift in chemical potential ($\Delta \mu$) in 
directions where the grooves develop. To show that, 
we analyse the shift in chemical potential ahead of 
the interface in the matrix. 
Fig.~\ref{subfig:phi_map} shows
the order parameter map in $(110)$ plane at time 
$t = 1100$, and the corresponding 
$\Delta \mu^*$ map is shown in 
Fig.~\ref{subfig:deltamu_map}. 
Fig.~\ref{subfig:contour} shows the contour map
at $\phi = 0.5$ in the one-fourth section of 
$(110)$ plane passing through the centre of the 
simulation box. We track the $\Delta \mu$ ahead of 
the interface in the matrix phase along the contour 
line shown in Fig.~\ref{subfig:contour} and plot 
the variation of $\Delta \mu$ in matrix along the 
contour as a function of arc-length $l_{\mathrm{arc}}$ as 
shown in Fig.~\ref{subfig:larc_vs_deltamu}. 
We obtain the arc-length by 
calculating the distance between the consecutive 
points on the contour in an anti-clockwise 
direction. The X-axis in the contour 
profile corresponds to $[\bar{1}10]$ direction, 
whereas the Y-axis corresponds to $[001]$. The tip of 
dendrite (directed along $[\bar{1}11]$) subtends an 
angle of $35.26\si{\degree}$ with X-axis. 
Fig.~\ref{subfig:larc_vs_deltamu} 
shows that regions along $[\bar{1}10]$ and $[001]$ 
have higher values of $\Delta \mu$ compared to other 
regions. This builds up the driving force for the solute 
to transfer from regions of higher $\Delta \mu$ to 
regions of lower $\Delta \mu$, i.e., from regions 
in the matrix along $[001]$ and $[\bar{1}10]$ 
directions to regions with lower values of 
$\Delta \mu$. Moreover, we track the evolution of 
the $\omega$ along $[001]$ 
direction at different time steps and evaluate the 
temporal evolution of jump in elastic energy 
normal to the interface. 
Fig.~\ref{subfig:fel_evolve} 
depicts $\omega$ profile at simulation 
times $t = 500$, $700$, $900$, and $1100$ 
along $[001]$ direction. The value of
$\omega$ in the precipitate increases with 
time which implies continuous temporal 
increase in the contribution of 
elastic energy to the chemical potential. 
Fig.~\ref{subfig:del_fel_evolve} shows 
the temporal evolution of jump in the 
$\omega$, $\Delta \omega$, 
which increases with time. The temporal 
increment of contribution of elastic 
energy over time results in the precipitate
dissolution along $[001]$ direction.

\begin{figure}[!htb]
    \centering
    \begin{subfigure}{0.5\textwidth}
    \centering
    \includegraphics[width=6.5cm]{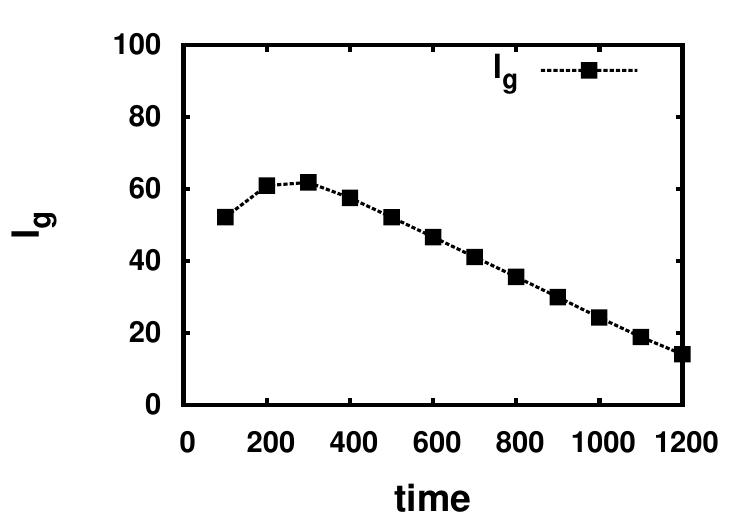}
    \caption{}
    \label{subfig:lg_evolve}
    \end{subfigure}%
    \begin{subfigure}{0.5\textwidth}
    \centering
    \includegraphics[width=6.5cm]{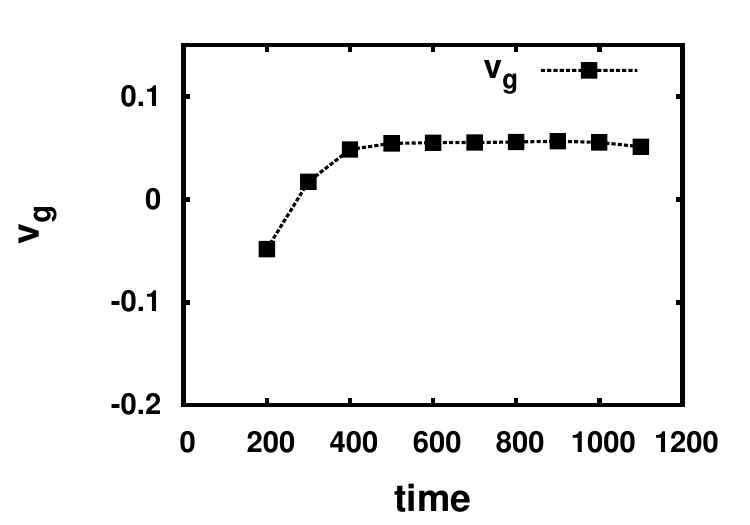}
    \caption{}
    \label{subfig:vg_evolve}
    \end{subfigure}
    \begin{subfigure}{0.5\textwidth}
    \centering
    \includegraphics[width=6.5cm]{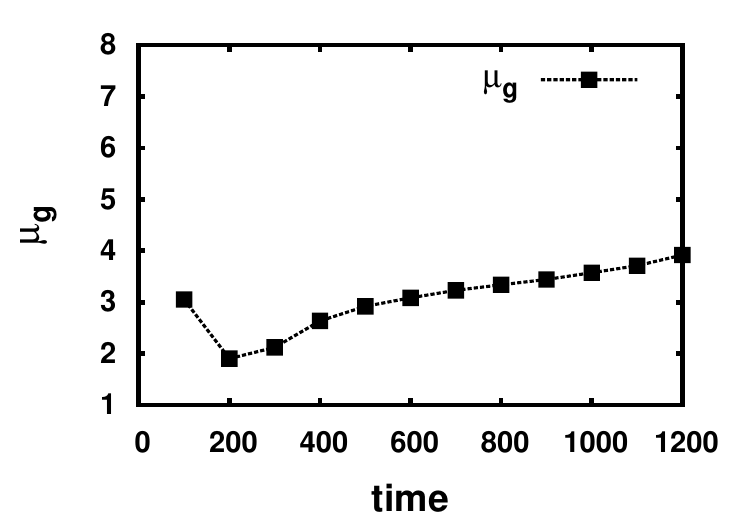}
    \caption{}
    \label{subfig:mug_evolve}
    \end{subfigure}
    \caption{
    (a) The evolution of distance of interface from 
    the centre of precipitate ($l_{\textrm{g}}$) along $[001]$ 
    direction. The $l_{\textrm{g}}$ decreases linearly with time.
    (b) The evolution of velocity $v_{\textrm{g}}$ of the interface 
    along $[001]$ direction as a function of time. The 
    $v_{\textrm{g}}$ reaches a steady-state over time before the pinch
    off takes place. (c) Temporal evolution of $\mu_{\textrm{g}}$ along
    $[001]$ direction. After the initial decrease, the value
    of $\mu_{\textrm{g}}$ tends to increase suggesting precipitate dissolution
    along $[001]$ direction.}
\end{figure}

\begin{figure}[!htb]
    \centering
    \begin{subfigure}{0.5\textwidth}
    \centering
    \includegraphics[width=6.5cm]{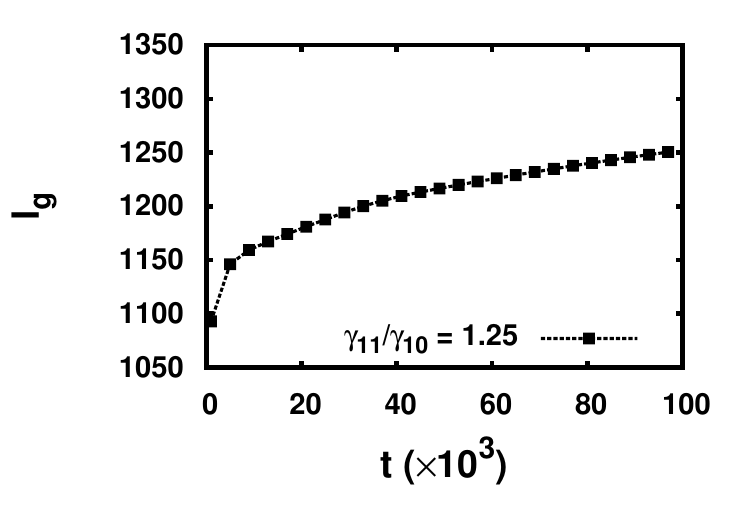}
    \caption{}
    \label{subfig:lg_evolve_int_aniso}
    \end{subfigure}%
    \begin{subfigure}{0.5\textwidth}
    \centering
    \includegraphics[width=6.5cm]{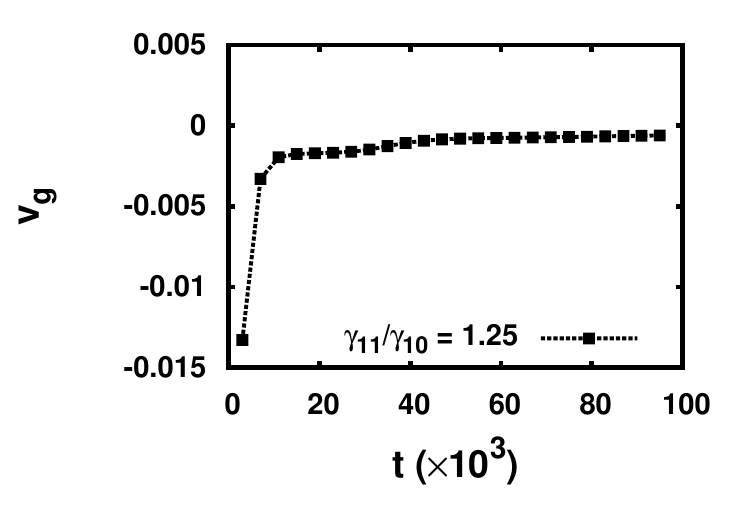}
    \caption{}
    \label{subfig:vg_evolve_int_aniso}
    \end{subfigure}
    \begin{subfigure}{0.5\textwidth}
    \centering
    \includegraphics[width=6.5cm]{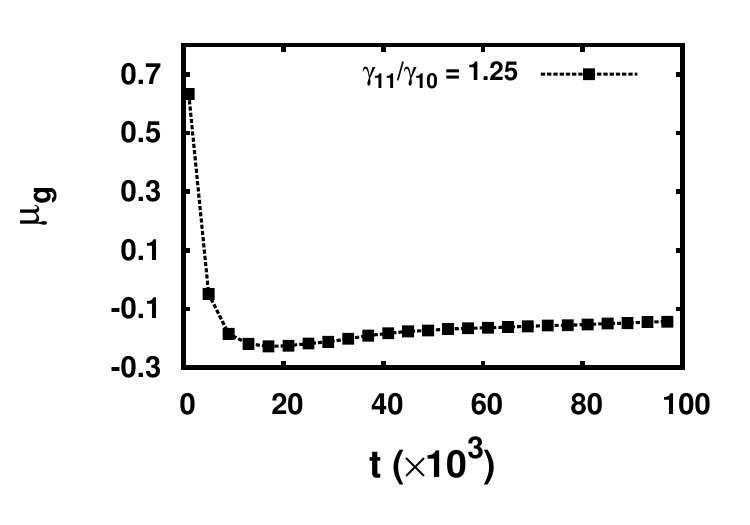}
    \caption{}
    \label{subfig:mug_evolve_int_aniso}
    \end{subfigure}
    \caption{
    (a) The evolution of distance of interface from 
    the center of precipitate ($l_{\textrm{g}}$) along $[01]$ 
    direction. The $l_{\textrm{g}}$ increases with time.
    (b) The evolution of velocity $v_{\textrm{g}}$ of the interface 
    along $[01]$ direction as a function of time. The 
    $v_{\textrm{g}}$ achieves a negative steady-state values 
    over time. (c) Temporal evolution of $\mu_{\textrm{g}}$ along
    $[01]$ direction. After the initial transient, the 
    $\mu_{\textrm{g}}$ tends to decrease with negative value which
    suggests advancement of interface along $[01]$ 
    direction away from the center of the precipitate.
    Here the ratio of interfacial energy along $[11]$
    direction to that along $[10]$ direction is $1.25$.}
\end{figure}

\begin{figure}[!htb]
    \centering
    \begin{subfigure}{0.5\textwidth}
    \centering
    \includegraphics[width=6.5cm]{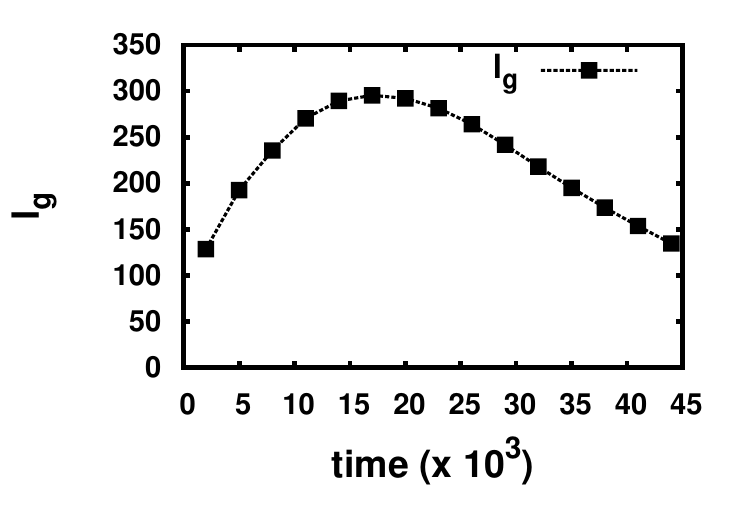}
    \caption{}
    \label{subfig:lg_evolve_2D}
    \end{subfigure}%
    \begin{subfigure}{0.5\textwidth}
    \centering
    \includegraphics[width=6.5cm]{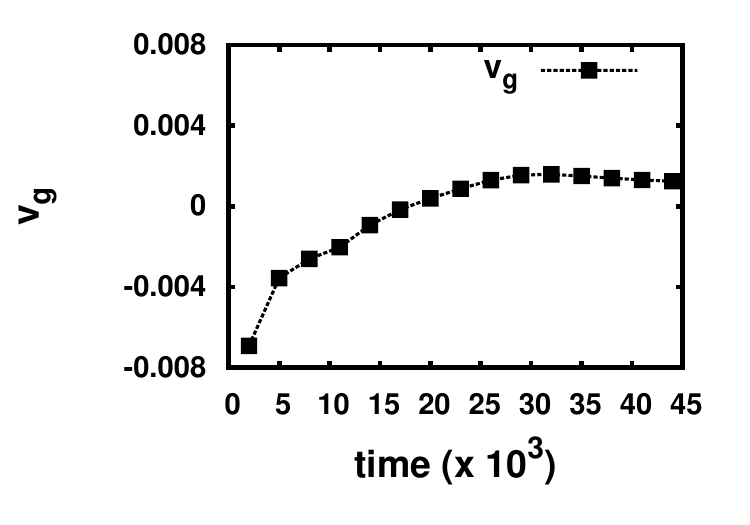}
    \caption{}
    \label{subfig:vg_evolve_2D}
    \end{subfigure}
    \begin{subfigure}{0.5\textwidth}
    \centering
    \includegraphics[width=6.5cm]{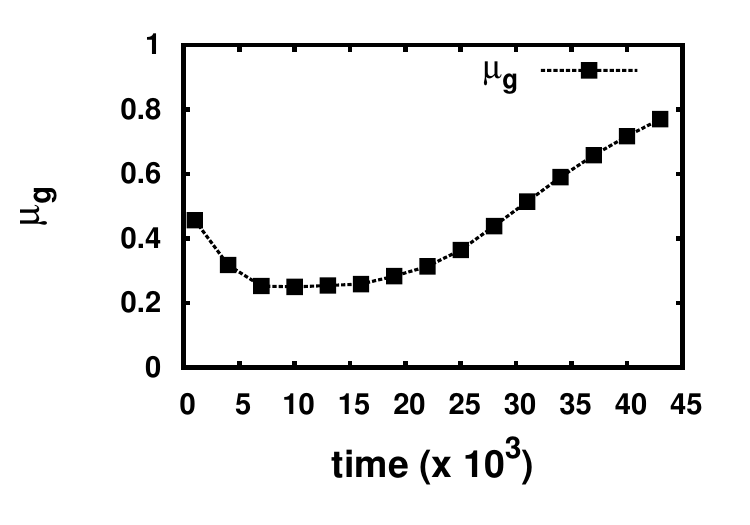}
    \caption{}
    \label{subfig:mug_evolve_2D}
    \end{subfigure}
    \caption{
    (a) The evolution of distance of interface from 
    the center of precipitate ($l_{\textrm{g}}$) along $[01]$ 
    direction. The $l_{\textrm{g}}$ decreases with time after achieving a 
    maximum value. (b) The evolution of velocity $v_{\textrm{g}}$ of the 
    interface along $[01]$ direction as a function of time. The 
    $v_{\textrm{g}}$ achieves a positive steady-state values 
    over time. (c) Temporal evolution of $\mu_{\textrm{g}}$ along
    $[01]$ direction. The 
    $\mu_{\textrm{g}}$ tends to increase which
    suggests advancement of interface along $[01]$ 
    direction towards the center of the precipitate.}
\end{figure}

In addition, we track the evolution of 
velocity of the interface $v_{\textrm{g}}$ 
along $[001]$ direction and 
$\Delta \mu$ in the matrix ahead of the 
interface along $[001]$ direction. 
Fig.~\ref{subfig:lg_evolve} 
shows the evolution of position of the 
interface $l_{\textrm{g}}$ 
along $[001]$ direction which increases 
initially and further linearly decreases 
until the pinch-off takes place. 
The linear decrease in $l_{\textrm{g}}$ 
occurs when the grooves develop and 
interface along $[001]$ advances towards 
the centre of the precipitate. As a 
result, the $v_{\textrm{g}}$ initially 
increases and reaches a steady state value 
when the interface starts advancing towards 
the centre of the precipitate. 
Fig.~\ref{subfig:mug_evolve} shows the 
evolution of $\mu_{\textrm{g}}$ as a 
function of time. The value of 
$\mu_{\textrm{g}}$ initially decreases and 
begin to increase 
as soon as the groove forms along the 
$[001]$ direction. The temporal increment 
of $\mu_{\textrm{g}}$ signifies 
the precipitate dissolution along $[001]$ 
direction. 
Two-dimensional simulation show similar 
trends of temporal evolution of 
$l_{\textrm{g}}$, $v_{\textrm{g}}$, and 
$\mu_{\textrm{g}}$ (see 
Figs.~\ref{subfig:lg_evolve_2D},~\ref{subfig:vg_evolve_2D}, 
and~\ref{subfig:mug_evolve_2D}). On the contrary, when only 
anisotropy in interfacial energy is present, the $l_{\textrm{g}}$ continues to increase with time as 
shown in Fig.~\ref{subfig:lg_evolve_int_aniso}. 
The velocity of the interface along $[01]$ direction 
reaches a negative steady state values (close to zero) 
which indicates the movement of interface along $[01]$ 
direction away from the centre of the precipitate 
(see Fig.~\ref{subfig:vg_evolve_int_aniso}). After the 
initial transient, $\mu_{\textrm{g}}$ continues to decrease 
having negative values (see 
Fig.~\ref{subfig:mug_evolve_int_aniso}).
The negative values of $\mu_{\textrm{g}}$ suggest growth of 
the interface along $[10]$ direction. 

\subsection{Effects of lattice misfit}
\label{sec:misfit_effect}
In this section, we discuss the effects of 
lattice misfit on the initiation of splitting 
instability. To understand the effect of lattice 
misfit, we choose different combinations of 
supersaturation ($c_{\infty} =$ $25\%$, $35\%$, 
$45\%$) and lattice misfit 
($\epsilon^* = $ $0.75\%$, $0.85\%$
and $1\%$). For all cases, the elastic 
energy anisotropy is $4$.
Table.~\ref{tab:effect_of_misfit} shows the precipitate 
morphologies at the same simulation time $t=10000$ for different levels 
of lattice misfit and supersaturations. At all given levels of 
lattice misfit and supersaturation, due to the point effect 
of diffusion, precipitate develops concavities, 
however, only at higher misfit and supersaturation 
concavities transform to grooves and lead to splitting of 
the precipitate. 

\begin{figure}[!htb]
    \centering
    \begin{subfigure}{0.3\textwidth}
    \centering
    \includegraphics[width=4cm]{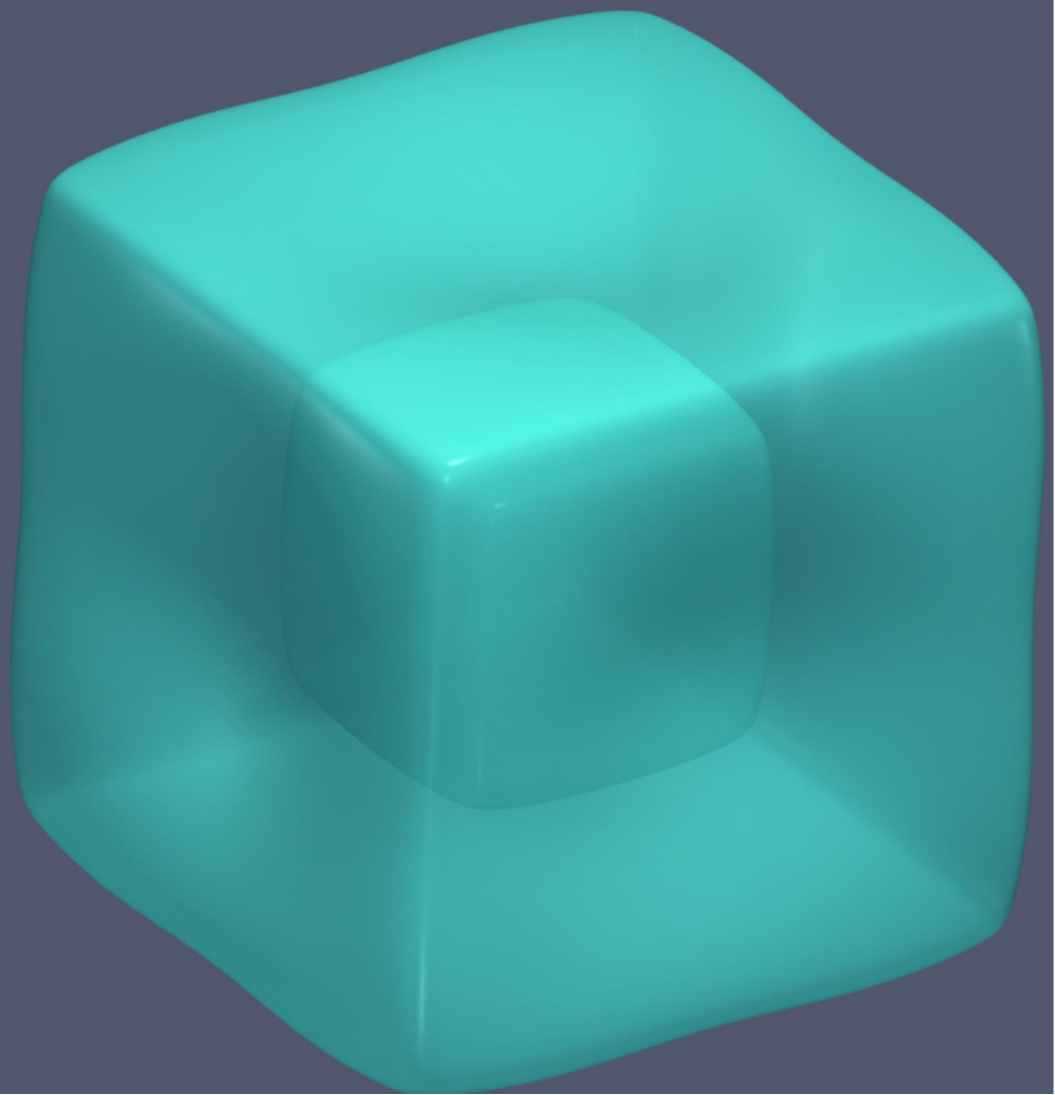}
    \caption{}
    \label{fig:eps0.0075_c0.45_iso}
    \end{subfigure}
    \begin{subfigure}{0.3\textwidth}
    \centering
    \includegraphics[width=4cm]{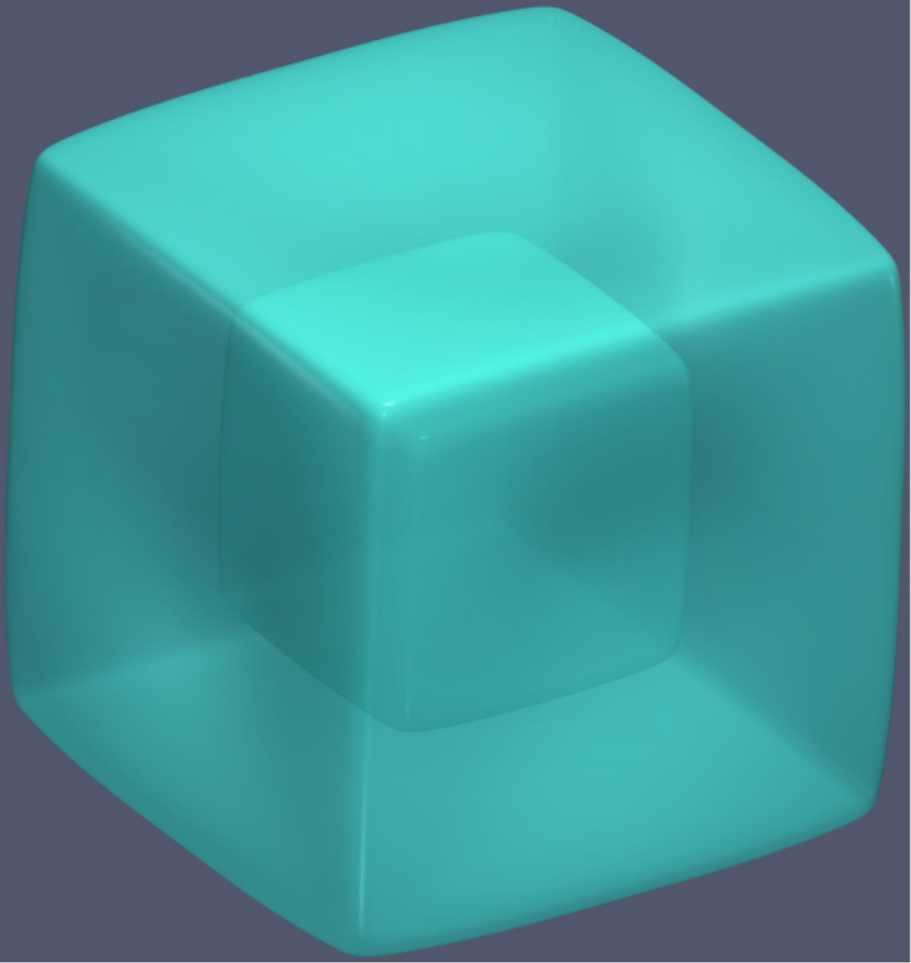}
    \caption{}
    \label{fig:eps0.01_c0.35_iso}
    \end{subfigure}
    \begin{subfigure}{0.3\textwidth}
    \centering
    \includegraphics[width=4cm]{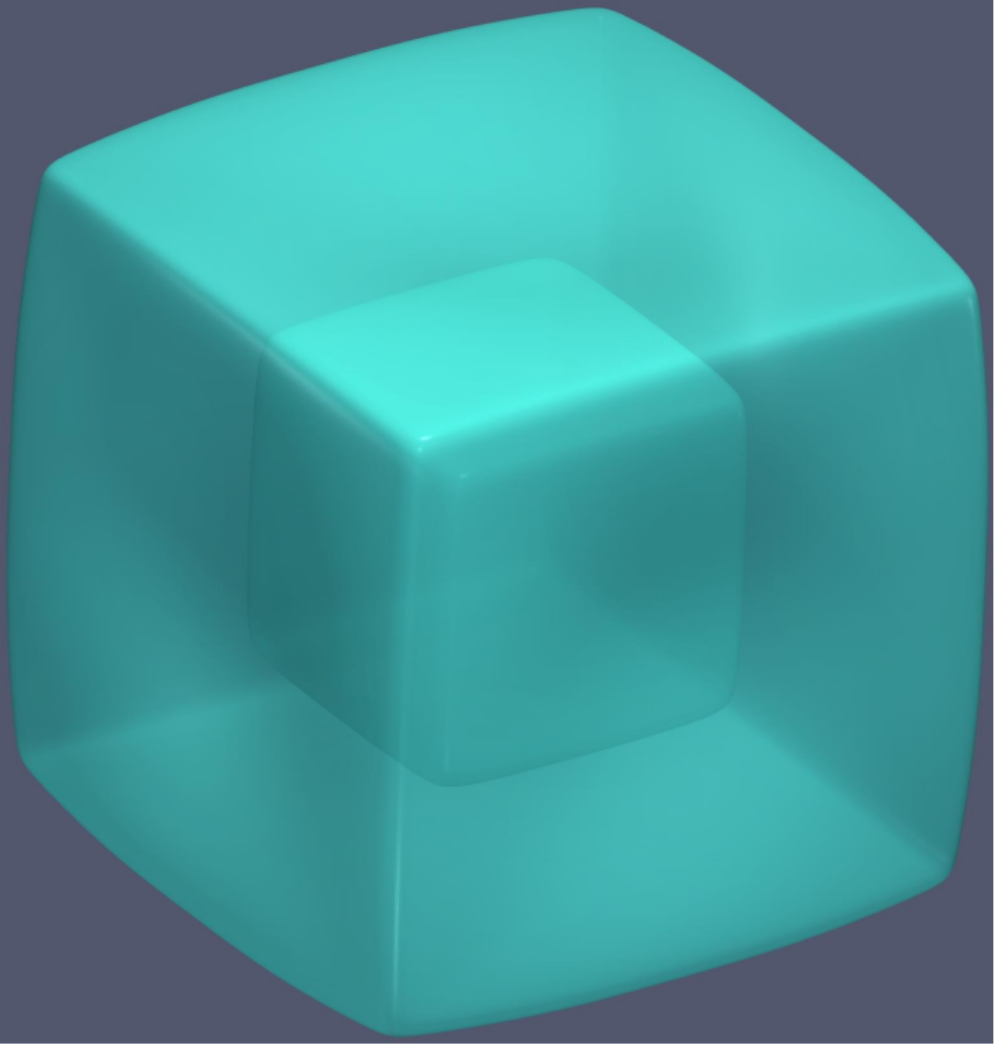}
    \caption{}
    \label{fig:eps0.01_c0.45_iso}
    \end{subfigure}%
    
    \begin{subfigure}{0.3\textwidth}
    \centering
    \includegraphics[width=4cm]{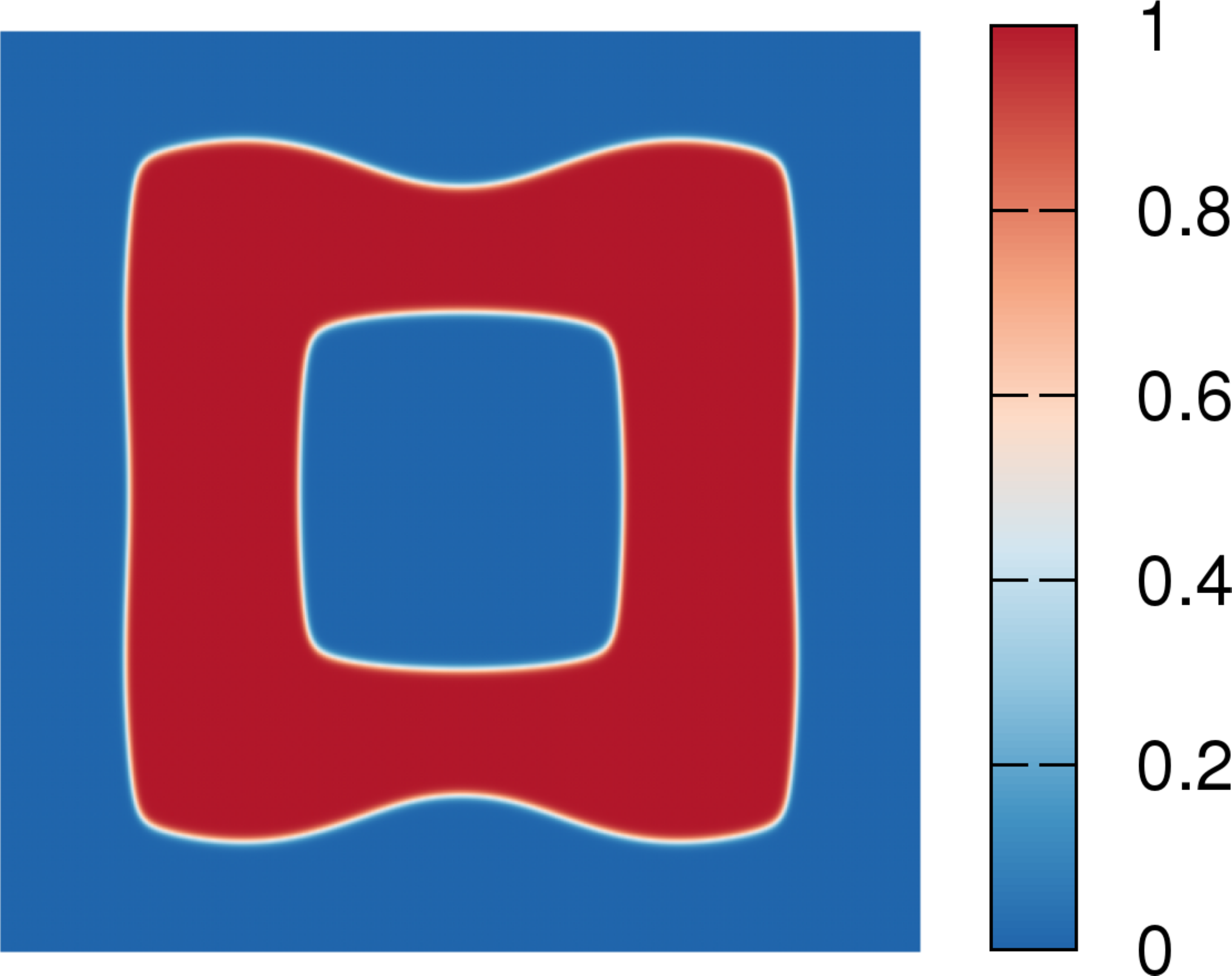}
    \caption{}
    \label{fig:eps0.0075_c0.45_2d}
    \end{subfigure}
    \begin{subfigure}{0.3\textwidth}
    \centering
    \includegraphics[width=4cm]{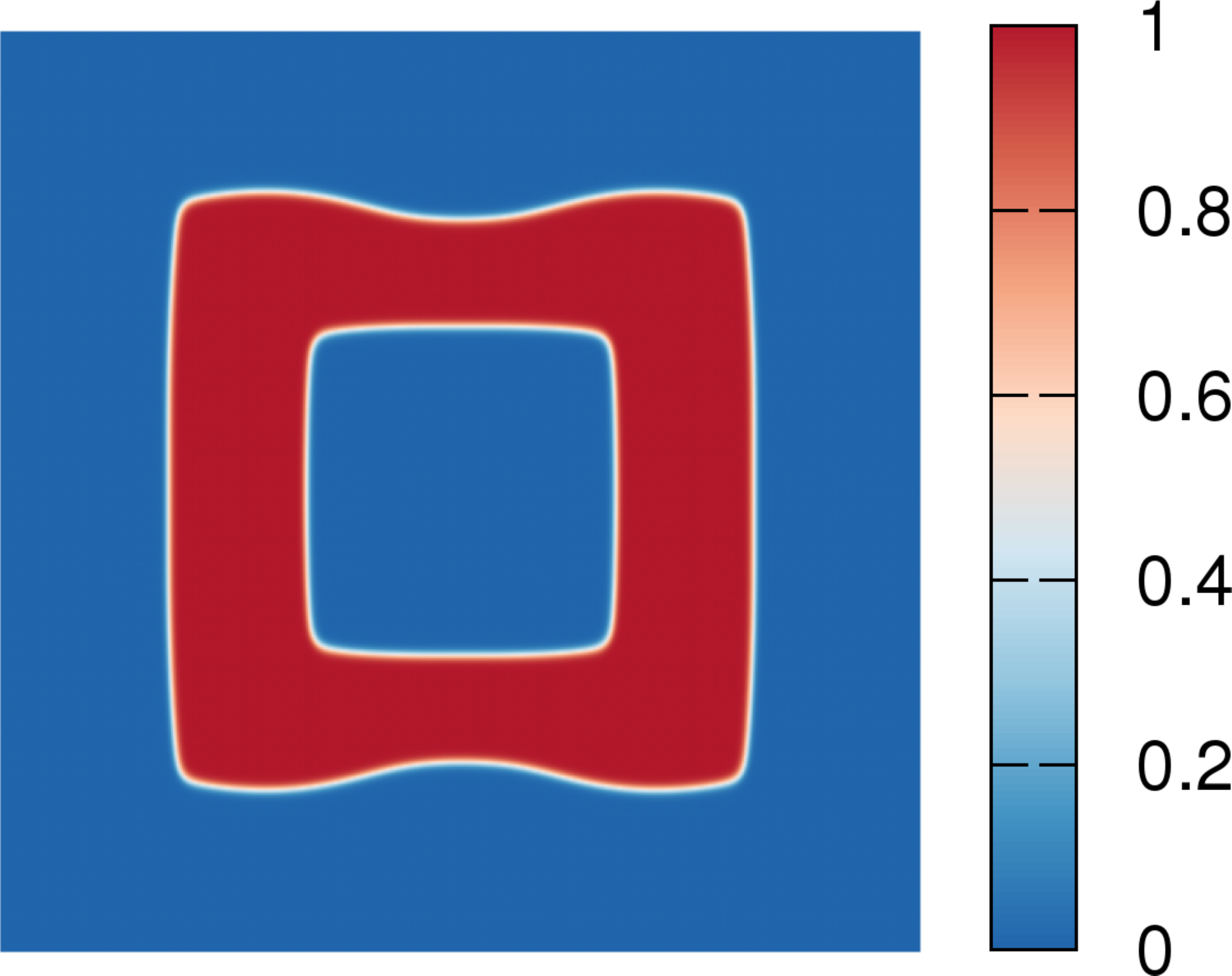}
    \caption{}
    \label{fig:eps0.01_c0.35_2d}
    \end{subfigure}
    \begin{subfigure}{0.3\textwidth}
    \centering
    \includegraphics[width=4cm]{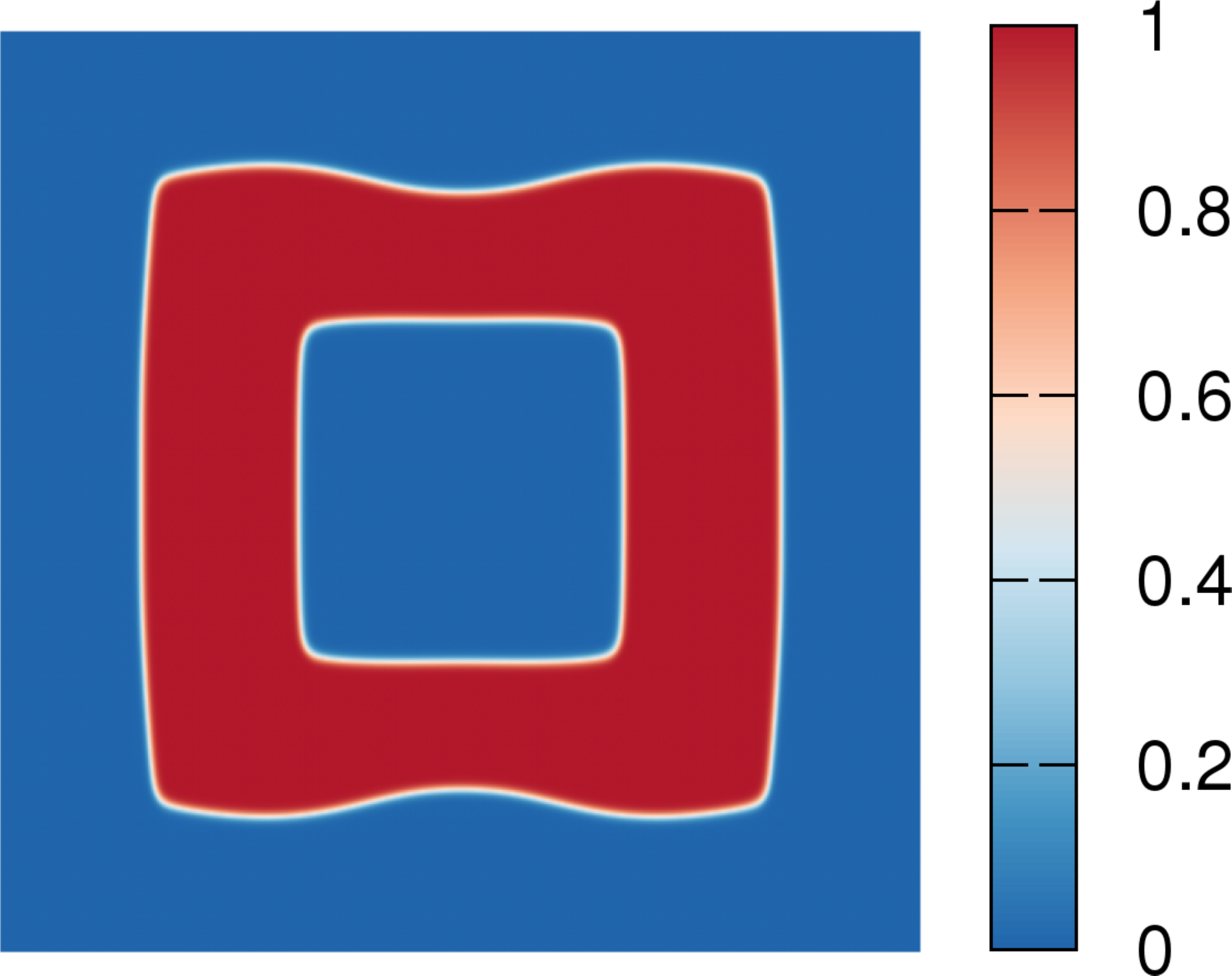}
    \caption{}
    \label{fig:eps0.01_c0.45_2d}
    \end{subfigure}%
    \caption{Isosurface representation at time $t = 10000$
     for (a) supersaturation $45\%$, $35\%$ and lattice misfit $0.75\%$ (b) supersaturation $35\%$ and lattice misfit $1\%$, and
    (c) supersaturation $45\%$ and lattice misfit $1\%$.}
    \label{fig:hollow_precipitate}
\end{figure}

\begin{table}[!htb]
    \centering
    \begin{tabular}{| M{1.75cm} | M{3cm}  M{3cm}  M{3cm} | }
         \hline
         & $c_{\infty} = 25\%$ & $c_{\infty} = 35\%$ & $c_{\infty} = 45\%$\\
         \hline
         & & & \\
         $\epsilon^* = 0.5\%$ & 
         \includegraphics[width=3cm]{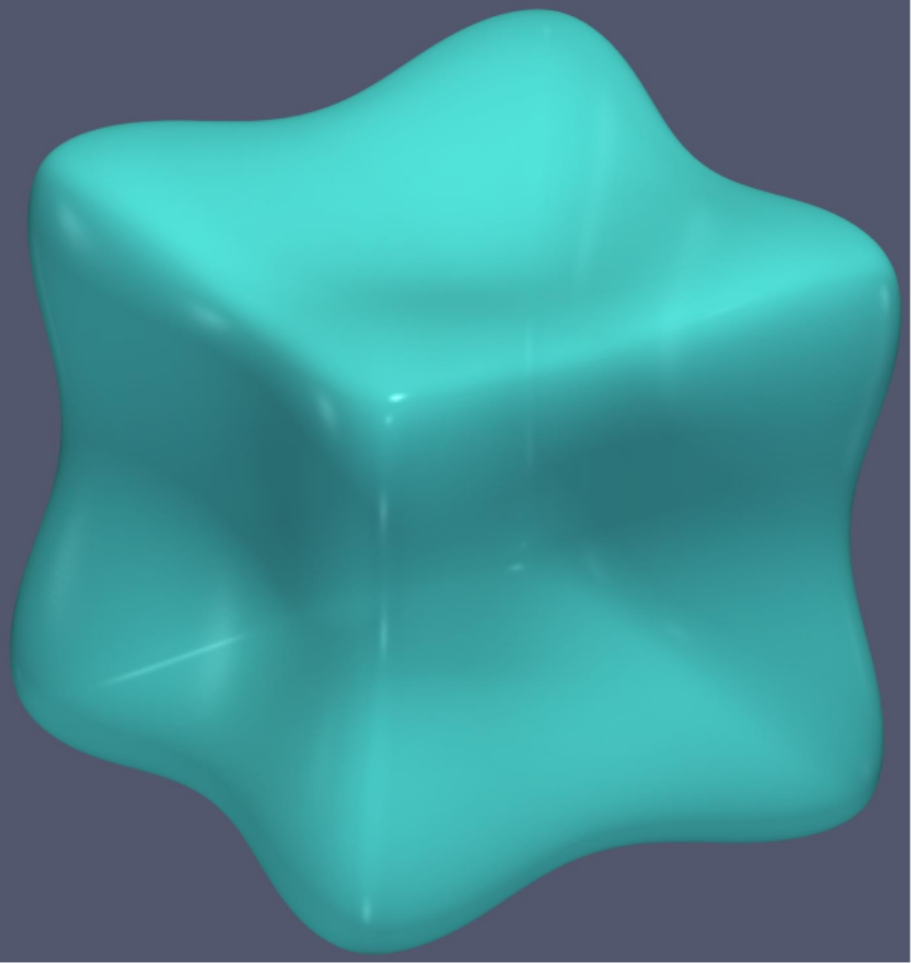}&
         \includegraphics[width=3cm]{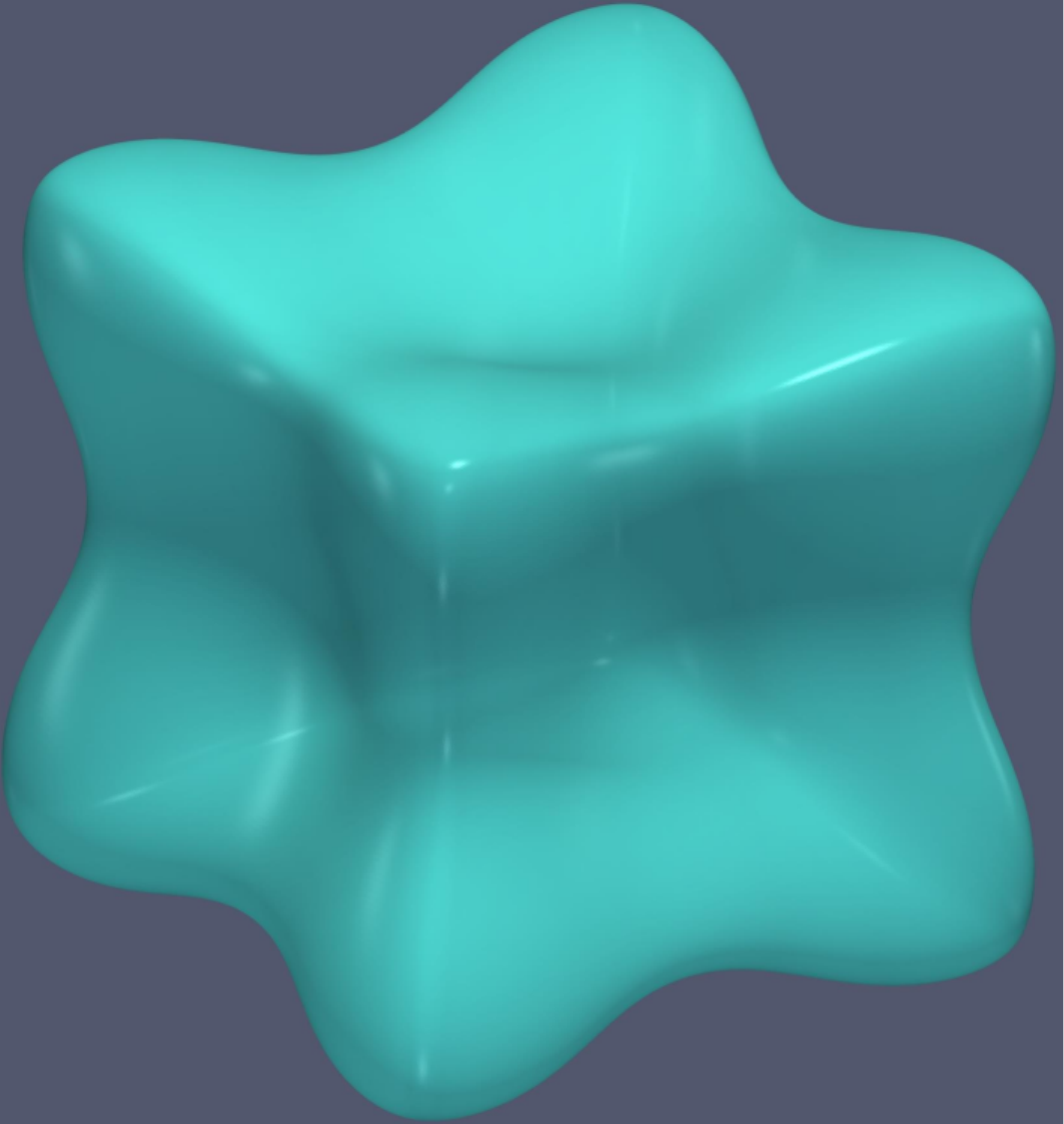}& 
         \includegraphics[width=3cm]{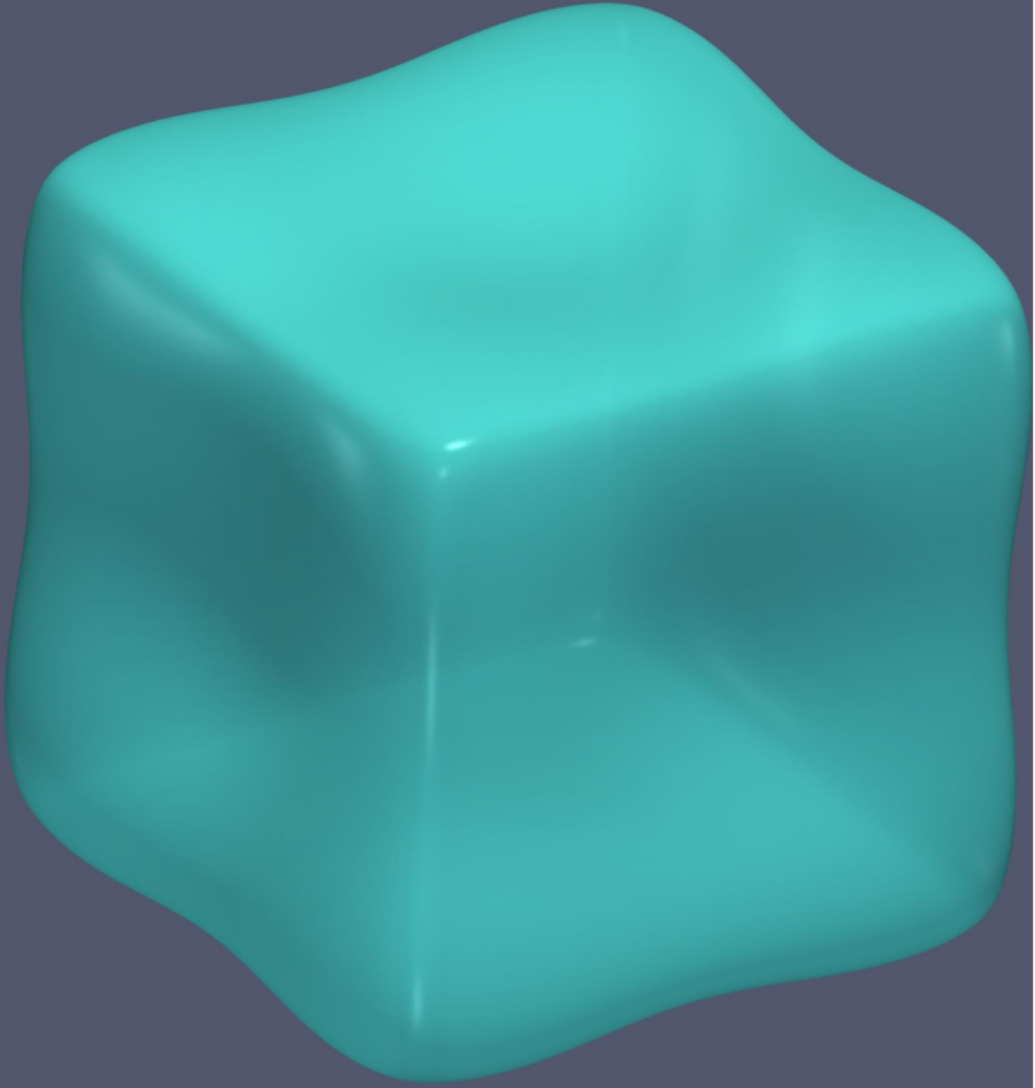}\\
         $\epsilon^* = 0.75\%$ & 
         \includegraphics[width=3cm]{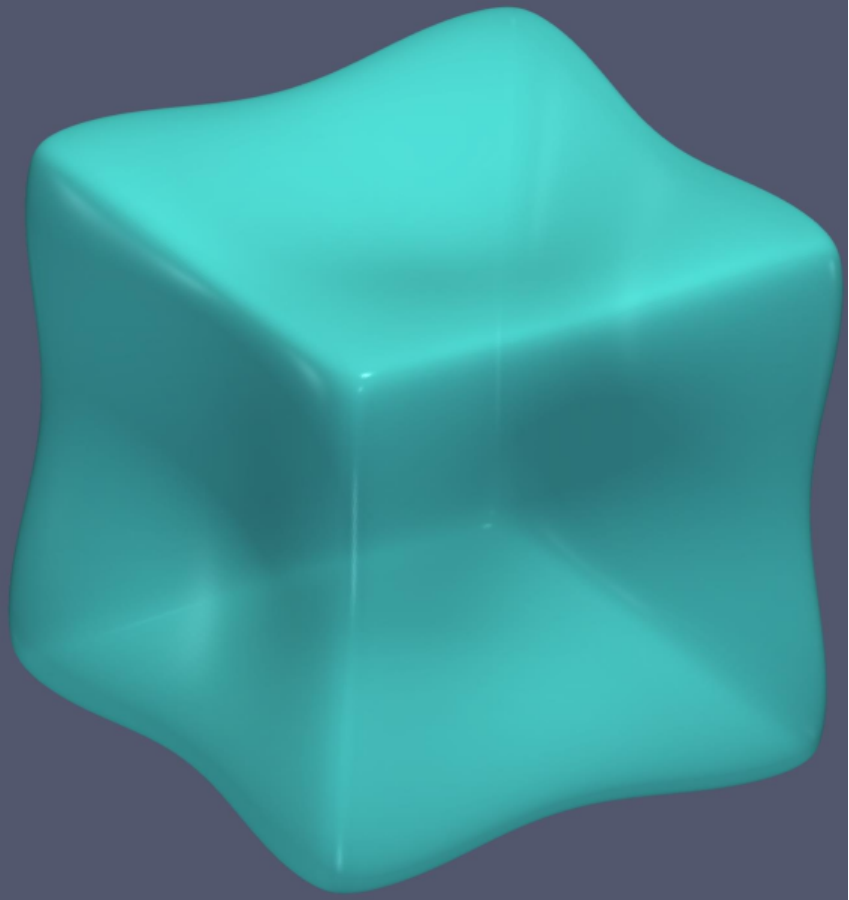}&
         \includegraphics[width=3cm]{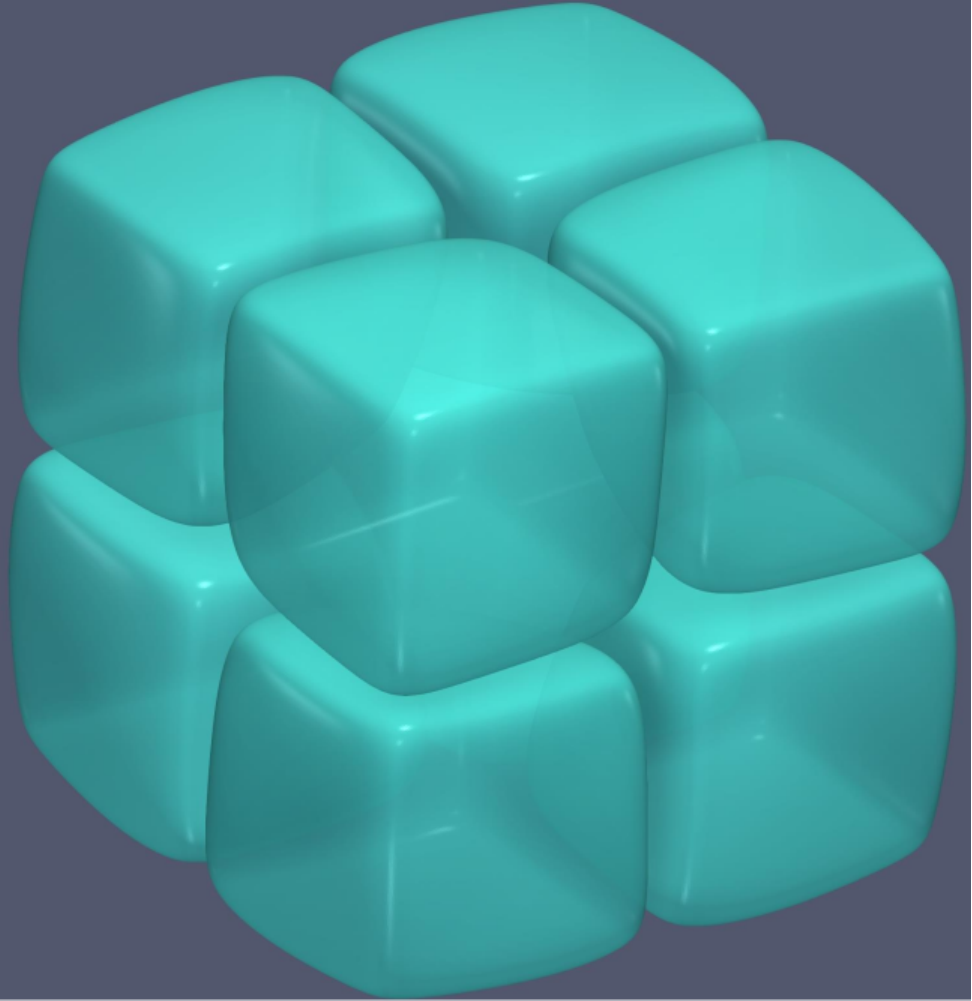}& 
         \includegraphics[width=3cm]{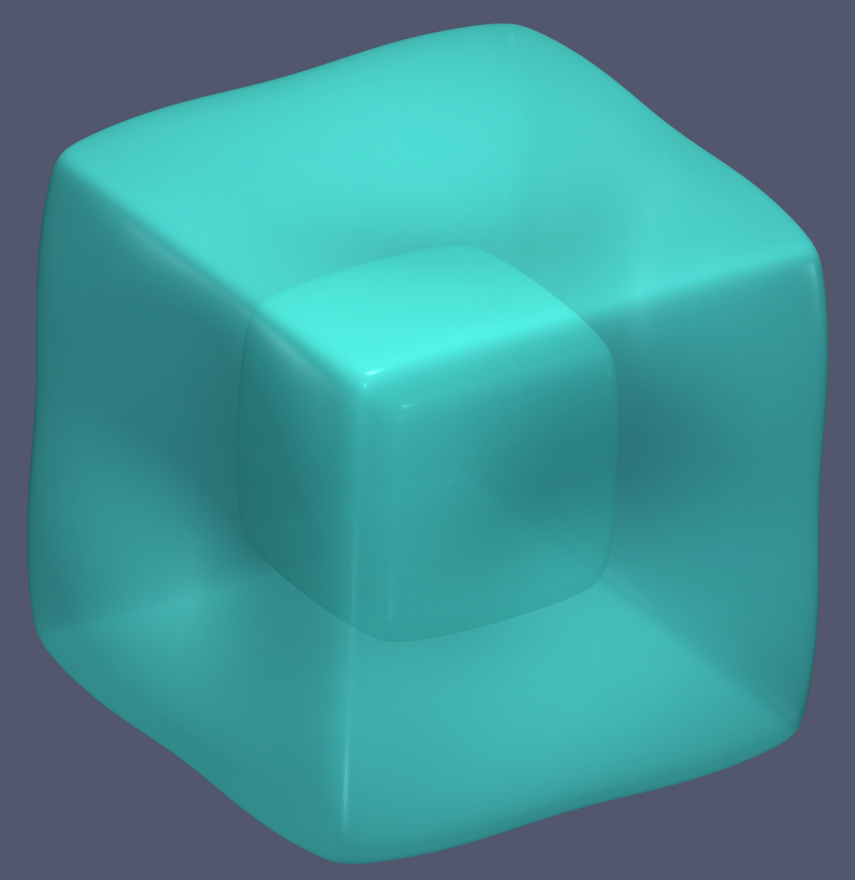}\\
         $\epsilon^* = 0.85\%$ & 
         \includegraphics[width=3cm]{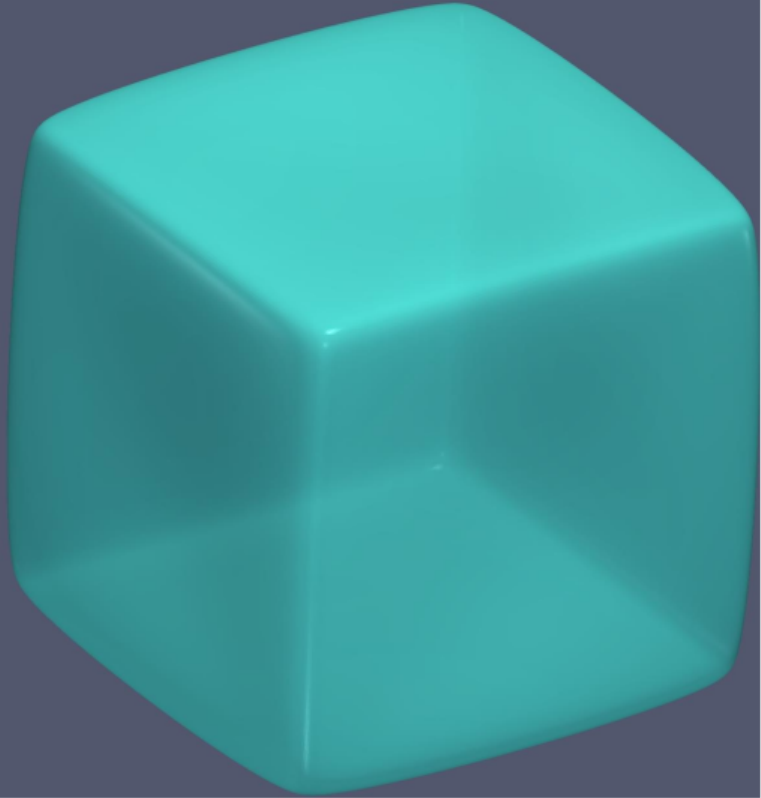}&
         \includegraphics[width=3cm]{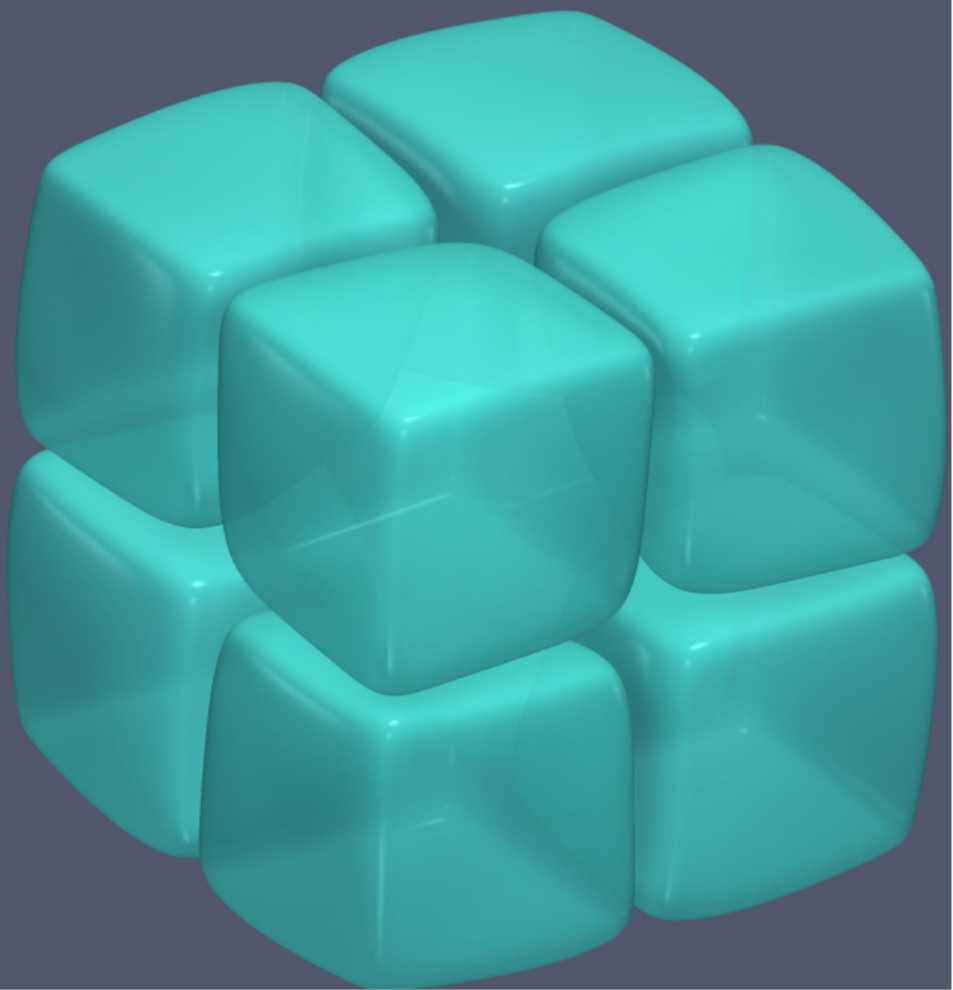}& 
         \includegraphics[width=3cm]{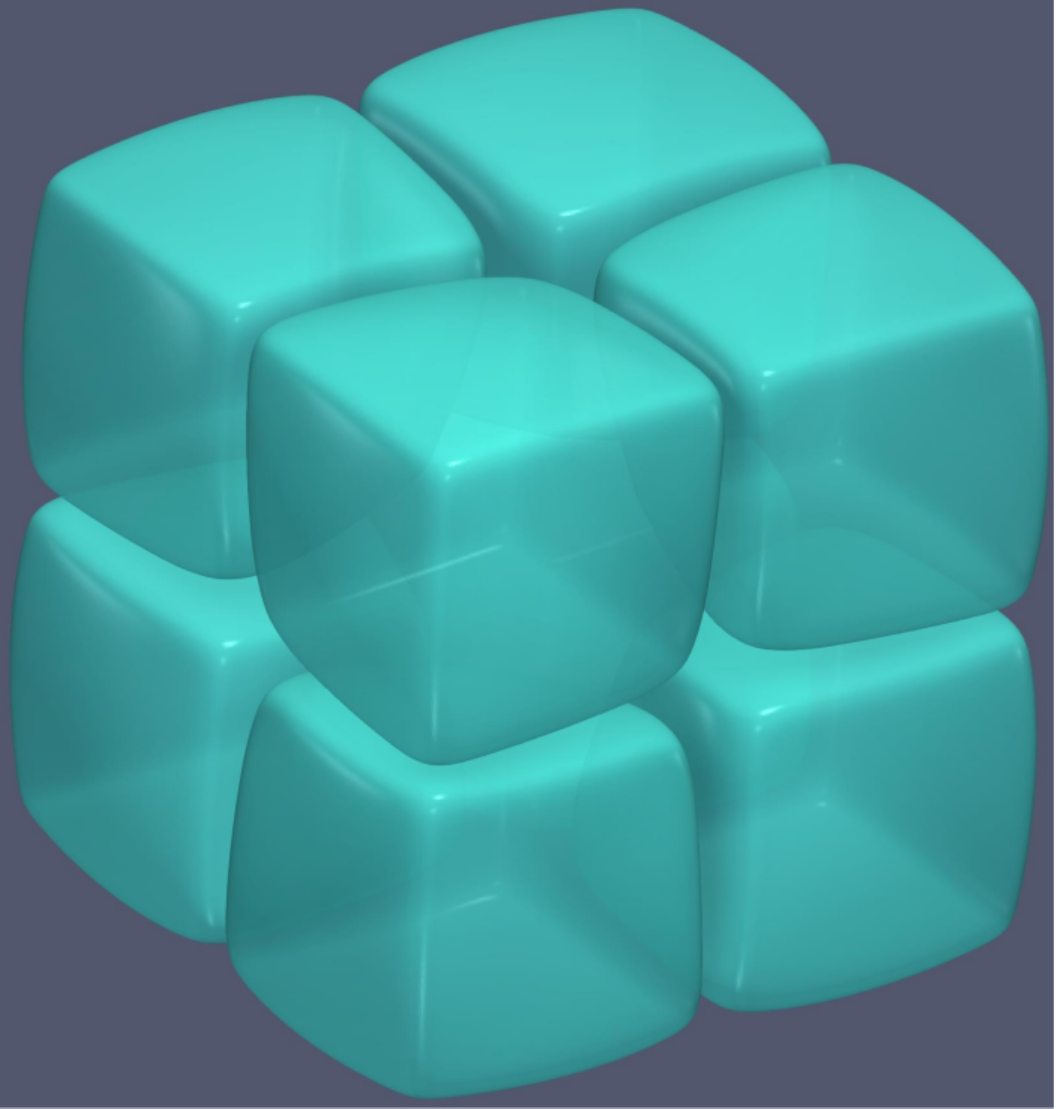}\\
         $\epsilon^* = 1\%$ & 
         \includegraphics[width=3cm]{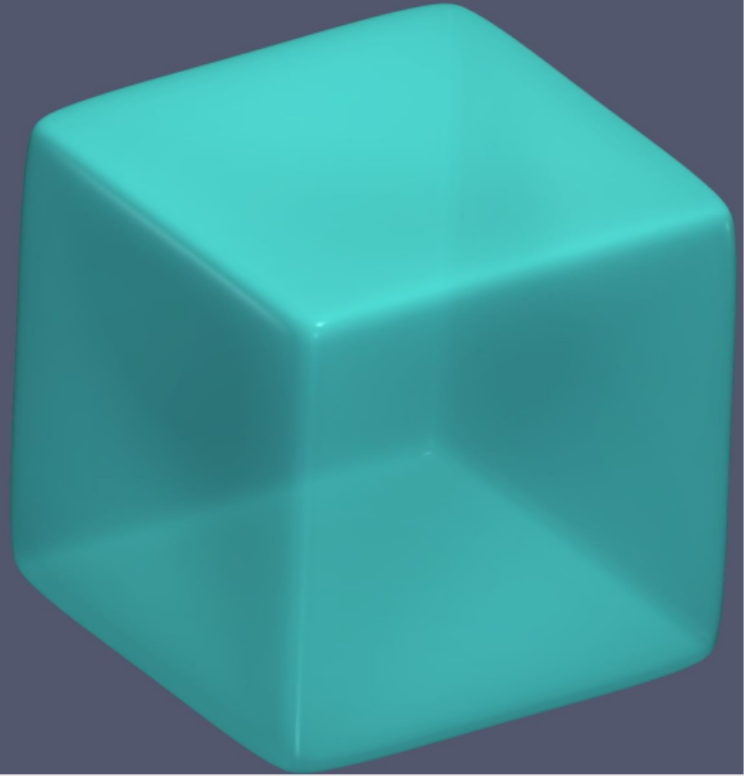}&
         \includegraphics[width=3cm]{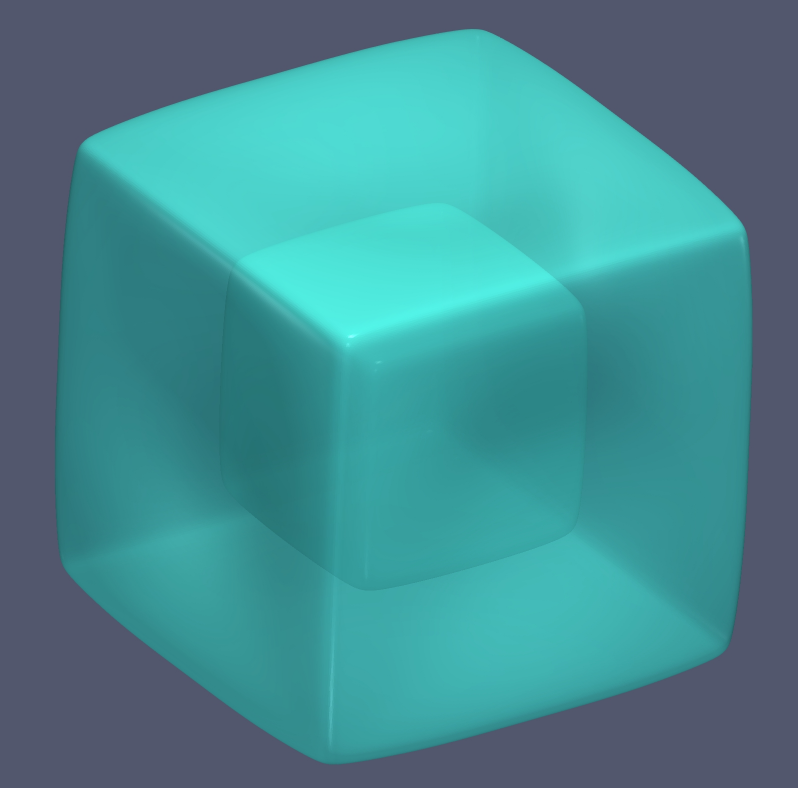}& 
         \includegraphics[width=3cm]{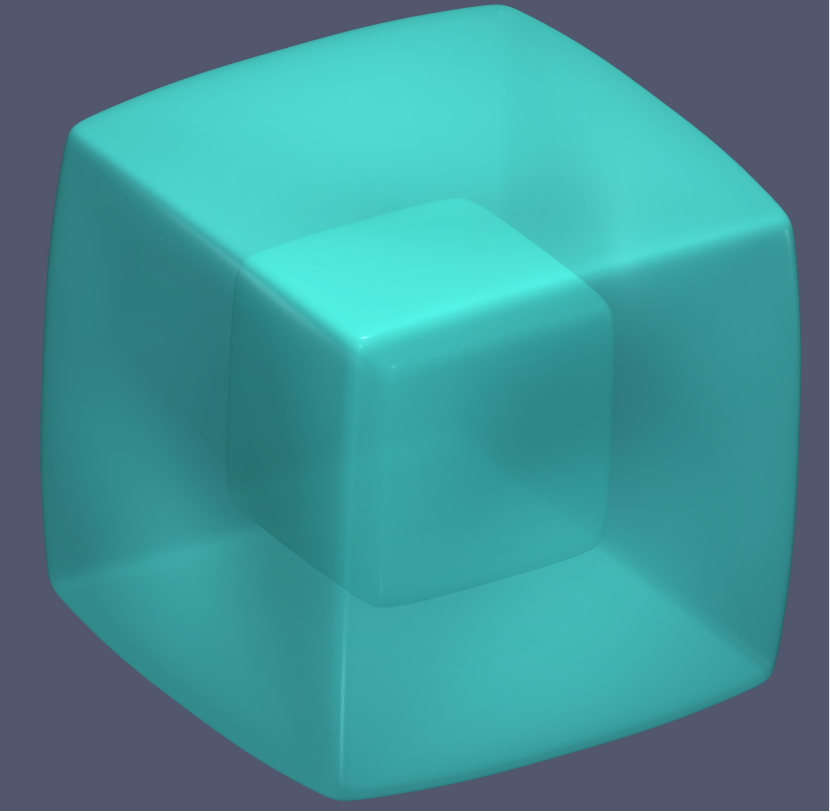}\\
         \hline
    \end{tabular}
    \caption{Precipitate morphologies at different 
    levels of lattice misfit $\epsilon^* = 0.5\%$, $0.75\%$, 
    $0.85\%$, $1\%$ and supersaturation $c_{\infty} = 25\%$, $35\%$, 
    $45\%$ at the same simulation time of $10000$. 
    Here, the Zener anisotropy parameter is $4$. A
    combination of higher misfit and supersaturation promotes 
    splitting instability}
    \label{tab:effect_of_misfit}
\end{table}
At $\epsilon^* = 0.5\%$,  for all given supersaturations,
the precipitate do not transform to form split pattern.
As the lattice misfit is increased to $0.75\%$ 
when the supersaturation is $45\%$, grooves develop 
and further precipitate coalesce to form a 
hollow precipitate with matrix phase at the core.
Similar phenomenon occurs when lattice misfit is $1\%$ 
and supersaturation is $c_{\infty} = 35\%, 45\%$.
Fig.~\ref{fig:eps0.0075_c0.45_iso},
~\ref{fig:eps0.01_c0.35_iso}, 
and~\ref{fig:eps0.01_c0.45_iso} show the 
isosurface representation of the precipitate morphology
at $c_{\infty} = 45$ and $\epsilon^* = 0.75\%$, 
$c_{\infty} = 35\%$ and $\epsilon^* = 1\%$, 
$c_{\infty} = 45\%$ and $\epsilon^* = 1\%$, respectively
at the same time $t = 10000$.
The corresponding plane section of precipitate 
morphology in $(110)$ plane clearly shows the matrix 
phase trapped inside the precipitate (see 
Fig.~\ref{fig:eps0.0075_c0.45_2d},~\ref{fig:eps0.01_c0.35_2d}, 
and~\ref{fig:eps0.01_c0.45_2d}). 

To understand how matrix phase get trapped inside the
precipitate, we track the temporal evolution of 
precipitate morphology in $(110)$ plane. 
Table~\ref{tab:HollowPrecipitateEvolve} 
shows the temporal evolution of precipitate morphology, 
where precipitate arms coalesce along
$\langle 110 \rangle$ directions due to which 
matrix phase gets trapped by the precipitate phase. 
The precipitate-matrix 
interface along $\langle110\rangle$ directions 
(one which is closer to the center of the precipitate) 
continues to advance towards the center of the 
precipitate leading to pinch-off of the primary arms. 
Later, the grooves along $\langle 100 \rangle$ directions
start closing, and the final morphology is a hollow 
precipitate with matrix phase at the center as shown in 
Fig.~\ref{fig:hollow_precipitate}. However, for lattice 
misfit $1\%$ and $c_{\infty} = 45\%$, although initially the
matrix phase appears at the center of the precipitate,
the coalescence of the arms takes place and 
entrapment of matrix occurs. Subsequently, the hollow 
precipitate forms where cuboidal matrix phase is trapped at 
the center.

When the 
lattice misfit is $0.85\%$ and $0.75\%$ for 
$c_{\infty} = 45\%$ and $c_{\infty} = 35\%$, respectively, 
the Euler characteristics are equal to two at the initial 
stages. Afterward, the Euler characteristics shoot up to 
$16$ which corresponds to the disjoint union of eight 
precipitates. On the other hand, for 
lattice misfit of $0.5\%$ and supersaturation of $25\%$, 
the Euler characteristics ($\chi = 2$) remains unchanged 
throughout the evolution. In other case ($c_{\infty} = 45\%$
and $\epsilon^* = 1\%$), initially the Euler characteristic 
of $4$ goes down to $-20$ and further jumps up to $-8$. The 
negative value of $\chi$ suggests the presence of packets of
matrix phase entrapped inside the precipitate. Although at 
the initial stage, the matrix phase forms at the 
center of precipitate due to the instability, during the 
process of groove running towards the center of precipitate,
the secondary arms of the precipitate impinge which results 
in entrapment of the matrix phases. In the late stage, 
the Euler characteristics achieve a value of four which 
indicates a morphology where cuboidal matrix phase is 
entrapped by the cuboidal precipitate (see 
table~\ref{tab:TopoPrecipitateEvolve}). 


\begin{table}[!htb]
    \centering
    \begin{tabular}{|M{1.8cm} | M{2.5cm} M{2.5cm} M{2.5cm} M{2.5cm} | }
    \hline
         &  $t=200$ & $t=400$ & $t=1800$ & $t=10000$\\
    \hline         
    & & & &\\
    Isosurface view     & 
    \includegraphics[width=2.5cm]{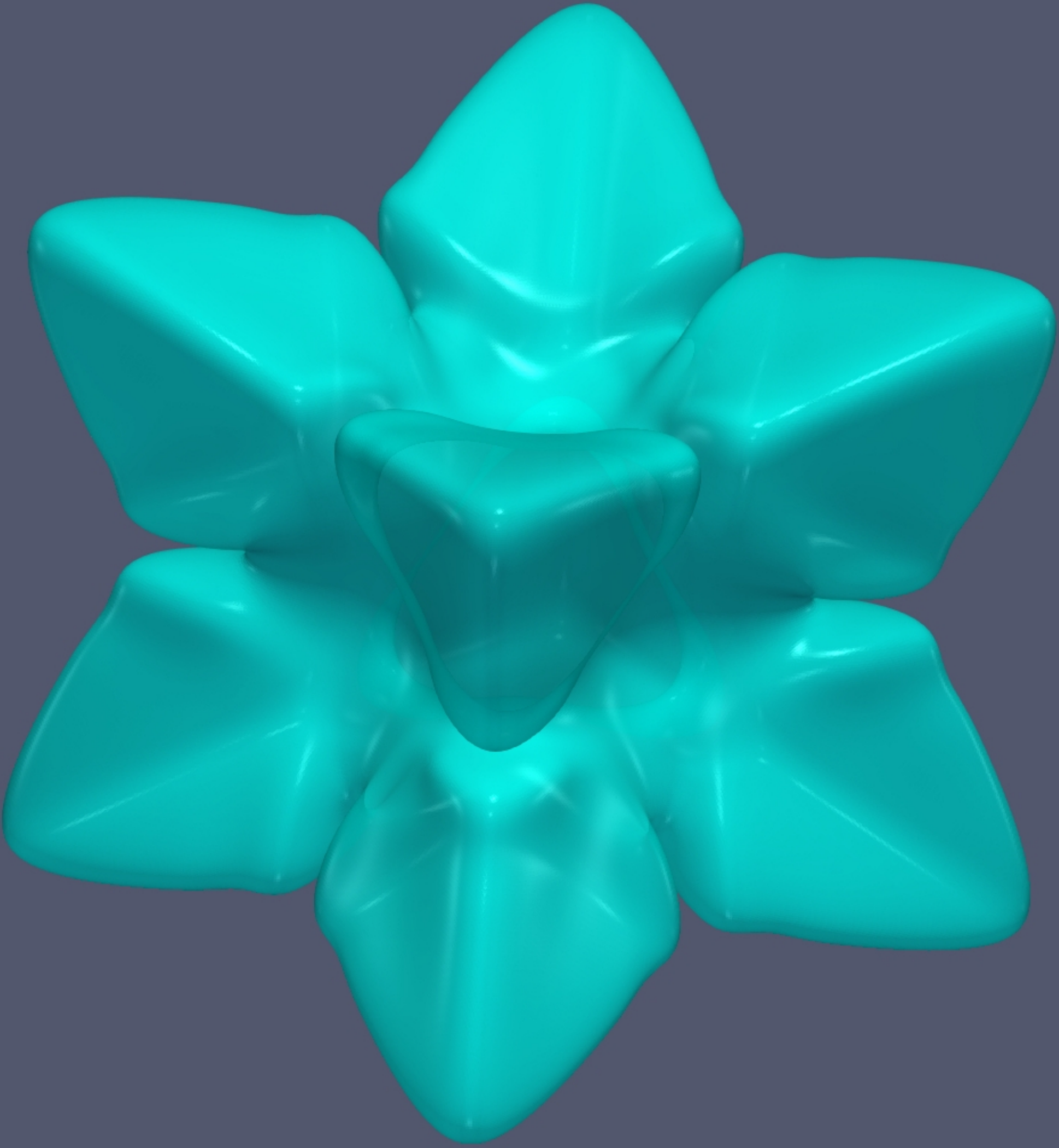} &
    \includegraphics[width=2.5cm]{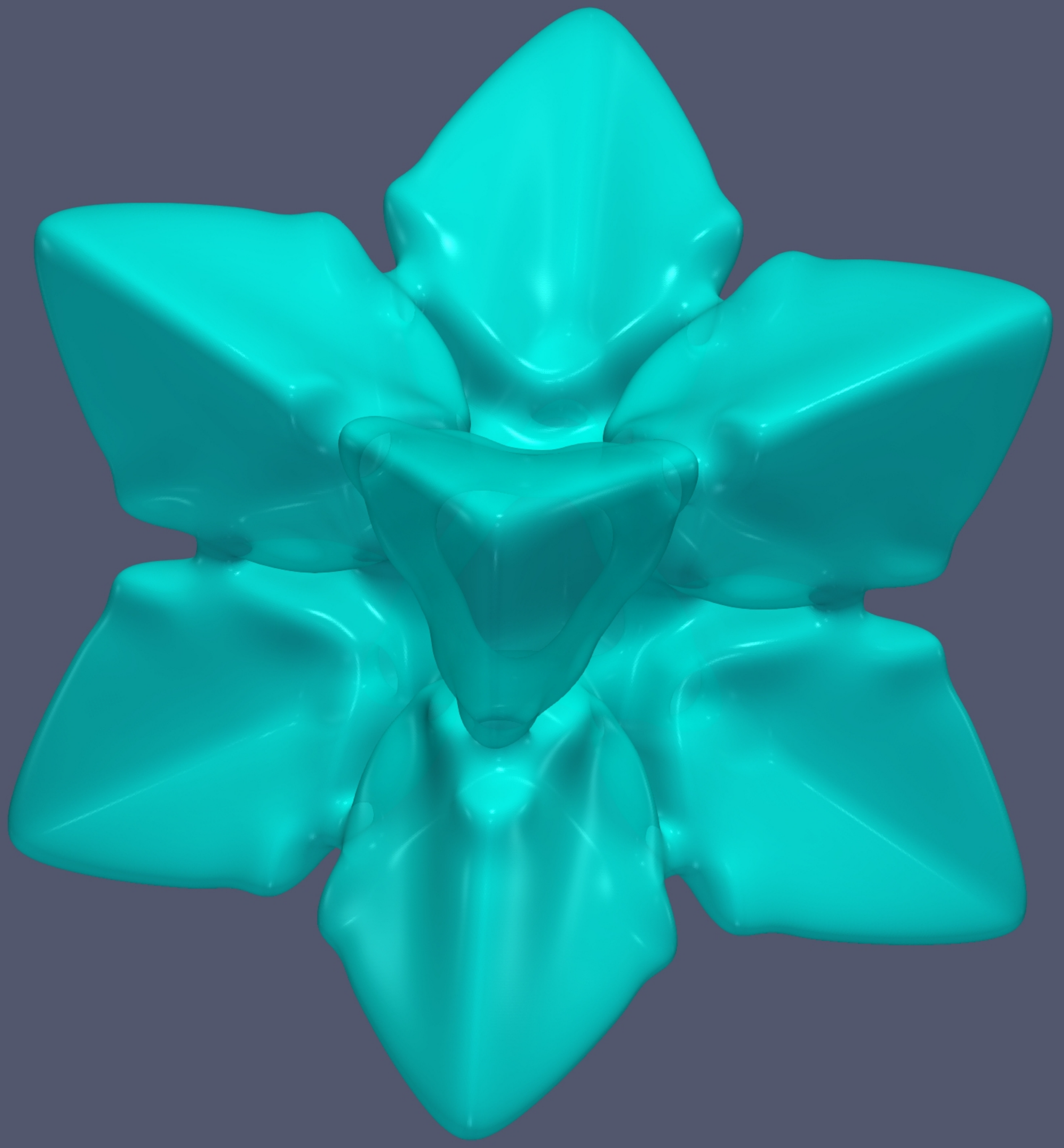} &
    \includegraphics[width=2.5cm]{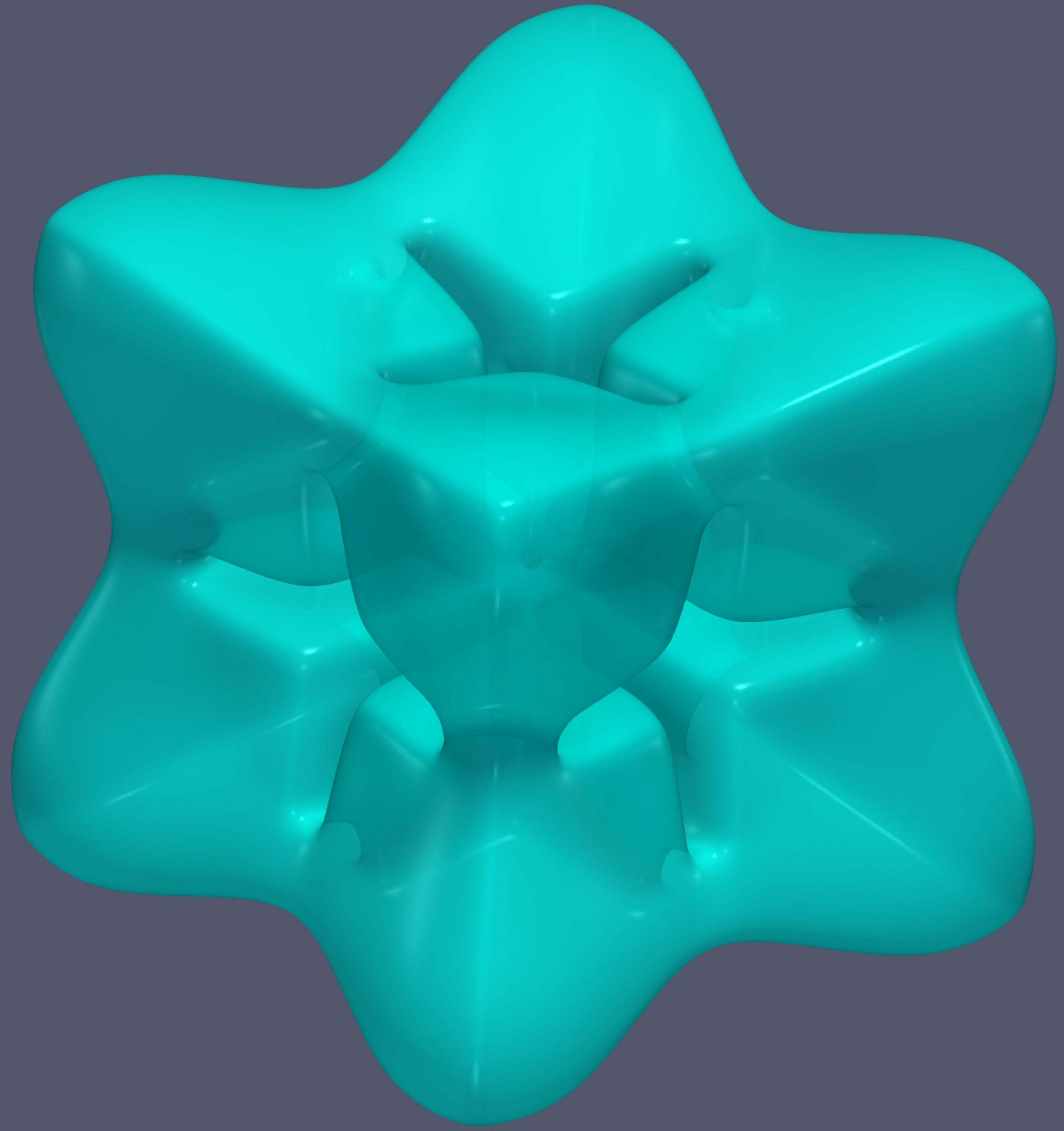} &
    \includegraphics[width=2.5cm]{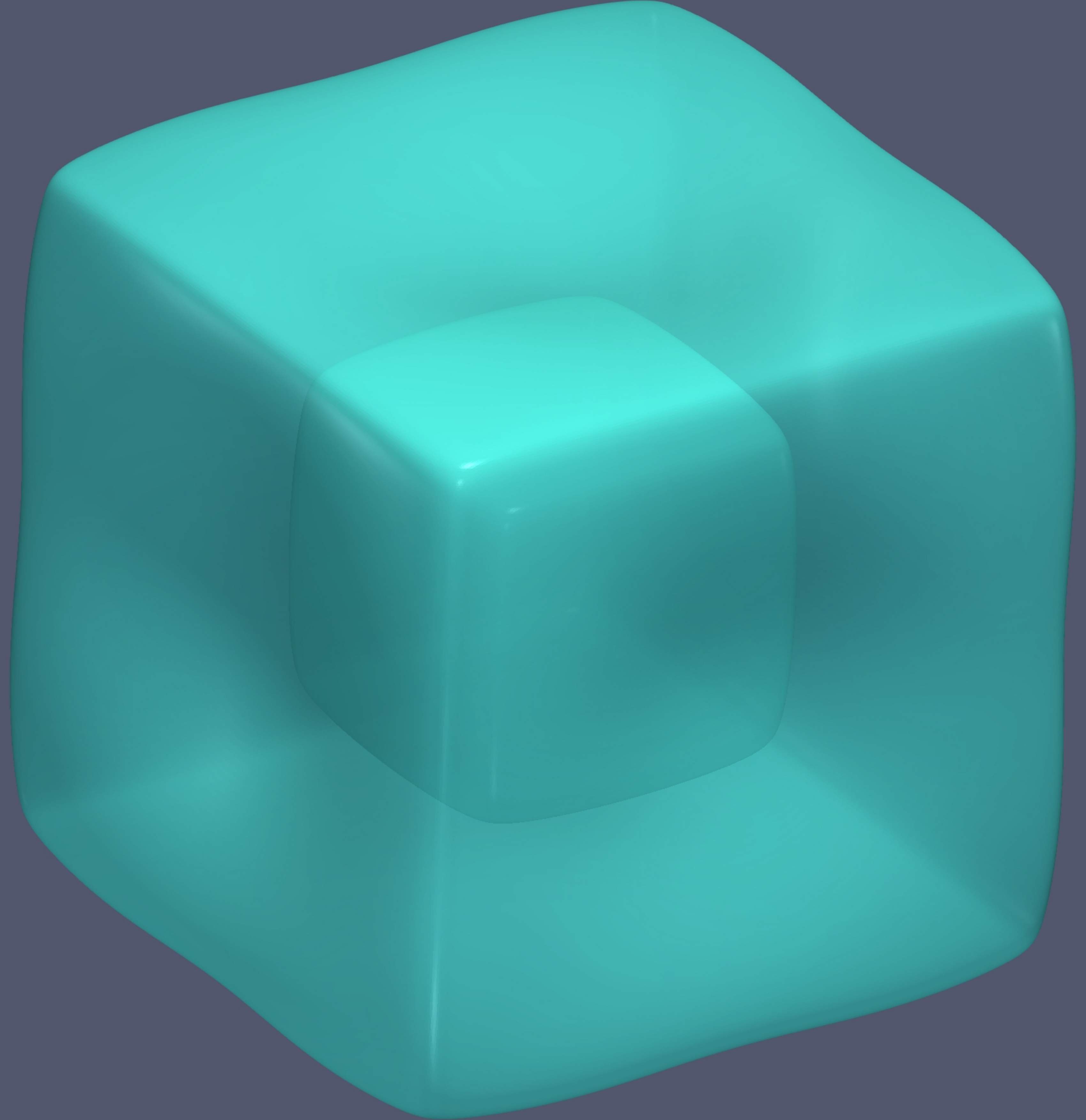}\\
    & & & &\\
    \hline
    & & & &\\
    Composition map &
    \includegraphics[width=2.5cm]{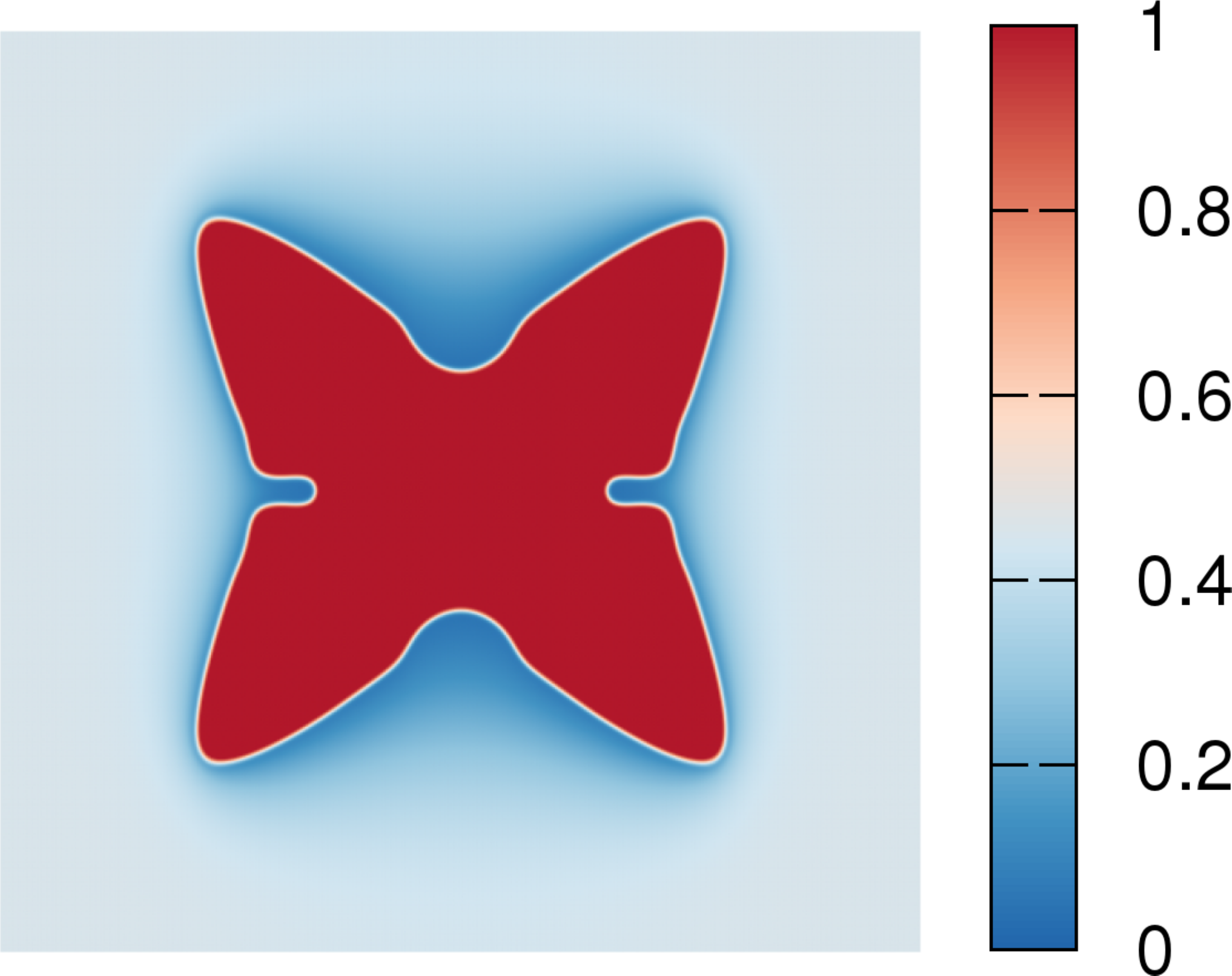} & 
    \includegraphics[width=2.5cm]{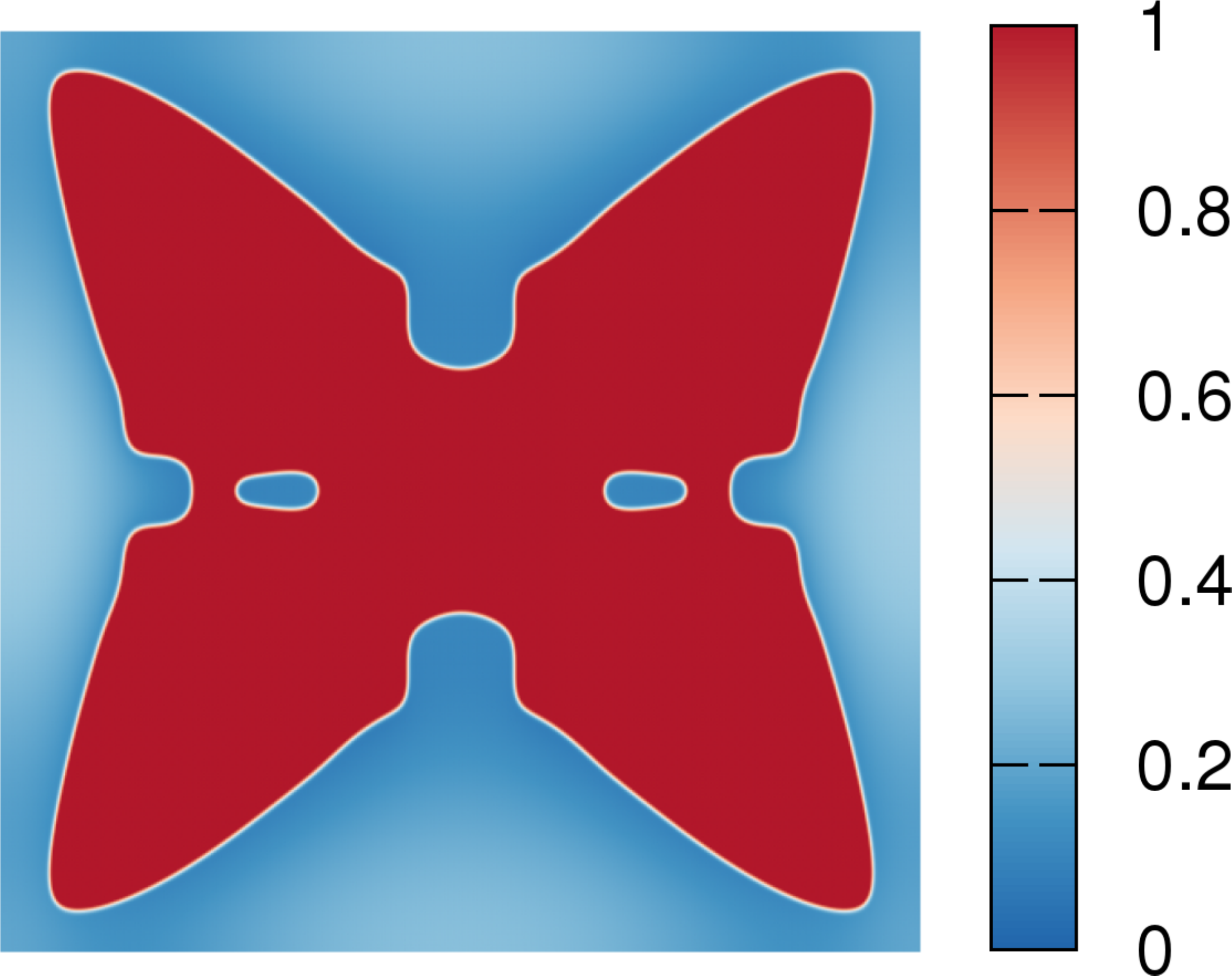} &
    \includegraphics[width=2.5cm]{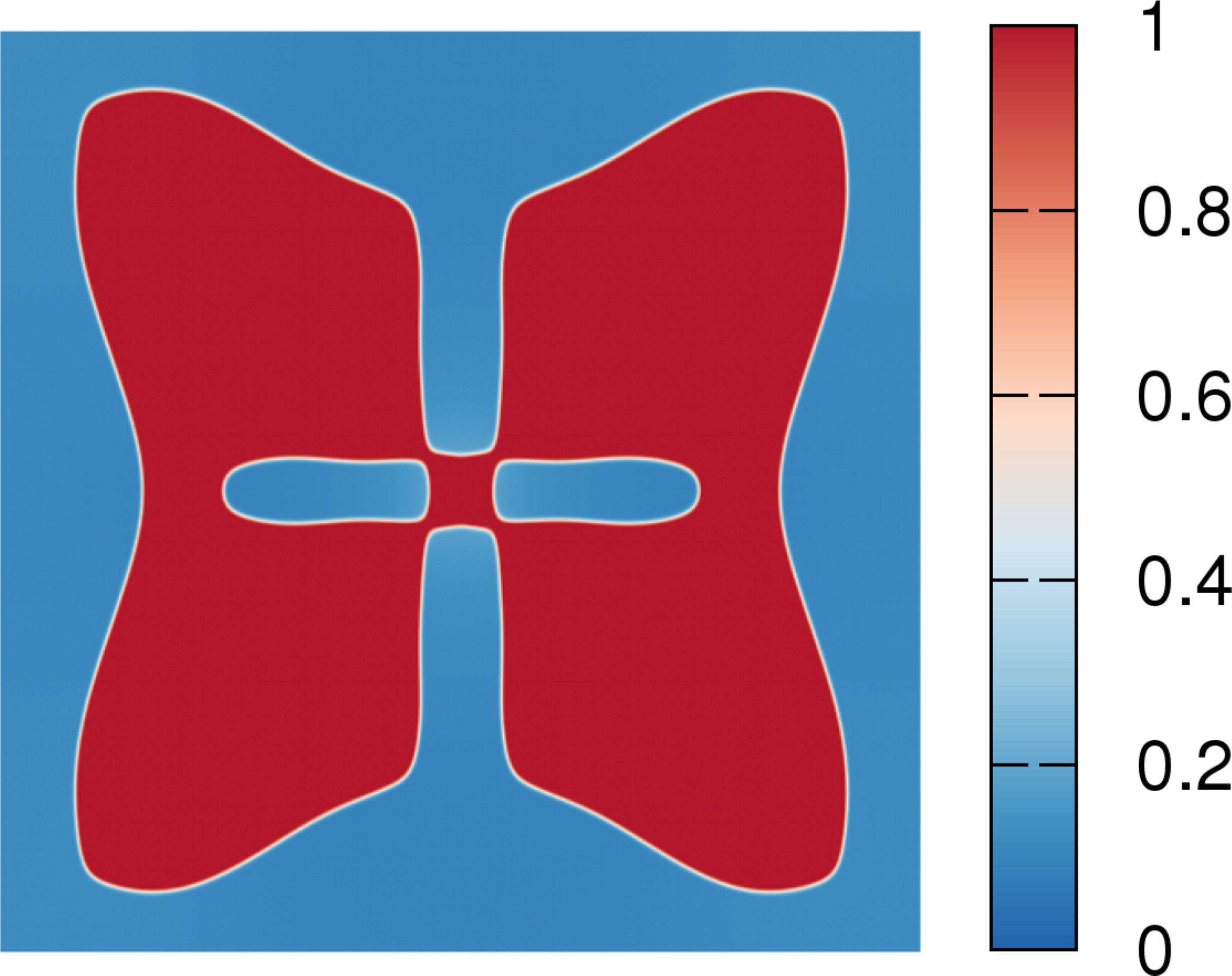} & 
    \includegraphics[width=2.5cm]{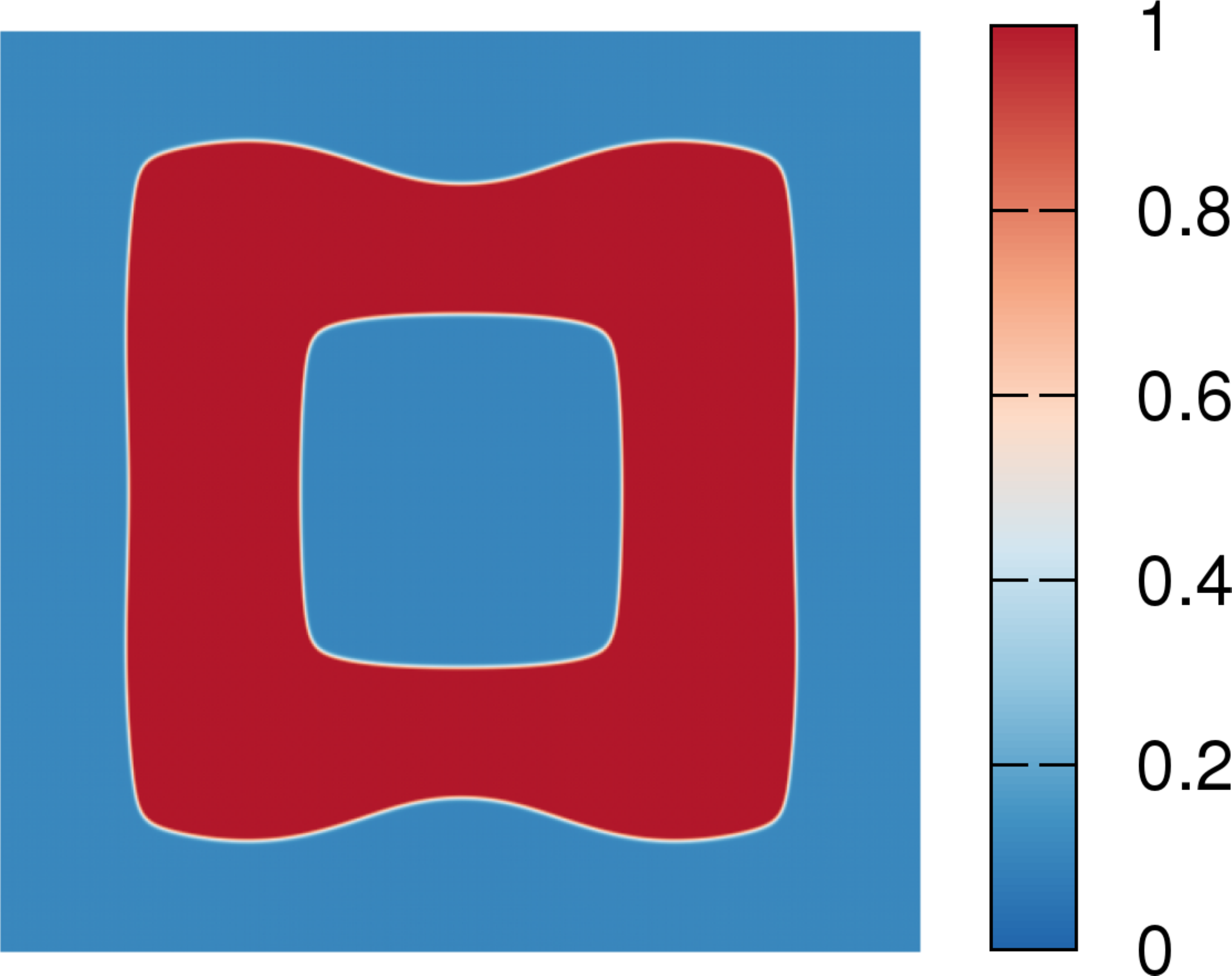}\\
    & & & &\\
    \hline
    & & & &\\
    Order Parameter map &
    \includegraphics[width=2.5cm]{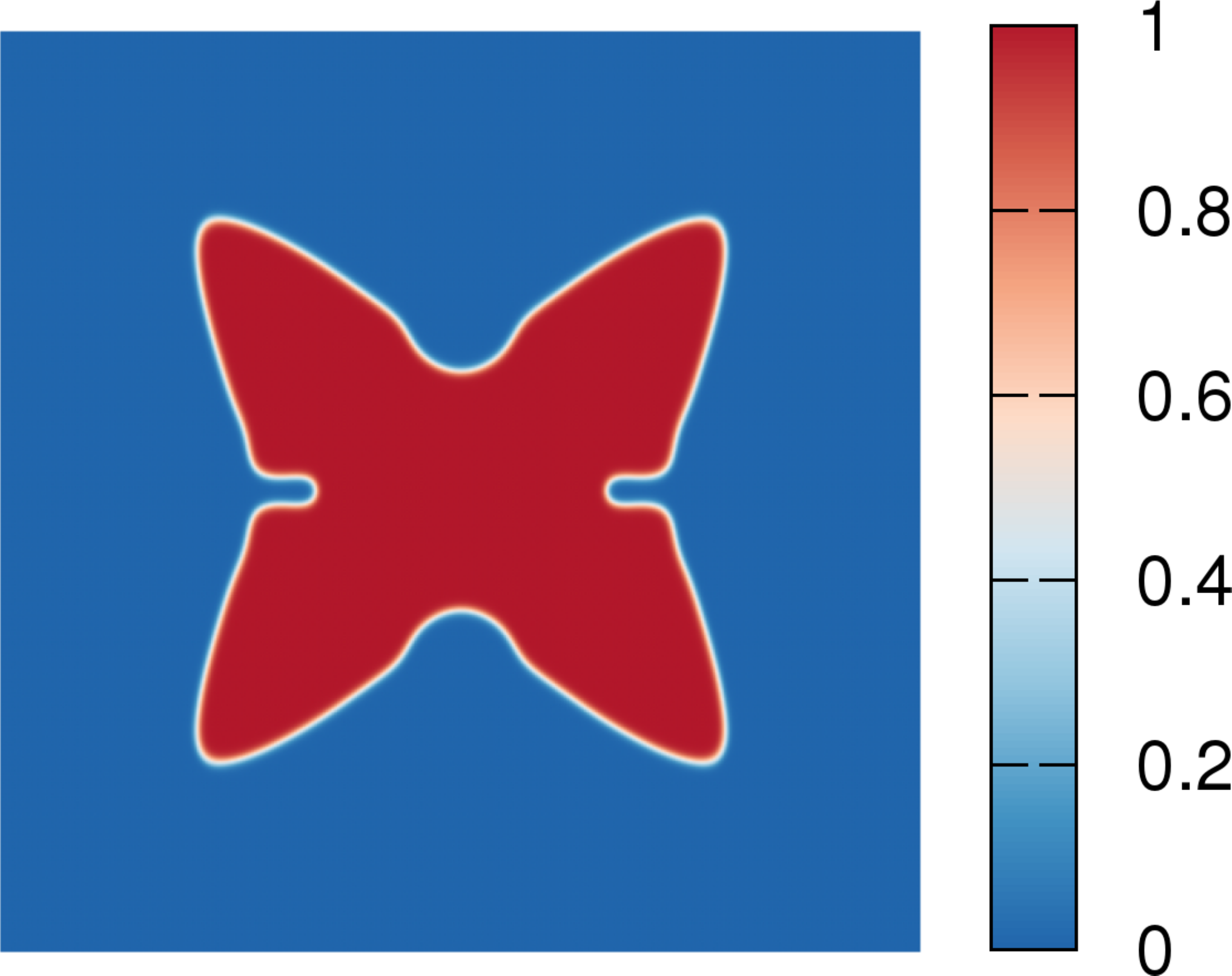} & 
    \includegraphics[width=2.5cm]{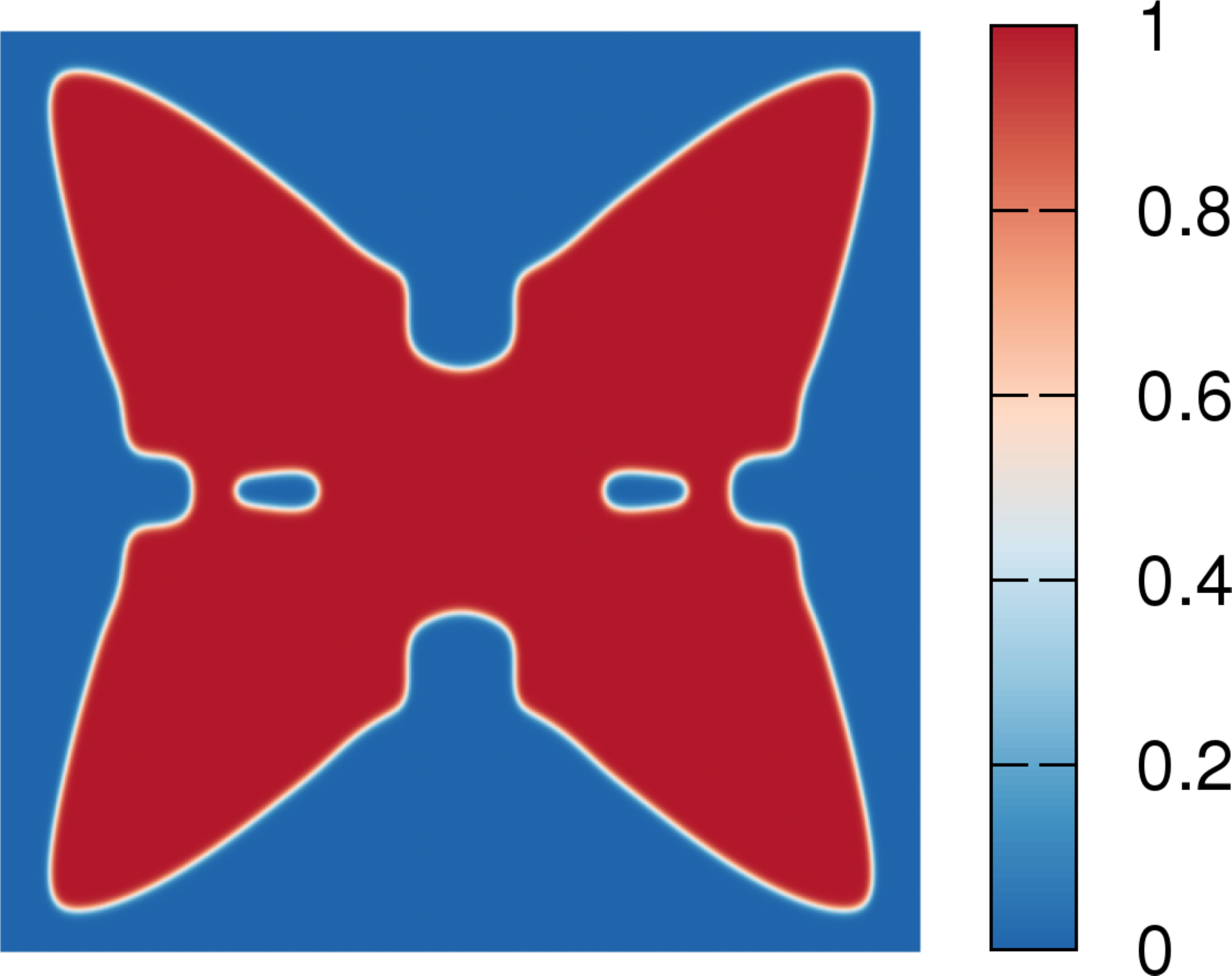} &
    \includegraphics[width=2.5cm]{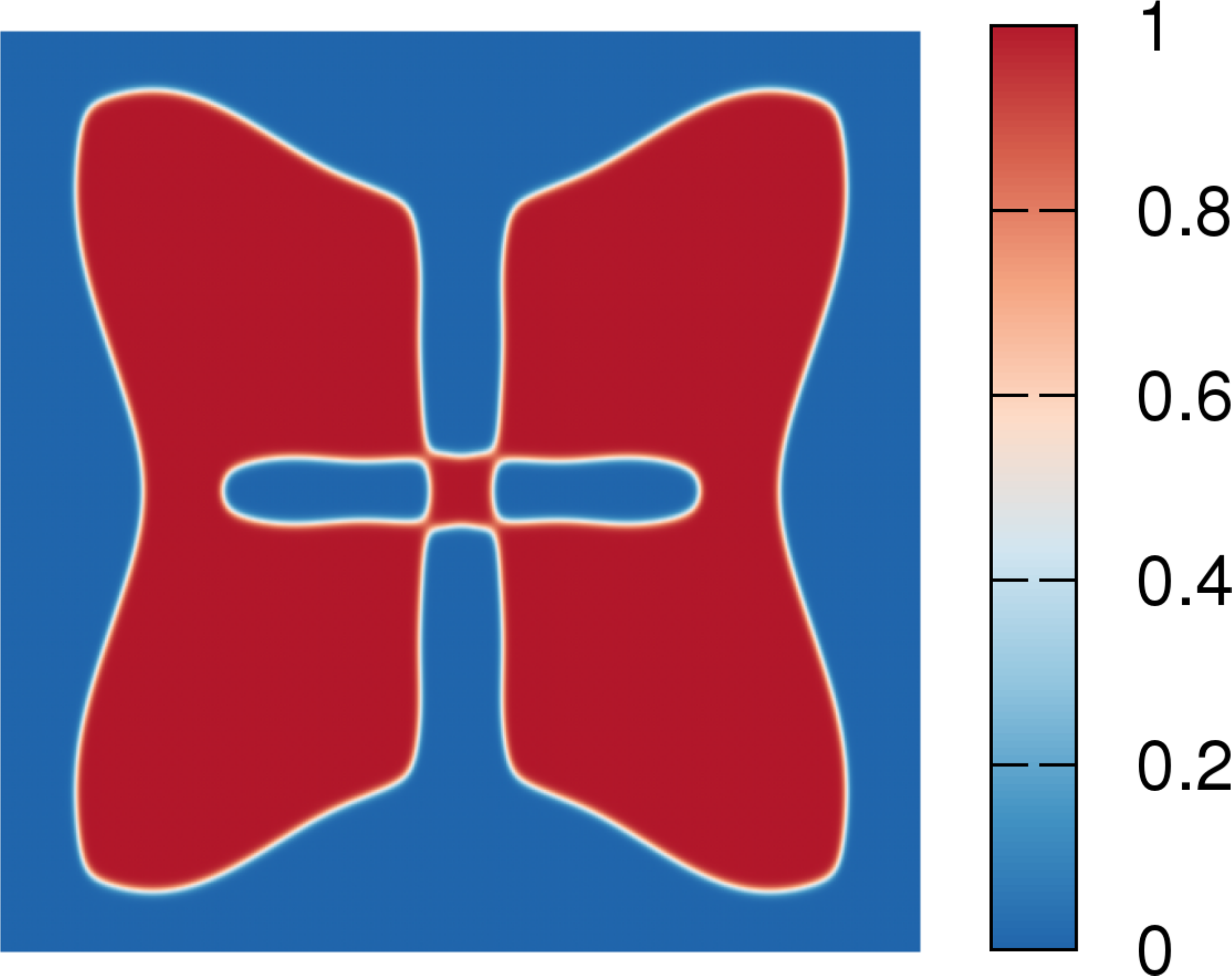} & 
    \includegraphics[width=2.5cm]{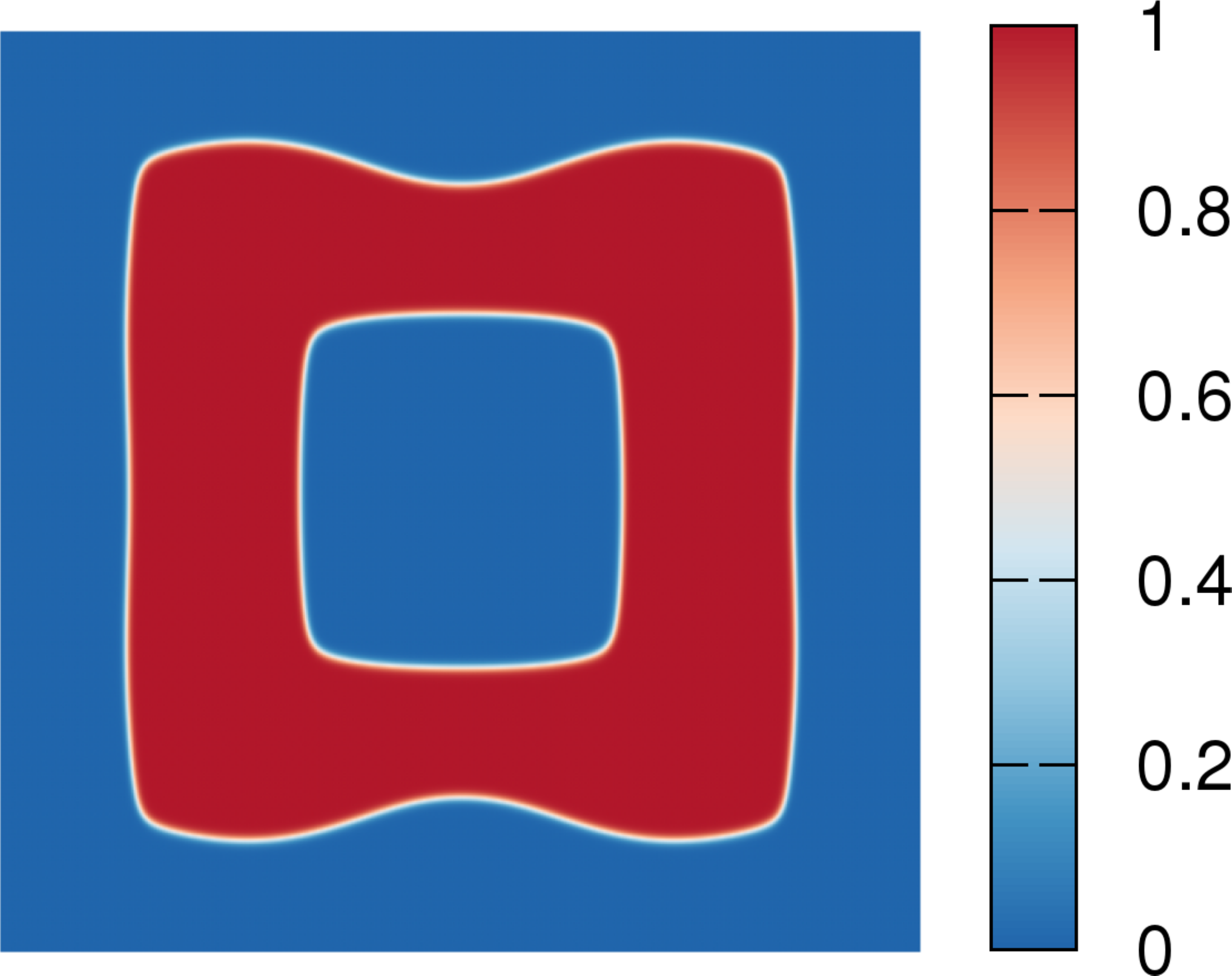}\\
    & & & &\\
    \hline
    \end{tabular}
    \caption{Temporal evolution of the precipitate morphology 
    at time $t = 200,\, 400,\, 1800,\, 10000$. First row shows the 
    isosurface representation of precipitate morphology. Second row 
    shows the composition map in $(110)$ plane passing through the 
    centre of the simulation box, and third row shows the corresponding 
    order parameter map. Here, $c_{\infty} = 45\%$ and 
    $\epsilon^* = 0.75\%$.}
    \label{tab:HollowPrecipitateEvolve}
\end{table}

\begin{figure}[!htb]
    \centering
    \begin{subfigure}{.5\textwidth}
    \centering
    \includegraphics[width=\linewidth]{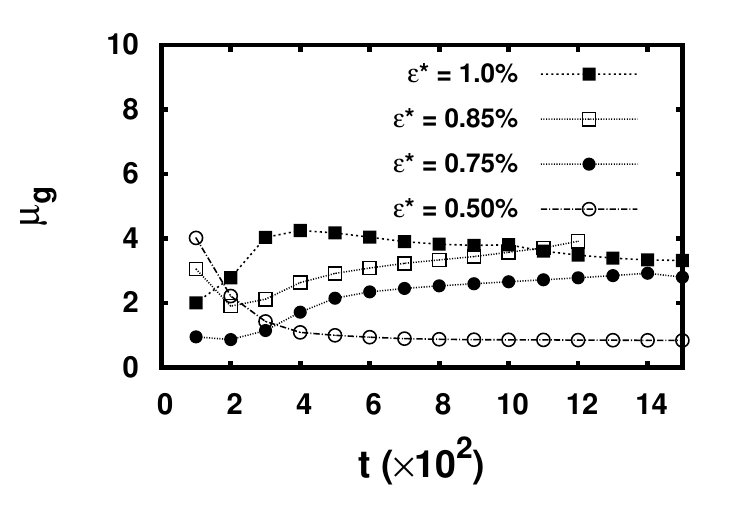}
    \caption{}
    \label{fig:mug_misfit_effect}
    \end{subfigure}%
    \begin{subfigure}{.5\textwidth}
    \centering
    \includegraphics[width=\linewidth]{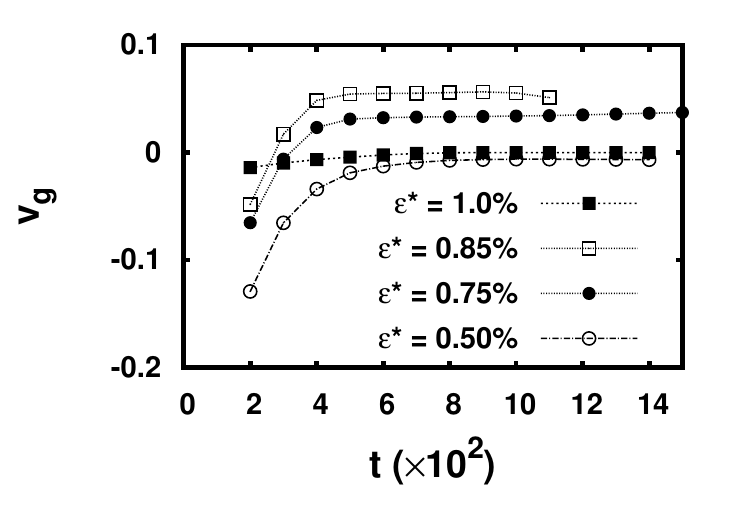}
    \caption{}
    \label{fig:vg_misfit_effect}
    \end{subfigure}
    \caption{(a) Temporal evolution of chemical potential ($\mu_{\textrm{g}}$) 
    of matrix ahead of interface along $[001]$ at different levels 
    of lattice misfit.  At higher lattice misfit 
    ($\epsilon^* = 0.85\%$, and $0.75\%$), $\mu_{\textrm{g}}$ continues to 
    increase once reached the lowest value. Whereas, at 
    $\epsilon^*=0.5\%$, $\mu_{\textrm{g}}$ continues to decrease. 
    The increase in $\mu_{\textrm{g}}$ suggests precipitate dissolution 
    along $[001]$. (b) Temporal evolution of velocity ($v_{\textrm{g}}$) 
    of the interface along $[001]$ direction for different levels of 
    lattice misfit. At lattice misfit of 0.5\%, $v_{\textrm{g}}$ 
    have negative steady state values, whereas other cases have
    positive steady state values. The positive values of $v_{\textrm{g}}$
    suggest movement of the interface towards the center of the 
    precipitate. Here, Zener anisotropy parameter is 4 and 
    supersaturation is 45\%.}
\end{figure}

Fig.~\ref{fig:mug_misfit_effect} shows the temporal 
evolution of the $\mu_{\textrm{g}}$ along $[001]$ direction 
for different lattice misfits $0.5\%$, $0.75\%$, and 
$0.85\%$ at $c_{\infty} = 45\%$. 
Fig.~\ref{fig:vg_misfit_effect} 
depicts the temporal evolution of $v_{\textrm{g}}$ along 
$[001]$ direction corresponding to the same levels of 
lattice misfit. The values of $\mu_{\textrm{g}}$ initially 
decrease at all levels of lattice misfit, however, for 
higher misfit ($\epsilon^* = 0.85\%$, and $0.75\%$) it 
continues to increase further with time. At the lattice 
misfit $0.5\%$, $\mu_{\textrm{g}}$ continues to decrease 
in contrast to the cases of lattice misfits $0.75\%$ and $0.85\%$. An 
increment in $\mu_{\textrm{g}}$ suggests precipitate 
dissolution along $[001]$ direction. For all the levels of 
lattice misfit, the $v_{\textrm{g}}$ achieves a steady 
state. However, higher lattice misfit 
($\epsilon^* = 0.85\%$ and $0.75\%$) have positive values 
of velocity at a steady state suggesting the movement of 
the interface towards the center of the precipitate. Whereas, 
in the case of lattice misfit $0.5\%$, the steady state 
value of velocity is negative which suggests the movement 
of the interface away from the centre of 
the precipitate along $[001]$ direction.

\begin{figure}[!htb]
    \centering
    \includegraphics[width=0.75\linewidth]{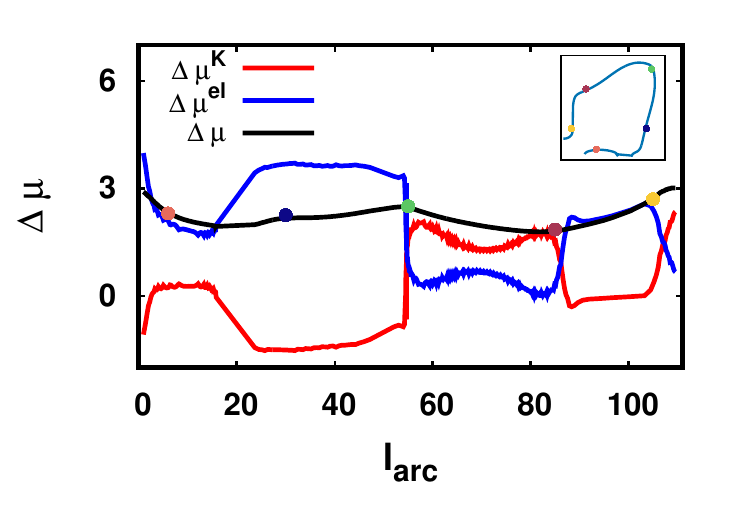}
    \caption{Variation of shift in chemical potential due to
    curvature ($\Delta \mu^{\textrm{K}}$) and elastic stresses 
    ($\Delta \mu^{\textrm{el}}$) as a function of arc-length at time $t = 1400$.
    Here, supersaturation is 45\% and lattice misfit is 0.75\%}
    \label{fig:mu_shift_eps_0.75}
\end{figure}

\begin{figure}[!htb]
    \centering
    \includegraphics[width=0.75\linewidth]{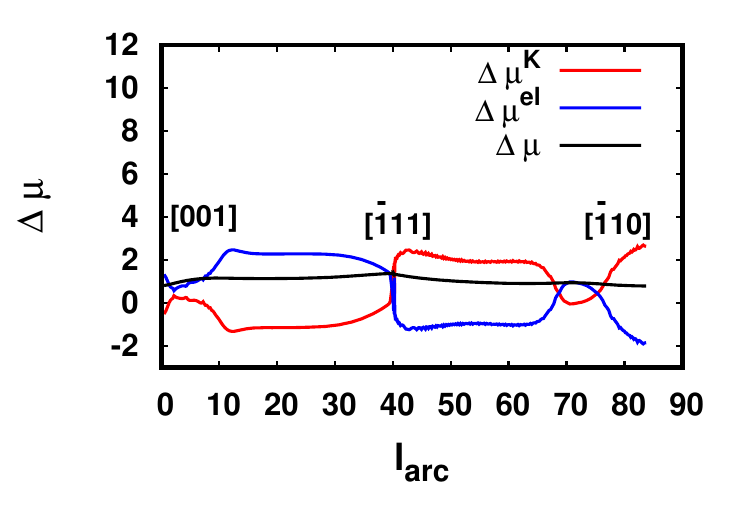}
    \caption{Variation of shift in chemical potential due to
    curvature ($\Delta \mu^{\textrm{K}}$) and elastic stresses 
    ($\Delta \mu^{\textrm{el}}$) as a function of arc-length at time $t = 1400$. 
    Here, supersaturation is 45\% and lattice misfit is 0.5\%}
    \label{fig:mu_shift_eps_0.5}
\end{figure}

Moreover, we plot the variation of elastic stress 
and curvature contribution to the shift in chemical 
potential as a function of arc-length. 
Figs.~\ref{fig:mu_shift_eps_0.75} 
and~\ref{fig:mu_shift_eps_0.5} 
represent the variation of $\Delta \mu^{\mathrm{el}}$ 
and $\Delta \mu^{\mathrm{K}}$ 
with arc-length for lattice misfits of 0.75\% and 0.5\% at 
simulation time $t = 1400$. Similar to the case of lattice 
misfit of $0.85\%$ (see Fig.~\ref{subfig:larc_vs_deltamu}), 
the values of $\Delta \mu^{\mathrm{el}}$ are higher in the 
region left to the red point and right to the yellow point 
(i.e. grooves oriented along $\langle 100 \rangle$ and
$\langle 110 \rangle$). On the hand, 
for the case of lattice misfit of 0.5\%, the $\Delta \mu$ 
value is the highest near the green dot due to which 
concavities present along $\langle 100 \rangle$ and 
$\langle 110 \rangle$ directions tend to vanish. 
With the decrease in misfit, the 
contribution of elastic stress to the shift in chemical 
potential decreases and thereby the point effect of 
diffusion subsides. As a result, 
the capillary forces kick in early during the 
evolution of precipitate morphology, and 
precipitate restores the cuboidal 
morphology by removing the concavities.

\subsection{Effects of anisotropy in elastic energy}
\label{sec:elast_anisotropy}
In this section, we discuss the effects of anisotropy 
in elastic energy on the initiation of particle splitting 
instability. We choose Zener anisotropy parameter of 
$2$, $3$, and $4$, lattice misfit of $0.5\%$, $0.75\%$, 
and $0.85\%$, and supersaturation of $45\%$. 
Table~\ref{tab:effect_of_Az} shows the precipitate 
morphologies at a same simulation time of $1000$ for
different levels of lattice misfit and Zener anisotropy 
parameter. At all combinations of lattice misfit and Zener 
anisotropy parameter, the dendritic instability grows, 
however, higher lattice misfit promotes particle splitting. 


\begin{table}[!htb]
    \centering
    \begin{tabular}{| M{1.75cm} | M{3cm}  M{3cm}  M{3cm} | }
         \hline
         & $A_{\textrm{z}} = 2$ & $A_{\textrm{z}} = 3$ & $A_{\textrm{z}} = 4$\\
         \hline
         & & & \\
         $\epsilon^* = 0.5\%$ & 
         \includegraphics[width=3cm]{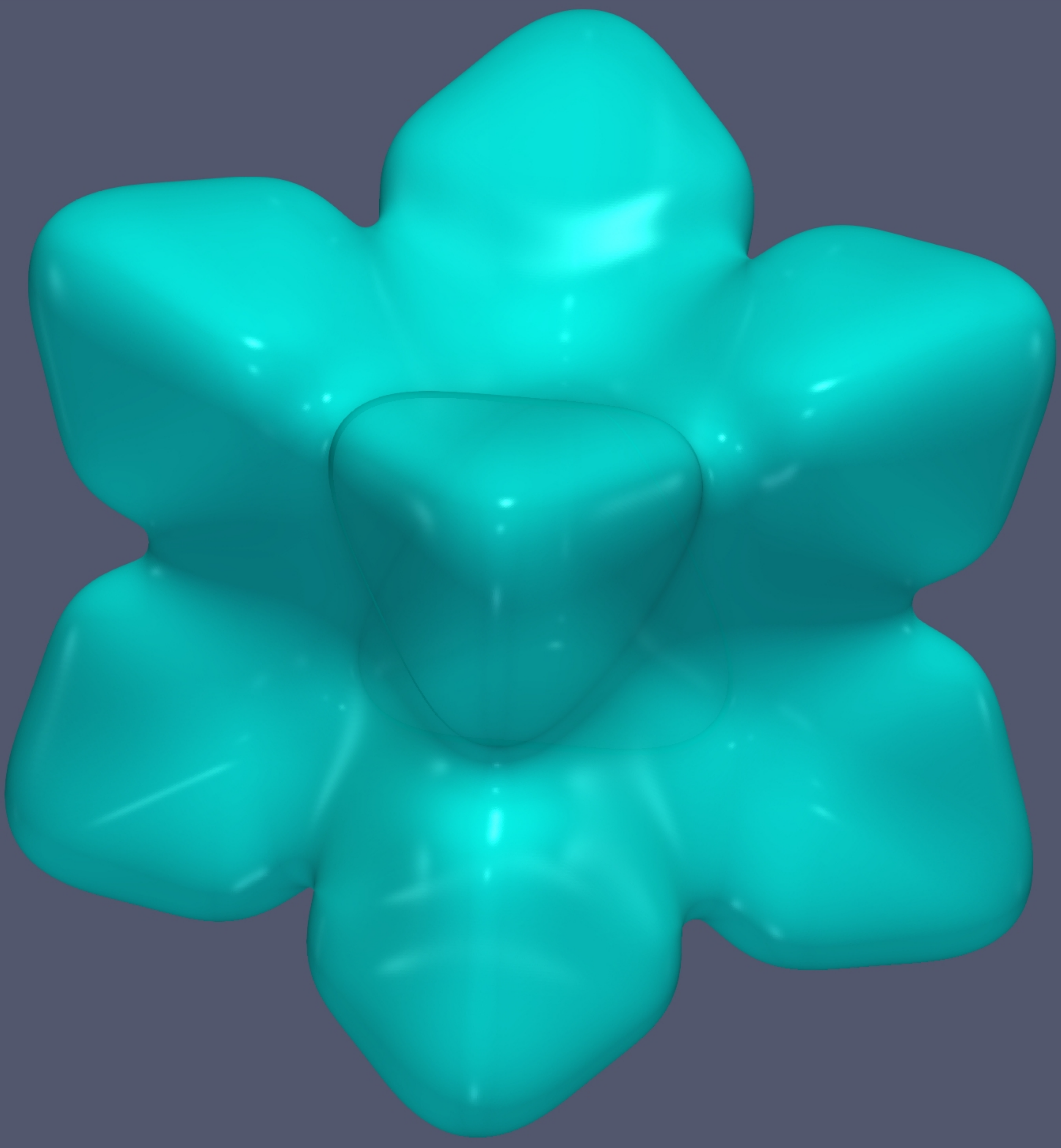}&
         \includegraphics[width=3cm]{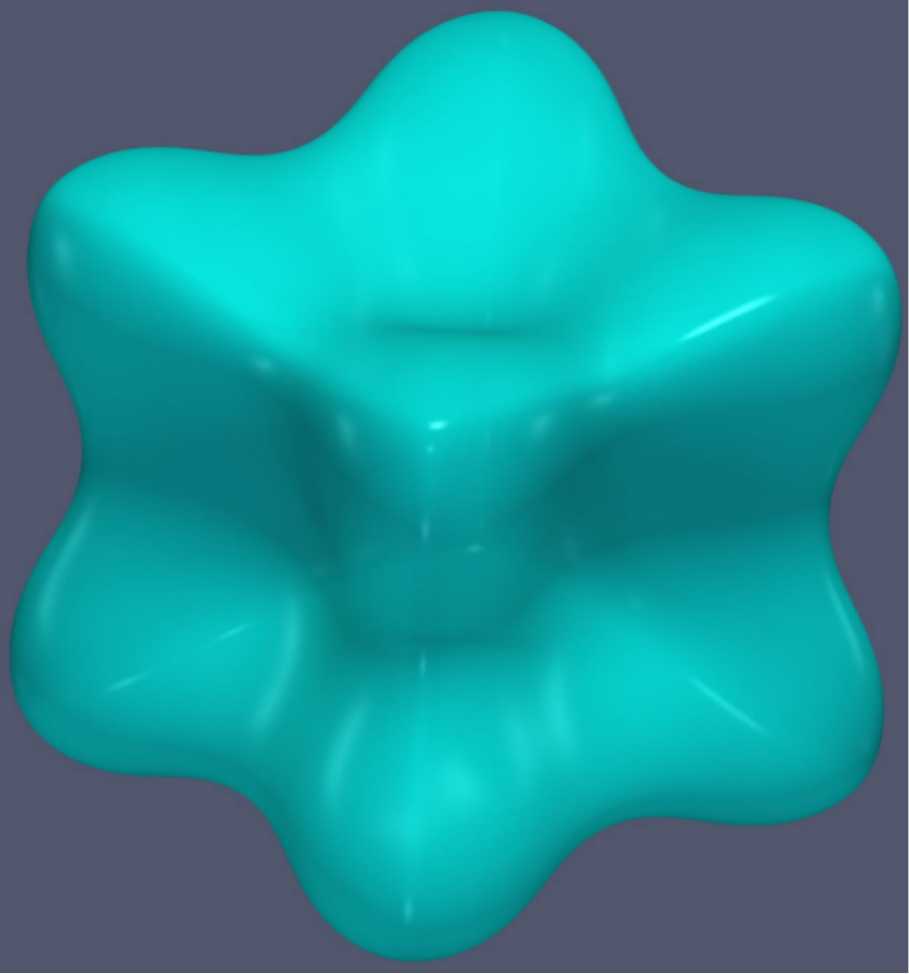}& 
         \includegraphics[width=3cm]{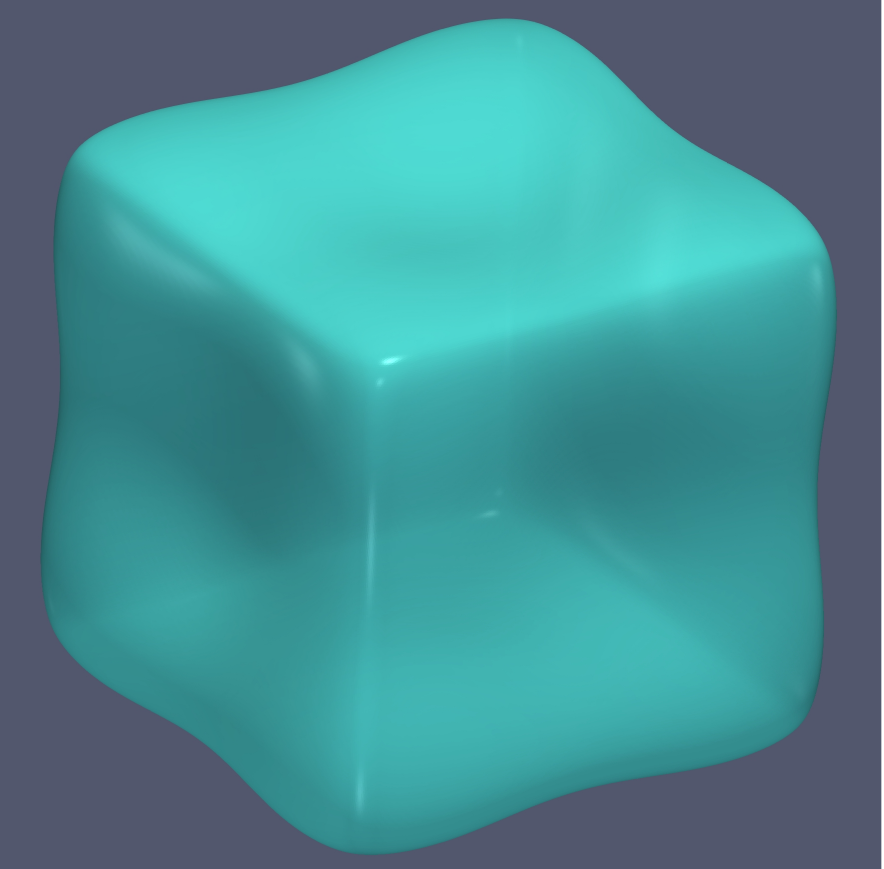}\\
         $\epsilon^* = 0.75\%$ & 
         \includegraphics[width=3cm]{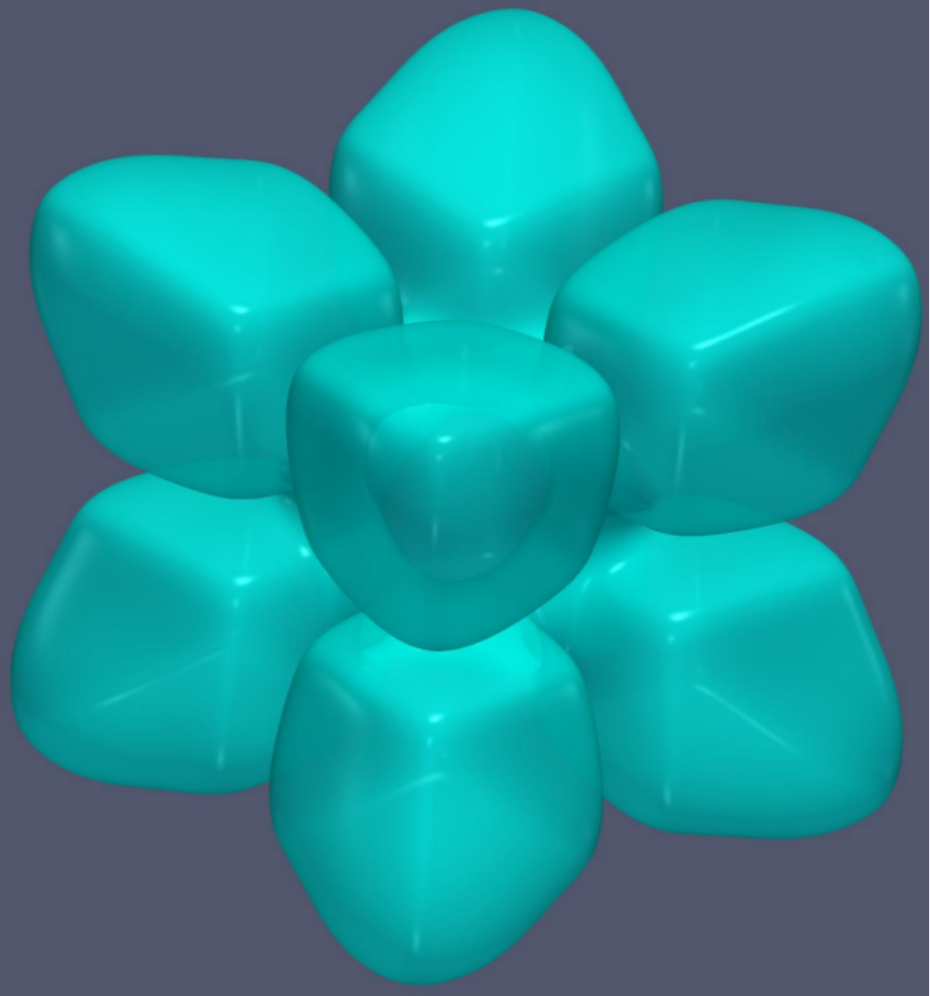}&
         \includegraphics[width=3cm]{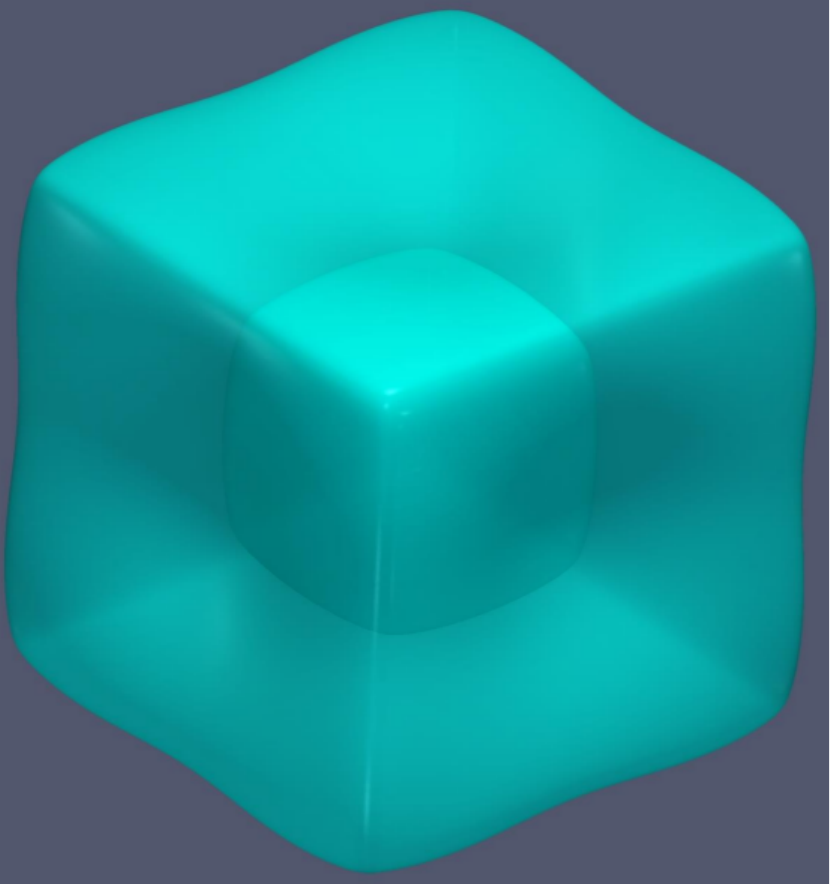}& 
         \includegraphics[width=3cm]{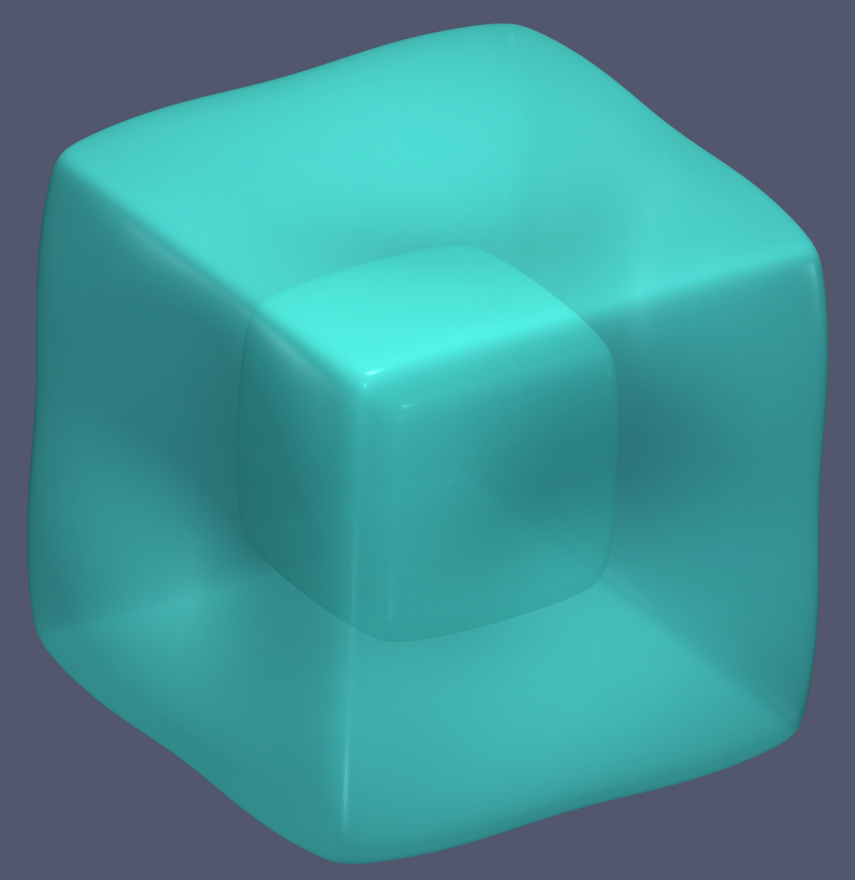}\\
         $\epsilon^* = 0.85\%$ & 
         \includegraphics[width=3cm]{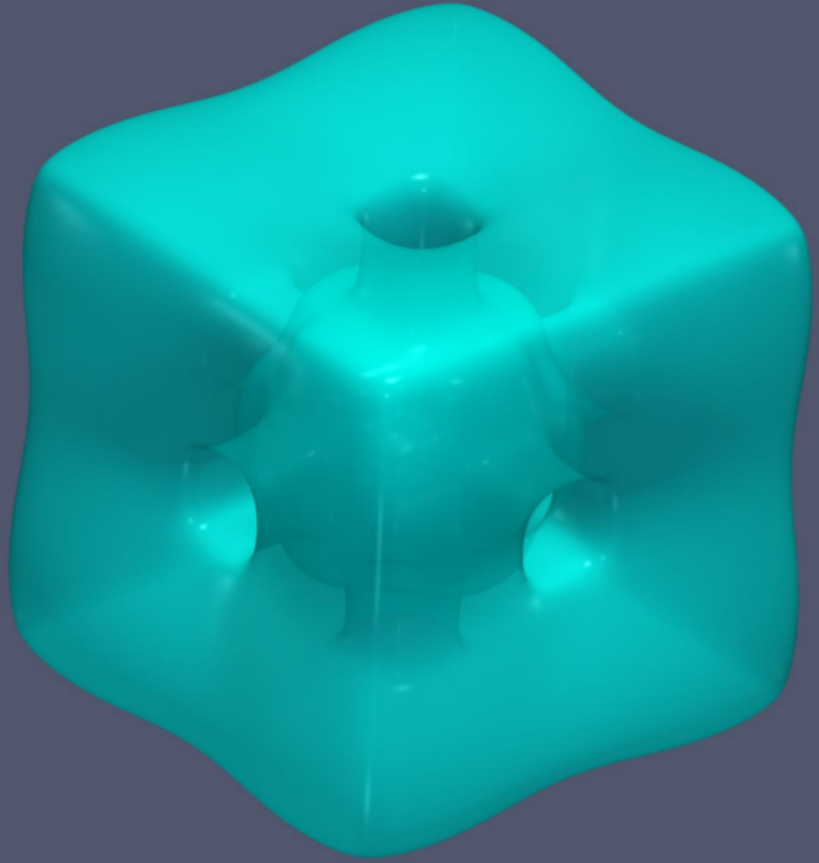}&
         \includegraphics[width=3cm]{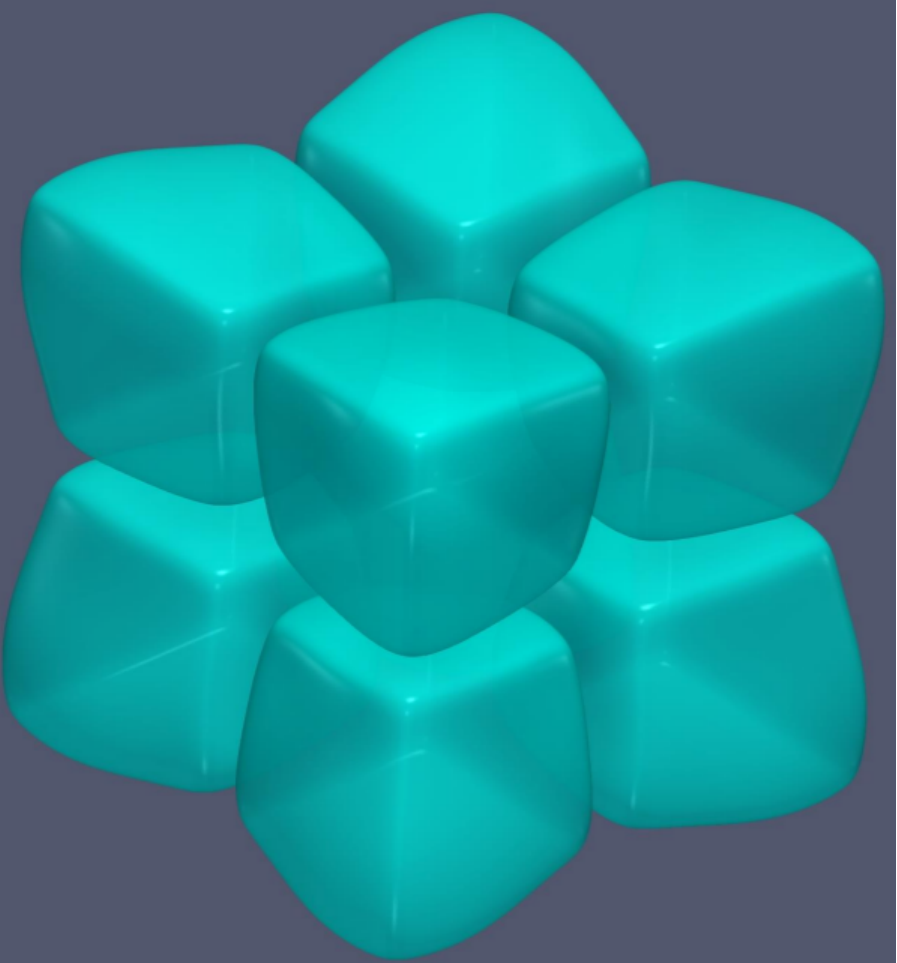}& 
         \includegraphics[width=3cm]{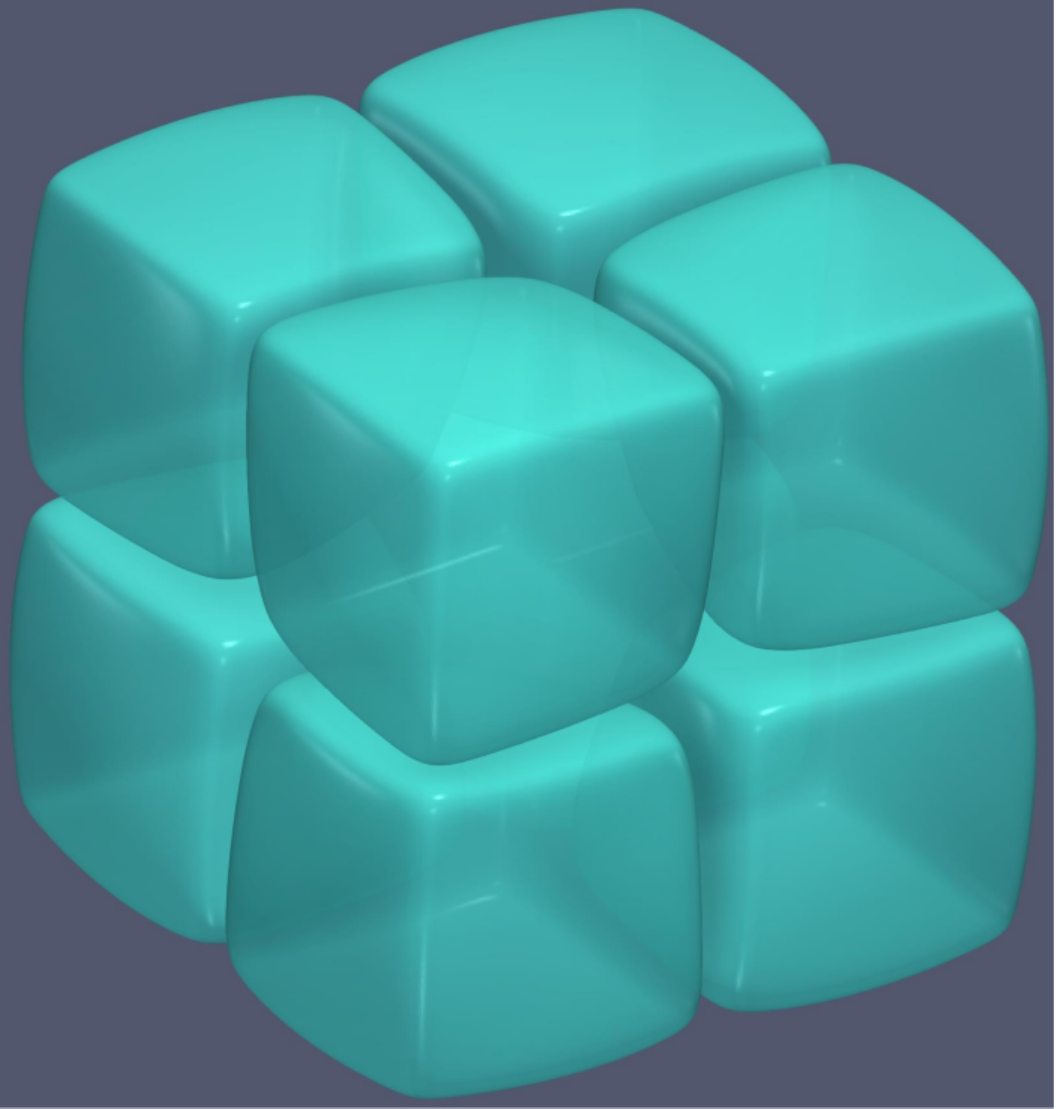}\\
         \hline
    \end{tabular}
    \caption{Precipitate morphologies for different levels of
    elastic misfit $\epsilon^* = 0.5\%, 0.75\%, 0.85\%$ and 
    elastic energy anisotropy $A_{\textrm{z}} = 2$, $3$, $4$ at time $t = 10000$.
    Here, supersaturation is 45\%. A higher value of lattice misfit 
    promotes splitting instability}
    \label{tab:effect_of_Az}
\end{table}

At a lattice misfit of 0.85\%, for $A_{\textrm{z}} = 4.0$ and $3.0$, the 
precipitate show split patterns, whereas for $A_{\textrm{z}} = 2.0$, 
hollow cuboidal precipitate forms. As discussed in the previous 
section, at $A_{\textrm{z}} = 2.0$, the precipitate coalesces along 
$\langle 110 \rangle$ directions; matrix phase gets trapped within 
precipitate phase and further pinch-off takes place. 

\begin{figure}[!htb]
    \centering
    \begin{subfigure}{0.5\textwidth}
    \includegraphics[width=\linewidth]{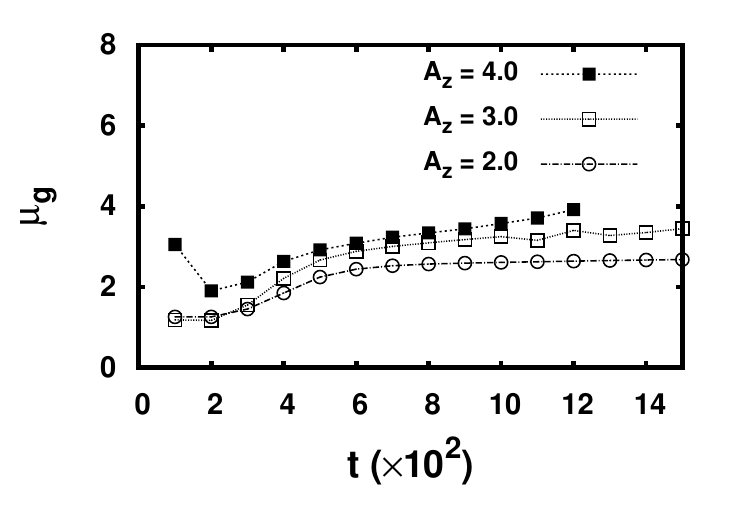}
    \caption{}
    \label{fig:mug_Az_effect}
    \end{subfigure}%
    \begin{subfigure}{0.5\textwidth}
    \includegraphics[width=\linewidth]{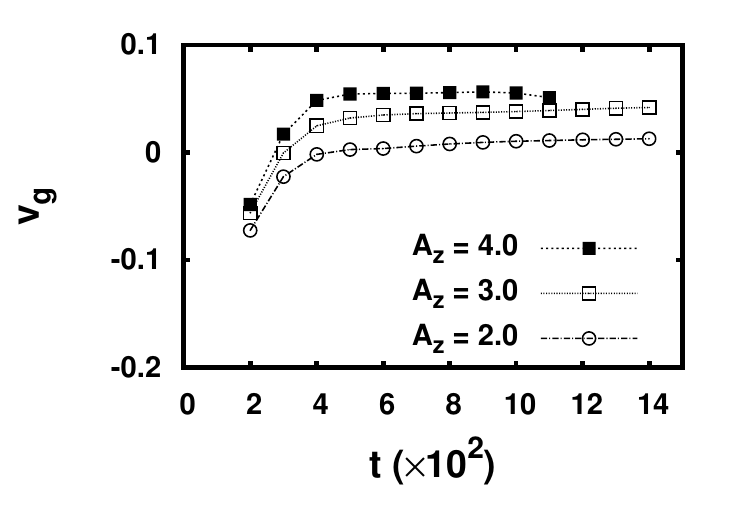}
    \caption{}
    \label{fig:vg_Az_effect}
    \end{subfigure}
    \caption{(a) Temporal evolution of chemical potential ($\mu_{\textrm{g}}$) 
    of matrix ahead of interface along $[001]$ at different levels 
    of Zener anisotropy parameter. At all levels of Zener anisotropy parameter, 
    $\mu_{\textrm{g}}$ tends to increase. The increase in $\mu_{\textrm{g}}$ suggests 
    precipitate dissolution along $[001]$ direction. (b)Temporal evolution of velocity 
    ($v_{\textrm{g}}$) of the interface along $[001]$ direction for different levels degree of 
    elastic energy anisotropy.  At all levels of Zener anisotropy parameter,
    $v_{\textrm{g}}$ achieve a positive steady state value. The positive values of 
    $v_{\textrm{g}}$ suggest movement of the interface towards the center of the 
    precipitate. Here, lattice misfit is 0.85\% and 
    supersaturation is 45\%.}
    
\end{figure}

Figs.~\ref{fig:mug_Az_effect} and~\ref{fig:vg_Az_effect} 
represents the temporal evolution of $\mu_{\textrm{g}}$ and $v_{\textrm{g}}$ 
along $[001]$ direction respectively. The values of 
$\mu_{\textrm{g}}$ tends to increase with time, whereas 
the values of $v_{\textrm{g}}$ achieve a positive steady state 
value for all levels of Zener anisotropy parameter at 
supersaturation of 45\% and lattice misfit of 0.85\%. 
As discussed in the previous section, the temporal 
increment in $\mu_{\textrm{g}}$ suggests precipitate dissolution 
along the $[001]$ direction.

    

\begin{figure}[!htb]
    \centering
    \includegraphics[width=0.6\linewidth]{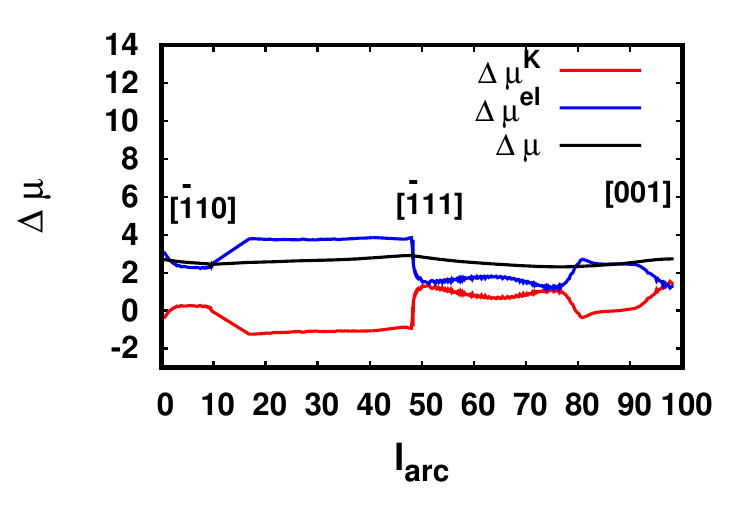}
    \caption{Variation of shift in chemical potential due to
    curvature ($\Delta \mu^{\textrm{K}}$) and elastic stresses 
    ($\Delta \mu^{\textrm{el}}$) as a function of arc-length 
    at time $t = 1400$ for $A_{\textrm{z}} = 3.0$. Here, supersaturation 
    is 45\% and lattice misfit is 0.85\%.}
    \label{fig:mu_shift_Az3}
\end{figure}

\begin{figure}[!htb]
    \centering
    \includegraphics[width=0.6\linewidth]{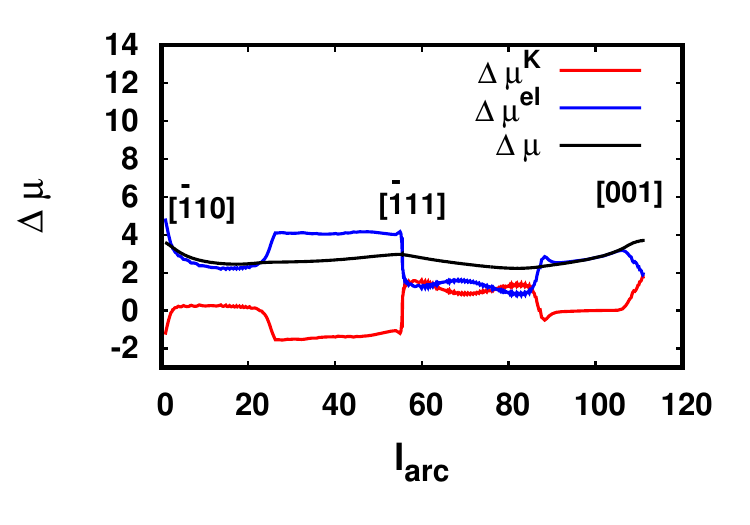}
    \caption{Variation of shift in chemical potential due to
    curvature ($\Delta \mu^{\textrm{K}}$) and elastic stresses 
    ($\Delta \mu^{\textrm{el}}$) as a function of arc-length at time $t = 1400$ for 
    $A_{\textrm{z}} =2.0$. Here, supersaturation is 45\% and lattice misfit is 0.5\%}
    \label{fig:mu_shift_Az2}
\end{figure}

Figs.~\ref{fig:mu_shift_Az3} and~\ref{fig:mu_shift_Az2} 
shows the variation of shift in chemical potential 
$\Delta \mu$ for $A_{\textrm{z}} = 3$ and $A_{\textrm{z}} = 2$ 
respectively at a same simulation time of $1400$. At $A_{\textrm{z}} = 3.0$, 
the shift in chemical potential are higher at regions along $[\bar{1}10]$,
$[\bar{1}11]$, and $[001]$ directions. Similar variation is 
observed for the case of $A_{\textrm{z}} = 2.0$.

\subsection{Effects of particle interactions}
\label{sec:part_interactions}
\begin{figure}[!htb]
    \centering
    \begin{subfigure}{0.33\textwidth}
    \centering
    \fbox{\includegraphics[width=3.25cm]{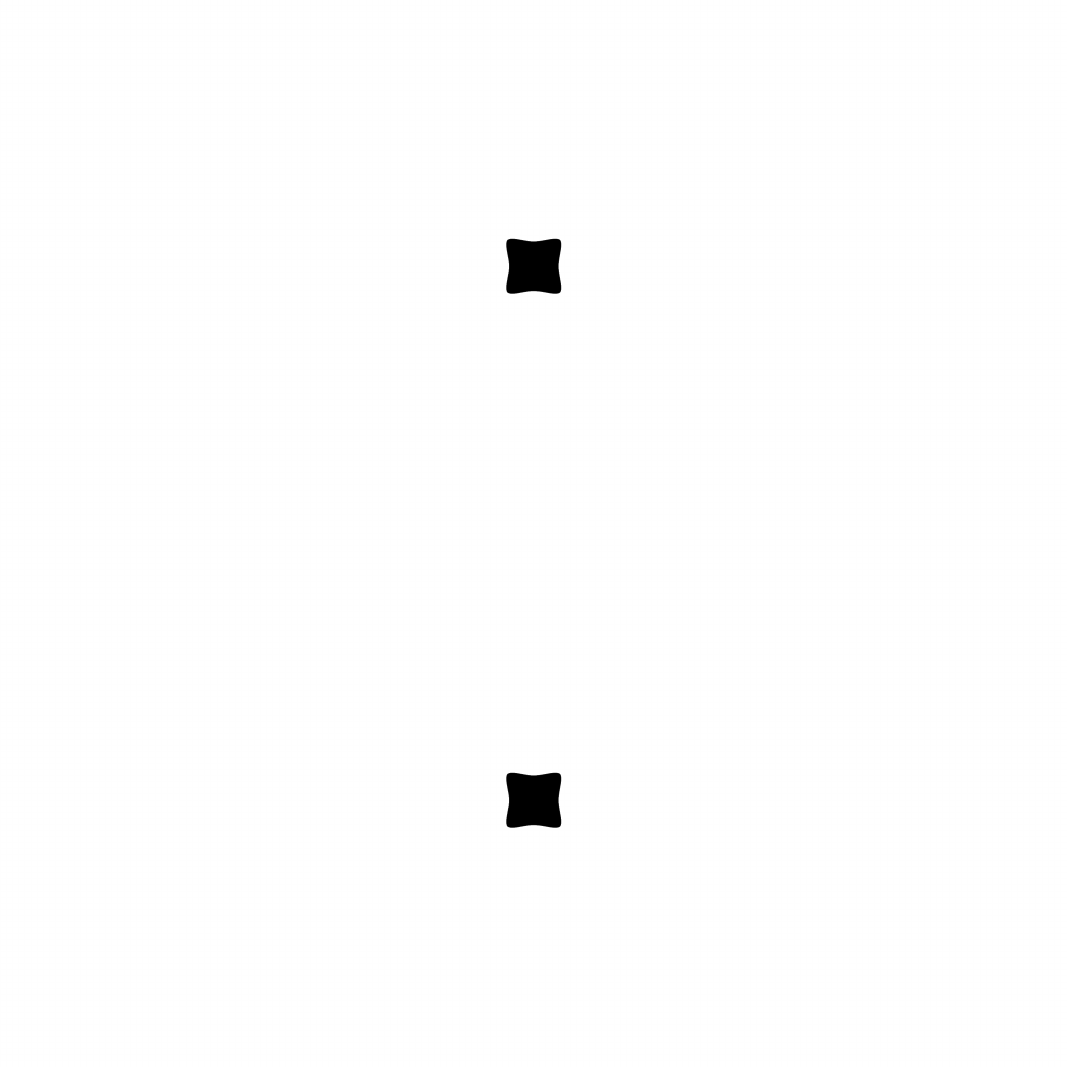}}
    \caption{t=1}
    \label{subfig:two-particle_t1}
    \end{subfigure}%
    \begin{subfigure}{0.33\textwidth}
    \centering
    \fbox{\includegraphics[width=3.25cm]{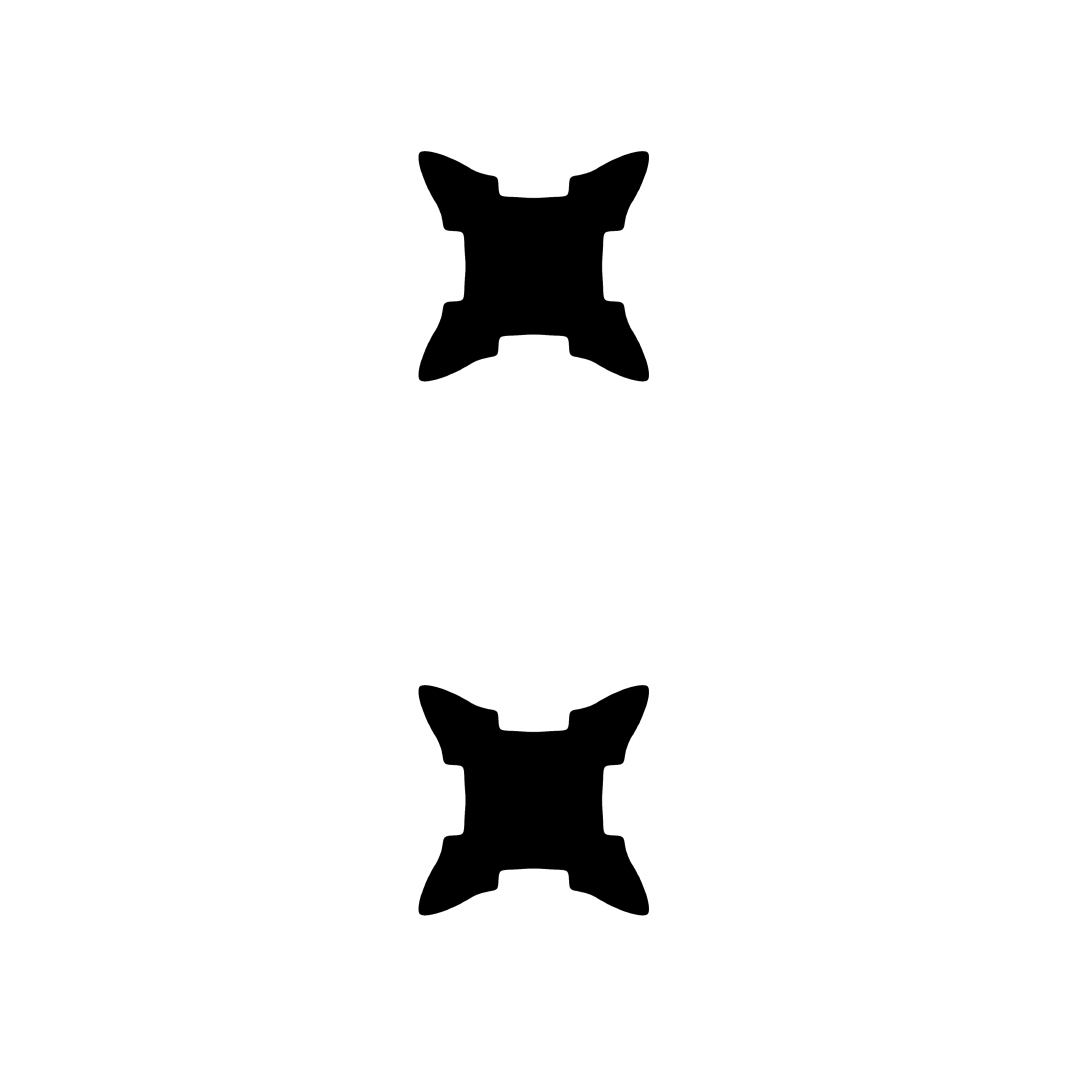}}
    \caption{t=10000}
    \label{subfig:two-particle_t10}
    \end{subfigure}%
    \begin{subfigure}{0.33\textwidth}
    \centering
    \fbox{\includegraphics[width=3.25cm]{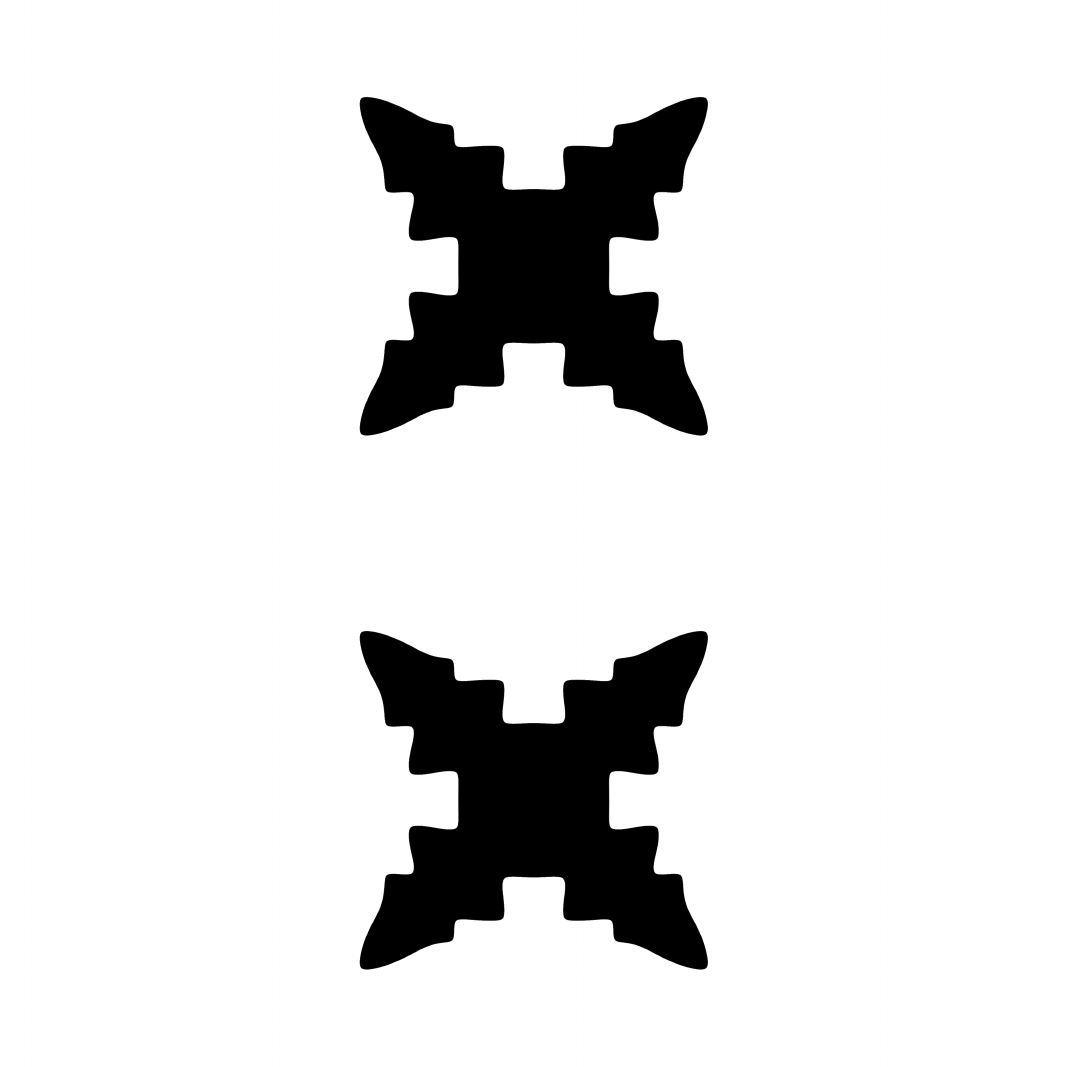}}
    \caption{t=20000}
    \label{subfig:two-particle_t20}
    \end{subfigure}
    
    \begin{subfigure}{0.33\textwidth}
    \centering
    \fbox{\includegraphics[width=3.25cm]{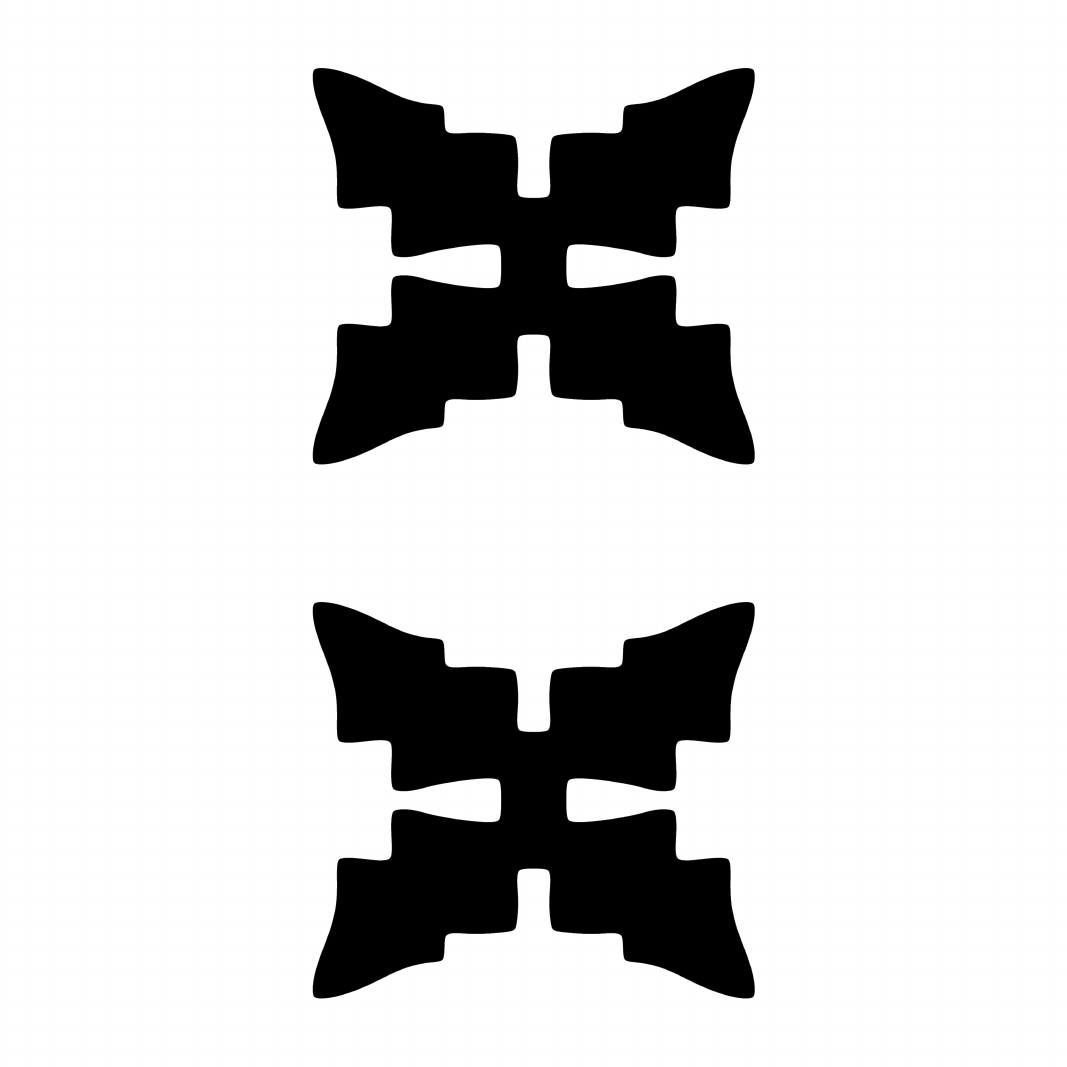}}
    \caption{t=40000}
    \label{subfig:two-particle_t40}
    \end{subfigure}%
    \begin{subfigure}{0.33\textwidth}
    \centering
    \fbox{\includegraphics[width=3.25cm]{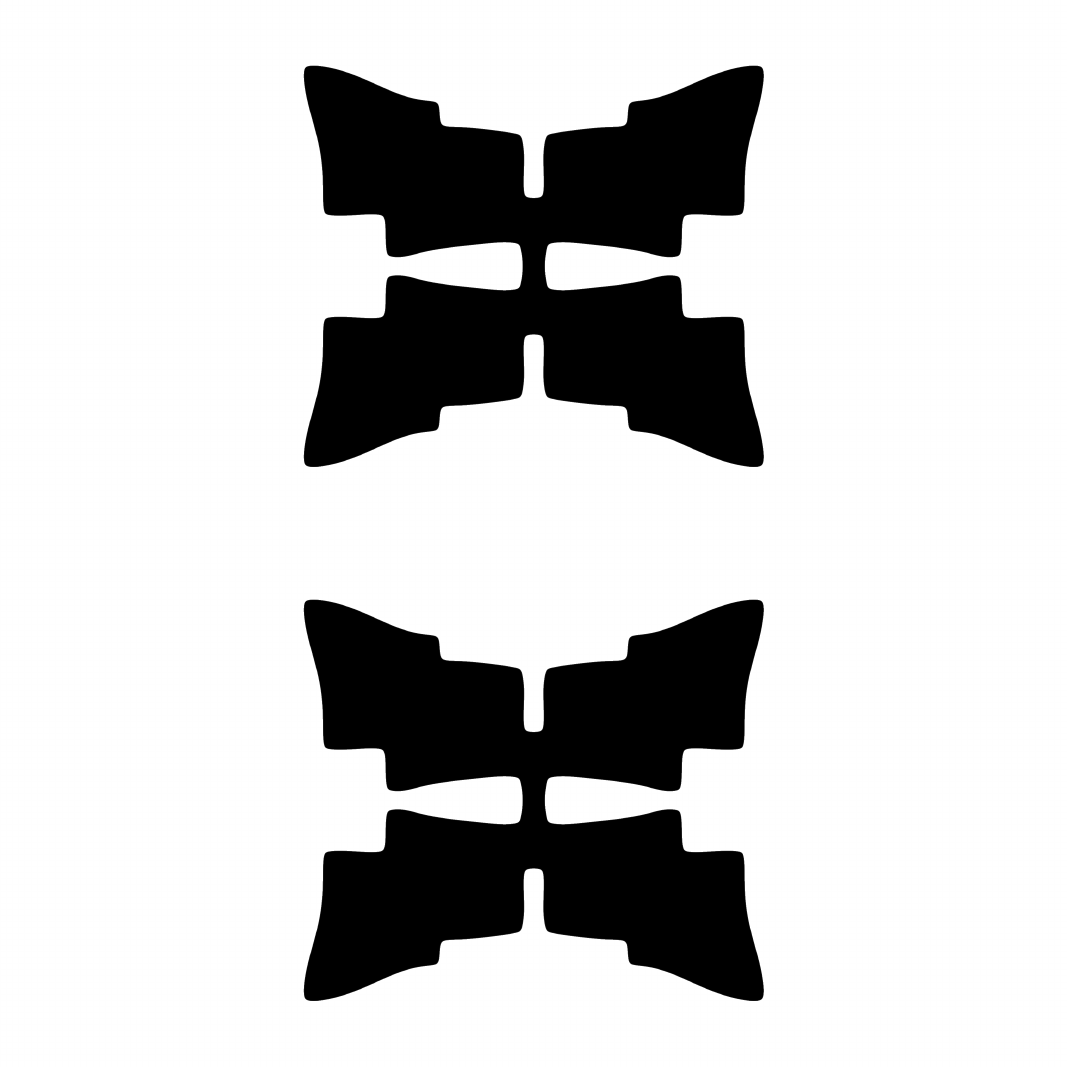}}
    \caption{t=50000}
    \label{subfig:two-particle_t80}
    \end{subfigure}%
    \begin{subfigure}{0.33\textwidth}
    \centering
    \fbox{\includegraphics[width=3.25cm]{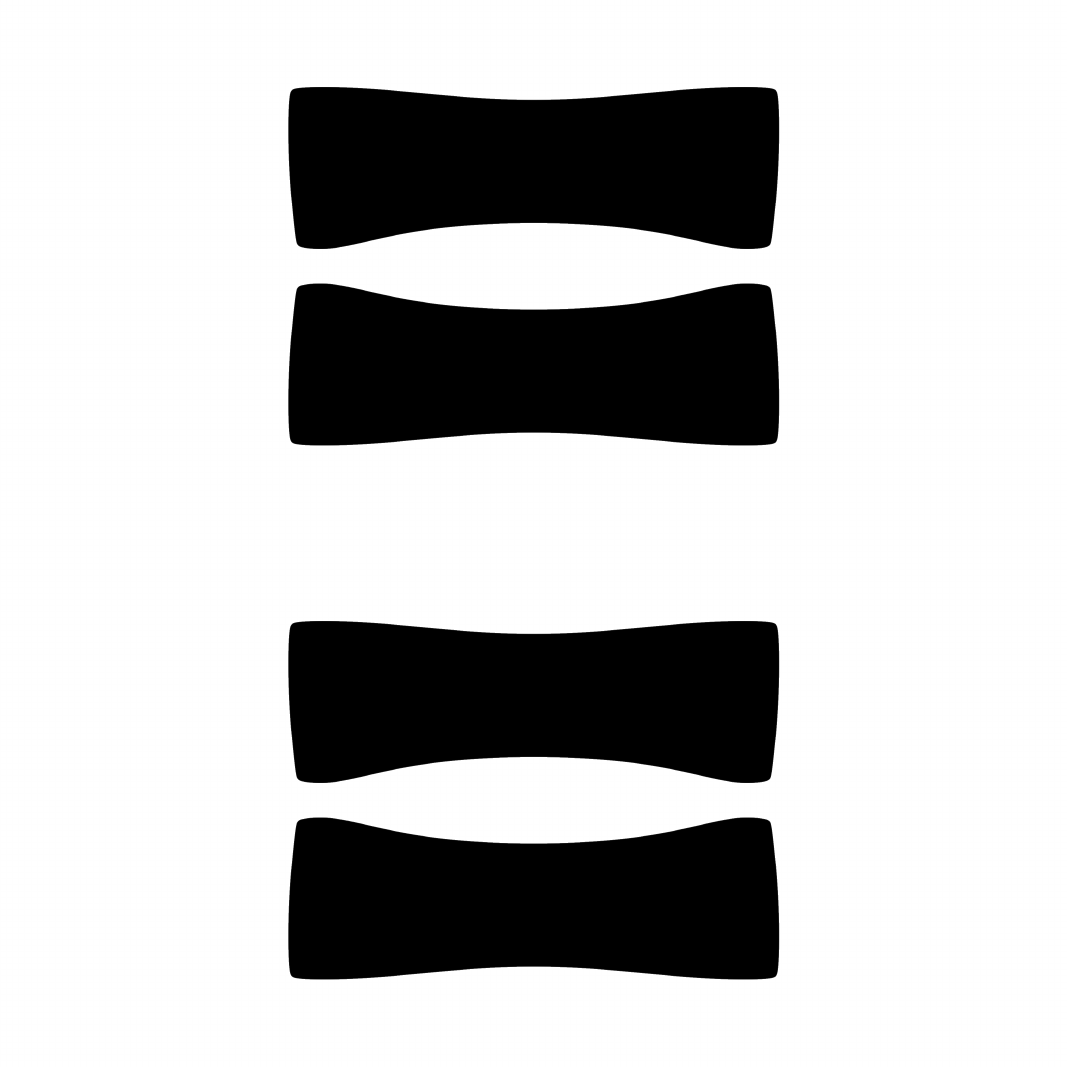}}
    \caption{t=200000}
    \label{subfig:two-particle_t200}    
    \end{subfigure}

    \begin{subfigure}{0.49\textwidth}
    \centering
    \includegraphics[width=7cm]{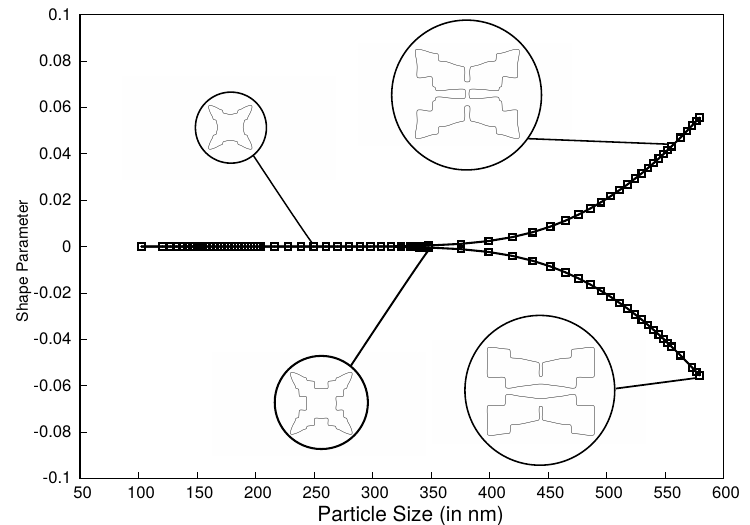}
    \caption{Bifurcation diagram showing T$\rightarrow$O transition}
    \label{subfig:Bifurcation Diagram}
    \end{subfigure}
    \begin{subfigure}{0.49\textwidth}
    \centering
    \includegraphics[width=7cm]{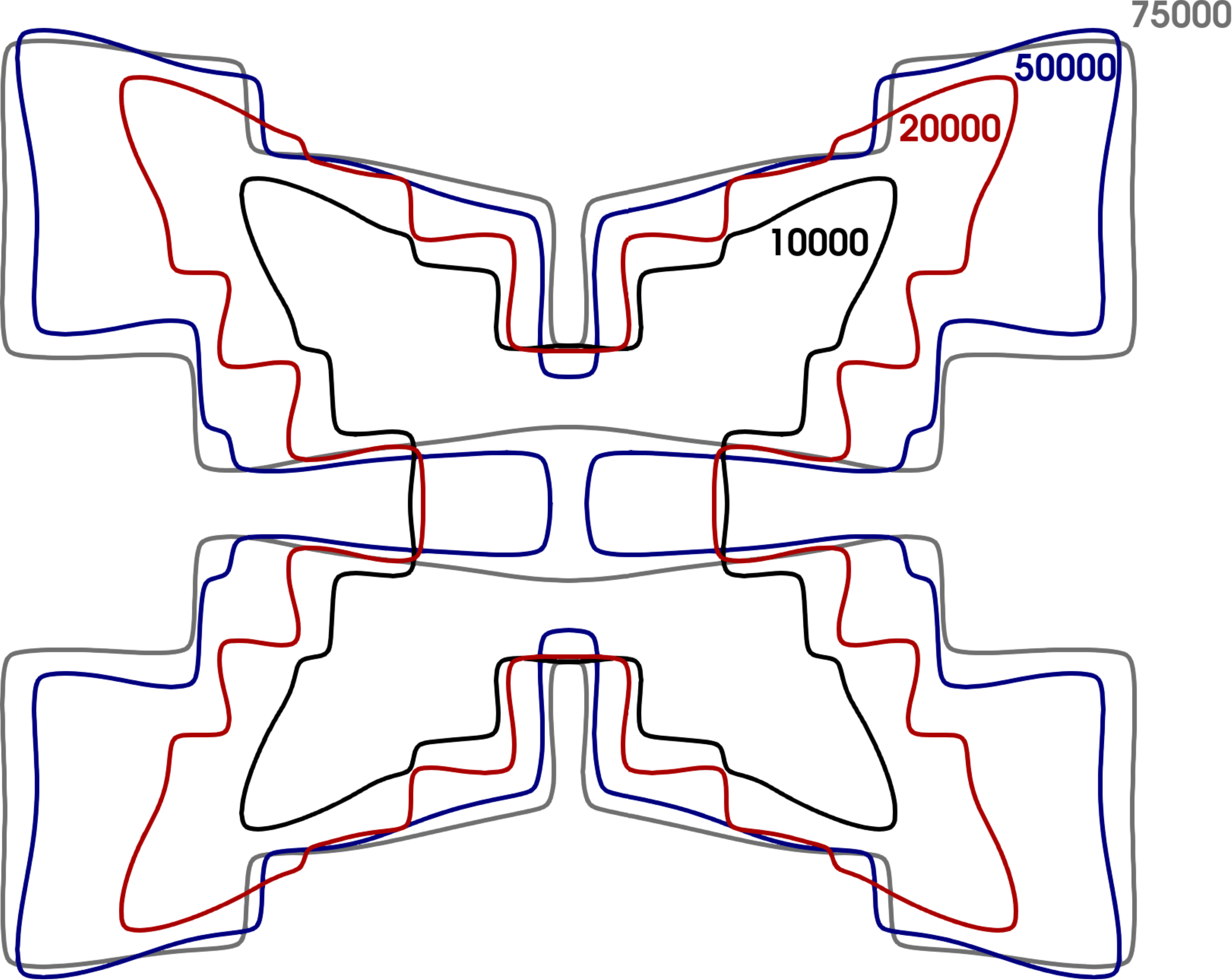}
    \caption{Contours of the particle at various timesteps}
    \label{subfig:Contour}
    \end{subfigure}
    
    \caption{Evolution of precipitates to the formation 
    of doublets due to interaction of two particles 
    oriented along $[01]$ direction. The particle splits 
    in the direction perpendicular to the direction of 
    particle orientation. Here, supersaturation is $25\%$ 
    and misfit is $0.85\%$.} 
    \label{fig:two-particle}
\end{figure}

Experimental reports show that, unlike solid-state 
dendrites, split patterns are observed even when particles 
are closely spaced during isothermal heat treatments~\cite{yoo1995effect}. 
Hence, we discuss the effects of particle interactions on 
the particle splitting instability. Here, we perform two-dimensional 
simulations with a box size having $4096 \times 4096$ grid points and 
grid spacing of $0.2$. We begin the simulation 
with two particles having centers at 
$\displaystyle \left(\frac{nx \times dx}{2},\frac{ny 
\times dy}{4}\right)$ 
and 
$\displaystyle \left(\frac{nx \times dx}{2},\frac{3ny 
\times dy}{4}\right)$. 
We assume supersaturation of 25\%, lattice misfit 
of 0.85\% and Zener anisotropy parameter of 4. 
Fig.~\ref{fig:two-particle} depicts the 
evolution of two particle microstructure which results 
in the formation of doublets. Initially, both the 
particles develop dendrite-like structure with predominant 
growth along $\langle11\rangle$ directions. Further, 
grooves develop on the particles along $\langle10\rangle$ 
directions. During the further evolution of microstructure, 
as the particles start interacting with each other, 
grooves along $[01]$ direction tends to vanish and 
grooves along $[10]$ are promoted to advance towards the 
centers of the precipitates. Later, the microstructure 
evolves to form the doublets oriented along 
$[10]$ direction.

\begin{figure}[!htb]
    \centering
    
    \begin{subfigure}{0.5\textwidth}
    \centering
    \includegraphics[width=6.2cm]{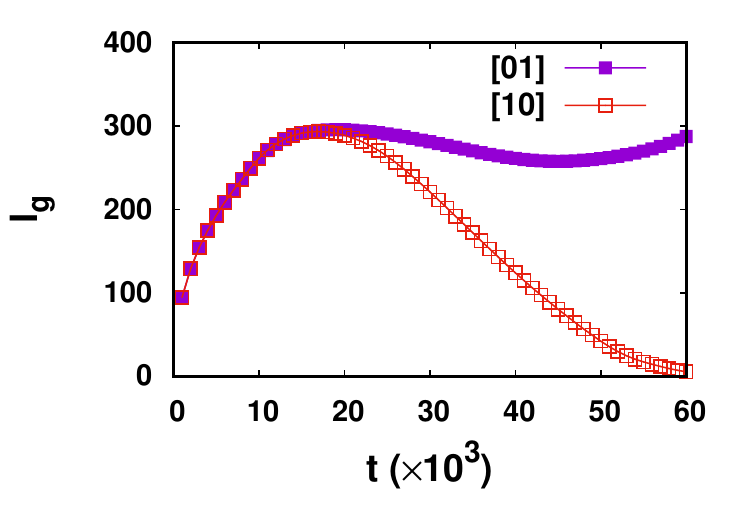}
    \caption{}
    \label{subfig:lg_evolve_two_particle}
    \end{subfigure}%
    \begin{subfigure}{0.5\textwidth}
    \centering
    \includegraphics[width=6.2cm]{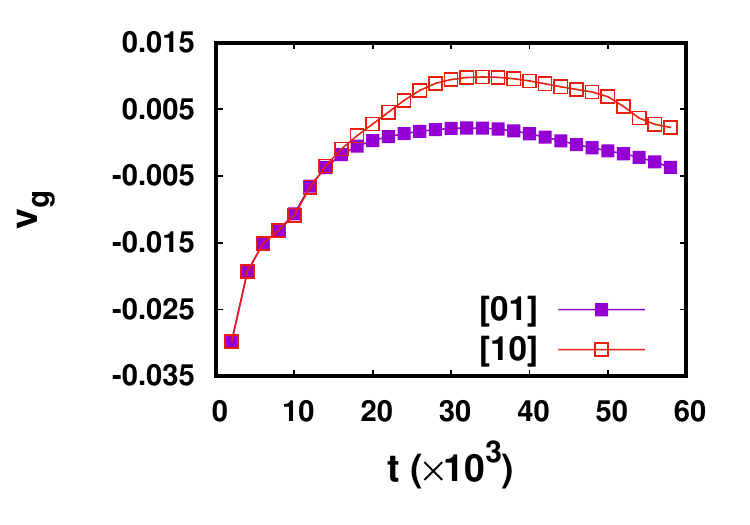}
    \caption{}
    \label{subfig:vg_evolve_two_particle}
    \end{subfigure}
    \begin{subfigure}{0.5\textwidth}
    \centering
    \includegraphics[width=6.2cm]{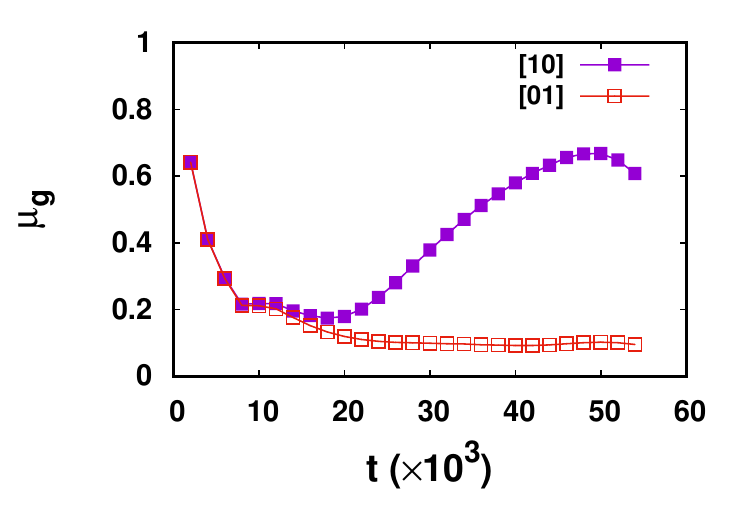}
    \caption{}
    \label{subfig:mug_evolve_two_particle}
    \end{subfigure}
    \caption{
    (a) Temporal evolution of $l_{\textrm{g}}$ along $[10]$ and 
    $[01]$ directions. The interface along $[10]$
    direction continues to advance towards the centre of 
    the precipitate. (b) Temporal evolution of $v_{\textrm{g}}$ along $[10]$ and 
    $[01]$ directions of the precipitate. Initially, 
    both the velocities are negative and equal. Further, the velocity along 
    $[10]$ achieves a positive value, whereas velocity along $[01]$ achieves
    a negative value. (c) Temporal evolution of $\mu_{\textrm{g}}$ along $[10]$ and 
    $[01]$ directions of the precipitate. After the initial equal values for 
    both directions, the $\mu_{\textrm{g}}$ along $[10]$ direction tends to grow with 
    time which suggests precipitate dissolution.}
\end{figure}

Fig.~\ref{subfig:vg_evolve_two_particle} 
shows the temporal evolution of 
$v_{\textrm{g}}$ along
$[10]$ and $[01]$ directions. 
Following the initial equal values of 
velocity for both the directions, 
$v_{\textrm{g}}$ along $[10]$ direction 
achieves a maximum value and decrease 
further before the split patterns form. On 
the other hand, $v_{\textrm{g}}$ along 
$[01]$ direction achieves a smaller 
positive value and further decreases to a 
negative value.

Fig.~\ref{subfig:mug_evolve_two_particle} 
depicts the temporal evolution of the 
chemical potential in the matrix ahead of 
the interface along
$[10]$ and $[01]$ directions. The initial 
same values of $\mu_{\textrm{g}}$
along $[10]$ and $[01]$ directions succeeds the increment in the 
values of $\mu_{\textrm{g}}$ along $[10]$ direction. On the other 
hand, the values of $\mu_{\textrm{g}}$ along $[01]$ direction 
keep on decreasing with time. A preference to increment in 
$\mu_{\textrm{g}}$ along $[10]$ direction results in doublets 
oriented along $[10]$ direction.



\subsection{Discussion}
We presented simulation results of precipitate growth at different levels of lattice misfit, supersaturation and elastic anisotropy in three dimensions. 
Simulation results indicate that the variation of lattice misfit, supersaturation, elastic anisotropy and interfacial energy anisotropy influences the formation of split patterns. 
Anisotropies in elastic energy and interfacial energy promote morphological instability, however, only anisotropy in elastic energy promotes the pinch-off instability. Thus, particle splitting is an elastically driven instability where dendritic structure is a prerequisite.

We observe that the precipitate develops primary dendritic arms along $\{111\}$ (in 3D) and $\{11\}$ directions (in 2D) before the precipitate splits into eight or four smaller precipitates.  
Three distinct precipitate morphologies are observed in three-dimensions, namely (i) No split pattern (NSP) (ii) Split pattern (SP) (iii) Trap split pattern (TSP). 
NSP represents the case where precipitate develops primary arms during the initial stages of growth, however the precipitate restores to cuboidal shape (e.g. refer to precipitate morphology for $c_{\infty} = 25\%$ and $\epsilon^* = 0.85\%$ in Table~\ref{tab:effect_of_misfit}). 
SP corresponds to the case where a single precipitate grows to form eight smaller cuboidal precipitates via formation of dendritic structure with primary arms along $\{111\}$ crystallographic directions and concavities along $\{100\}$ and $\{110\}$ crystallographic directions (refer to precipitate morphology for $c_{\infty} = 35\%$ and $\epsilon^* = 0.75\%$ in Table~\ref{tab:effect_of_misfit}). 
Trap split pattern represents a precipitate morphology where precipitate phase entraps the matrix phase (refer to precipitate morphology $c_{\infty} = 45\%$ and $\epsilon^* = 1\%$ in Table~\ref{tab:effect_of_misfit}).
In case of two-dimensional simulations, we only observe split patterns.
In his classical monograph, Doi et al.~\cite{doi1992Coarsening} explained the splitting phenomenon
based on bifurcation theory wherein elastic interaction energy modulates the precipitate to form smaller octet or quartet split
patterns. By taking into consideration the work of Doi et al., we rationalize our results by plotting elastic free-energy density and interfacial free-energy density. 
\begin{figure}[htb]
    \centering
    \begin{subfigure}{0.5\textwidth}
    \includegraphics[width=\textwidth]{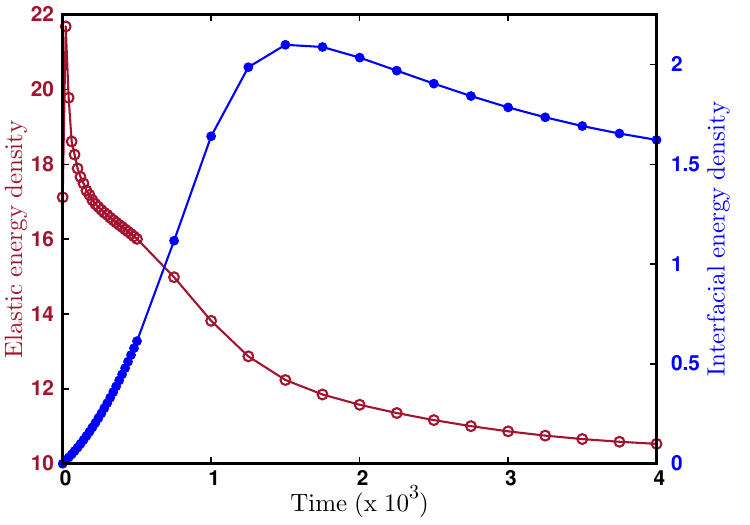}
    \caption{}
    \label{subfig:elast_int_energy_c25_3D}%
    \end{subfigure}%
    \begin{subfigure}{0.5\textwidth}
    \includegraphics[width=\textwidth]{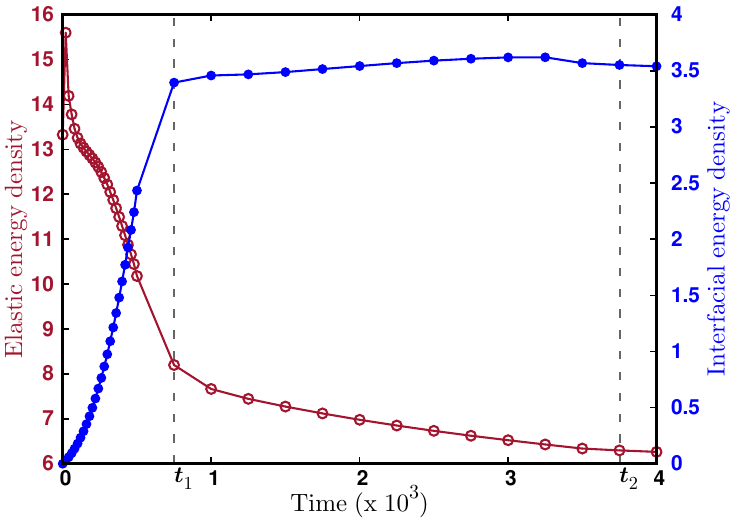}
    \caption{}
    \label{subfig:elast_int_energy_c35_3D}%
    \end{subfigure}
    \begin{subfigure}{0.5\textwidth}
    \includegraphics[width=\textwidth]{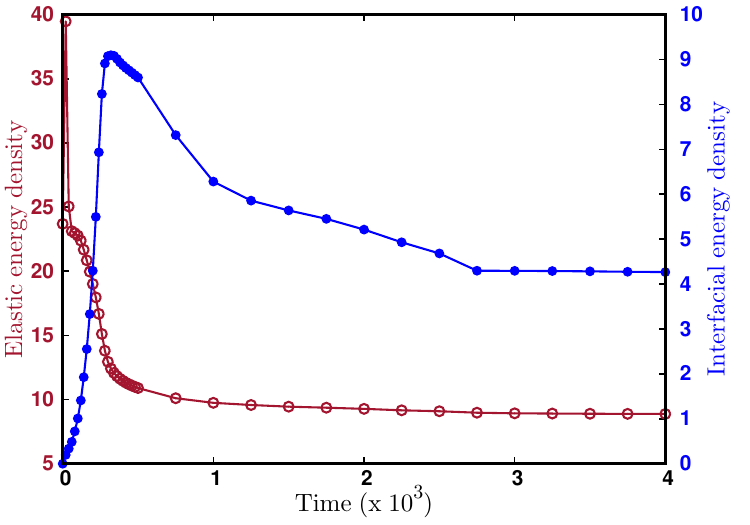}
    \caption{}
    \label{subfig:elast_int_energy_c45_3D}%
    \end{subfigure}%
    \begin{subfigure}{0.5\textwidth}
    \includegraphics[width=\textwidth]{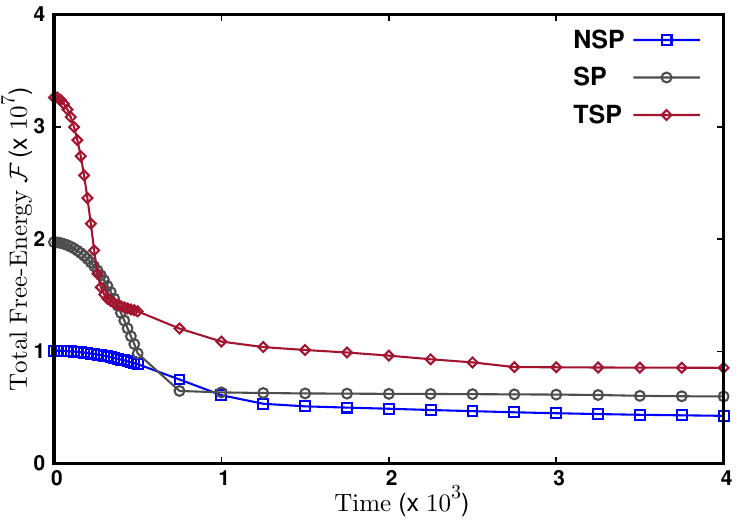}
    \caption{}
    \label{subfig:tot_energy_3D}%
    \end{subfigure}
    \caption{(a) Temporal variation of elastic energy density and 
    interfacial energy density for (a) $c_{\infty} = 25\%$, $A_{\textrm{z}} = 4$ and
    $\epsilon^* = 0.85\%$. (b) $c_{\infty} = 35\%$, $A_{\textrm{z}} = 4$ and
    $\epsilon^* = 0.75\%$ (c) $c_{\infty} = 45\%$, $A_{\textrm{z}} = 4$ and
    $\epsilon^* = 1\%$. (d) Corresponding temporal evolution of total 
    free-energy $\mathcal{F}$ }
    \label{fig:elenergy_and_intenergy_3D}%
\end{figure}
\begin{figure}[htb]
    \centering
    \begin{subfigure}{0.5\textwidth}
    \includegraphics[width=\textwidth]{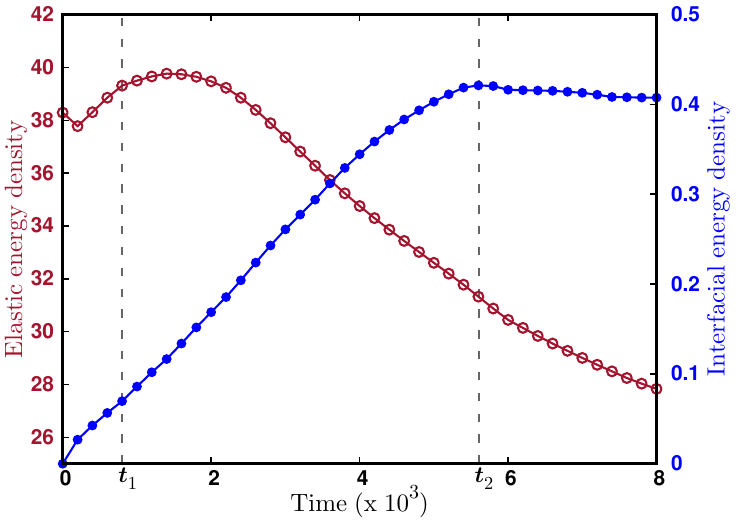}
    \label{subfig:elast_int_energy_2D}%
    \caption{}
    \end{subfigure}%
    \begin{subfigure}{0.5\textwidth}
    \includegraphics[width=\textwidth]{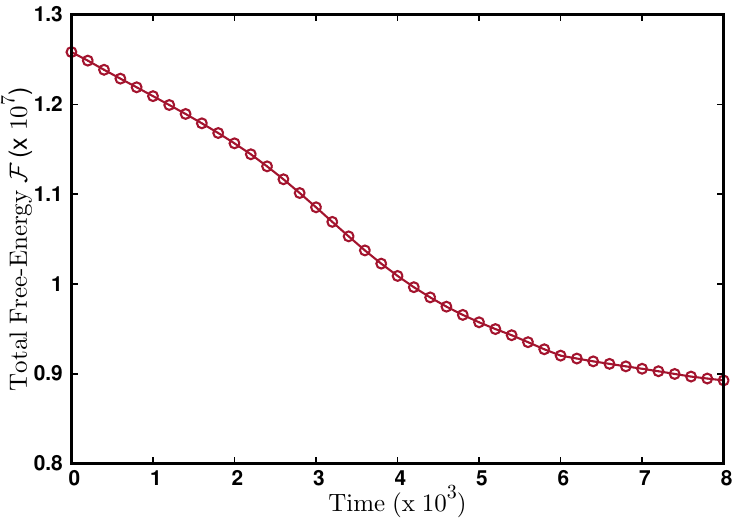}
    \label{subfig:tot_energy_2D}%
    \caption{}
    \end{subfigure}
    \caption{(a) Temporal variation of elastic energy density and 
    interfacial energy density for $c_{\infty} = 25\%$, $A_{\textrm{z}} = 4$ and
    $\epsilon^* = 1\%$. $t_1$ represents the initiation of grooving, and $t_2$ denotes the 
    time of pinch-off. (b) Corresponding temporal evolution of total free-energy $\mathcal{F}$ }
    \label{fig:elenergy_and_intenergy_2D}
\end{figure}

Fig.~\ref{subfig:elast_int_energy_c25_3D}, \ref{subfig:elast_int_energy_c35_3D} and \ref{subfig:elast_int_energy_c45_3D} depict 
the evolution of elastic free-energy density and interfacial free-energy density for NSP, SP and TSP respectively.
Fig.~\ref{subfig:tot_energy_3D} represents the temporal evolution of total free-energy for NSP, SP and TSP cases.
For all cases, the total energy of the system decreases 
monotonically. However, within time intervals $t_1$ and $t_2$ the interfacial energy density continuous to increase, whereas 
elastic energy density after attaining a peak value continually drop (see Fig.~\ref{subfig:elast_int_energy_c35_3D}). Likewise, 
we observe inverse behavior of elastic free-energy density and interfacial free-energy density for 2D simulations of split 
patterns (see Fig.~\ref{fig:elenergy_and_intenergy_2D}).
Thus, reduction in the elastic free-energy density driven by increasing elastic interaction energy contribution concomitant with 
the increase in interfacial free-energy density promotes the process of grooving which finally results in split pattern.
On the contrary, the concomitant decrement in the elastic free-energy density and interfacial free-energy density promotes the 
elimination of concavities or grooves which results in single cuboidal morphology (NP).  TSP is a peculiar case where 
the process of elimination of concavities by merging of primary arms and pinch-off at the center of precipitate occur.
Initially elastic free-energy density and interfacial free-energy density have inverse relationship until interfacial energy 
attains a peak. Post peak value of interfacial energy suggests a process of merging of primary arms, whereas grooves continues to 
advance towards the center of the precipitate. The decrease in the interfacial free-energy is not continuous, rather with 
intermittent temporal change in slope. 

\begin{figure}[!htb]
    \centering
    \begin{subfigure}{0.5\textwidth}
        \includegraphics[scale=0.45]{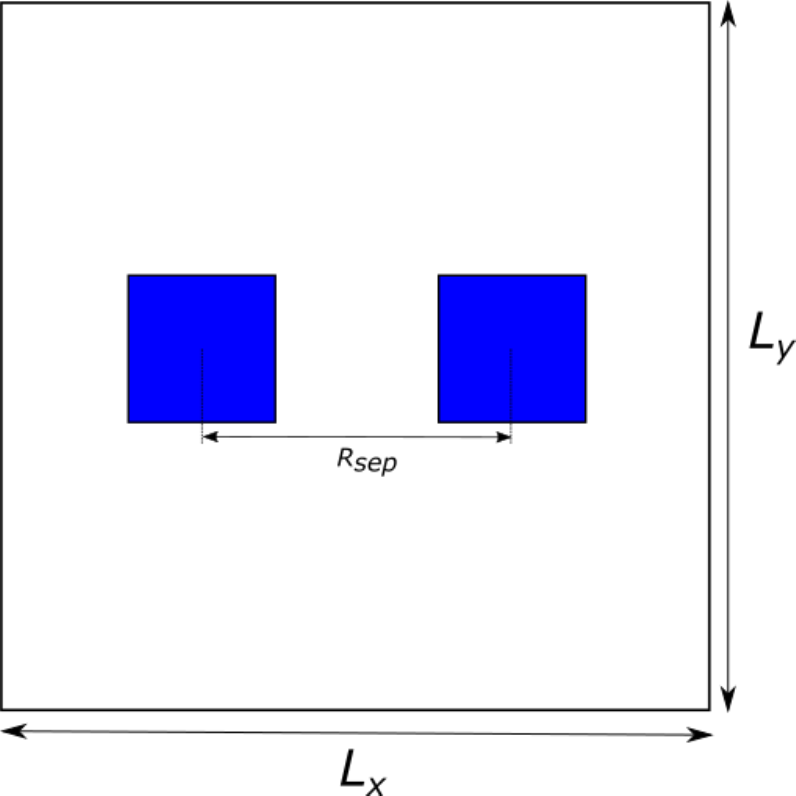}
        \caption{}
         \label{subfig:schematic_F_el_int}
    \end{subfigure}%
    \begin{subfigure}{0.5\textwidth}
        \includegraphics[scale=0.55]{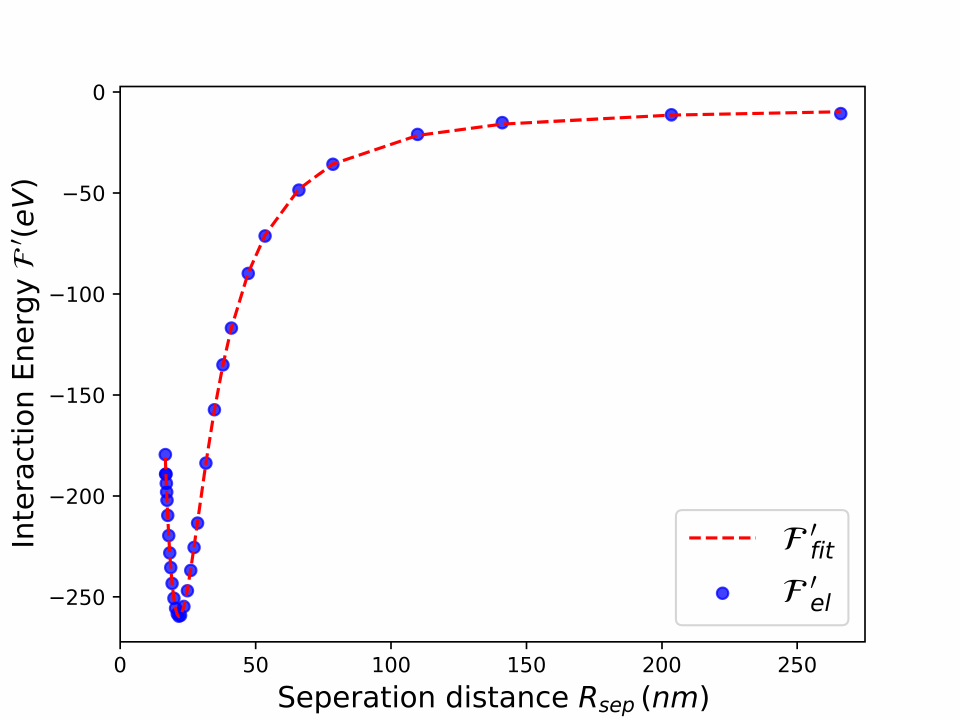}
        \caption{}
        \label{subfig:F_el_int_vs_r}
    \end{subfigure}
    \caption{(a) Schematic representatin of configuration showing arrangement of pair of cuboidal 
    misfitting precipitate in a matrix (b) Variation of elastic interaction energy
    $F_{el}^{int}$ as a function of separation distance ($R_{sep}$) along $[100]$ direction.}
\end{figure}

Doi ~\cite{doi1992Coarsening,DOI199679} calculated the elastic interaction 
energy for a pair of ellipsoidal $\gamma'$ particles as a function of distance 
between them and alignment directions $\langle hkl \rangle$. The analysis calculation 
the negative minimum for elastic interaction energy when two particles are aligned along $\langle 100 \rangle$.   
On the similar manner, we evaluated the elastic interaction energy for a configuration 
in two-dimensions as shown in Fig.~\ref{subfig:schematic_F_el_int}. The configuration has 
a pair of cuboidal $\gamma'$ precipitates of equal size in a box of dimensions $L_x \times L_y$ such that 
$L_x, L_y >> R_{sep}$, where $R_{sep}$ denotes the distance between centers of $\gamma'$ 
precipitates. The elastic interaction energy $\mathcal{F}'_{el}$ for a pair of precipitates 
is evaluated as
\begin{equation}
    \mathcal{F}'_{el} = \mathcal{F}_{el} - 2\mathcal{F}_{self},
\end{equation}
where $\mathcal{F}_{el}$ is the total elastic energy and $\mathcal{F}_{self}$ 
is the self elastic energy of a single precipitate.  Fig.~\ref{subfig:F_el_int_vs_r} 
shows the variation of elastic interaction energy for a pair of 
cuboidal precipitates aligned along $[100]$ direction as a function of $R_{sep}$. 
We notice that the variation of $\mathcal{F}'_{el}$ shows a minimum at 
$R_{sep} ^{min} \approx 25 nm$ which suggests the tendency for a pair of precipitates 
to remain separated at $R_{sep}^{min}$ and aligned along $\langle 100 \rangle$.
We fit the elastic interaction energy data points to a even polynomial in terms of $1/R_{sep}$ of degree twelve:
\begin{equation}
    \mathcal{F}'_{fit} = \sum_{i=0}^{i=6}\left(\frac{a_{i}}{R_{sep}^{2i}}\right), 
\end{equation}
where $a_i$ represent the coefficients of the polynomial. The fitted curve of the polynomial is presented in 
Fig.~\ref{subfig:F_el_int_vs_r}. 
%
    

\section{Conclusion}
\label{sec: conclusion}

We present a diffuse-interface model which 
can recover the conditions of interfacial 
equilibrium for a curved elastically
stressed interface. We have performed 
three-dimensional as well as 
two-dimensional phase-field simulations of 
particle splitting where dendrite-like 
structure leads to particle splitting. 
Our results show that the
presence of elastic anisotropy and lattice 
misfit is necessary for splitting. 
Anisotropy in interfacial energy do not 
lead to splitting of precipitate.
Hence, particle splitting is an 
elastically induced phenomenon. The 
splitting occurs when the contribution of 
elastic stresses is higher as compared to 
curvature along $\langle 110 \rangle$ and 
$\langle 100 \rangle$ directions 
($\langle 10 \rangle$ in two-dimensions). 
Particles in proximity can restrict the growth 
of grooves in certain direction and doublet 
can form. Finite size effects result in 
formation of plate shaped precipitates which 
compare well with the bifurcation theories.

\section*{Acknowledgement}
We gratefully acknowledge the financial 
support from Defence Metallurgical Research 
Laboratory (DMRL), India under through the 
sanction number DMRL/DMR-309/01/TC.  

\newpage 

\bibliography{sample}

\begin{thebibliography}{10}
\expandafter\ifx\csname url\endcsname\relax
  \def\url#1{\texttt{#1}}\fi
\expandafter\ifx\csname urlprefix\endcsname\relax\def\urlprefix{URL }\fi
\expandafter\ifx\csname href\endcsname\relax
  \def\href#1#2{#2} \def\path#1{#1}\fi

\bibitem{doi1992Coarsening}
M.~Doi, Coarsening behaviour of coherent precipitates in elastically
  constrained systems —with particular emphasis on gamma-prime precipitates
  in nickel-base alloys—, Materials Transactions Jim 33 (1992) 637--649.

\bibitem{DOI199679}
M.~Doi, Elasticity effects on the microstructure of alloys containing coherent
  precipitates, Prog. in Mater. Sci. 40~(2) (1996) 79 -- 180.

\bibitem{su1996dynamics}
C.-H. Su, P.~W. Voorhees, The dynamics of precipitate evolution in elastically
  stressed solids—i. inverse coarsening, Acta Mater. 44~(5) (1996)
  1987--1999.

\bibitem{gururajan2016elastic}
M.~Gururajan, L.~Arka, Elastic stress effects on microstructural instabilities,
  J. of the Indian Inst. of Sci. 96~(3) (2016) 199--234.

\bibitem{yeon2005phase}
D.-H. Yeon, P.-R. Cha, J.-H. Kim, M.~Grant, J.-K. Yoon, A phase field model for
  phase transformation in an elastically stressed binary alloy, Model. and
  Simul. in Mater. Sci. and Eng. 13~(3) (2005) 299.

\bibitem{fratzl1999modeling}
P.~Fratzl, O.~Penrose, J.~L. Lebowitz, Modeling of phase separation in alloys
  with coherent elastic misfit, J. of Stat. Phys. 95~(5-6) (1999) 1429--1503.

\bibitem{johnson1984elastically}
W.~Johnson, J.~Cahn, Elastically induced shape bifurcations of inclusions, Acta
  metallurgica 32~(11) (1984) 1925--1933.

\bibitem{johnson1984elastic}
W.~Johnson, On the elastic stabilization of precipitates against coarsening
  under applied load, Acta Metall. 32~(3) (1984) 465--475.

\bibitem{johnson1990coarsening}
W.~Johnson, T.~Abinandanan, P.~W. Voorhees, The coarsening kinetics of two
  misfitting particles in an anisotropic crystal, Acta Metall. et Mater. 38~(7)
  (1990) 1349--1367.

\bibitem{VOORHEES20041}
P.~Voorhees, W.~C. Johnson, The Thermodynamics of Elastically Stressed
  Crystals, Vol.~59 of Solid State Physics, Academic Press, 2004, Ch.~1, pp. 1
  -- 201.

\bibitem{westbrook1958precipitation}
J.~Westbrook, Precipitation of {$\mathrm{Ni_3Al}$} from nickel solid solution
  as ogdoadically diced cubes, Z. f{\"u}r Krist.-Cryst. Mater. 110~(1-6) (1958)
  21--29.

\bibitem{miyazaki1982formation}
T.~Miyazaki, H.~Imamura, T.~Kozakai, The formation of ``$\gamma^{\prime}$
  precipitate doublets” in {Ni-Al} alloys and their energetic stability,
  Mater. Sci. and Eng. 54~(1) (1982) 9--15.

\bibitem{doi1984effect}
M.~Doi, T.~Miyazaki, T.~Wakatsuki, The effect of elastic interaction energy on
  the morphology of $\gamma^{\prime}$ precipitates in nickel-based alloys,
  Mater. Sci. and Eng. 67~(2) (1984) 247--253.

\bibitem{doi1985effects}
M.~Doi, T.~Miyazaki, T.~Wakatsuki, The effects of elastic interaction energy on
  the $\gamma^{\prime}$ precipitate morphology of continuously cooled
  nickel-base alloys, Mater. Sci. and Eng. 74~(2) (1985) 139--145.

\bibitem{Kaufman1989}
M.~J. Kaufman, P.~W. Voorhees, W.~C. Johnson, F.~S. Biancaniello, An
  elastically induced morphological instability of a misfitting precipitate,
  Metall. Trans. A 20~(10) (1989) 2171–2175.

\bibitem{yoo1995effect}
Y.~Yoo, D.~Yoon, M.~Henry, The effect of elastic misfit strain on the
  morphological evolution of $\gamma^{\prime}$-precipitates in a model
  {Ni-base} superalloy, Metal. and Mater. 1~(1) (1995) 47--61.

\bibitem{qiu1996retarded}
Y.~Qiu, Retarded coarsening phenomenon of $\gamma^{\prime}$ particles in
  {Ni-based} alloy, Acta Mater. 44~(12) (1996) 4969--4980.

\bibitem{qiu1998splitting}
Y.~Qiu, The splitting behavior of $\gamma^{\prime}$ particles in {Ni-based}
  alloys, J. of Alloys and Compd. 270~(1-2) (1998) 145--153.

\bibitem{grosdidier1998precipitation}
T.~Grosdidier, A.~Hazotte, A.~Simon, Precipitation and dissolution processes in
  $\gamma$/$\gamma^{\prime}$ single crystal nickel-based superalloys, Mater.
  Sci. and Eng.: A 256~(1-2) (1998) 183--196.

\bibitem{doi1996transmission}
M.~Doi, Transmission electron microscopy observations of the splits of
  {$\mathrm{D0_3}$} and {$\mathrm{B2}$} precipitates in fe-based alloys, Phil.
  Mag. Lett. 73~(6) (1996) 331--336.

\bibitem{calderon1997coarsening}
H.~Calderon, G.~Kostorz, Y.~Qu, H.~Dorantes, J.~Cruz, J.~Cabanas-Moreno,
  Coarsening kinetics of coherent precipitates in {Ni-Al-Mo} and {Fe-Ni-Al}
  alloys, Mater. Sci. and Eng.: A 238~(1) (1997) 13--22.

\bibitem{castro1998isothermal}
M.~Castro, R.~Romero, Isothermal $\gamma$ precipitation in a $\beta$
  {Cu--Zn--Al} alloy, Mater. Sci. and Eng.: A 255~(1-2) (1998) 1--6.

\bibitem{yamabe2002formation}
Y.~Yamabe-Mitarai, H.~Harada, Formation of a `splitting pattern' associated
  with {$\mathrm{L1_2}$} precipitates in {Ir-Nb} alloys, Phil. Mag. Lett.
  82~(3) (2002) 109--118.

\bibitem{cornish2007overview}
L.~Cornish, et~al., Overview of the development of new {Pt-based} alloys for
  high temperature application in aggressive environments, J. of the South.
  Afr. Inst. of Min. and Metall. 107~(11) (2007) 697--711.

\bibitem{ricks1983growth}
R.~Ricks, A.~Porter, R.~Ecob, The growth of $\gamma^{\prime}$ precipitates in
  nickel-base superalloys, Acta Metall. 31~(1) (1983) 43--53.

\bibitem{leo1989effect}
P.~H. Leo, R.~Sekerka, The effect of elastic fields on the morphological
  stability of a precipitate grown from solid solution, Acta Metall. 37~(12)
  (1989) 3139--3149.

\bibitem{Bhadak2020formation}
B.~Bhadak, T.~Jogi, S.~Bhattacharya, A.~Choudhury, Formation of solid-state
  dendrites under the influence of coherency stresses: A diffuse interface
  approach, Unpublished work.

\bibitem{wang1995shape}
Y.~Wang, A.~Khachaturyan, Shape instability during precipitate growth in
  coherent solids, Acta Metall. et Mater. 43~(5) (1995) 1837--1857.

\bibitem{wang1993kinetics}
Y.~Wang, L.-Q. Chen, A.~Khachaturyan, Kinetics of strain-induced morphological
  transformation in cubic alloys with a miscibility gap, Acta Metall. et Mater.
  41~(1) (1993) 279--296.

\bibitem{zhang_li_chen_1997}
J.~D. Zhang, D.~Y. Li, L.~Q. Chen, Shape evolution and splitting of a single
  coherent particle, MRS Proc. 481 (1997) 243.

\bibitem{li1998shape}
D.~Li, L.~Chen, Shape evolution and splitting of coherent particles under
  applied stresses, Acta Mater. 47~(1) (1998) 247--257.

\bibitem{liu2016elastic}
L.~Liu, Z.~Chen, Y.~Wang, Elastic strain energy induced split during
  precipitation in alloys, J. of Alloys and Compds. 661 (2016) 349--356.

\bibitem{liu2017split}
L.~Liu, Z.~Chen, Y.~Wang, M.~Zhang, The split of dendritic precipitates with
  interfacial anisotropy in solid transformations in alloys, J. of Alloys and
  Compds. 703 (2017) 321--329.

\bibitem{banerjee1999formation}
D.~Banerjee, R.~Banerjee, Y.~Wang, Formation of split patterns of
  $\gamma^{\prime}$ precipitates in {Ni-Al} via particle aggregation, Scr.
  Mater. 41~(9) (1999) 1023--1030.

\bibitem{calderon2000direct}
H.~Calderon, J.~Cabanas-Moreno, T.~Mori, Direct evidence that an apparent
  splitting pattern of gamma particles in {Ni} alloys is a stage of
  coalescence, Phil. Mag. Lett. 80~(10) (2000) 669--674.

\bibitem{Calderon2005high}
H.~A. Calderon, G.~Kostorz, L.~Calzado-Lopez, C.~Kisielowski, T.~Mori,
  High-resolution electron-microscopy analysis of splitting patterns in {Ni}
  alloys, Phil. Mag. Lett. 85~(2) (2005) 51--59.

\bibitem{kisielowski2007statistical}
C.~Kisielowski, T.~Mori, H.~Calderon, Statistical analysis of $\gamma^{\prime}$
  quartet split patterns in $\gamma$--$\gamma^{\prime}$ ni alloys revealed by
  high resolution electron microscopy, Phil. Mag. Lett. 87~(1) (2007) 33--40.

\bibitem{mehl2017aflow}
M.~J. Mehl, D.~Hicks, C.~Toher, O.~Levy, R.~M. Hanson, G.~Hart, S.~Curtarolo,
  The aflow library of crystallographic prototypes: part 1, Comput. Mater. Sci.
  136 (2017) S1--S828.

\bibitem{cha2005phase}
P.-R. Cha, D.-H. Yeon, S.-H. Chung, Phase-field study for the splitting
  mechanism of coherent misfitting precipitates in anisotropic elastic media,
  Scr. Mater. 52~(12).

\bibitem{leo2001elastically}
P.~Leo, J.~Lowengrub, Q.~Nie, On an elastically induced splitting instability,
  Acta Mater. 49~(14) (2001) 2761--2772.

\bibitem{maheshwari1992elastic}
A.~Maheshwari, A.~J. Ardell, Elastic interactions and their effect on gamma
  prime precipitate shapes in aged dilute {Ni-Al} alloys, Scr. Metall. et
  Mater. 26.

\bibitem{zhao2013effects}
X.~Zhao, R.~Duddu, S.~P. Bordas, J.~Qu, Effects of elastic strain energy and
  interfacial stress on the equilibrium morphology of misfit particles in
  heterogeneous solids, J. of the Mech. and Phys. of Solids 61~(6) (2013)
  1433--1445.

\bibitem{cool2017359}
T.~Cool, P.~Voorhees, The evolution of dendrites during coarsening:
  Fragmentation and morphology, Acta Mater. 127 (2017) 359 -- 367.

\bibitem{herlach2001grain}
D.~Herlach, K.~Eckler, A.~Karma, M.~Schwarz, Grain refinement through
  fragmentation of dendrites in undercooled melts, Mater. Sci. and Eng.: A 304
  (2001) 20--25.

\bibitem{choudhury2012grand}
A.~Choudhury, B.~Nestler, Grand-potential formulation for multicomponent phase
  transformations combined with thin-interface asymptotics of the
  double-obstacle potential, Phys. Rev. E 85~(2) (2012) 021602.

\bibitem{choudhury2013quantitative}
A.~N. Choudhury, Quantitative phase-field model for phase transformations in
  multi-component alloys, KIT Scientific Publishing, 2013.

\bibitem{kim1999}
S.~G. Kim, W.~T. Kim, T.~Suzuki, Phase-field model for binary alloys, Phys.
  Rev. E 60 (1999) 7186--7197.

\bibitem{kim2007phase}
S.~G. Kim, A phase-field model with antitrapping current for multicomponent
  alloys with arbitrary thermodynamic properties, Acta Mater. 55~(13) (2007)
  4391--4399.

\bibitem{abinandanan2001extended}
T.~Abinandanan, F.~Haider, An extended cahn-hilliard model for interfaces with
  cubic anisotropy, Phil. Mag. A 81~(10) (2001) 2457--2479.

\bibitem{allen1979microscopic}
S.~M. Allen, J.~W. Cahn, A microscopic theory for antiphase boundary motion and
  its application to antiphase domain coarsening, Acta Metall. 27~(6) (1979)
  1085--1095.

\bibitem{khachaturyan2013theory}
A.~G. Khachaturyan, Theory of structural transformations in solids, Dover
  Publication, 2008.

\bibitem{chen1998applications}
L.~Q. Chen, J.~Shen, Applications of semi-implicit fourier-spectral method to
  phase field equations, Comput. Phys. Commun. 108~(2-3) (1998) 147--158.

\bibitem{hu2001phase}
S.~Hu, L.~Chen, A phase-field model for evolving microstructures with strong
  elastic inhomogeneity, Acta Mater. 49~(11) (2001) 1879--1890.

\bibitem{bhattacharyya2012spectral}
S.~Bhattacharyya, T.~W. Heo, K.~Chang, L.-Q. Chen, A spectral iterative method
  for the computation of effective properties of elastically inhomogeneous
  polycrystals, Commun. in Comput. Phys. 11~(3) (2012) 726--738.

\bibitem{johnson1987precipitate}
W.~C. Johnson, Precipitate shape evolution under applied
  stress—thermodynamics and kinetics, Metall. and Mater. Trans. A 18~(2)
  (1987) 233--247.

\bibitem{laraia1988growth}
V.~Laraia, W.~C. Johnson, P.~Voorhees, Growth of a coherent precipitate from a
  supersaturated solution, J. of Mater. Res. 3~(2) (1988) 257--266.

\bibitem{LandoltBornstein1990}
A.~LeClaire~(auth.), G.~Neumann~(auth.), H.~Mehrer~(eds.), Diffusion in Solid
  Metals and Alloys, 1st Edition, Landolt-Börnstein - Group III Condensed
  Matter 26 : Condensed Matter, Springer-Verlag Berlin Heidelberg, 1990.

\bibitem{schmidt1997equilibrium}
I.~Schmidt, D.~Gross, The equilibrium shape of an elastically inhomogeneous
  inclusion, J. of the Mech. and Phys. of Solids 45~(9) (1997) 1521--1549.

\bibitem{cufft}
{Nvidia Corporation}, \href{https://developer.nvidia.com/cufft}{cuFFT library,
  CUDA Toolkit}.
\newline\urlprefix\url{https://developer.nvidia.com/cufft}

\end{thebibliography}

\appendix
\section{Gradient energy density}
\label{app:grad_energy_density}
The gradient free energy density $f_{\textrm{grad}}$ 
of the system is written as:
\begin{equation}
f_{\textrm{grad}}(\nabla \phi) = \varepsilon^2 |\nabla \phi|^2 + 
\Gamma_{\langle hkl \rangle} |\Delta \phi|^2,
\label{eqn:grad_energy}
\end{equation}
where $\varepsilon^2$ is the gradient free energy coefficient, 
$\Gamma_{\langle hkl \rangle}$ is the fourth-rank gradient 
tensor which incorporates anisotropy in the interfacial 
energy.
The fourth-rank gradient tensor reads as
\begin{equation}
    \Gamma_{\langle hkl \rangle} = \Gamma_{\textrm{A}} (h^4 + k^4 + l^4) + \Gamma_{\textrm{I}},
\end{equation}
where $\Gamma_{\textrm{A}}$ and $\Gamma_{\textrm{I}}$ denote the 
anisotropic and isotropic components of the fourth-rank gradient 
energy tensor, and $h$,$k$,$l$ represent the direction cosines of 
normal to the interface.
The interfacial energy of the system must be non-negative, as
a result the following constraints must be valid:
\begin{eqnarray}
 \Gamma_{\textrm{I}} &\geq& \lvert \Gamma_{\textrm{A}} \rvert \quad \forall \; \; \Gamma_{\textrm{A}} \leq 0, \\
 -\Gamma_{\textrm{I}} &\leq&  \frac{\Gamma_{\textrm{A}}}{3} \; \quad  \forall \; \; \Gamma_{\textrm{A}} > 0.
\end{eqnarray}
When $\Gamma_{\textrm{A}} < 0$, the values of 
$\Gamma_{\langle hkl \rangle}$ are maximum along 
$\langle 111 \rangle$ directions and minimum along 
$\langle 100 \rangle$ directions. Consequently,
interfacial energy along $\langle 100 \rangle$ will be lower than 
that along $\langle 111 \rangle$ directions. Hence, the Wulff 
shape will have facets along $\langle 100 \rangle$ directions. On 
the other hand, when $\Gamma_{\textrm{A}} > 0$, 
$\Gamma_{\langle hkl \rangle}$ possesses maximum values along 
$\langle 100 \rangle$ directions and minimum along 
$\langle 111 \rangle$ directions. Hence, the Wulff shape will
exhibit facets along $\langle 111 \rangle$ directions. We choose
the earlier case for our study to show the effect of 
interfacial energy anisotropy.
\section{Elastic driving force}
\label{app:elast_driv_force}
The elastic-free energy density is expressed as
\begin{eqnarray}
    f^{\mathrm{el}} & = & \frac{1}{2} 
    C_{ijkl}(\mathbf{r})
    [\delta \epsilon_{ij}(\mathbf{r}) + 
    \Bar{\epsilon}_{ij} -
    \epsilon_{ij}^*(\mathbf{r})] 
    [\delta \epsilon_{kl}(\mathbf{r}) + 
    \Bar{\epsilon}_{kl} -
    \epsilon_{kl}^*(\mathbf{r})] \nonumber \\
    & = & \frac{1}{2} \left[ C_{ijkl}^0 +
    C_{ijkl}^{\prime} \left(2h(\phi) - 1 \right) 
    \right] \left[\delta \epsilon_{ij}(\mathbf{r}) + 
    \Bar{\epsilon}_{ij} -
    \epsilon^0 \delta_{ij} h(\phi)\right] 
    \left[\delta \epsilon_{kl}(\mathbf{r}) + 
    \Bar{\varepsilon}_{kl} -
    \epsilon^0 \delta_{ij} h(\phi)\right]
\end{eqnarray}
Taylor expansion of elastic free-energy density
about $f^{\mathbf{el}}$ derives as
\begin{eqnarray}
    f^{\mathrm{el}} + \delta f^{\mathrm{el}}  = 
    \frac{1}{2} & \left[ C_{ijkl}^0 +
    C_{ijkl}^{\prime} \left(2h(\phi) - 1 \right) 
    + 2 C_{ijkl}^{\prime} h^{\prime}(\phi) 
    \delta \phi \right] \nonumber \\ 
    & \left[ \delta 
    \epsilon_{ij}(\mathbf{r}) + 
    \Bar{\varepsilon}_{ij} -
    \epsilon^0 \delta_{ij} h(\phi) - 
    \epsilon^0 \delta_{ij} h^{\prime}(\phi) 
    \delta \phi \right] \nonumber \\
    & \left[ \delta \epsilon_{kl}(\mathbf{r}) 
    + \Bar{\epsilon}_{kl} -
    \epsilon^0 \delta_{kl} h(\phi) - 
    \epsilon^0 \delta_{kl} h^{\prime}(\phi) 
    \delta \phi \right]
\end{eqnarray}
Further simplification leads to
\begin{eqnarray}
    \delta f^{\mathrm{el}} =  
& & - \frac{1}{2} C_{ijkl} (\mathbf{r}) 
    \left[ 
    \delta \epsilon_{ij}(\mathbf{r}) + 
    \bar{\epsilon}_{ij} -
    \epsilon^0 \delta_{ij} h(\phi) 
    \right] \epsilon^0 \delta_{kl} 
    h^{\prime}(\phi) \delta \phi \nonumber \\ 
& & - \frac{1}{2} C_{ijkl} (\mathbf{r}) 
    \left[ 
    \delta \epsilon_{kl} (\mathbf{r}) +
    \bar{\epsilon}_{kl} -
    \epsilon_{kl}^0 h(\phi)
    \right] \epsilon^0 \delta_{ij} 
    h^{\prime}(\phi) \delta \phi \nonumber \\
& & + \frac{1}{2} C_{ijkl} (\mathbf{r}) 
    \left[ \epsilon^{0} \epsilon^{0} 
    \delta_{ij} \delta_{kl} 
    (h^{\prime}(\phi))^2 (\delta \phi)^2  
    \right] \nonumber \\ 
& & + C^{\prime}_{ijkl} 
    h^{\prime}(\phi) \delta \phi 
    \left[\delta \epsilon_{ij}(\mathbf{r}) + 
    \bar{\epsilon}_{ij} - \epsilon^0 \delta_{ij}
    h(\phi)
    \right] 
    \left[ 
    \delta \epsilon_{kl}(\mathbf{r}) + 
    \bar{\epsilon}_{kl} - \epsilon^0 \delta_{kl} 
    h(\phi)
    \right] \nonumber \\
& & - C_{ijkl}^{\prime} 
    h^{\prime}(\phi) \delta \phi  
    \left[ \delta \epsilon_{ij}(\mathbf{r}) +
    \bar{\epsilon}_{ij} -
    \epsilon^0 \delta_{ij} h(\phi) \right] \epsilon^0 \delta_{kl} h^{\prime}(\phi) 
    \delta \phi \nonumber \\
& & - C^{\prime}_{ijkl} 
    h^{\prime}(\phi) \delta \phi 
    \left[ \delta \epsilon_{kl} (\mathbf{r}) + 
    \bar{\epsilon}_{kl} - 
    \epsilon^0 \delta_{kl} h(\phi)
    \right] \epsilon^0 \delta_{ij} h^{\prime}(\phi) \delta \phi \nonumber \\
& & + C^{\prime}_{ijkl} 
    h^{\prime}(\phi) \delta \phi 
    \left[  
    \epsilon^0 \epsilon^0 (h^{\prime}(\phi))^2 (\delta \phi)^2 
    \right] \delta_{ij} \delta_{kl}
\end{eqnarray}
Considering only first-order terms in 
$\delta \phi$ and neglecting 
higher-order terms in $\delta \phi$,
\begin{eqnarray}
    \delta f^{\mathrm{el}} = \frac{1}{2} \Big[
    & 2 C^{\prime}_{ijkl} h^{\prime}(\phi) 
    \delta \phi
    \left[ 
    \delta \epsilon_{ij}(\mathbf{r}) +
    \bar{\epsilon}_{ij} - 
    \epsilon^0 \delta_{ij}  h(\phi)
    \right] 
    \left[ 
    \delta \epsilon_{kl}(\mathbf{r})
    \bar{\epsilon}_{kl} - 
    \epsilon^0 \delta_{kl} h(\phi)
    \right] \nonumber \\
& - C_{ijkl} \left[ 
    \delta \epsilon_{ij}(\mathbf{r}) +
    \bar{\epsilon}_{ij} -
    \epsilon^0 \delta_{ij} h(\phi) 
    \right] \epsilon^0 \delta_{kl} 
    h^{\prime}(\phi) \delta \phi \nonumber \\
& - C_{ijkl} \left[ 
    \delta \epsilon_{kl} (\mathbf{r}) +
    \bar{\epsilon}_{kl} -
    \epsilon^0 \delta_{kl} h(\phi)
    \right] \epsilon^0 \delta_{ij} 
    h^{\prime}(\phi) \delta \phi \Big]
\end{eqnarray}
Invoking symmetries of $C_{ijkl}$, the 
elastic driving force is expressed as
\begin{eqnarray}
    \frac{\delta f^{\mathrm{el}}}{\delta \phi}
    & = & \frac{1}{2} C^{\prime}_{ijkl} 
    h^{\prime}(\phi)
    \left[  
    \delta \epsilon_{ij}(\mathbf{r}) + 
    \bar{\epsilon}_{ij} - 
    \epsilon^0 \delta_{ij} h(\phi)
    \right] 
    \left[ 
    \delta \epsilon_{kl}(\mathbf{r}) +
    \bar{\epsilon}_{kl} - 
    \epsilon^0 \delta_{kl} h(\phi)
    \right] \nonumber \\
&  & - C_{ijkl}
    \left[
    \delta \epsilon_{ij}(\mathbf{r}) +
    \bar{\epsilon}_{ij} -
    \epsilon^0 \delta_{ij} h(\phi) 
    \right] \epsilon^0 \delta_{kl} 
    h^{\prime}(\phi)
\end{eqnarray}

\section{Non-dimensionalization}
\label{app:nondimen}
The governing equations Eqs.~\eqref{eqn:compeqn} and~\eqref{eqn:phieqn} 
are presented in non-dimensional
form. Using characteristics energy $\mathcal{E}$, characteristics length $\mathcal{L}$, and characteristics time $\mathcal{T}$, 
all quantities are rendered non-dimensional.
In the following, we discuss the non-dimensionalization scheme in detail.
Here, all the quantities with asterisk represent dimensional quantities.

\begin{eqnarray}
    \mathcal{F}^{*} = \int_{V^{*}} \left[  
     f^{*}(\phi,c^{*}) + 
    {\epsilon^{*}}^2 | \nabla^{*} \phi (\boldsymbol{r})|^2 \; + \; 
    \gamma_{\langle hkl \rangle}^* |{\boldsymbol{\nabla}^{*}}^2 
    \phi (\boldsymbol{r})|^2 + 
    \frac{1}{2} \boldsymbol{\upvarepsilon}^{\mathrm{el}}:
    \mathbb{C}^*:\boldsymbol{\upvarepsilon}^{\mathrm{el}}
    \right] \mathrm{d}V^{*}
    \label{eqn:dimen_free_energy}
\end{eqnarray}

Since the total energy $\mathcal{F}^{*}$ has a unit of \si{\joule}, $f^{*}$ has a unit of \si{\joule \per \meter \cubed}, 
${\epsilon^{*}}^2$ has a unit of \si{\joule \per \meter}, and $\gamma_{\langle hkl \rangle}^{*}$ has a unit of \si{\joule \meter}. Moreover,
$\mathbb{C}^{*}$ has a unit of \si{\joule \per \meter \cubed}.

The scaled composition $c$ is obtained as 

\begin{equation}
    c = \frac{c^{*}-c_{m}^{*}}{c_{p}^{*}-c_{m}^{*}},
\end{equation}

where $c_{m}^{*}$ and $c_{p}^{*}$ represent local
compositions of matrix and precipitate phases in dimensional form, respectively. 
We follow the procedure shown below to non-dimensionalize other physical quantities:

\begin{eqnarray}
    f &=& \frac{f^{*}\mathcal{L}^3}{\mathcal{E}},\nonumber \\
    \epsilon^2 &=& \frac{{\epsilon^{*}}^2 \mathcal{L}}{\mathcal{E}},\nonumber \\
    \gamma_{\langle hkl \rangle} &=& 
    \frac{\gamma_{\langle hkl 
    \rangle}^{*}}{\mathcal{E}\mathcal{L}},\nonumber \\
    \mathbb{C} &=& 
    \frac{\mathbb{C}^{*}\mathcal{L}^3}{\mathcal{E}},\nonumber \\
    V &=& \frac{V^{*}}{\mathcal{L}^3}
\end{eqnarray}

The non-dimensional form of Eqn.~\eqref{eqn:dimen_free_energy} is given as
\begin{eqnarray}
    \mathcal{F} & = &\frac{\mathcal{F}^{*}}{\mathcal{E}},\\
    & = &\frac{\mathcal{L}^3}{\mathcal{E}} \mathlarger{\int_V} \Bigg[  
    f(\phi(\boldsymbol{r}),c) \frac{\mathcal{E}}{\mathcal{L}} + 
    {\epsilon}^2 \frac{\mathcal{E}}{\mathcal{L}} \frac{1}{\mathcal{L}^2}
    | \boldsymbol{\nabla} \phi (\boldsymbol{r})|^2 + \nonumber \\ 
    & &\mathbb{H}\mathcal{E} \mathcal{L} \frac{1}{\mathcal{L}^4}|{\nabla}^2 \phi (\boldsymbol{r})|^2 + 
    \frac{1}{2} C_{ijkl} \frac{\mathcal{E}}{\mathcal{L}^3} 
    \varepsilon^{\mathrm{el}}_{ij} 
    \varepsilon^{\mathrm{el}}_{kl} 
    \Bigg] \mathrm{d}V \\
    & = &\mathlarger{\int_V} \Bigg[  
     f(\phi(\boldsymbol{r}),c) + 
     \epsilon^2 | \boldsymbol{\nabla} 
     \phi (\boldsymbol{r})|^2 + 
     \nonumber \\
    & & \gamma_{\langle hkl \rangle} |{\nabla}^2 \phi (\boldsymbol{r})|^2 + 
    \frac{1}{2} \boldsymbol{\upvarepsilon}^{\mathrm{el}}:\mathbb{C} : \boldsymbol{\upvarepsilon}^{\mathrm{el}}
    \Bigg] \mathrm{d}V
\end{eqnarray}

The non-dimensional form of Allen-Cahn equation (interfacial anisotropy disregarded) is given as
\begin{eqnarray}
\frac{\partial \phi}{\partial t^{*}} &=& -L^{*} \frac{\delta \mathcal{F}^{*}}{\delta \phi} 
\\
\frac{1}{\mathcal{T}}\frac{\partial \phi}{\partial t} &=& L^{*} 
\left(
{\epsilon^{*}}^{2} {\nabla^{*}}^2 \phi - 
\omega^{*} \frac{\partial f_{\mathrm{dw}}(\phi(\boldsymbol{r}))}{\partial \phi} 
\right)\\
\frac{\partial \phi}{\partial t} &=& L^{*} \mathcal{T} \left(\epsilon^2 \frac{\mathcal{E}}{\mathcal{L}} \frac{1}{\mathcal{L}^2}
\nabla^2 \phi - 
\omega \frac{\mathcal{E}}{\mathcal{L}^3} \frac{\partial f_{\mathrm{dw}}(\phi(\boldsymbol{r}))}{\partial \phi}\right)\\
&=& \frac{L^{*} \mathcal{E} \mathcal{T}}{\mathcal{L}^3}  
\left(\epsilon^2 \nabla^2 \phi - 
\omega \frac{\partial f_{\mathrm{dw}}(\phi(\boldsymbol{r}))}{\partial \phi}
\right)
\end{eqnarray}
Here, $L = \displaystyle \frac{L^{*}\mathcal{E}\mathcal{L}^3}{\mathcal{T}}$ 
represents non-dimensional form for the relaxation coefficient, $L^{*}$.

The diffusion equation in non-dimensional form is given as
\begin{eqnarray}
\frac{\partial c^{*}}{\partial t^{*}} &=& 
\nabla^{*} \cdot \frac{D^{*}}{f_{cc}^{*}} \nabla^{*} 
\mu^{*}\\
\frac{c_p^* - c_m^*}{\mathcal{T}}
\frac{\partial c}{\partial t} &=& 
 \frac{\mathcal{L}^3 }{\mathcal{E}} \nabla \cdot \frac{D^{*}}{f_{cc}} \frac{\mathcal{E}}{\mathcal{L}^5} \nabla \mu\\
\frac{\partial c}{\partial t} &=& \frac{D^{*} \mathcal{T}}{(c_p^* - c_m^*)\mathcal{L}^2} 
\nabla \cdot \frac{1}{f_{cc}} \nabla \mu
\end{eqnarray}
Here, $D = \displaystyle \frac{D^{*}\mathcal{T}}{(c_p^* - c_m^*)\mathcal{L}^2}$ denote the 
non-dimensional form of diffusion coefficient.

\section*{Characteristic quantities}

To justify the degree of reality, we show that parameters used in simulations and 
values from CALPHAD databases show good match. We choose Ni-17Al (at $\%$) alloy which 
is used by Kaufmann et al.~to present the results of particle splitting at temperature 
$T=1000$ \si{\celsius}. At temperature $T=1000$ \si{\celsius}, we find 
parameters for Ni-17Al (at$\%$) alloy and compare them with simulation parameters. 
Here, the characteristic energy is $\mathcal{E} = 10^{-20}$ \si{\joule}, 
characteristic time is $\mathcal{T} = 3 \times 10^{-3}$ \si{\second}, and
characteristic length is $\mathcal{L}= 0.25$ \si{\nano\meter} 
(calculated from dimensional molar volume). 
Using values of characteristic quantities, we derive non-dimensional 
values. Thus, a non-dimensional time-step $dt^*=0.1$ corresponds to $dt = 3 \times 
10^{-4}$ \si{\second}. The following table shows the comparison between dimensional 
parameters from simulation and Thermo-Calc databases.   
\begin{table}[!htb]
    \centering
    \begin{tabular}{|p{3.5cm}|p{4cm}|p{4cm}|p{4cm}|}
    \hline
    \hline
    Parameter  & Non-Dimensional values & Dimensional values & Values from TCNI9 \\
    \hline
    \hline
    Free energy curvature & $2.24$ & $1.5\times10^{-19}$
    \si{\joule \per \atom} & $6\times10^{-19}$ \si{\joule \per \atom} \\
    \hline
    Lattice misfit & $0.5-1\%$ & $0.5-1\%$ & $0.615\%$ at $T = 1000$ \si{\celsius} for 
    Ni-17Al (at \%) \\
    \hline
    Interfacial energy & $0.12$ & $20$ \si{\milli\joule \per \meter \squared} & $20-60$ \si{\milli\joule\per\meter \squared} at $T=1000$ \si{\celsius} \\
    \hline
    Diffusivity & $1$ & $10^{-15}$ \si{\meter \squared \per \second} & $8 \times 10^{-15}$ 
    \si{\meter \squared \per \second} at $T = 1000$ \si{\celsius}\\
    \hline
    Compositions ($c_{\upalpha}^{\textrm{e}}$,$c_{\upbeta}^{\textrm{e}}$) & $(0,1)$ &
    $-$ & $(0.16,0.23)$ \\
    \hline
    \hline
    \end{tabular}
    \caption{Comparison of simulation parameters to parameters from Thermo-Calc databases.
    Simulation parameters correspond to Ni-17Al (at.~\%) which is used by Kaufman 
    et al.~to report splitting of $\gamma'$ precipitates.}
    \label{tab:sim_data_and_TC}
\end{table}
\end{document}